\documentclass[10pt]{report}
\newcommand{\doublespaced}{\renewcommand{\baselinestretch}{2}\normalfont}
\newcommand{\singlespaced}{\renewcommand{\baselinestretch}{1}\normalfont}
\newcommand{\draftspaced}{\doublespaced} %for final version
\usepackage{latexsym}\usepackage{revsymb}\usepackage{amssymb}\usepackage{amsmath}\usepackage{amscd}\usepackage{amsthm}%\usepackage{apalike}
\usepackage[numbers, sort&compress]{natbib} 
\usepackage{comment}
\usepackage{enumerate}\usepackage{setspace}
\usepackage{multirow}\usepackage{array}%\usepackage{rotating}
\usepackage[dvips]{epsfig}\usepackage{float}
\usepackage[cmtip,arrow]{xy}\usepackage{pb-diagram,pb-xy}
\usepackage{color}
\numberwithin{equation}{section}
\usepackage{bm}        % for math

\usepackage{hyperref} %clickable references.
\hypersetup{
    colorlinks,
    citecolor=blue,
    filecolor=black,
    linkcolor=blue,
    urlcolor=blue
}

\DeclareMathOperator{\Tr}{Tr}
\newcommand{\defnn}[2]{\noindent {\underline {\it #1}} -- #2\\}
\newcommand{\defn}[3]{\noindent {\underline {\it #1}} (#2) -- #3\\}

\newcommand{\abs}[1]{\left| #1 \right|} % for absolute value
\newcommand{\avg}[1]{\left< #1 \right>} % for average
\newcommand{\pd}[2]{\frac{\partial #1}{\partial #2}} 
\newcommand{\pdd}[2]{\frac{\partial^2 #1}{\partial #2^2}} 
\newcommand{\ket}[1]{\left| #1 \right>} % for Dirac bras
\newcommand{\bra}[1]{\left< #1 \right|} % for Dirac kets
\newcommand{\braket}[2]{\left< #1 \vphantom{#2} \right|
 \left. #2 \vphantom{#1} \right>} % for Dirac brackets
\newcommand{\matrixel}[3]{\left< #1 \vphantom{#2#3} \right|
 #2 \left| #3 \vphantom{#1#2} \right>} % for Dirac matrix elements
\renewcommand{\v}[1]{\ensuremath{\mathbf{#1}}} % for vectors
 
\newcommand{\lstar}[0]{\ensuremath{\ell^*}}

\newcommand{\Nccut}[0]{\ensuremath{N_\text{c}^\text{cut}}}
\newcommand{\Ncextra}[0]{\ensuremath{N_\text{c}^\text{extra}}}
\newcommand{\keff}[0]{\ensuremath{k_\text{eff}}}
\newcommand{\p}[0]{\ensuremath{^\prime}}
\newcommand{\ZmZiso}[0]{Z-Z_\mathrm{iso}}
\newcommand{\Ziso}[0]{Z_\mathrm{iso}}

\newcommand{\eq}[1]{\begin{align} #1 \end{align}}
\newcommand{\eqs}[1]{\begin{align} \begin{split} #1 \end{split}\end{align}}

\doublespaced
\def\thetitle{UNEARTHING THE ANTICRYSTAL: CRITICALITY IN THE LINEAR RESPONSE OF DISORDERED SOLIDS}
\def\theauthor{Carl P.\ Goodrich}\def\theauthorlegalname{Carl P.\ Goodrich}   
\def\theadvisor{Andrea J.\ Liu}

\def\theyear{2015}

\begin{document}

\pagenumbering{roman}
\doublespaced
%\onehalfspace
%\large\newlength{\oldparskip}\setlength\oldparskip{\parskip}\parskip=.2in
\thispagestyle{empty}

\begin{center}
%\vspace*{10mm}
%\Large\thetitle
%
%\Large\theauthor

\large\thetitle

\large\theauthor

A DISSERTATION

in 

Physics and Astronomy

Presented to the Faculties of the University of Pennsylvania

in

Partial Fulfillment of the Requirements for the

Degree of Doctor of Philosophy 
\end{center}

%\noindent\singlespaced%\normalsize
%Presented to the Faculties of the University of Pennsylvania in Partial Fulfillment of the Requirements for the Degree of Doctor of Philosophy

\singlespaced\large
\begin{center}
\theyear
\end{center}

\vspace*{10mm}
\noindent\makebox[0in][l]{\rule[2ex]{3in}{.3mm}}\singlespaced
\theadvisor, Professor of Physics and Astronomy\\
Supervisor of Dissertation

\vspace*{10mm}
\noindent\makebox[0in][l]{\rule[2ex]{3in}{.3mm}}\singlespaced
Sidney R.\ Nagel, Professor of Physics (University of Chicago, Chicago, IL)\\
Co-Supervisor of Dissertation

\vspace*{10mm}
\noindent\makebox[0in][l]{\rule[2ex]{3in}{.3mm}}\singlespaced
Marija Drndic, Professor of Physics and Astronomy\\
Graduate Group Chairperson

\vspace*{10mm}

\noindent\onehalfspace
Dissertation Committee\\
Douglas J.\ Durian, Professor of Physics and Astronomy\\
Eleni Katifori, Assistant Professor of Physics and Astronomy\\
Tom C.\ Lubensky, Professor of Physics and Astronomy\\
%Sidney Nagel, Professor of Physics (University of Chicago, Chicago, IL) \\
Rob Riggleman, Assistant Professor of Chemical and Biomolecular Engineering \\
David Srolovitz, Professor of Materials Science and Engineering \\

%Christopher A. Hunter, Professor of Pathobiology\\
%Randall D. Kamien, Professor of Physics and Astronomy\\
%Michael A. Lampson, Professor of Biology\\
%Tom C. Lubensky, Professor of Physics and Astronomy\\
%Phillip C. Nelson, Professor of Physics and Astronomy\\

%\vspace*{\fill}

%\normalsize\parskip=\oldparskip

\newpage
\large\newlength{\oldparskip}\setlength\oldparskip{\parskip}\parskip=.2in
\doublespaced
\thispagestyle{empty}
\vspace*{\fill}
%\begin{center}
\noindent
	\thetitle\\%\vspace{\fill}
  COPYRIGHT\\
  \theyear \\
  \theauthorlegalname\\

%\end{center}
\vspace*{\fill}

\newpage 
\vspace{2cm}
\begin{center} Dedication \\
\textit{For my mother, whom I miss every day.}
\end{center}

\newpage
\doublespaced

\chapter*{Acknowledgements}

This dissertation would never have been possible without the tremendous support and guidance I have received from countless individuals.
%I am extremely grateful for the tremendous support and guidance I have received from countless individuals, without which this dissertation would not be possible. 
First and foremost, I thank my advisors, Andrea Liu and Sid Nagel, for their unwavering support, advice, patience, encouragement, and confidence. More than anything, Andrea and Sid have taught me what it means to be a scientist, and I am forever humbled to have been able to work with and learn from them. 
Thank you for the countless opportunities, and for challenging me every day.
I am also very grateful for the considerable financial support they have provided for me to attend numerous conferences and schools, which has been an essential part of my education and development.

I have been extremely lucky not only to have two outstanding advisors but to be in an ideal environment for a graduate student to learn and grow.
%In addition, the soft matter theory group at Penn has been an ideal environment in which to learn and grow. 
I sincerely thank Randy Kamien for challenging me each and every day, for making DRL a fun and enjoyable work environment, and for always being generous with his time. It is an understatement to say that Randy has had a significant and profound effect on me both as a scientist and a person.
Furthermore, I thank Tom Lubensky for many stimulating discussions and interactions over the years. %regarding the subtleties and nuances of isostaticity and for numerous other constructive interactions. 
I am also grateful to Doug Durian, Eleni Katifori, Allison Sweeney, and Arjun Yodh for countless interactions that have been invaluable to my education. 

%I am also grateful to the entire soft matter group at Penn.
I have also had the privilege to work alongside many amazingly talented, insightful and kind students and postdocs at Penn. I sincerely thank the entire soft matter theory group, particularly
Ed Banigan, for teaching me to love biophysics and for always explaining things in a clear and concise way so that even I could understand;
%Bryan Chen, for being one of the smartest, kindest, and most humble people I know;
Bryan Chen, for many invaluable discussions regarding isostaticity, the ``pebble game," and the like;
Wouter Ellenbroek, for taking me under his wing when I was first learning about jamming; 
Timon Idema, for teaching me, both directly and by example, more about science and about life than I can possibly put in words;
Anton Souslov, for helping me think clearly about criticality and jamming;
and finally my jamming brethren Sam Schoenholz and Daniel Sussman, for numerous discussions and for their constant insight, feedback and support. 
I also owe a tremendous amount to others at Penn that are outside of the soft matter theory group. 
I thank Dave Srolovitz and Spencer Thomas for teaching me how to generate polycrystals and for other invaluable discussions.
I thank the entire DISCONAP community, especially John Crocker, Jennifer Rieser, and Rob Riggleman, as well as the entire Yodh lab, especially Matt Gretale, Matt Lohr, Tim Still, and Peter Yunker. 

%Andy Mastbaum,
%Pat Vora

%I also thank, and apologize to, my office mates over the years: 
%Ed Banigan, 
%Dan Beller,
%Tom Haxton, 
%Sabetta Matsumoto, 
%Jason Rocks, 
%and Lisa Tran.

%I have also had the privilege of many significant and extremely rewarding interactions with 
In addition, I have benefitted greatly from numerous interactions with researchers outside of Penn. I especially thank
Lisa Manning,
Brian Tighe,
and Martin van Hecke
for countless discussions that never failed to teach me something new. I am also grateful for many fruitful interactions with, among others,
Justin Burton, 
Patrick Charbonneau
Eric Corwin, 
Simon Dagois-Bohy,
Eric DeGiuli,
Amy Graves,
Silke Henkes,
Stan Leibler,
Edan Lerner, 
Michael Mitchell,
Corey O'Hern, 
Ken Schweizer,
Jen Schwartz,
Jim Sethna,
Mark Shattuck,
Mike Thorpe, 
Tsvi Tlusty,
Sal Torquato,
Vincenzo Vitelli, 
Matthieu Wyart, 
Ning Xu, 
and Zorana Zeravcic.

Finally, I thank my fathers Bill and Scot, and my sister Kirstin. I could not have done any of this without your unyielding love and support. Thank you for always being there for me. I also thank my cousin Elaine and her wonderful family for giving me a home away from home.

\newpage
\doublespaced

%\newpage
%\vspace*{\fill}
\begin{center}
  ABSTRACT\\
\thetitle\\
\vspace{.5in}
  \theauthor\\
  \theadvisor \\
  Sidney R. Nagel
\end{center}
\doublespaced
\noindent

The fact that a disordered material is not constrained in its properties in the same way as a crystalline one presents significant and yet largely untapped potential for novel material design. However, unlike their crystalline counterparts, disordered solids are not well understood. Though currently the focus of intense research, one of the primary obstacles is the lack of a theoretical framework for thinking about disorder and its relation to mechanical properties. To this end, we study a highly idealized system composed of frictionless soft spheres at zero temperature that, when compressed, undergos a jamming phase transition with diverging length scales and clean power-law signatures. This critical point is the cornerstone of a much larger ``jamming scenario" that has the potential to provide the essential theoretical foundation that is sorely needed to develop a unified understanding of the mechanics of disordered solids. 
We begin by showing that jammed sphere packings have a valid linear regime despite the presence of a new class of ``contact nonlinearities," demonstrating that the leading order behavior of such solids can be ascertained by linear response. We then investigate the critical nature of the jamming transition, focusing on two diverging length scales and the importance of finite-size effects. 
Next, we argue that this jamming transition plays the same role for disordered solids as the idealized perfect crystal plays for crystalline solids. Not only can it be considered an idealized starting point for understanding the properties of disordered materials, but it can even influence systems that have a relatively high amount of crystalline order. As a result, the behavior of solids can be thought of as existing on a spectrum, with the perfect crystal at one end and the jamming transition at the other. Finally, we introduce a new principle for disordered solids wherein the contribution of an individual bond to one global property is independent of its contribution to another. This principle allows the different global responses of a disordered system to be manipulated independently of one another and provides a great deal of flexibility in designing materials with unique, textured and tunable properties.

\vspace*{\fill}

\newpage
\begin{onehalfspace}
\tableofcontents
\end{onehalfspace}

%\newpage
%\begin{singlespaced}
%\listoftables
%\end{singlespaced}
%

\newpage
\begin{singlespaced}
\listoftables
\end{singlespaced}

\newpage
\begin{singlespaced}
\listoffigures
\end{singlespaced}

%\chapter*{Preface}

\newpage
\draftspaced
\pagenumbering{arabic}

\chapter{Introduction} \label{introductionchapter}

\section{Ordered and disordered solids}

Anyone who has made ice cubes in the freezer knows that liquids can transition into solids when cooled below a certain temperature. In many cases, this occurs when the constituent atoms or molecules order themselves into a periodically repeating array, called a crystal. The differences between the disordered fluid and the ordered solid are not only obvious but also well understood, and this understanding comes directly from the presence of long-range crystalline order in the solid.

However, many fluids solidify without forming a crystalline lattice, instead remaining in a disordered, or amorphous, state. 
The most common example of this is the glass transition, which occurs when a glass-forming liquid is cooled below some temperature, $T_g$, where the relaxation time diverges. 
Disordered solids can form in other ways as well. 
Foams flow when stressed but hold their shape when the stress is lowered below the yield stress, and colloidal suspensions flow only when the density is sufficiently low. In all of these examples, the fluid transitions into a solid without any obvious change in the microscopic structure, and our understanding of the liquid-crystal transition that occurs in our freezer is clearly insufficient since it is based on the formation of structural order. 
Indeed, understanding of the glass transition has defied physicists for decades, and other disordered materials like foams and suspensions are also 
currently the focus of considerable research.

However, one can attempt to understand the material properties of a solid independently from the associated fluid-solid phase transition. In crystals, for example, the presence of a periodically repeating array has a profound effect on the solid's mechanical and thermal properties. For over a century, physicists have understood these properties by first considering an idealized {\it perfect} crystal ({\it i.e.} an infinite crystal with no defects or imperfections of any kind), whose high symmetry makes analytical calculations possible~\cite{Ashcroft:1976ud}. Real crystals, which always have defects, can then be treated by perturbing around this idealized starting point.

Although they lack this symmetry, many disordered solids exhibit common features, including a characteristic temperature dependence of the heat capacity and thermal conductivity~\cite{Phillips:1981um} and brittle response to mechanical load~\cite{Maloney:2007en}. 
These commonalities suggests that an underlying physical principle could be at work, influencing the behavior of a vast range of systems that appear to share only a dearth of crystalline order. %A possible explanation for this is 
To this end, the so-called ``jamming scenario"~\cite{Liu:1998up,OHern:2003vq,Liu:2010jx} stipulates that the onset of rigidity caused by, for example, lowering the temperature of a glass-forming liquid, lowering the stress applied to a foam, or increasing the density of a colloidal suspension are manifestations of an umbrella concept called jamming. The utility of this approach is that there is an ideal case where the rigidity transition is controlled by a critical point. While it remains to be seen whether or not this critical point influences the glass transition itself, there is mounting evidence that the properties of the glass, as well as that of other disordered solids, can indeed be understood from within the context of the jamming scenario.

\section{Jamming as a unifying concept}
\begin{figure}
	\centering
	\includegraphics[width=0.7\linewidth]{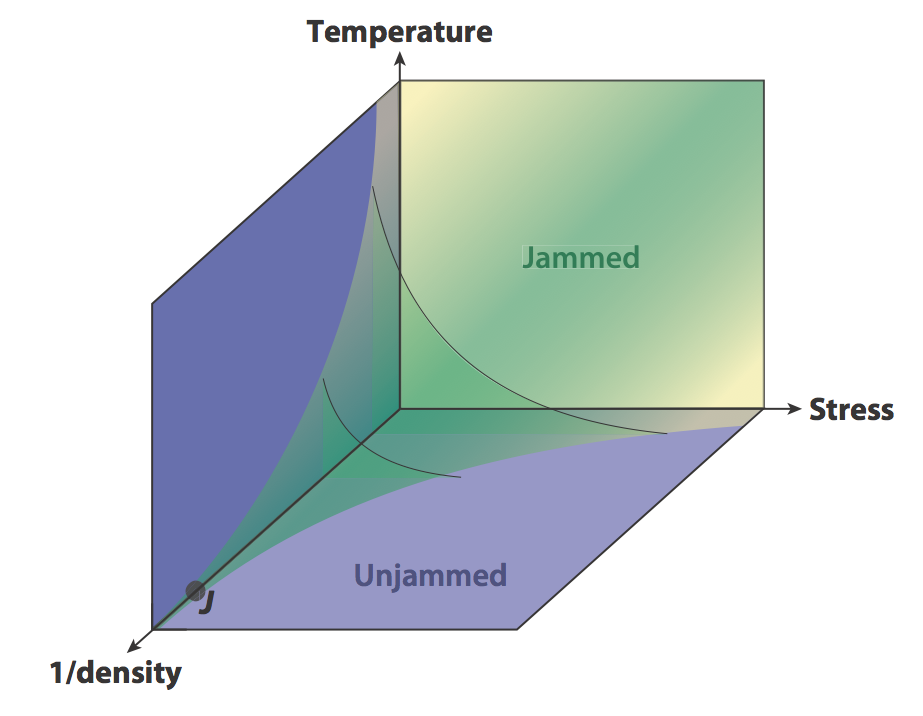}
	\caption[The jamming phase diagram.]{The jamming phase diagram -- reproduced with permission from Ref.~\cite{Liu:2010jx}. Outside the shaded green region, at high temperature $T$, applied shear stress $\Sigma$, and high inverse density $1/\phi$, the system is unjammed and can flow; inside the shaded green region, the system is jammed. The lines in the ($T-1/\phi$) and ($1/\phi - \Sigma$) planes represent the generic dynamical glass transition and yield stress, respectively. The point ``J'' marks the jamming transition for ideal spheres at zero temperature and applied stress.}
	\label{fig:jamming_phase_diagram}
\end{figure}

In 1998, Liu and Nagel proposed~\cite{Liu:1998up} that the onset of rigidity in many different disordered systems may be viewed as manifestations of a single concept: jamming. The connection is made through the so-called {\it jamming phase diagram}, which is reproduced in Fig.~\ref{fig:jamming_phase_diagram} with permission from Ref.~\cite{Liu:2010jx}. In this paradigm, a disordered system is unjammed (behaves like a fluid) when it is at high temperature, high applied stress and low density; lowing the temperature, lowering the applied stress, or increasing the density can then cause the system to become rigid and jam (behave like a solid). 
For example, the glass transition exists on the zero stress plane, when a glass-forming liquid in the unjammed phase is cooled at constant density, and a jammed suspension in the zero temperature plane flows (unjams) when sheared past its yield stress. %While the details of this phase diagram of course depend on the 

Furthermore, the jamming phase diagram for systems composed of frictionless soft spheres has a critical point, called the jamming point or point J, on the zero temperature and zero stress axis~\cite{Liu:1998up,OHern:2002bs,OHern:2003vq,Liu:2010jx}. This realization sparked tremendous enthusiasm because the presence of a critical point allows jamming to be studied in its purest form, and soft sphere packings have since been referred to as ``the `Ising model' for jamming"~\cite{vanHecke:2009go}. If we can develop a theory for the jamming of frictionless soft spheres, then it may be possible to understand real disordered materials in terms of how they differ from this ideal case. 

Frictionless soft spheres are described by a short-ranged, purely repulsive pairwise interaction that is described in detail in Sec.~\ref{sec:def_of_system}. The essential characteristics are 1) spheres have a well defined radii, 2) two non-overlapping spheres do not interact with each other, and 3) spheres are allowed to overlap (as opposed to hard sphere models) but  repel each other when they do. There are no other forces or interactions in the system. Thus, at zero temperature and zero applied stress, force balance must be satisfied on every particle and the entire system sits in a local minimum of the total energy. 

At low packing fractions, $\phi$, there is enough room for all the particles to avoid making contact with each other. With no interactions at all, the system is in a ``mechanical vacuum;"\footnote{This phrase is attributed to Martin van Hecke.} the energy is zero and the energy landscape is locally flat. 
Bulk quantities like the pressure and elastic constants vanish -- the system is not jammed. 
As the packing fraction is increased to a critical $\phi_c$, there is a sudden rigidity transition that is marked by the simultaneous onset of a number of quantities, including the energy, pressure, bulk modulus, and shear modulus~\cite{OHern:2003vq}, indicating that the system has indeed gained rigidity. Conceptually, this occurs when particles can no longer avoid each other and are forced to overlap. From the perspective of the energy landscape, this corresponds to the local minimum transitioning from being highly degenerate (flat) to being point-like.\footnote{The energy landscape is still locally flat in a few directions, corresponding to the motion of rattlers.} 

The presence of particle-particle contacts is quantified by the contact number, $Z$, which is the average number of contacts at each particle. At $\phi_c$, there is a sharp jump in $Z$ from zero to the isostatic contact number, $\Ziso$. The notion of isostaticity dates back to Maxwell and refers to a balance between the number of constraints (contacts) and number of degrees of freedom (particle positions), so that $\Ziso$ is approximately twice the dimensionality of space (see Chapter~\ref{chapter:finite_size} for a thorough analysis of this that includes finite-size corrections to $\Ziso$). 

%\setlength\extrarowheight{6pt}
%\begin{table}
%\centering
%\begin{tabular}{ | c | c | c |} %{| M{4cm} | M{4cm} | M{4cm} | N}
% \hline
%%\begin{tabular}{| c | c | c |} \hline
%quantity & scaling & comments \\ \hline
%pressure & $p \sim (\phi-\phi_c)^{\alpha-1}$ & \\
%contact number & $\ZmZiso \sim (\phi-\phi_c)^{1/2}$ & \\
%bulk modulus & $B \sim (\phi-\phi_c)^{\alpha-2}$ & approaches constants as $\phi \rightarrow \phi_c^+$ for $\alpha=2$ \\
%shear modulus & $G \sim (\phi-\phi_c)^{\alpha-1.5}$ &  \\
%frequency scale & $\omega^* \sim (\phi-\phi_c)^{1/2}$ NOT CORRECT!!! & derived as $\omega^* \sim 1/\lstar$ \\
%rigidity length & $\lstar \sim (\phi-\phi_c)^{-1/2}$ & derived as $\lstar \sim (\ZmZiso)^{-1}$ \\
%transverse length & $\ell_\text{T} \sim  (\phi-\phi_c)^{-1/4}$ & \\
%\hline
%\end{tabular}
%\caption{\label{table:scaling_summary}... All of these power-law scalings are the same for both two and three dimensional systems, and are observed numerically over many decades in $\phi-\phi_c$.}
%\end{table}

\setlength\extrarowheight{6pt}
\begin{table}
\centering
\begin{tabular}{ | c | c |} %{| M{4cm} | M{4cm} | M{4cm} | N}
 \hline
%\begin{tabular}{| c | c | c |} \hline
quantity & scaling  \\ \hline
contact number & $\ZmZiso \sim (\phi-\phi_c)^{1/2}$ \\
pressure & $p / \keff \sim (\phi-\phi_c)$ \\
bulk modulus & $B / \keff \sim (\phi-\phi_c)^0$ \\
shear modulus & $G / \keff \sim (\phi-\phi_c)^{1/2}$ \\
frequency scale & $\omega^*/\sqrt{\keff} \sim (\phi-\phi_c)^{1/2}$ \\
rigidity length & $\lstar \sim (\phi-\phi_c)^{-1/2}$ \\
transverse length & $\ell_\text{T} \sim  (\phi-\phi_c)^{-1/4}$ \\
\hline
\end{tabular}
\caption[Power-law scalings near the jamming transition.]{\label{table:scaling_summary} Power-law scalings near the jamming transition. All scalings are independent of dimensionality and are observed numerically over many decades in $\phi-\phi_c$.
 %All of these power-law scalings are the same for both two and three dimensional systems, and are observed numerically over many decades in $\phi-\phi_c$. 
 All effects associated with the form of the interaction potential are included in the effective spring constant, $\keff$, which scales as $\keff \sim (\phi - \phi_c)^{\alpha-2}$ (see Eq.~\eqref{eq:pairwise_iteraction_potential} details).}
\end{table}

\subsubsection{Scaling above $\phi_c$}

\begin{figure}
	\centering
	\includegraphics[width=0.8\linewidth]{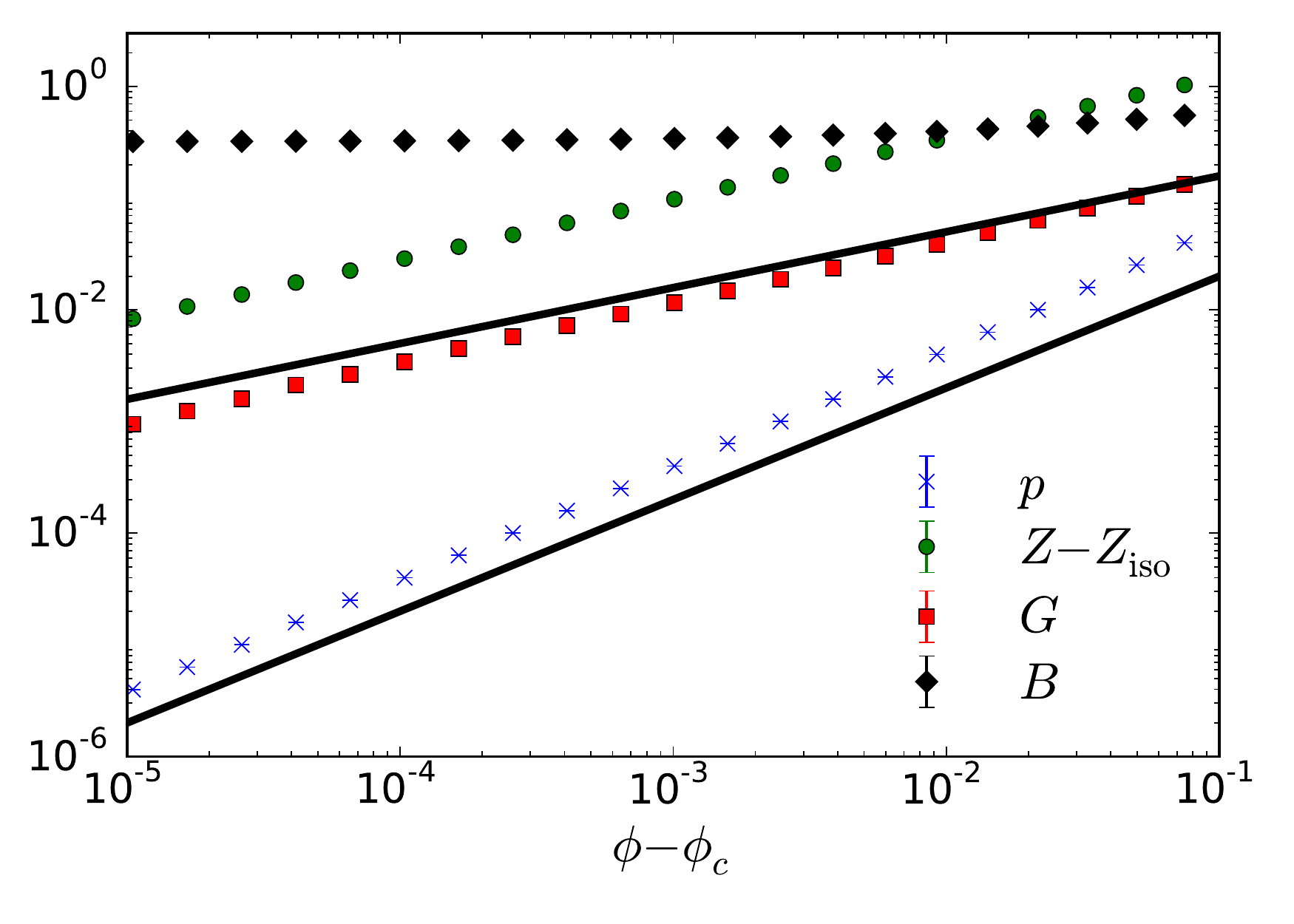}
	\caption[The pressure, excess contact number, shear modulus, and bulk modulus as a function of the excess packing fraction.]{The pressure, excess contact number, shear modulus, and bulk modulus as a function of the excess packing fraction for packings of $N=8192$ polydisperse disks with harmonic interactions ($\alpha=2$) in two dimensions. Here, $\phi_c$ is found to be $\phi_c \approx 0.8451$. The solid black lines have slopes $1/2$ and $1$.}
	\label{fig:basic_scalings}
\end{figure}

Numerous quantities exhibit clean power-law scaling as the packing fraction is increased above $\phi_c$.
%, numerous quantities exhibit clean power-law scaling over many decades in $\phi-\phi_c$. 
Many of the exponents are found to be very close to simple rational numbers, but it is possible that some or all of them actually deviate slightly. 
Here, we will review the scaling in the infinite size limit and assume that all exponents are simple rational numbers. 
As the packing fraction increases past $\phi_c$, particles in contact begin to overlap and new contacts are formed. The {\it excess} contact number, $\ZmZiso$, increases with the excess packing fraction as $\ZmZiso \sim (\phi - \phi_c)^{1/2}$, for small $\phi-\phi_c$~\cite{Durian:1995eo,OHern:2003vq}. Additionally, the pressure increases from zero as $p/\keff \sim (\phi - \phi_c)$~\cite{Durian:1995eo,OHern:2003vq}, 
where $\keff$ is the average effective spring constant and accounts for trivial differences in scaling caused by the exact form of the interaction potential (see Eq.~\ref{eq:pairwise_iteraction_potential}). Most of this thesis will focus on harmonic interactions where $\keff \sim \text{const}$ and does not affect the scaling. 
%where $\alpha$ characterizes the particular form of the interaction potential (see Eq.~\ref{eq:pairwise_iteraction_potential}) and typically takes values of $\alpha=2$ (corresponding to Hookian spring-like interactions) and $\alpha=2.5$ (corresponding to Hertzian interactions). 
The bulk modulus, $B$, and the shear modulus, $G$, scale as $B/\keff \sim (\phi-\phi_c)^0 \sim \text{const}$ and $G/\keff \sim (\phi-\phi_c)^{1/2}$, respectively, so that $G/B \sim(\phi-\phi_c)^{1/2}$~\cite{Durian:1995eo,OHern:2003vq}. % For $\alpha=2$, the bulk modulus approaches a constant at the transition while the shear modulus vanishes, and the ratio $G/B \sim (\phi-\phi_c)^{1/2}$ is independent of the form of the potential. 
These scalings are summarized in Table~\ref{table:scaling_summary} and shown in Fig.~\ref{fig:basic_scalings}.

For finite-size systems, these relations are complicated by the fact that $\phi_c$ fluctuates from one system to another~\cite{OHern:2003vq}. In fact, the observed values of $\phi_c$ form a distribution whose width scales as $N^{-1/2}$, where $N$ is the number of particles. To account for this, one approach would be to subtract the $\phi_c$ for each system individually, but this is difficult because when a system at its $\phi_c$ is compressed, the particles will rearrange and the relevant $\phi_c$ for the compressed system is not necessarily the same as the original $\phi_c$. It has therefore become common practice to treat $p/\keff$ as a proxy for $\phi - \phi_c$. This is especially simple for harmonic interactions where $\phi - \phi_c \sim p$. Since work presented in this thesis uses harmonic interactions almost exclusively, data will repeatedly be shown as a function of pressure. 

\subsubsection{Length scales and the density of states}
\begin{figure}
	\centering
	\begin{tabular}{cc}
	\includegraphics[width=0.45\linewidth]{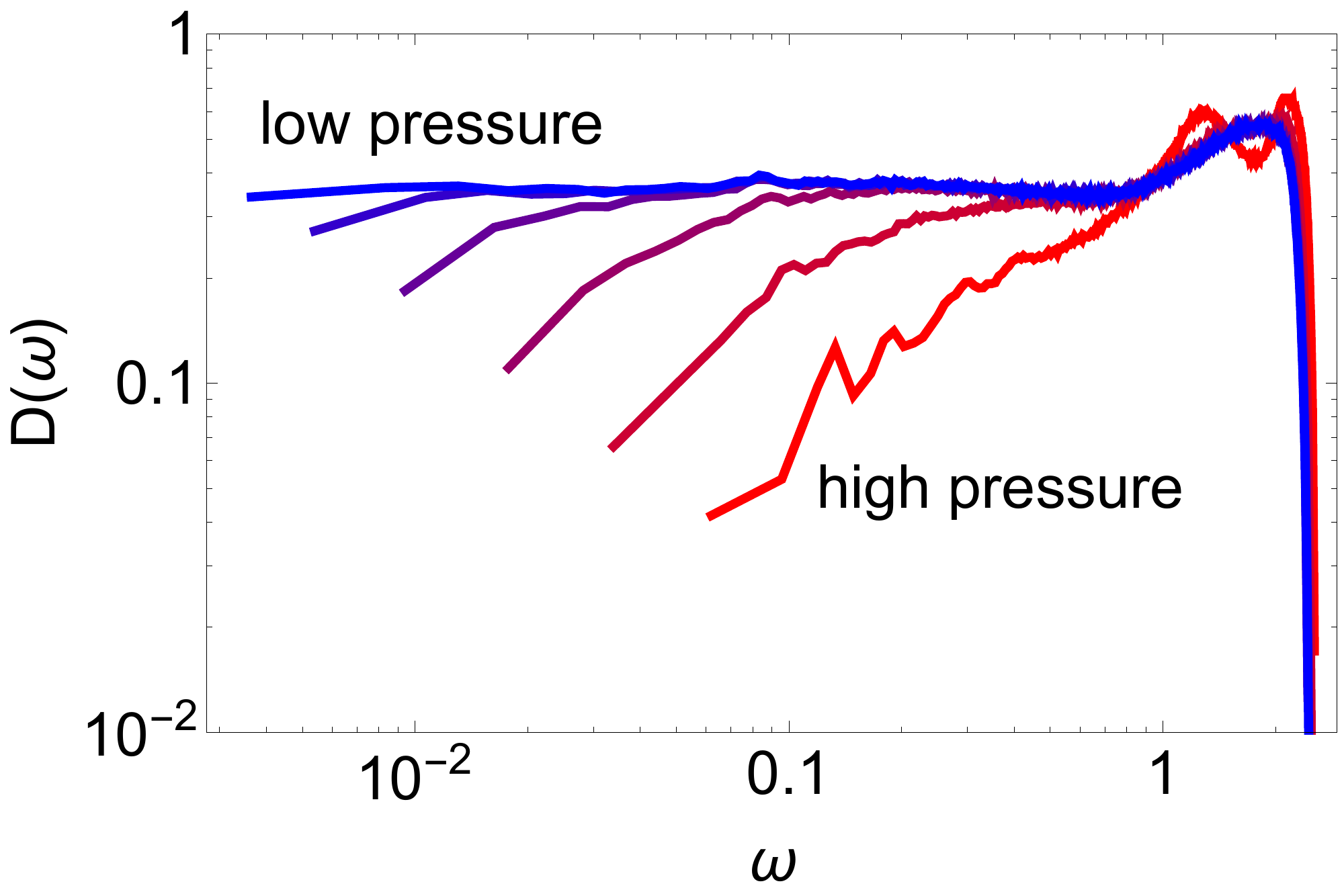}&
	\includegraphics[width=0.45\linewidth]{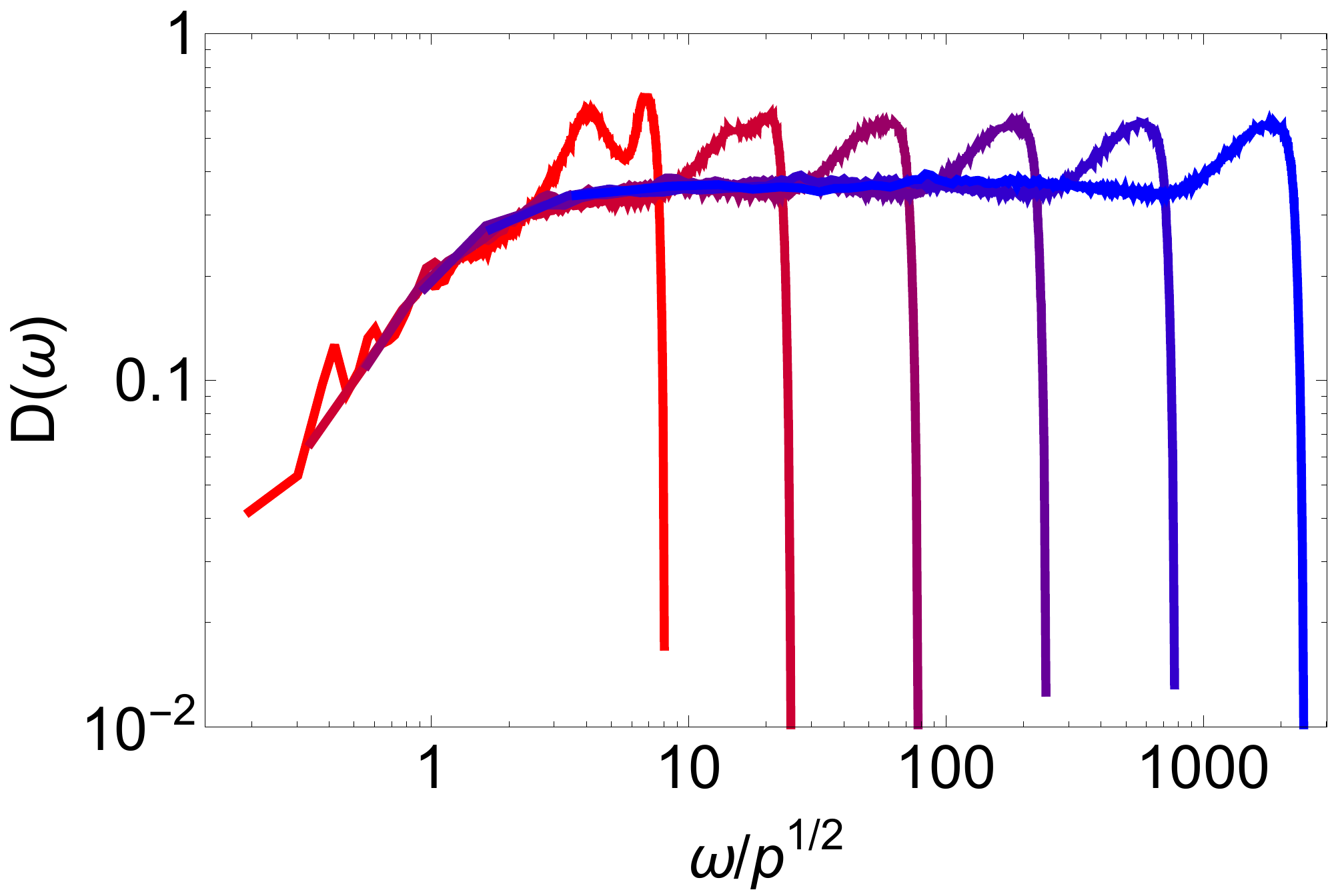}
	\end{tabular}
	\caption[Density of states.]{Left: Density of states, $D(\omega)$, for $2d$ packings of $N=2048$ harmonic disks at pressures of $10^{-6}$ (blue), $10^{-5}$, $10^{-4}$, $10^{-3}$, $10^{-2}$, and $10^{-1}$ (red). Right: The density of states collapses at low frequency when $\omega$ is divided by $p^{1/2}$, suggesting a frequency scale $\omega^* \sim p^{1/2} \sim (\phi - \phi_c)^{1/2}$.
	}
	\label{fig:dos_intro}
\end{figure}

One of the most important quantities to understand when studying a solid is the density of vibrational states, $D(\omega)$, particularly at low frequency $\omega$. For crystals, low frequency modes are plane waves, and their density scales as $D(\omega) \sim \omega^{d-1}$ in $d$ dimensions. This is the canonical {\it Debye scaling} that is observed regardless of crystal structure. Sphere packings above the jamming transition, however, exhibit a new class of vibrational modes that are disordered and completely overwhelm plane waves~\cite{OHern:2003vq,Silbert:2005vw}. These excess modes result in a plateau in the density of states at low frequency (see Fig.~\ref{fig:dos_intro}) that extends down to a characteristic frequency called $\omega^*$. This characteristic frequency vanishes at the jamming transition and scales as $\omega^*/\sqrt{\keff} \sim (\phi-\phi_c)^{1/2}$, so that at the jamming transition $D(\omega) \sim \mathrm{const}$ in the limit $\omega \rightarrow 0^+$~\cite{OHern:2003vq,Silbert:2005vw}. 

%, even in the limit $\omega \rightarrow 0$. Indeed, the low frequency density of states is flat, $D(\omega) \sim 1$, all the way down to zero frequency. For systems above the jamming transition, $D(\omega) \sim 1$ only down 

The density of states and the scaling of $\omega^*$ can be understood in terms of a length scale, called $\lstar$, that diverges at the jamming transition and scales as $\lstar  \sim (\phi-\phi_c)^{-1/2}$~\cite{Wyart:2005wv}. The so-called ``cutting argument" that predicts this length and rationalizes the $\omega^*$ scaling is reviewed in detail in Sec.~\ref{sec_cutting_argument_review}. A second length scale was also identified~\cite{Silbert:2005vw} from the transverse component of the Fourier transform of modes at $\omega^*$. This ``transverse length" also diverges at the jamming transition but is smaller than $\lstar$, scaling as $\ell_\text{T} \sim (\phi-\phi_c)^{-1/4}$~\cite{Silbert:2005vw}. Chapter~\ref{chapter:LengthScales} discusses these length scales and shows that they can both be understood in terms of rigidity against different types of boundary perturbations.

\section{Outline}
The remainder of this dissertation focuses on two primary goals: 1) to further develop the theory of the jamming transition for ideal soft spheres, and 2) to understand how this theory fits in to, enhances, and develops our overall understanding of solids. 
It is organized as follows. 
Chapter~\ref{chapter:numerical_methods} gives significant details regarding the numerical procedures used throughout this thesis. Later chapters will still contain enough information to be understood on their own, but this chapter provides much more detail. A major motivation for this chapter is to document precisely the protocol used to generate sphere packings at very low pressure, as well as include ``tricks of the trade" that have been accumulated over the years in the hope that someone may find them useful in the future.

Chapters~\ref{chapter:linear_response}-\ref{chapter:finite_size} discuss studies into the critical nature of jamming.
Chapter~\ref{chapter:linear_response}, whose contents appear in Refs.~\cite{Goodrich:2014jw} and \cite{Goodrich:2014dm}, discusses the validity and utility of the harmonic approximation in the study of the jamming transition, which was recently called into question by Refs.~\cite{Schreck:2011kl,Bertrand:2014ig,Schreck:2013gg}. We show that above the transition ($\phi > \phi_c$), the harmonic approximation is valid for any system size, which is important because our understanding of jammed solids, like that of crystalline solids, is based to a large degree on the theory of linear elasticity and the density of states, both of which are defined in the linear regime. This result validates the use of these quantities in later chapters.

Chapter~\ref{chapter:LengthScales}, whose contents appear in Refs.~\cite{Goodrich:2013ke} and \cite{Schoenholz:2013jv}, discusses the two diverging length scales $\lstar$ and $\ell_\mathrm{T}$. First, we re-derive $\lstar$ in terms of rigid clusters, which overcomes a number of difficulties in the original ``cutting argument"~\cite{Wyart:2005wv,Wyart:2005jna} and allows us to directly measure the length and verify its scaling numerically.  We then show that $\ell_\text{T}$ also arises naturally as a rigidity length scale, but here the rigidity is with respect to a more subtle class of boundary deformations that does not affect the connectivity of the packing. 

Chapter~\ref{chapter:finite_size}, whose contents appear in Refs.~\cite{Goodrich:2012ck} and \cite{Goodrich:2014iu}, discusses the effects of finite size on the jamming transition. Since rigidity is a global property (either the entire system is jammed or the entire system is unjammed), finite size has subtle and nontrivial effects on even the very definition of jamming. Various jamming criteria are discussed, along with corresponding ensembles that fulfill these requirements. We also present an extensive analysis of finite-size scaling in the excess contact number and elastic constants. This includes an analysis of anisotropy, which is inherently a finite-size effect because jammed packings are isotropic in the thermodynamic limit. 

In Chapter~\ref{chapter:anticrystal}, whose contents appear in Ref.~\cite{Goodrich:2014fl}, we take a step back and discuss the role of jamming in our overall understanding of solids. We propose that the physics of the jamming transition represents an extreme limit of mechanical behavior, which we dub the ``anticrystal." In this paradigm, the anticrystal represents a pole that sits opposite to the perfect crystal and provides an equally valid starting point for the study of solids. Ordinary matter should then be thought of as existing somewhere on the continuum between these two idealized limits: in some cases it makes sense to describe its properties starting from the perspective of a crystal, while in other cases it is more appropriate to describe it from the perspective of a jammed solid.

Finally, Chapter~\ref{chapter:independent_response}, whose contents appear in Ref.~\cite{Goodrich:2015vy}, shows that one of the fundamental features of jamming, that $G/B \rightarrow 0$ at the transition, is really a special case of a more fundamental principle for disordered solids: independent bond-level response. This new principle states that the way an individual bond contributes to one global property (such as the shear modulus) is independent of how it contributes to another (such as the bulk modulus). Therefore, using a simple and experimentally relevant procedure to remove specific bonds from a disordered spring network, we can tune $G/B$ to either the incompressible limit, $G/B \rightarrow 0$, or the auxetic limit, $G/B \rightarrow \infty$. 
%One of the fundamental features of jamming is that the ratio of the shear modulus to bulk modulus vanishes at the transition with the excess contact number, $G/B \sim \ZmZiso$. We show that the ability to independently tune two bulk properties (in this case $G$ and $B$) is a result of a generic property of disordered networks called independent bond-level response. The connectivity in spring networks is not determined from the node geometry as it is for sphere packings, and this flexibility allows us to drastically and efficiently tune $G/B$ to either the incompressible limit $G/B \rightarrow 0$ or the auxetic limit $G/B \rightarrow \infty$ through a simple and experimentally relevant bond-pruning procedure. 
This principle of independent bond-level response allows the different global responses of a disordered system to be manipulated independently of one another and provides a great deal of flexibility in designing materials with unique, textured and tunable properties.

\chapter{General numerical methods}
\label{chapter:numerical_methods}

%This chapter is somewhat of an outlier in that it does not present any novel research. Instead, 
This chapter serves a few purposes. 
First, it provides complete and precise definitions for many standard quantities that are considered throughout this thesis. It is my experience that in the literature, quantities can fluctuate by factors of $2$ or $N/V$ (number of particles per unit volume), so my hope is that this will help make my results more reproducible. 
In addition, much of my work uses sphere packings that were created at logarithmically spaced pressure intervals, and therefore very closely approach the jamming transition. This chapter will describe the nontrivial protocol used to create an extensive database of such states, including detailed numerical procedures and ``tricks of the trade." 
Finally, we will briefly discuss linear response, including normal modes of vibration and the linear elastic constants, with an emphasis on numerical procedure. 

My hope is that someone wishing to reproduce my results will find this chapter extremely useful, but my expectation is that a more casual reader can use it as a reference or skip it entirely. 
%it can be skipped by the more casual reader since each ensuing chapter is self contained. 

\section{Packings of soft spheres\label{sec:def_of_system}}

Consider a system of $N$ particles of mass $M$ in a $d$ dimensional simulation box with linear length $L$ and periodic boundary conditions in each direction. Let $\v r_m$ and $R_m$ be the position and radius of particle $m$, respectively. Two particles, $m$ and ${m\p}$, interact with a pairwise potential of the form
\eq{ U_{m{m\p}} = \left \{ \begin{array}{l l}
	\frac \epsilon \alpha \left(1 - \frac {r_{m{m\p}}}{R_m+R_{m\p}} \right)^\alpha \;\;\;\;\;\;  & \text{if $r_{m{m\p}} < R_m + R_{m\p}$ } \\
	0 & \mathrm{otherwise}, \\
\end{array} \right. \label{eq:pairwise_iteraction_potential}}
%\eq{ U_{ij} = \frac \epsilon \alpha \left(1 - \frac {r_{ij}}{R_i+R_j} \right)^\alpha \Theta\left(1- \frac {r_{ij}}{R_i+R_j} \right). }
where $\epsilon$ sets the energy scale, $\alpha \geq 2$, and 
\eq{	r_{m{m\p}} \equiv \left| \v r_{m\p} - \v r_m \right| \label{eq:rij_definition}} 
is the distance between the centers of the two particles. Equation~\eqref{eq:pairwise_iteraction_potential} represents a one-sided repulsive spring-like interaction, in which particles repel when they overlap but do not otherwise interact. This is called the ``soft sphere" class of potentials. The exponent $\alpha$ has typical values of $\alpha=2$ (leading to ``harmonic" or ``Hookian" interactions) and $\alpha=2.5$ (leading to ``Hertzian" interactions). 

Note that Eq.~\eqref{eq:rij_definition} implicitly accounts for the periodic boundary conditions; a particle will never interact with multiple images of another particle provided that $L > 4\max(R_m)$, which we will always enforce. The units of length, mass, and energy are $D_\text{avg}$, $M$, and $\epsilon$ respectively, where $D_\text{avg}\equiv N^{-1}\sum_m 2R_m$ is the average particle diameter. The units of all quantities in this thesis are given by the appropriate combination of these three.

\subsection{Additional definitions}

\defn{Total energy}{$U$}{
The only interactions in the system come from Eq.~\eqref{eq:pairwise_iteraction_potential}, and the total energy is given by 
\eq{	U = \sum_{\left<m,{m\p}\right>} U_{m{m\p}}. \label{eq:total_energy_definition} } 
$\left<m,{m\p}\right>$ means that the sum is over all pairs of interacting particles. 
}

\defn{Packing fraction}{$\phi$}{The packing fraction is the ratio of the volume of the spheres to the total volume, and is given by 
\eq{	\phi = \frac 1V \sum_m V_m,}
where $V=L^d$ is the volume and $V_m$ is the volume of sphere $m$ ({\it i.e.} $V_m = \pi R_m^2$ in $2d$ and $V_m = \tfrac 43 \pi R_m^3$ in $3d$). Note that when two spheres overlap, this definition double counts the overlapping region and it is therefore possible for $\phi$ to be greater than 1. This is different from the ``experimental" definition of the packing fraction that does not double count. 
}

\defn{Stress tensor}{$\sigma_{ij}$}{
The $d$ by $d$ stress tensor is given by 
\eq{	\sigma_{ij} = \frac 1V \sum_{\left<m,{m\p}\right>} \left(\pd {U}{r}\right) \frac{r_i r_j}{\abs{r}}, }
where the subscripts on $U_{mm\p}$ and $r_{mm\p}$ are implied, and $i$ and $j$ index spatial dimensions.
}

\defn{Pressure}{$p$}{
 The pressure is related to the trace of the stress tensor and is given by 
\eq{	p = -\frac{1}{d} \Tr \sigma.}
}

\defnn{Rattler}{
In a jammed packing, a rattler is a particle that is not part of the rigid, jammed backbone of the system. Any contacts it might have are marginal (meaning the particles are {\it just} touching and have negligible overlap), and the particle does not contribute at all to the overall rigidity or other mechanical properties of the system. Since a non-rattler must have at least $d+1$ contacts to be geometrically constrained, we define a rattler as a particle that is in contact with fewer than $d+1$ non-rattlers.\footnote{Some people also include a geometrical requirement that there is no $d$ dimensional hemisphere without a contact. My experience is that this additional requirement has little to no effect and that the $d+1$ requirement is sufficient, though it is important to keep this in mind.}
}

\defn{Number of ``participating" particles}{$N_{pp}$}{
This is just the number of particles minus the number of rattlers. It is often convenient to ignore the rattlers in static calculations, and this is the effective number of particles. However, it is quite common (both in the literature and in this thesis) for $N$ to implicitly mean the number of ``participating" particles. 
% Many times throughout this thesis and in the literature to simply use $N$ when this is what is really meant.
}

\defn{Number of contacts}{$N_c$}{
The total number of contacts in the system, not including any (marginal) contacts associated with rattlers. 
}

\defn{Contact number}{$Z$}{
Also known as the coordination number, the contact number is the average number of contacts per particle, ignoring rattlers. This quantity is only sensible when the interaction potential has a well-defined and sharp cutoff, as is the case for soft spheres. %If $N_c$ is the total number of contacts in the system, then the contact number is
The contact number is given by
\eq{	Z = \frac{2N_c}{N_{pp}}.}
%Note that $N$ here is the total number of non-rattlers. 
}

\defn{Fabric tensor}{$F_{ij}$}{
The fabric tensor is a measure of structural anisotropy and is given by 
\eq{	F_{ij} = \frac 2{N_{pp}} \sum_{\left<m,{m\p}\right>} \frac{r_ir_j}{\abs{r}^2}. }
Note that $Z = \Tr F$.
}

\section{Preparation of state databases}

\subsection{Energy minimization details\label{sec:energy_minimization_details}}
At zero temperature, a system of soft spheres always sits at a local energy minimum so that force balance is satisfied on all particles. 
The energy is minimized by running the so-called FIRE algorithm~\cite{Bitzek:2006bw}. In the past I have used a number of standard nonlinear optimization algorithms, including Conjugate Gradient (CG) and L-BFGS, but given its speed, ease of implementation, and sensitivity to shallow features in the energy landscape, I now recommend the exclusive use of the FIRE algorithm. Reference~\cite{Bitzek:2006bw} suggests values for all of the parameters in the algorithm except for the baseline MD simulation time step, $\Delta t_{MD}$. Based on trial and error to maximize efficiency, I use $\Delta t_{MD} = 0.02$. 

Any minimization algorithm will run until a certain stopping condition is met. The stopping condition I use is
\eq{	\abs{ f_\alpha} < f_\text{tol} \label{eq:minimization_stopping_condition}}
for all $\alpha$, where $\alpha \in [1,dN]$, $f_\alpha$ is a component of the $Nd$ dimensional vector of forces on particles, and $f_\text{tol}$ is some small tolerance.\footnote{In principle, you want the residual net forces in the system to be much smaller than the typical force scale in the system, which in our case is proportional to the pressure. In practice, however, we are limited by machine precision, and our approach (Eq.~\eqref{eq:minimization_stopping_condition}) is to do ``as good as can reasonably be expected.'' One can then decide if the residual forces in a configuration at a particular pressure are sufficiently small for his or her purposes.} In other words, the minimization algorithm will continue to run until Eq.~\eqref{eq:minimization_stopping_condition} is satisfied for all $\alpha$. I use a tolerance of $f_\text{tol} = 10^{-13}$. 

Note that many optimizations algorithms use changes in energy as a stopping condition, meaning that the algorithm is considered to have converged when the energy stops decreasing. Due to rounding errors in the calculation of the energy, however, changes in energy typically reach machine precision well before Eq.~\eqref{eq:minimization_stopping_condition} is satisfied. Therefore, one can do a much better job of obtaining a mechanically stable state by only looking at the gradient of the energy.

\subsection{Procedure for generating configurations\label{sec:procedure_for_generating_configurations}}
The goal here is to generate configurations at specific pressures with logarithmic spacing. Consider a set of target pressures, for example: $p_\text{target} = 10^{-1}$, $10^{-1.2}$, $10^{-1.4}$, $...$, $10^{-8}$. Configurations of $N$ total particles are generated at each of these pressures as follows. 
\begin{enumerate}
\item Using a uniquely seeded random number generator, place all the particles at random in the simulation box and set the packing fraction to be $\phi = 1.0$,\footnote{Recall that our definition of the packing fraction double counts the volume of overlapping regions, meaning that $\phi$ can be made arbitrarily large.} which is a crude approximation for the packing fraction at the largest target pressure. 
\item Minimize the energy and calculate the pressure ($p$) and the bulk modulus ($B$). See Sec.~\ref{sec:energy_minimization_details} for details on energy minimization and Sec.~\ref{sec:numerical_procedures_linear_elasticity} for details on calculating the elastic moduli.
\item Set $p_\text{target}$ to be the largest of the target pressures ({\it e.g.} $p_\text{target} = 10^{-1}$). 
\item \label{step:target_pressure_loop} With the current configuration, repeat the following steps until the pressure satisfies 
\eq{	\abs{\frac{p_\text{target} - p}{p_\text{target}}} < p_\text{tol}, \label{eq:target_pressure_test} }
where we use a tolerance of $p_\text{tol} = 0.01$.\footnote{This typically only needs to be repeated 1-2 times. If the pressure has not converged in 6 cycles, the loop is exited and the configuration is not included in the database.}
\begin{enumerate}
	\item Change the packing fraction of the system according to the difference between the pressure and the target pressure. Assuming the harmonic approximation, set
%	This is done via the harmonic approximation:
	\eq{	\phi \rightarrow \phi + \eta \left[ \frac{\phi + (p_\text{target} - p)}{B} \right]	,	}
	where the constant $\eta$ is initially set to $1$. 
	\item Re-minimize the energy and calculate the pressure ($p$) and the bulk modulus ($B$).
	\item If the system is not jammed\footnote{There are numerous signatures of the jamming transition that can be used to determine if a system is jammed. However, many of these ({\it e.g.} the energy, pressure, shear modulus, or bulk modulus for non-harmonic interactions) require an arbitrary threshold to distinguish zero from non-zero. To avoid this ambiguity, I only check that 1) not every particle is a rattler and 2) the number of contacts, $N_c$, satisfies the relation $N_c - d(N_{pp}-1) \geq 1$. While these tests may not technically be sufficient to determine if a system is jammed (depending on one's preferred definition of jamming), they are sufficient for the algorithm to proceed.} return the system to the previous state and set $\eta \rightarrow 0.5 \eta$.
	\item If the system is jammed but does not satisfy Eq.~\eqref{eq:target_pressure_test}, set $\eta \rightarrow 1$.
\end{enumerate}
\item If this is successful, we now have a mechanically stable configuration whose pressure is sufficiently close to the target pressure. Now change the target pressure to the next largest ({\it e.g.} $10^{-1.2}$) and repeat the procedure in Step~\ref{step:target_pressure_loop}. Repeat this for all of the target pressures.
\end{enumerate}

\subsection{Database details}
In the procedure described above, there are a number of parameters that describe the setup of the system. These are the number of particles ($N$), the potential ({\it e.g.} harmonic, Hertzian), the dimensionality of space (2d or 3d), and the distribution of particle radii ({\it e.g.} mono-disperse, bi-disperse, etc.). In addition, there is the choice of the seed for the random number generator, which allows for many independent configurations with otherwise identical parameters. A {\it database set}, which is the set of all generated states with the same $N$, potential, dimension and radii distribution, is created as follows.

%The procedure described in Sec.~\ref{sec:procedure_for_generating_configurations} is run many times with different initial seeds. For each seed, 
For a given seed and set of parameters, the procedure described in Sec.~\ref{sec:procedure_for_generating_configurations} generates $N_{p_{\text{target}}}$ configurations, one for every target pressure. A number of checks are then performed on each of the $N_{p_{\text{target}}}$ states, and if any of these tests fail for any configuration, then all $N_{p_{\text{target}}}$ configurations are thrown out. Otherwise, they are added to the database. Therefore, there is a well-defined correspondence between states at different pressures.\footnote{For example, in a particular database set, the 43rd configuration at $p_\text{target}=10^{-6.2}$ will be the configuration that was directly derived from the 43rd configuration at $p_{\text{target}}=10^{-6}$.} This is then repeated many times with different seeds until there is a predetermined number ({\it e.g.} 5000) of ``successful" configurations at each pressure. 

The tests are:
\begin{enumerate}
\item Not all particles are rattlers: $N_{pp} > 0$.
\item The number of contacts satisfies the theoretical lower bound: $N_c - d(N_{pp}-1) \geq 1$.
\item The pressure is sufficiently close to the target pressure: see Eq.~\ref{eq:target_pressure_test}.
\item Energy minimization successfully converged.\footnote{Note: After minimization, Eq.~\eqref{eq:minimization_stopping_condition} is satisfied for every degree of freedom $\alpha$. However, rattlers are always removed before doing any calculations. While most rattlers have zero contacts, some have up to $d$ contacts, all of which can cary loads of order $f_\text{tol}$. Therefore, removing rattlers (and all associated interactions) can affect the net force on neighboring particles by order $f_\text{tol}$. It is therefore possible that Eq.~\eqref{eq:minimization_stopping_condition} is satisfied before rattlers are removed but not after. To account for this, I calculate the force vector after removing rattlers and require that 
\eq{	\abs{ f_\alpha} < 10 f_\text{tol}. }
%If this is not satisfied, then the system is thrown out. 
}
\item All standard quantities are reproducible, meaning the configuration can be successfully saved and reloaded. This is primarily an i/o consistency check. 
\end{enumerate}

\subsubsection{Distribution of particle radii}
There are a number of ways we can choose the radius of each particle. The simplest is to have them all be equal (monodisperse), but in two-dimensional packings this leads to a high degree of local crystalline order. Instead, it is common to frustrate the system by introducing some polydispersity. A typical example is to have a mixture of 50\% small and 50\% large particles, where $R_\text{large}/R_\text{small} = 1.4$. While this is a very common choice, it still leads to a non-negligible amount of crystallinity, as discussed in Sec.~\ref{sec:JammingCriteria}. Therefore, I have sometimes used a polydisperse distribution where the radii are evenly spaced with the ratio of the largest divided by the smallest is $1.4$.

%\subsubsection{Table of database sets}
%
%\begin{table}[h!]
%\centering
%\begin{tabular}{| c | c | c | c |}\hline
% & \multicolumn{3}{c|}{radii distribution} \\ \hline
%N & mono & bi & poly\_uniform \\ \hline
%32 & & 5000 & 5000 \\
%64 & & 5000 & 5000 \\
%128 & & 5000 & 5000 \\
%256 & & 5000 & 5000 \\
%512 & & 5000 & 5000 \\
%1024 & & 5000 & 5000 \\
%2048 & & 5000 & 5000 \\
%4096 & & 5000 & 5000 \\
%8192 & & & 2000 \\
%16384 & & & \\ \hline
%\end{tabular}
%\caption{Harmonic interactions in 2 dimensions.}
%\end{table}

\section{The harmonic approximation and linear response}
Let $\{u_\alpha\}$, where $\alpha \in [1,dN]$, be some set of collective particle displacements. A Taylor expansion of the total energy about $u=0$ gives
\eq{ U = U^0 - F^0_\alpha u_\alpha + \tfrac 1{2} D^0_{\alpha\beta}u_\alpha u_\beta + \tfrac 1{3!}T^0_{\alpha\beta\gamma}u_\alpha u_\beta u_\gamma + ... \label{eq:energy_expansion}}
Here, $U^0$ is the energy of the initial system and $\vec F^0$ gives the net force component on every particle, $F^0_\alpha = - \pd{U}{u_\alpha}\big|_{u=0}$. %, which vanishes if the system is mechanically stable. 
The dynamical matrix $D^0$ (also referred to as the ``Hessian") is given by the second derivative of the energy, $D^0_{\alpha\beta} = \pd{^2U}{u_\alpha \partial u_\beta}\big|_{u=0}$,\footnote{Note that the $1/M$ term that typically appears in the dynamical matrix is omitted here because we are explicitly setting $M=1$ for all particles.} and the tensors $T^0$, etc. are given by higher-order derivatives. The ``0" superscripts emphasize that the derivatives are evaluated at $u=0$, although going forward this will be dropped to avoid unnecessary clutter. 

If the system is mechanically stable, then $F_\alpha = 0$ for all $\alpha$. The harmonic approximation is obtained by truncating Eq.~\eqref{eq:energy_expansion} after the second order term:
\eq{ U - U^0 \approx \tfrac 1{2} D^0_{\alpha\beta}u_\alpha u_\beta. \label{eq:harmonic_approximation_def}}
Chapter~\ref{chapter:linear_response} provides an in-depth analysis of when this approximation is justified. 
Under the harmonic approximation, the system behaves as a many body harmonic oscillator. 

\subsection{Normal modes of vibration}
The linearized equations of motion are 
\eq{	\ddot u = -Du.	 \label{eq:linearized_eq_of_motion}}
Equation~\eqref{eq:linearized_eq_of_motion} can be solved by projecting onto the $n$th eigenvector $e_n$ of $D$:
\eq{	e_n^T \ddot u = - \lambda_n (e_n^T u),	}
where $\lambda_n$ is the $n$th eigenvalue. This has the solution
\eq{	e_n^T u(t) = e_n^T u(0) e^{i t \sqrt{\lambda_n}}. }
Thus, the eigenvectors of $D$ are the normal modes of vibration of the system and the eigenvalues give the square of the corresponding characteristic frequencies. If a system is mechanically stable, then all the eigenvalues should be strictly positive, except for the $d$ modes that correspond to global translation\footnote{As we are considering systems with periodic boundary conditions, we do not get zero modes associated with global rotation.} and have eigenvalue zero.\footnote{This statement assumes that rattlers have been removed. Otherwise additional zero modes would exist.} The density of normal modes as a function of frequency is referred to as the density of states, $D(\omega)$ (not to be confused with the dynamical matrix, $D$), and is of fundamental interest.

Numerical diagonalization of large matrices, such as the dynamical materix, is a notoriously difficult problem; computation time is typically $\mathcal O(N^3)$ and the $\mathcal O(N^2)$ memory required to store the matrix and eigenvectors can become prohibitive. I predominantly use the numerical package ARPACK (and the C++ wrapper ARPACK++), which has two primary advantages. 1) Due to the short range nature of the interaction potential in Eq.~\eqref{eq:pairwise_iteraction_potential}, the dynamical matrix is extremely sparse, with only order $N$ non-zero values. ARPACK uses algorithms that are designed for very large and sparse matrices. 2) It is often the case that we only care about the smallest eigenvalues and their corresponding eigenvectors. ARPACK allows you to calculate, for example, only the 100 eigenvectors with the smallest eigenvalue. Since the size of the matrix can be up $\sim 10^5$ or more, this is \emph{significantly} more efficient. Although getting ARPACK to work can be troublesome, I highly recommend it for anyone needing to diagonalize large sparse matrices.\footnote{ARPACK can have some difficulty when it encounters highly degenerate eigenvalues. For example, there have been times when I knew precisely how many zero eigenvalues the matrix had, but if I requested precisely that many eigenvectors, ARPACK would miss some of the zero modes and instead give me eigenvectors with positive eigenvalue. However, if I requested, say, 100 additional eigenvectors, then it would correctly find all the zero modes. As a rule of thumb, I always request at least 100 modes, and if I am interested in a degenerate spectrum, I always request at least 100 more than my estimate for the size of the degeneracy.}$^\text{,}$\footnote{ARPACK has a ``shift and invert" mode that supposedly can be significantly faster and allow you to find modes anywhere in the spectrum. However, I have had some difficulty getting this to work properly. While this may reflect more on me than on ARPACK itself, I nonetheless leave this as a warning.}

\subsection{Hooke's law}
If an external force $f^\text{ext}$ is applied to the system, force balance will no longer be satisfied. The resulting response of the system $u$ is obtained to linear order by solving\footnote{Technical note: $D$ is positive semi-definite, not positive definite, because it has zero frequency modes that correspond to global translations (as well as to motion of rattlers if they have not been removed). Therefore, inverting $D$ to solve for the response is ill-defined. Rattlers have no affect on linear response and can be completely ignored. To suppress the translational modes, I simply add $\varepsilon \mathcal{I}$ to the dynamical matrix, where $\mathcal{I}$ is the identity and $\varepsilon$ is order $10^{-12}$. This effectively adds a very weak tether to each particles and makes $D$ positive definite, raising each zero mode to $\varepsilon$. Numerically, this suppresses drift along the zero modes but has essentially zero effect on the non-trivial response.}
\eq{	Du = f^\text{ext}. \label{eq:Hookes_law}}
I solve Eq.~\eqref{eq:Hookes_law} numerically using UMFPACK (part of the SuiteSparse package\footnote{See http://faculty.cse.tamu.edu/davis/suitesparse.html.}) which efficiently computes an LU decomposition on large sparse matrices. Of its many advantages, this allows you to solve Eq.~\eqref{eq:Hookes_law} for many different $f^\text{ext}$ without having to recalculate the LU decomposition. 

\subsection{Linear elasticity\label{sec:numerical_procedures_linear_elasticity}}
Consider a global deformation ({\it e.g.} compression or shear) that is given to lowest order by the linearized symmetric strain tensor $\epsilon_{ij}$. To second order in $\epsilon$, the change in energy $\Delta U$ is given by 
\eq{	\frac{\Delta U}{V} = \sigma_{ij}\epsilon_{ij} + \frac 12 c_{ijkl}\epsilon_{ij}\epsilon_{kl} + ...}
where $V$ is the volume, $\sigma_{ij}$ is the initial stress tensor and $c_{ijkl}$ is the elastic modulus tensor. It is convenient to express this in terms of the enthalpy-like quantity $H \equiv U -   \sigma_{ij} \, \epsilon_{ji} \, V$:
\eq{ \frac{\Delta H}{V} = \frac 12 c_{ijkl}\epsilon_{ij}\epsilon_{kl} + ...}
The symmetries of $\epsilon_{ij}$ ({\it i.e.} $\epsilon_{ij}=\epsilon_{ji}$) imply the following symmetries in the elastic modulus tensor:
\eq{	c_{ijkl} = c_{jikl} = c_{ijlk} = c_{klij}.	}
When no further symmetries are assumed, this reduces the number of independent elastic constants from $d^4$ to $\tfrac 18 d(d+1)(d^2+d+2)$, which is 6 in 2 dimensions and 21 in 3 dimensions. Note that for an isotropic system, this reduces further to just the bulk modulus and the shear modulus. We can calculate the components of $c_{ijkl}$ by imposing various strain tensors and calculating the change in energy to linear order.

A linearized symmetric strain tensor $\epsilon_{ij}$ is imposed by transforming all distance vectors $\v r$ affinely according  to
\eq{ r_i \rightarrow r_i + \sum_j \epsilon_{ij} r_j. \label{strain_tensor_transformation}	}
When a disordered system undergoes such a transformation, force balance will no longer be satisfied and the particles will exhibit a secondary response. If $f^\text{ext}$ is the $Nd$ dimensional vector of net forces on the particles, then we can use Eq.~\eqref{eq:Hookes_law} to solve for the secondary, non-affine displacements. By adding the affine and non-affine displacements, it is straight forward to calculate the total change in energy and back out $c_{ijkl}$.

\chapter{Contact nonlinearities and the validity of linear response}
\label{chapter:linear_response}

\newcommand{\denl}[0]{\ensuremath{\delta_\mathrm{enl}^*}}

\section{Introduction}
The harmonic approximation of an energy landscape is the foundation of much of solid state physics~\cite{Ashcroft:1976ud}. Calculations that invoke this simplifying assumption are said to be in the linear regime and are responsible for our understanding of many material properties such as sound propagation and the elastic or vibrational response to small perturbations~\cite{Ashcroft:1976ud,Landau1986}. While the harmonic approximation is not exact and breaks down for large perturbations, the existence of a linear regime is essential to our understanding of ordered solids.

While the lack of any periodic structure has long made amorphous materials difficult to study, the past decade has seen significant progress towards uncovering the origin of commonality in disordered solids by way of the jamming scenario~\cite{Liu:2010jx}. Specifically, numerous studies of the jamming transition of athermal soft spheres have exploited the harmonic approximation to reveal a non-equilibrium phase transition~\cite{Liu:1998up,OHern:2003vq,Liu:2010jx,Wyart:2005wv,Silbert:2005vw,Ellenbroek:2006df,Xu:2007dd,Xu:2010vd,Goodrich:2012ck}. Near this jamming transition, the shape of the landscape near each minimum is essentially the same within the harmonic approximation -- for example, the distribution of curvatures around the minimum, which is directly related to the density of normal modes of vibration, is statistically the same for the vast majority of energy minima.  As a result, linear response properties such as the elastic constants can be characterized by a single property of the minimum, such as its energy, pressure or contact number, which quantifies the distance from the jamming transition for that state.  This powerful property forms the basis of the jamming scenario, which has been shown to explain similarities in the mechanical and thermal properties of many disordered solids. 

However, the jamming scenario is based on systems with finite-ranged potentials. It was pointed out by Schreck {\it et al.}~\cite{Schreck:2011kl} that for such potentials, breaking and forming contacts are a source of nonlinearity and they concluded that the harmonic approximation should not be valid for disordered sphere packings in the large-system limit even for infinitesimal perturbations.  Without a valid linear regime, quantities like the density of vibrational modes and elastic constants are ill defined. Thus, their claim calls into question much of the recent progress that has been made in understanding the nature of the jammed solid.

In this chapter, we examine the effect of nonlinearities in jammed sphere packings.  
As we discuss in Sec.~\ref{sec:nonlinearities}, there are two distinct classes of nonlinearities: \emph{expansion nonlinearities} are those that can be understood from the Taylor expansion of the total energy about the local minima, while \emph{contact nonlinearities} are those arising from particles coming in and out of contact. Hentschel {\it et al.}~\cite{Hentschel:2011cba} recently asked whether expansion nonlinearities destroy the linear regime in the thermodynamic limit. By considering carefully the proximity of the system to a plastic rearrangement, which is often preceded by a vibrational mode with vanishing frequency, they concluded that the elastic moduli are indeed well defined.  Here, we provide a detailed analysis of the effect of \emph{contact} nonlinearities; just as Hentschel {\it et al.} found that expansion nonlinearities do not invalidate linear response in the thermodynamic limit, we find that the same is true for contact nonlinearities. Our main results are presented in Sec.~\ref{sec:thermodynamic_limit}, where we show that packings at densities above the jamming transition have a linear regime in the thermodynamic limit despite an extensive number of altered contacts.  We then discuss finite-amplitude vibrations in Sec.~\ref{sec:finite_amplitude_discussion}, and conclude in Sec.~\ref{sec:discussion} with a discussion of our results and their implications for the jamming scenario. 

One somewhat counterintuitive result is that for intrinsically anharmonic potentials such as the Hertzian potential, contact nonlinearities do not affect the harmonic approximation in the limit of small displacements.  Such nonlinearities only pose a danger for Hookian repulsions, but even in that case, there is a well-defined linear regime in the thermodynamic limit for any density above the transition, contrary to the conclusions of Ref.~\cite{Schreck:2011kl}.  Thus, our results show that the harmonic approximation is on footing that is as firm for disordered solids as it is for ordered solids.  

\section{The harmonic approximation and its leading nonlinear corrections\label{sec:nonlinearities}}

We consider athermal packings of $N$ soft spheres in $d$ dimensions that interact with the pair potential
\eq{	
	V_{mn}(r) = \left\{ 
	\begin{array}{l l}
		\frac \epsilon \alpha \left(1-\frac {r}{\sigma}\right)^\alpha & \quad \mbox{if  $r<\sigma$}\\
		0 & \quad \mbox{if $r\geq \sigma$.}\\ 
	\end{array} 
	\right. \label{pair_potential}
}
Here, $r$ is the center-to-center distance between particles $m$ and $n$, $\sigma$ is the sum of the particles' radii, $\epsilon\equiv 1$ sets the energy scale, and $\alpha \ge 2$ determines the power law of the interactions. Such packings jam when the packing fraction $\phi$ exceeds a critical density $\phi_c$, and we will use the excess packing fraction, $\Delta \phi \equiv \phi-\phi_c$, as a measure of the distance to jamming.

The harmonic approximation is obtained from the expansion of the total energy:
\eq{
	U &\equiv \sum_{m,n} V_{mn}(r) \label{total_energy} \\
	&= U^0 - F^0_i u_i + \tfrac 1{2} D^0_{ij}u_i u_j + \tfrac 1{3!}T^0_{ijk}u_iu_ju_k + ... \label{energy_expansion}
}
where the indices $i$, $j$, $k$ run from 1 to $dN$ and index the $d$ coordinates of each of the $N$ particles, and the $dN$-dimensional vector $\vec{u}$ represents some collective displacement about the initial positions. It will be useful to denote the magnitude of $\vec{u}$ as $\delta$ and the direction as $\hat{u}$, so that $\vec{u}=\hat{u}\delta$.
$U^0$ is the energy of the initial system. $\vec F^0$ gives the net force component on every particle, $F^0_i = - \pd{U}{u_i}\big|_{\vec u=0}$, which vanishes if the system is mechanically stable. The dynamical matrix $D^0$ is given by the second derivative of the energy, $D^0_{ij} = \pd{^2U}{u_i \partial u_j}\big|_{\vec u=0}$, and the tensors $T^0$, etc. are given by higher-order derivatives. The ``0" superscripts emphasize that the derivatives are evaluated at $\vec u=0$.

The mechanical response of an athermal system of particles is governed by the equations of motion,
\eq{	m_i \ddot u_i = F_i(\vec u), \label{eq_of_motion}	}
where $m_i$ is the particle mass and $\vec F(\vec u)$ is the vector of instantaneous forces, {\it i.e.}, evaluated at $\vec u$. Since $D_{ij}(\vec u) =- \pd{F_i(\vec u)}{u_j}$, where $D(\vec u)$ is the instantaneous dynamical matrix, this force is generically given by
\eq{	F_i(\vec u) = -\int D_{ij}(\vec u) du_j,	\label{integral_of_dynamical_matrix} }
where the integral follows the trajectory of the particles from the mechanically stable minima at $\vec u = 0$ to the current configuration.

A mechanically stable system is said to be in the linear regime if the harmonic approximation 
\eq{	U-U^0 \approx \tfrac 12 D^0_{ij}u_iu_j	\label{harmonic_approx}}
is accurate enough to describe the phenomenon of interest. Under this assumption, the dynamical matrix is constant and the equations of motion become linear:
\eq{	m_i \ddot u_i = -D^0_{ij}u_j. \label{linearized_eq_of_motion}	}
The solutions to Eq.~\eqref{linearized_eq_of_motion} are called the normal modes of vibration and are among the most studied quantities in solid state physics.

\subsection{Microscopic vs. bulk response}
Importantly, the extent of the linear regime depends on the quantity one wishes to measure; Eq.~\eqref{linearized_eq_of_motion} might accurately describe one phenomena but fail to describe another. Thus, it is important to clarify the quantities of interest~\cite{Goodrich:2014jw}. 
For crystalline solids, the linear approximation is often used to calculate bulk thermal and mechanical properties, such as the elastic moduli and thermal conductivity. However, it is typically \emph{not} used to predict exact microscopic details over long times. If one were to perturb a system along one of its vibrational modes, for example, the linear equations of motion predict simple oscillatory motion confined to the direction of that mode. However, this is not what happens, since even very slight nonlinearities can cause energy to gradually leak into other modes~\cite{Ashcroft:1976ud}. 

Clearly, the linear theory fails to describe such microscopic details, except for the very special case where the harmonic approximation is \emph{exact}, and one would not expect disordered sphere packings to be an exception. However, linear response has had tremendous success in predicting the \emph{bulk} mechanical and thermal properties of crystals. It is these bulk linear quantities, such as the normal modes and elastic moduli, and not the details of microscopic response that are central to the theory of jamming and will thus be the focus of the remainder of this chapter.

We will primarily be concerned with determining whether the harmonic approximation is valid in the limit of infinitesimal displacements, $\delta$.  In other words, we will be asking whether $\delta$ can be made small enough so that Eq.~\eqref{linearized_eq_of_motion} accurately describes bulk response. 
If so, then linear quantities such as the density of states or the elastic constants are well-defined. While experimental measurements in real systems necessarily involve nonzero displacements, our focus on the limit $\delta\rightarrow 0$ will reveal whether the lowest-order behavior can be ascertained from the harmonic approximation.   

To understand the breakdown of the harmonic approximation, it is useful to separate nonlinear corrections into two distinct classes, as outlined below.

\begin{figure}
	\centering	
	\epsfig{file=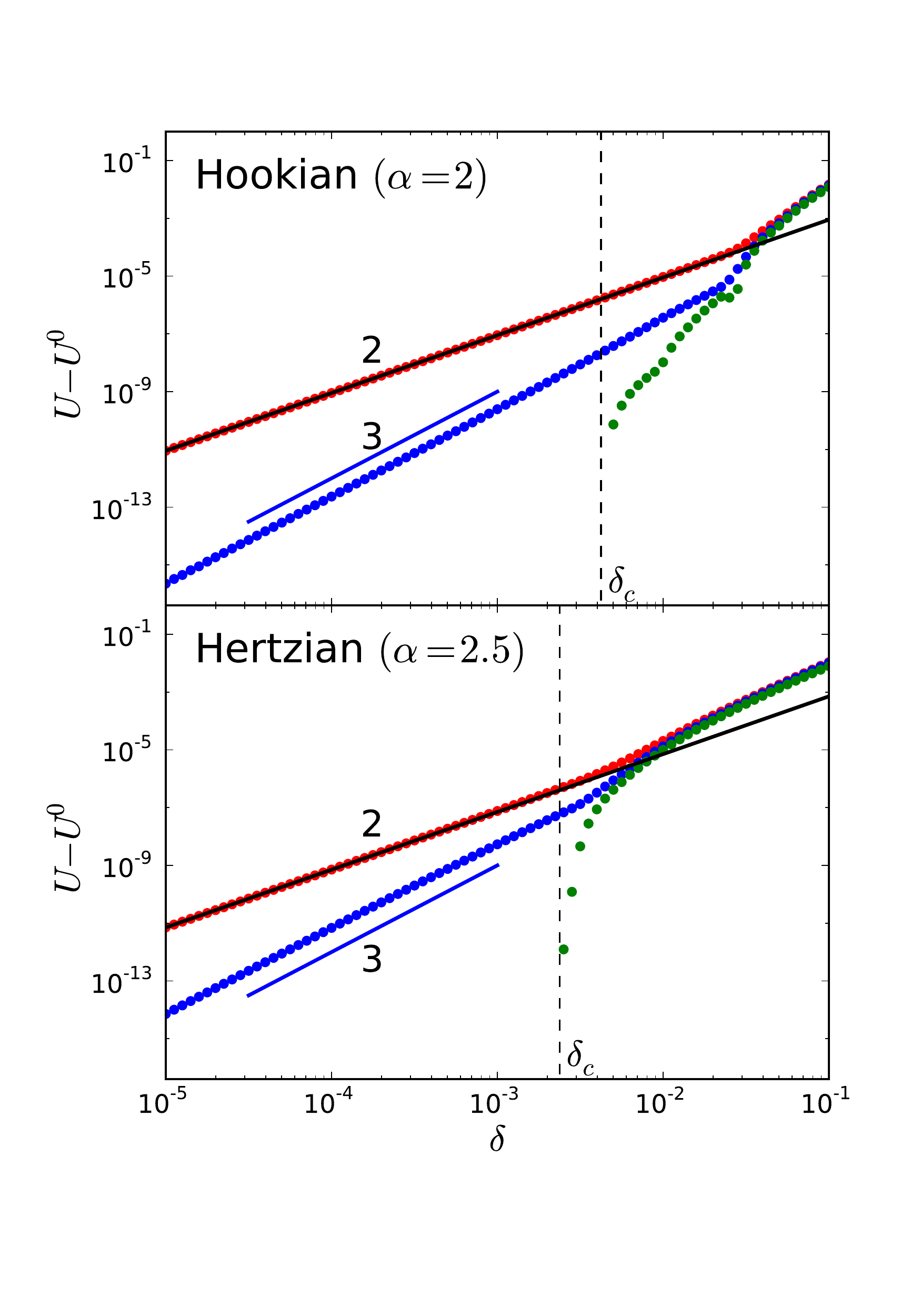,width=0.5\linewidth,viewport=20 90 480 650, clip}
	\caption[Illustration of nonlinearities for packings with Hookian and Hertzian interactions.]{Illustration of nonlinearities for packings of $N=64$ particles with Hookian spring-like interactions ($\alpha=2$, top) and Hertzian interactions ($\alpha=2.5$, bottom). Both systems are at a pressure of $10^{-2}$. The red data shows the total energy $U-U^0$ as the system is displaced by an amount $\delta$ along its lowest non-zero mode. The black line gives the prediction of the harmonic approximation, see Eq.~\eqref{harmonic_approx}. The difference between $U-U^0$ and the predicted energy is shown by the blue data, and the blue line has a slope of $3$. The vertical dashed line represents the value of $\delta$ where the contact network first changes, and the green data gives the magnitude of the change in energy due to contact changes.}
	\label{fig:expansion_nl}
\end{figure}

\subsection{Expansion nonlinearities}
\emph{Expansion nonlinearities} are those which are described by the higher order terms in Eq.~\eqref{energy_expansion}, and can thus be understood from derivatives of the total energy at the energy minimum. However, provided the quadratic term $\tfrac 12 D^0_{ij}u_iu_j$ is positive in all directions, $\delta$ can always be made small enough so that the higher order terms become negligible.\footnote{It is not necessary for $\tfrac 12 D^0_{ij}u_iu_j$ to be positive in the directions of global translation because the energy landscape in these directions is completely flat.} At the jamming transition, {\it i.e.} $\Delta \phi = 0$, the quadratic term vanishes in some directions in configurational space,\footnote{At $\Delta \phi =0$, the density of states is constant in the limit $\omega \rightarrow 0^+$, see Fig.~\ref{fig:dos_intro}, so $\lim_{\epsilon\rightarrow 0} \frac 1\epsilon \int_0^\epsilon d\omega D(\omega) > 0$.} so the harmonic approximation fails.  Away from the jamming transition, {\it i.e.} $\Delta \phi>0$, however, the quadratic term is indeed positive in all directions.  (In all our calculations, we remove rattlers, which correspond to zero-frequency modes, so that the dynamical matrix only contains particles that are part of the jammed network.)  Thus, although expansion nonlinearities can be important and even dominate certain phenomena, they cannot prevent a system from having a linear regime provided $\Delta \phi>0$. 

An easy way to observe expansion nonlinearities is to displace a system by an amount $\delta$ in some direction $\hat u$ and measure the energy as a function of $\delta$. $U-U^0$ can then be compared to the prediction of the harmonic approximation given by Eq.~\eqref{harmonic_approx}. A typical example of this is shown in Fig.~\ref{fig:expansion_nl} for jammed packings of particles with Hookian spring-like interactions ($\alpha=2$, top) and Hertzian interactions ($\alpha=2.5$, bottom). 
The corrections to the harmonic approximation have clear cubic behavior at small $\delta$.  Note that they are present when $\alpha=2$: even a spring network has expansion nonlinearities in dimension $d>1$.  This can be seen from Eq.~\eqref{energy_expansion} by writing the tensor $T^0_{ijk}$ as
\eq{
	T^0_{ijk} \equiv{}& \sum_{m,n} \pd{^3V_{mn}(r)}{r_i \partial r_j \partial r_k} \nonumber \\
	={}&\sum_{m,n} t \left(\pd{r}{r_i} \pd{r}{r_j} \pd r {r_k}\right)     - f \left(\pd{^3r}{r_i \partial r_j \partial r_k}\right) \label{d3Vij}     \\
		&+ k \left( \pd{^2r}{r_i \partial r_k} \pd r{r_j} +\pd{^2r}{r_j \partial r_k} \pd r{r_i}  + \pd{^2r}{r_i \partial r_j} \pd r{r_k}\right), \nonumber
}
where $V_{mn}(r)$ is the pair interaction potential of Eq.~\eqref{pair_potential}, $f\equiv -\pd{V_{mn}(r)}{r}$, $k\equiv \pdd{V_{mn}(r)}{r}$, and $t\equiv \pd{^3V_{mn}(r)}{r^3}$, and the terms $\pd r{r_i}$, $\pd{^2r}{r_i \partial r_j}$ and $\pd{^3r}{r_i \partial r_j \partial r_k}$ are generically nonzero.

Clearly, expansion nonlinearities are more dangerous when the harmonic term is small. At the jamming transition ($\Delta \phi=0$), for example, there exist vibrational modes with arbitrarily low frequency (see Fig.~\ref{fig:dos_intro}) that are thus highly susceptible to expansion nonlinearities. Additionally, it is well known that plastic rearrangements in athermal amorphous solids are preceded by a mode frequency going to zero.\footnote{A notable exception is packings of particles with one-sided Hookian interactions.} Since the density of plastic rearrangements increases with system size, so too does the likelihood that a mode exists with arbitrarily low frequency that is thus highly susceptible to expansion nonlinearities. The effect that this has on the elastic response was studied in detail by Hentschel {\it et al.}~\cite{Hentschel:2011cba}, who showed that the shear modulus is well defined over a small linear regime. For the remainder of this chapter, we will focus on the second class on nonlinearities, the contact nonlinearities introduced in Ref.~\cite{Schreck:2011kl}, which we now discuss.

\subsection{Contact nonlinearities\label{sec:contact_nonlinearities}}
Unlike a true spring network, contacts in a sphere packing are allowed to form and break. Since the total energy is a sum over particles in contact, nonlinearities arise when the contact network is altered. Such \emph{contact nonlinearities} cannot be understood from derivatives of the energy at the minimum. For pair interactions of the form of Eq.~\eqref{pair_potential}, the energy expansion of Eq.~\eqref{energy_expansion} is not analytic when contacts form or break and the second derivative is discontinuous if $\alpha\leq 2$. 

For the two systems in Fig.~\ref{fig:expansion_nl}, the green data show the magnitude of the change in energy due to altered contacts.
The vertical dashed lines indicate the minimum displacement magnitude, $\delta_c$, required to change the contact network.  While the value of $\delta_c$ varies greatly depending on the realization and displacement direction, Schreck {\it et al.}~\cite{Schreck:2011kl} showed that $\avg{\delta_c}\rightarrow 0$ in two important limits. As the number of particles increases, so too does the number of contacts and thus the probability that \emph{some} contact is on the verge of forming or breaking must also increase. Similarly, all contacts are on the verge of breaking in a marginally jammed system at $\Delta \phi = 0$. Therefore, the onset amplitude $\delta_c$ of contact nonlinearities vanishes as either $N\rightarrow \infty$ or $\Delta \phi \rightarrow 0$~\cite{Schreck:2011kl}.

\subsection{Important limits}
Due to the existence of a phase transition at the jamming point, the limits $N\rightarrow \infty$ and $\Delta\phi \rightarrow 0$ are of particular interest. When studying the leading order mechanical properties of a solid, one also considers the limit of infinitesimal displacements, {\it i.e.} $\delta \rightarrow 0$. However, the order at which these limits are taken is important. For example, Schreck {\it et al.} showed that $\delta_c>0$ for finite $\Delta \phi$ and $N$~\cite{Schreck:2011kl}, so there is a perfectly well-defined linear regime if $\delta\rightarrow 0$ is the first limit taken.  This is the standard order of limits taken, for example, in the harmonic theory of crystalline solids~\cite{Ashcroft1976}.  

We already saw that expansion nonlinearities can occur if one considers taking the limit $\Delta \phi \rightarrow 0$ before $\delta \rightarrow 0$, and the importance of these nonlinearities is emphasized in, for example, Refs.~\cite{OHern:2003vq,Xu:2010fa,Gomez:2012ji,Ikeda:2013gu}. Furthermore, Schreck {\it et al.}~\cite{Schreck:2011kl,Schreck:2013gg} showed that contact nonlinearities will also be present in this case, regardless of system size. 
Thus, there is no linear regime at $\Delta \phi=0$. This result was generalized to finite temperatures by Ikeda {\it et al.}~\cite{Ikeda:2013gu} and Wang and Xu~\cite{Wang:2013gy}, who independently showed that the linear regime breaks down above a temperature $T^*$ when $\Delta \phi > 0$.

Finally, for athermal systems above the jamming transition ($\Delta \phi>0$), contact nonlinearities are unavoidable if  we take the limit $N\rightarrow \infty$ before $\delta \rightarrow 0$.   Nevertheless, we will show next that there is still a well-defined linear regime in this case.

\section{The linear regime in the thermodynamic limit\label{sec:thermodynamic_limit}}
In this section, we will show that there is always a well-defined linear regime in the thermodynamic limit whenever $\Delta \phi>0$. We will assume that $T=0$ and that the limit $N\rightarrow \infty$ is taken before the limit $\delta \rightarrow 0$ so that any infinitesimal displacement $\delta\ket{\hat u}$ changes the contact network, leading to contact nonlinearities. 
As discussed above, we will primarily be concerned with establishing the existence of a linear regime for bulk quantities, such as the elastic constants or heat capacity. Since these quantities are described by the density of vibrational modes, $D(\omega)$, it will suffice to show that $D(\omega)$ is insensitive to nonlinear corrections in the limit $\delta \rightarrow 0$. This is not the case for microscopic quantities, such as the precise time evolution following a particular perturbation to a particle or group of particles, which can be highly sensitive to microscopic details that have no noticeable bulk effect.

We will first present a perturbation argument to show that changes to $D(\omega)$ due to contact nonlinearities vanish in the thermodynamic limit as $N^{-1}$~\cite{Goodrich:2014jw}. This result is independent of potential and shows that linear response holds for bulk quantities. We will then present a far simpler argument, based on the continuity of the dynamical matrix for potentials with $\alpha >2$, that shows a clear linear regime for \emph{both} bulk and microscopic quantities~\cite{Goodrich:2014jw}. Our results can be reconciled with those of Schreck {\it et al.}~\cite{Schreck:2011kl,Schreck:2013gg} because they only look at microscopic quantities of relatively small packings close to the transition.

\subsection{Validity of bulk linear response\label{sec:perturbation_argument}}
Here, we will construct a perturbation theory to describe the effect of contact nonlinearities on the vibrational modes and their corresponding frequencies.  We will begin by considering only a single altered contact and then extend the results to the case of many altered contacts. We will assume that $N^{-1}\ll \delta\ll 1$ so that contact nonlinearities are unavoidable but all expansion nonlinearities can be ignored. 

Let $\Delta D$ be the change in the dynamical matrix as a result of the change of a single contact, so that the new dynamical matrix is $\tilde D = D^0 + \Delta D$. Note that $\Delta D$ is highly sparse with only $4d^2$ non-zero elements, where $d$ is the dimensionality. We now consider the effect of the perturbation $\Delta D$ on the eigenmodes of $D^0$ ({\it i.e.}, the original normal modes of vibration).

\subsubsection{Extended modes}
Let $\ket{\hat e_n}$ and $\omega_n^2$ be the $n$th eigenmode and eigenvalue of $D^0$, respectively. If a normalized mode is extended, then every component will be of order $N^{-1/2}$.  For now, we will assume that all modes are extended; the extension of the argument to localized modes is discussed below. The change in the $n$th eigenvalue of $D^0$ can be described by the expansion
\eq{	
	\Delta \omega_n^2 & \equiv \tilde \omega_n^2 - \omega_n^2 \nonumber \\
	&= \matrixel{\hat e_n}{\Delta D}{\hat e_n} + \sum_{m \neq n} \frac{\abs{\matrixel{\hat e_m}{\Delta D}{\hat e_n}}^2}{\omega_n^2 - \omega_m^2} + ...\label{eigenvalue_perturbation_2order}
}
where $\tilde \omega_n^2$ is the eigenvalue of $\tilde D$ and $\matrixel{\hat e_n}{\Delta D}{\hat e_n} \sim \matrixel{\hat e_m}{\Delta D}{\hat e_n} \sim N^{-1}$ because the modes are extended and $\Delta D$ is highly sparse. While the first-order term clearly scales as $N^{-1}$, the higher-order terms depend on the mode spacing as well. Since the probability distribution of eigenvalues does not depend on $N$, the average eigenvalue spacing is proportional to $N^{-1}$. If we assume that 
\eq{	\abs{\omega_n^2-\omega_m^2} > N^{-1}, \label{nondegenerate_criteria}}
then all higher order terms in Eq.~\eqref{eigenvalue_perturbation_2order} are at most proportional to $N^{-1}$. 

However, just because the average mode spacing is of order $N^{-1}$ does not mean that \emph{all} modes are separated by $N^{-1}$. To account for the possibility of, for example, two nearly degenerate modes, $\ket{\hat e_n}$ and $\ket{\hat e_m}$, that do not satisfy Eq.~\eqref{nondegenerate_criteria}, we explicitly solve the degenerate perturbation problem given by
\eq{	
	V \equiv \left( 
	\begin{array}{c c}
		\matrixel{\hat e_n}{\Delta D}{\hat e_n} &\matrixel{\hat e_m}{\Delta D}{\hat e_n} \\
		\matrixel{\hat e_n}{\Delta D}{\hat e_m} &\matrixel{\hat e_m}{\Delta D}{\hat e_m}
	\end{array}
	\right)
	\label{V_extended_only}
}
that treats the coupling between $\ket{\hat e_n}$ and $\ket{\hat e_m}$.
The eigenvalues of $V$ give the full corrections to $\omega_n^2$ and $\omega_m^2$ from their mutual interaction with the perturbation $\Delta D$. The coupling with the other modes is given by Eq.~\eqref{eigenvalue_perturbation_2order}, where terms involving the two nearly degenerate modes are omitted. 

Since the elements of $V$ are all proportional to $N^{-1}$, so too are its eigenvalues. We have already shown that the non-degenerate effect is at most order $N^{-1}$, so the full effect of the perturbation on all eigenvalues $\omega_n^2$ must vanish in the thermodynamic limit. 

We can construct a similar expansion for the eigenvectors. The non-degenerate case is given by
\eq{	\ket{\tilde{e}_n} = \ket{\hat e_n} + \sum_{m\neq n} \frac{\matrixel{\hat e_m}{\Delta D}{\hat e_n}}{\omega_n^2 - \omega_m^2} \ket{\hat e_m} + ...	 \label{eigenvector_expansion}}
while the coupling between nearly degenerate modes that do not satisfy Eq.~\eqref{nondegenerate_criteria} is given by the eigenvectors of $V$. As should be expected, the eigenvectors of $V$ can cause considerable mixing between the nearly degenerate modes. 
Furthermore, the coefficients in front of $\ket{\hat e_m}$ in Eq.~\eqref{eigenvector_expansion} do not vanish when $\abs{\omega_n^2-\omega_m^2}$ is of order $N^{-1}$. Thus, an eigenmode can mix with the few modes nearest in frequency, but the eigenvalue difference between such modes vanishes as $N^{-1}$. In the thermodynamic limit, modes that are able to mix must already be degenerate, so distinguishing between them is meaningless. It is clear that the mode mixing caused by the perturbation $\Delta D$ cannot change the spectral density in the thermodynamic limit.

\subsubsection{Localized modes}
We will now consider the effect of localized modes. We will show that localized modes that overlap with the altered contact can change substantially, but their presence does not affect the extended modes. Furthermore, since the number of modes that are localized to a given region cannot be extensive, the total density of states will be unaffected. Although we will consider modes that are completely localized to a few particles, the arguments can be easily applied to quasi-localized modes by including appropriate higher-order corrections. 

If a localized mode does not overlap with the altered contact, then the matrix elements in Eqs.~\eqref{eigenvalue_perturbation_2order} and \eqref{eigenvector_expansion} involving that mode are zero. In this trivial case, the localized mode is unchanged and does not couple to any other modes. However, if a localized mode \emph{does} overlap with the altered contact, then the matrix elements coupling it to an extended mode are proportional to $N^{-1/2}$ (not $N^{-1}$, as it is for the extended modes).

In this case, we cannot use the non-degenerate perturbation theory of Eqs.~\eqref{eigenvalue_perturbation_2order} and \eqref{eigenvector_expansion}. For instance, there is always a $k$th order term in Eq.~\eqref{eigenvalue_perturbation_2order} that is proportional to $N^{-1}/\abs{\omega_n^2-\omega_m^2}^{k-1}$ and does not converge unless $\abs{\omega_n^2-\omega_m^2} \gg \mathcal{O}\left(N^{-1/(k-1)}\right)$. Therefore, we must treat the interaction between the localized mode and all nearby extended modes by solving the degenerate problem. 

Let $\omega_l^2$ be the the eigenvalue of a localized mode, and let the indices $s$ and $t$ run over the set of $\rho N$ modes that satisfy
\eq{	\abs{\omega_{s,t}^2 - \omega_l^2} < c \nonumber}
where $c$ is some small constant. Note that the localized mode is among those spanned by $s$ and $t$.
To diagonalize the symmetric perturbation matrix
\eq{ V_{st} \equiv \matrixel{\hat e_s}{\Delta D}{\hat e_t},}
note that the dynamical matrix can be written~\cite{Pellegrino:1993uu} as
\eq{ D = AF^{-1}A^T.  \nonumber}
Here, $A$ is the equilibrium matrix and has $dN$ rows and $N_c$ columns, where $N_c$ is the number of contacts, $N$ is the number of particles and $d$ is the dimensionality. $F$ is the diagonal flexibility matrix and has $N_c$ elements $F_{ii} = 1/k_i$, where $k_i$ is the stiffness of the $i$th contact. When $N_c=1$, as is the case for our perturbation matrix $\Delta D$, the equilibrium matrix becomes a vector, $A\rightarrow\ket{A}$, and the flexibility matrix becomes the scalar $1/k$. We can now write the matrix elements as
\eq{
	V_{st} &= \matrixel{\hat e_s}{\Big(\ket{A}k \bra{A}\Big)}{\hat e_t} \nonumber \\
	&= a_s a_t,
	\label{V_reduced}
}
where $a_s = k^{1/2}\braket{A}{\hat e_s}$ and $\braket{A}{\hat e_s}$ is simply the projection of the original eigenvector $\ket{\hat e_s}$ onto the broken contact. Note that for extended modes, the magnitude of $a_s$ scales as
\eq{	a_s  \sim N^{-1/2} 	}
while for localized modes
\eq{	a_l  \sim 1.}
The eigenvalues and eigenvectors of the $\rho N$ by $\rho N$ matrix $V_{st}$ can be solved exactly, with the following results.

A matrix of the form of Eq.~\eqref{V_reduced} has only one non-zero eigenvalue,
\eq{	\Delta \omega_l^2 = a_l^2 + \sum_{s\neq l} a_s^2,	}
which gives the change in energy of the localized mode and does not vanish in the thermodynamic limit. This is not surprising given the drastic overlap between the mode and the altered contact. Similarly, the corresponding eigenvector gives the coupling from the extended modes:
\eq{	\ket{\tilde e_l} = \ket{\hat e_l} + \sum_{s\neq l} \frac{a_s}{a_l}\ket{\hat e_s}.	}
From the scalings of $a_s$ and $a_l$, there is a $N^{-1/2}$ contribution to $\ket{\tilde e_l}$ from each of the $\rho N$ extended modes, the elements of which also scale like $N^{-1/2}$. Therefore, $\ket{\tilde e_l}$ becomes at least partially extended if $\sum_{s\neq l} a_s^2 > 0$.

Thus, forming or breaking a single contact can significantly change the eigenvalue of localized modes that happen to overlap with the altered contact. However, since the density of such modes vanishes as $N^{-1}$, the effect on the density of states in negligible. Furthermore, note that if the initial displacement $\ket{u}$ is along a localized mode, then there is always a finite displacement amplitude $\delta_c$ before the first contact change and so contact nonlinearities can be avoided. 

To understand the effect of localized modes on the extended modes, we see that all other eigenvalues of $V_{st}$ are zero:
\eq{	\Delta \omega_s^2 = 0 \;\;\;\; \text{for all } s\neq l	}
This implies that the frequency of an extended mode does not change due to the presence of a localized mode. However, there is a small correction to the mode itself,
\eq{	\ket{\tilde e_s} &= \ket{\hat e_s} - \frac {a_s}{a_l} \ket{\hat e_l}, \;\;\;\; \text{for all } s\neq l	}
but this correction vanishes in the thermodynamic limit.

Thus, we have shown that even for Hookian springs, altering a single contact in the thermodynamic limit cannot change the density of states~\cite{Goodrich:2014jw}. For extended modes, eigenvalues can change by at most order $N^{-1}$ and mode mixing is allowed only between modes whose eigenvalue spacing is less than $N^{-1}$. While localized modes that overlap with the altered contact can have a non-negligible change in eigenvalue and mix with a large number of extended modes, the density of such localized states vanishes as $N^{-1}$.

\subsubsection{Breaking many contacts}
So far we have considered the effect of changing a single contact. We can find an upper bound for the total number of contacts that can change, $\Delta N_c$, by considering the probability distribution $P(r - \sigma)$, where $r$ is the center-to-center distance between two particles and $\sigma$ is the sum of their radii. $P(x)$ measures the likelihood that two particles are a distance $x$ away from \emph{just} touching, and is conceptually very similar to the radial distribution function (the two are identical for monodisperse packings). Since a contact can only change if $\left| r-\sigma\right| \lesssim \delta$, where $\delta$ is again the perturbation amplitude, we can approximate $\Delta N_c$ by integrating $P(x)$ from $-\delta$ to $\delta$ and multiplying by the system size. For finite $\Delta \phi$, $P(r-\sigma)$ is finite at $r=\sigma$, so the total number of altered contacts is
\eq{	\Delta N_c \sim N \delta  \qquad \text{ for $\Delta \phi > 0$}. \label{eq:DeltaNc}}

Although this diverges when the limit $N\rightarrow \infty$ is taken before $\delta \rightarrow 0$, the density of altered contacts $\Delta N_c/N$ vanishes. Using the above result that each altered contact affects the density of states by at most $N^{-1}$, we see that the net effect of altering $\Delta N_c$ contacts is proportional to $\delta$. Thus, even when an extensive number of contacts are altered in the thermodynamic limit, the effect on the density of states vanishes as $\delta \rightarrow 0$ and we conclude that the linear regime is well defined.

Finally, we can use this result to estimate how the size of the linear regime vanishes in the limit $\Delta \phi \rightarrow 0$. Here, we will only consider the effect of contact nonlinearities; see Appendix~\ref{sec:linear_response_appendix} for a rough estimation of when expansion nonlinearities become important. 
Like the radial distribution function, $P(r-\sigma)$ forms a $\delta$ function at $r=\sigma$ when $\Delta\phi=0$. This means that even for arbitrarily small perturbations, a macroscopic number of contacts change, implying that the above argument does not hold when the limit $\Delta \phi \rightarrow 0$ is taken before $\delta \rightarrow 0$. 
For small but finite $\Delta \phi$, the peak in $P(r-\sigma)$ shifts slightly and its height is proportional to $\Delta \phi^{-1}$~\cite{OHern:2003vq}. Therefore, we can overestimate the above integral by assuming $P(r-\sigma) \sim \Delta\phi^{-1}$ over the range of integration. Equation~\eqref{eq:DeltaNc} becomes $\Delta N_c \sim N \delta / \Delta \phi$, so the net effect of contact nonlinearities on the density of states is proportional to $\delta / \Delta \phi$. Setting this to a constant, determined by the ``acceptable" amount of deviation from linear behavior, we see that the displacement amplitude at which contact nonlinearities become important vanishes linearly with $\Delta\phi$.

\begin{figure}
	\centering	
	\epsfig{file=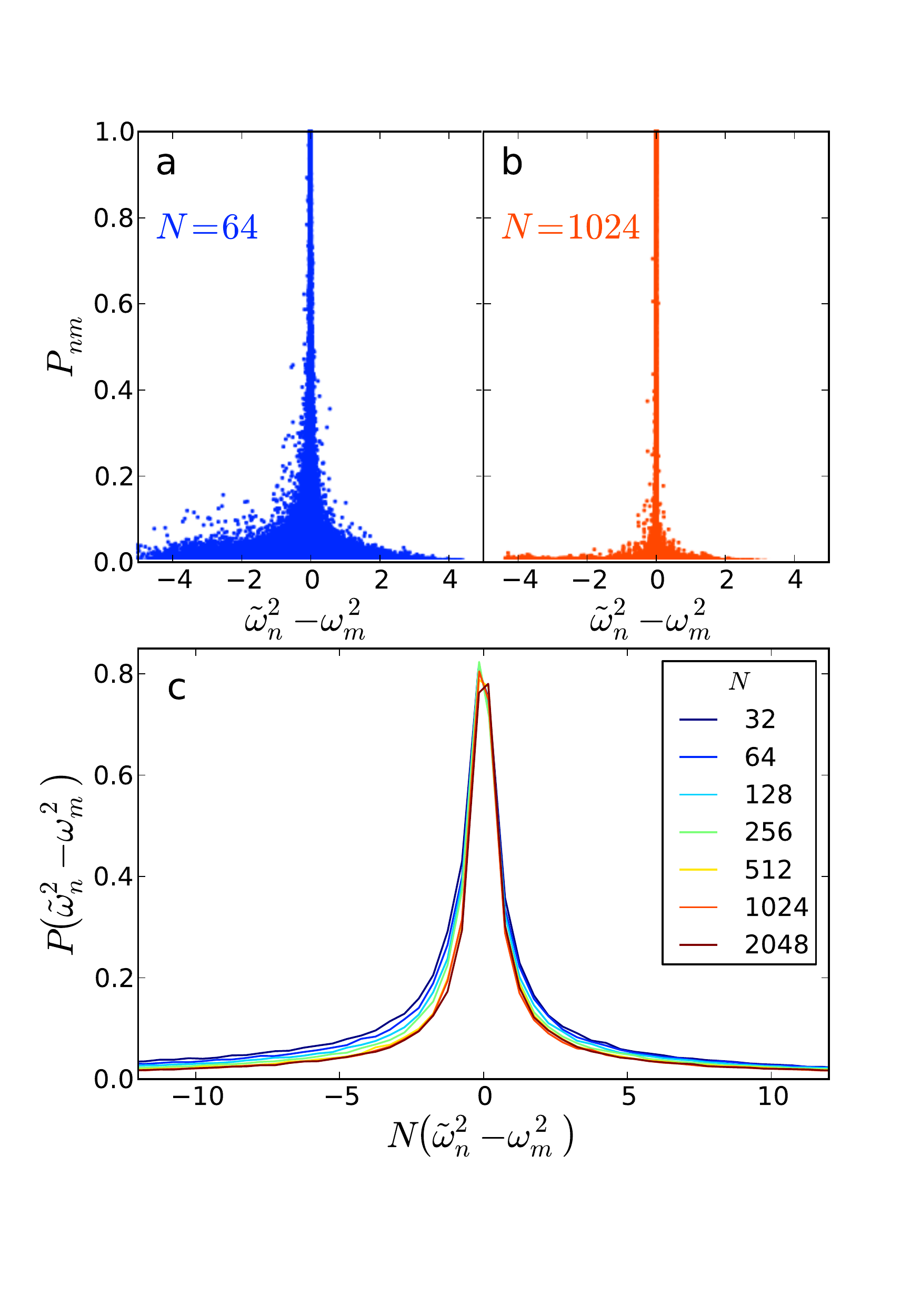,width=0.5\linewidth,viewport=20 80 480 655, clip}
	\caption[The projection of a perturbed mode onto the original modes.]{The projection of a perturbed mode $\ket{\tilde e_n}$ onto the original modes $\ket{\hat e_m}$ as a function of the eigenvalue difference. a) 16 realizations of $N=64$ particle systems. b) 1 realization of a $N=1024$ particle system. All systems have Hookian interactions ($\alpha=2$) in 2 dimensions and are at a pressure of $10^{-2}$. c) The average projection as a function of $N\left(\tilde \omega_n^2-\omega_m^2\right)$ for various system sizes. The range of $\tilde \omega_n^2-\omega_m^2$ over which the projection is relevant vanishes slightly faster than $N^{-1}$.}
	\label{fig:mode_projection}
\end{figure}

\subsubsection{Numerical Verification}
We now provide numerical evidence to support the analytical result that changing a single contact has a $N^{-1}$ effect on the linear vibrational properties. To do this, we generate mechanically stable 2-dimensional packings of disks that interact according to Eq.~\eqref{pair_potential} with $\alpha=2$. For each mechanically stable system, we first obtain the normal modes of vibration by diagonalizing the dynamical matrix $D^0$. We then perturb the system by removing the weakest contact without actually displacing any particles. This perturbed system no longer corresponds to a sphere packing but allows us to isolate contact nonlinearities without considering expansion nonlinearities. The diagonalization of the resulting dynamical matrix $\tilde D$ gives the normal modes of vibration for the perturbed system.

We compare the vibrational modes in Fig.~\ref{fig:mode_projection} by projecting each mode $\ket{\tilde e_n}$ of the perturbed system onto each mode $\ket{\hat e_m}$ of the unperturbed system. The projection 
\eq{ P_{nm} \equiv \braket{\tilde e_n}{\hat e_m} }
quantifies how close a perturbed mode is to an unperturbed mode. Fig.~\ref{fig:mode_projection}a shows a scatter plot of $P_{nm}$ as a function of the difference in eigenvalue $\tilde \omega_n^2 - \omega_m^2$ for 16 systems of $N=64$ particles at a pressure of $10^{-2}$. Fig.~\ref{fig:mode_projection}b shows similar data but for a system of $N=1024$ particles. As expected, the projection has a sharp peak at $\tilde \omega_n = \omega_m$, because mode mixing is stronger among modes of the same frequency. 

The width of the peak in the projection is clearly smaller for the larger system. The $N$-dependence of the width is quantified in Fig.~\ref{fig:mode_projection}c, which shows the average projection, $P\left(\tilde{\omega}_n^2 - \omega_m^2\right)$, as a function of $N\left(\tilde{\omega}_n^2 - \omega_m^2\right)$. By comparing the width of $P\left(\tilde{\omega}_n^2 - \omega_m^2\right)$ at different system sizes, we see that it vanishes slightly faster than $N^{-1}$, confirming the fundamental result of our perturbation calculation above. 

\begin{figure}
	\centering	
	\epsfig{file=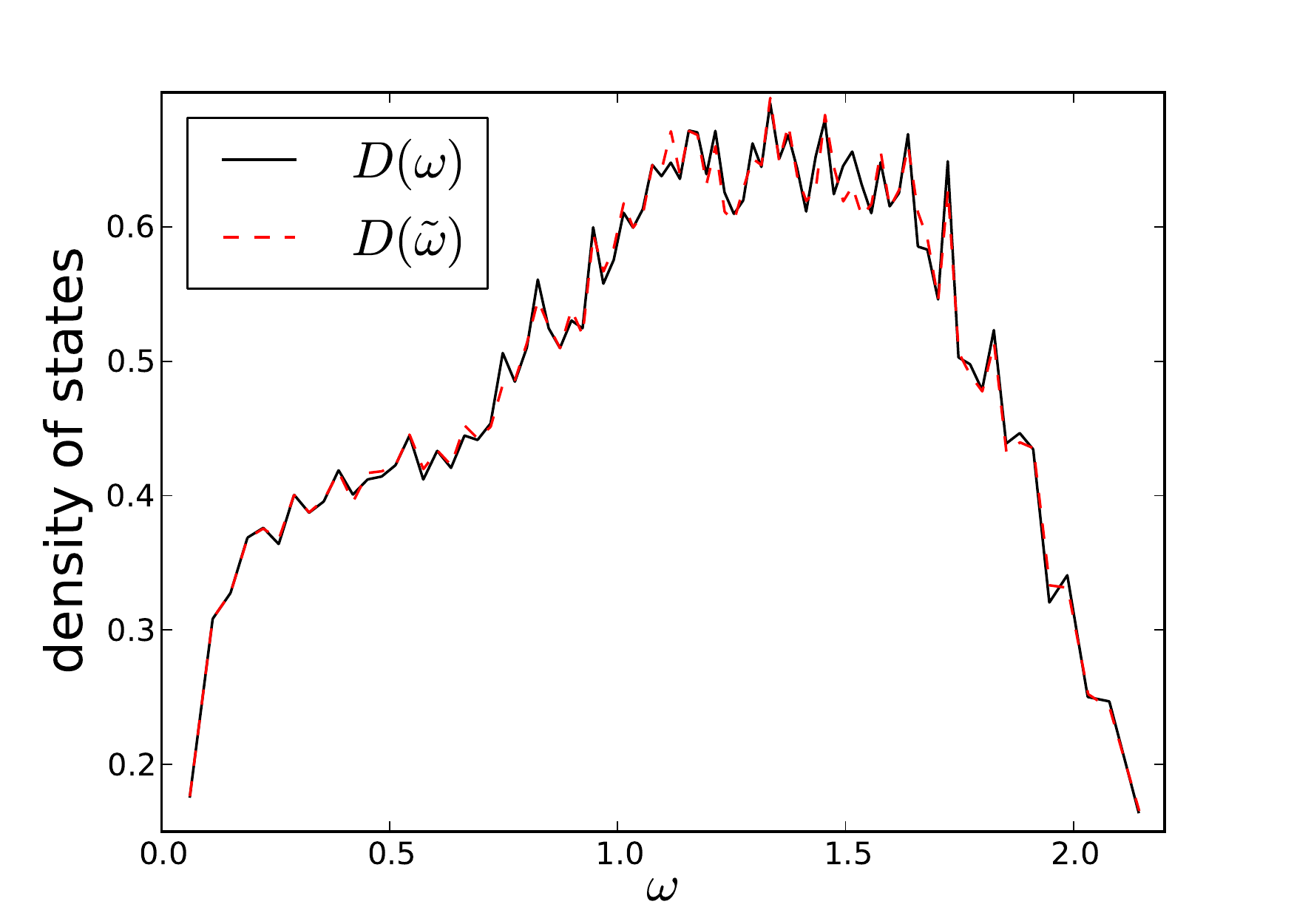,width=0.5\linewidth,viewport=0 0 480 330, clip}
	\caption[Change in the density of states.]{Density of states for a single mechanically stable system ($D(\omega)$, solid black curve) and an identical system with a single contact removed ($D(\tilde \omega)$, dashed red curve). The $N=2048$ particle system has Hookian interactions and is at a pressure of $10^{-2}$. The difference between the two density of states is well within the inherent fluctuations.}
	\label{fig:dos_change}
\end{figure}

We can also measure the shift in vibrational frequency due to the removal of a single contact. The solid black curve in Fig.~\ref{fig:dos_change} shows the density of vibrational states, $D(\omega)$, for a single mechanically stable $N=2048$ particle system. The dashed red curve shows the density of states, $D(\tilde \omega)$, for the corresponding perturbed system. While there is a small change, this difference is not systematic and is much smaller than the fluctuations inherent in the measurement. Since the density of states remains non-zero down to a characteristic frequency $\omega^*$, which is related to the number of contacts above jamming, $\Delta Z$~\cite{Wyart:2005wv,Silbert:2005vw}, changes to the contact network could have a more drastic effect if $\Delta Z \lesssim N^{-1}$. In the thermodynamic limit, however, this only occurs when $\Delta \phi \sim \Delta Z^{2} \rightarrow 0$, {\it i.e.}, at the jamming transition, where nonlinear effects are known to dominate.

\subsection{Continuity of the dynamical matrix for $\alpha>2$}
In the above perturbation argument, we looked at the effect of forming or breaking a contact on the eigenmodes and frequencies of the system. We exploited the sparsity of the perturbation matrix $\Delta D$ to show that the effects scale like $N^{-1}$, but we allowed the non-zero values of $\Delta D$ to be finite in magnitude. If these non-zero elements were vanishingly small, however, then the effect of altering the contact would be negligible and the above perturbation argument would not be necessary. 
We will see that this is the case for potentials where the dynamical matrix is a continuous function of the particle positions.  In that case, the forming or breaking of a contact has a negligible effect on the response in the limit of small displacements. 

We begin by considering the difference between the linear and nonlinear equations of motion (Eqs.~\eqref{eq_of_motion} and \eqref{linearized_eq_of_motion}). If we define $F^\text{harm}_i(\vec u)\equiv -D^0_{ij}u_j$, then the quantity
\eq{	\Delta F_i(\vec u) \equiv \left|F_i^\text{harm}(\vec u)- F_i(\vec u)\right|/\left|F_i^\text{harm}(\vec u)\right|	}
measures the relative error associated with the linearized equations of motion. If $\Delta F_i(\vec u)$ remains ``suitably" small, which again depends on the quantity being measured, then the harmonic approximation is justified and there is a valid linear regime.

Note that $\left|F_i^\text{harm}(\vec u)\right|$ is clearly proportional to $\delta$. Furthermore, using Eq.~\eqref{integral_of_dynamical_matrix} we can write $F_i^\text{harm}(\vec u)- F_i(\vec u) = \int  \left( D_{ij}(\vec u)-D_{ij}^0\right) du_j$, where $D_{ij}(\vec u)$ are the elements of the instantaneous dynamical matrix at displacement $\vec u$. If all elements $ D_{ij}(\vec u)-D_{ij}^0$ vanish in the limit $\delta \rightarrow 0$, then $F_i^\text{harm}(\vec u)- F_i(\vec u)$ must vanish \emph{faster} than $\delta$ and $\Delta F_i(\vec u) \rightarrow 0$ as $\delta \rightarrow 0$.

Also note that the instantaneous dynamical matrix can generically be written as
\eq{
	D_{ij}(\vec u) \equiv{}& \sum_{\text{contacts}} \pd{^2V(r)}{r_i \partial r_j} \nonumber \\
	={}&\sum_{\text{contacts}} k(r) \pd{r}{r_i} \pd{r}{r_j} -f(r) \pd{^2r}{r_i \partial r_j}, \label{D_ij}
}
where $f\equiv -\pd{V(r)}{r}$ and $k\equiv \pdd{V(r)}{r}$ are the force and stiffness of each contact, respectively, evaluated at $\vec u$. Therefore, we see that if $f(r)$ and $k(r)$ are continuous functions of the distance $r$ between two particles, then $D_{ij}(\vec u)$ is a continuous function of particle positions, which implies that $D_{ij}(\vec u)-D_{ij}^0$ vanishes for small $\delta$ and there is a valid linear regime.

Now, for one-sided interaction potentials of the form of Eq.~\eqref{pair_potential}, $f(r)$ is given by  
\eq{
	f(r) & \equiv \pd{V(r)}{r} = \left\{
		\begin{array}{l l}
			\frac{\epsilon}{\sigma} \left(1-\frac r\sigma\right)^{\alpha-1} & \quad \mbox{if $r<\sigma$}\\
			0 & \quad \mbox{if $r \geq \sigma$}
		\end{array} \right. 
}
and $k(r)$ is given by 
\eq{
	k(r) & \equiv \pdd{V(r)}{r} \nonumber \\
	&= \left\{
		\begin{array}{l l}\frac{\epsilon(\alpha-1)}{\sigma^2} \left(1-\frac r\sigma\right)^{\alpha-2} & \quad \mbox{if $r<\sigma$}\\
		0 & \quad \mbox{if $r \geq \sigma$}.
		\end{array} \right. 
	\label{k_of_r}
}
$f(r)$ and $k(r)$ are both continuous when $r<\sigma$ and when $r>\sigma$; it is the point of contact ($r=\sigma$) that poses a potential problem. Discontinuities do indeed arise when the exponent $\alpha$ is less than or equal to 2, but $f(r)$ and $k(r)$, and thus $D_{ij}(\vec u)$, are clearly continuous whenever $\alpha > 2$. Thus, there is always a valid linear regime for interaction potentials with $\alpha>2$~\cite{Goodrich:2014jw}.

We can calculate a lower bound for the size of the linear regime by requiring that the change in any element of $D_{ij}(\vec u)$ never exceeds some $\Delta D_\text{max}$. This is satisfied if the change in $k(r)$ of any bond never exceeds $\Delta D_\text{max}$. From Eq.~\eqref{k_of_r}, we see that the maximum change in contact length, $\Delta r_\text{max}$, is given by
\eq{	\frac{\Delta r_\text{max}}{\sigma} = \left( \frac{\Delta D_\text{max} \sigma^2}{\epsilon \left(\alpha-1\right)}\right)^{1/(\alpha-2)}. \label{Deltar_max}	}
Therefore, if $\delta \braket{\hat u}{\!r}$ is the projection of the displacement onto the bond length $r$, then the system has a well-defined linear regime for $\delta < \delta_0$~\cite{Goodrich:2014jw}, where
\eq{	\delta_0 = \Delta r_\text{max} / \braket{\hat u}{\!r}.}

This statement is valid for any potential $\alpha>2$ and is independent of the number of contacts that change or the system size. Importantly, the limit $\alpha \rightarrow 2^+$ is still well-behaved so that it is only in the case $\alpha=2$ that $\delta_0=0$ and we must resort to the perturbation argument presented above in Sec.~\ref{sec:thermodynamic_limit}. 

In Ref.~\cite{Schreck:2013gg}, Schreck {\it et al.} showed that for systems with Hertzian interactions, contact nonlinearities have a smooth effect on the spectral density as $\delta$ increases above $\delta_c$, the minimum displacement magnitude required to change the contact network.  This implies that although $\delta_c \rightarrow 0$ in the thermodynamic limit, the harmonic approximation should still describe small amplitude perturbations, in complete agreement with our results. The smooth onset of contact nonlinearities for $\alpha>2$ also implies that, provided the time scale of the measurement is suitable, low amplitude \emph{microscopic} measurements, for which $\delta < \delta_0$ , can also be described by linear response. The issue of time scales is important and is discussed in the next section.

\section{Nonzero-amplitude vibrations and time scales\label{sec:finite_amplitude_discussion}}
So far, by considering infinitesimal vibrations, we have shown that linear response can accurately describe the lowest-order behavior of bulk quantities. However, since experiments must study nonzero-amplitude perturbations, it is also important to understand how nonlinearities affect the response.  

First, consider a perturbation in the direction of one of the normal modes of vibration with frequency $\omega_1$. The motion is determined by the position dependent dynamical matrix $D_{ij}(\vec u)$ according to Eqs.~\eqref{eq_of_motion} and \eqref{integral_of_dynamical_matrix}. Ignoring expansion nonlinearities, $D_{ij}(\vec u)$ is constant if contacts do not change and the system undergoes oscillatory motion with a $\delta$-function in the Fourier transform at $\omega_1$. The dynamical matrix changes when a contact forms (or breaks), so when represented by the eigenvectors of the new dynamical matrix the trajectory becomes smeared out over multiple modes. Motion along these modes will evolve at different frequencies and so the system will be in a slightly different position when the contact reopens (or reforms). This leads to mode mixing and a broadening in time of the Fourier transform. 

References~\cite{Schreck:2011kl} and \cite{Schreck:2013gg} showed that for small systems close to jamming, the effect of mode mixing is particularly sudden and dramatic as soon as a contact forms or breaks, {\it i.e.} when $\delta > \delta_c$.
However, mode mixing can also occur without a change to the contact network because expansion nonlinearities cause $D_{ij}(\vec u)$ to change for any $\vec u$.
Indeed, mode mixing is a generic feature of finite-amplitude vibrations even for systems without contact nonlinearities ({\it e.g.} with Lennard-Jones interactions) and for which there is a clear linear regime ({\it e.g.} a crystal)~\cite{Ashcroft:1976ud}. Although mode mixing is ubiquitous, the effect might not be noticeable over short times when $\delta < \delta_c$, as demonstrated by Refs.~\cite{Schreck:2011kl} and \cite{Schreck:2013gg}. 

Therefore, an important factor is the time scale over which a measurement is made. To understand this time scale for a particular system, one must know how the eigenvectors and associated frequencies of $D_{ij}(\vec u)$ change as the system evolves. For example, if the initial trajectory only projects onto nearly degenerate modes, then the broadening of the Fourier transform will be very slow whereas if the trajectory projects onto modes with very different frequencies, then the broadening will occur quickly. Also of relevance is the amount of time during the oscillation for which $D_{ij}(\vec u)$ differs from $D_{ij}(\vec 0)$.

While this time scale can be important, for example in phonon scattering, it has little relevance to understanding bulk response to leading order, which is the focus of this dissertation. Importantly, the presence of mode mixing is \emph{not} an indication that there is no linear regime~\cite{Ashcroft:1976ud}. For large amplitude vibrations ($\delta \gg \delta_c$), Refs.~\cite{Schreck:2011kl} and \cite{Schreck:2013gg} showed that the Fourier transform differs greatly from the harmonic prediction. However, it is important to distinguish this from the density of normal modes, which is \emph{defined} from the dynamical matrix and the harmonic approximation. 
We note that while we have focused on the harmonic approximation of the potential energy, one can also think of normal modes of the free energy ({\it e.g.} of hard sphere glasses~\cite{Brito:2009ed}), though they are still defined within the harmonic approximation. 
If one assumes that the harmonic approximation is valid, then there are a variety of ways to calculate the density of states, including also the velocity autocorrelation function and the displacement covariance matrix~\cite{Chen:2010gb,Ghosh:2010hu}. While these approaches are often much more feasible, especially in experimental systems, they only measure the density of states provided the systems remains in the linear regime.

\section{Discussion\label{sec:discussion}}
We have shown that jammed soft sphere packings always have a well-defined linear regime regardless of system size whenever $\Delta \phi > 0$, thus providing sound justification for the use of the harmonic approximation in the study of bulk response. Although Ref.~\cite{Schreck:2011kl} showed that $\delta_c$, which marks the onset of contact nonlinearities, vanishes as $N\rightarrow \infty$, individual contact nonlinearities have a vanishing effect on bulk response in the thermodynamic limit. When measuring microscopic quantities like the evolution over time after a specific perturbation, Schreck {\it et al.}~\cite{Schreck:2011kl,Schreck:2013gg} showed that nonlinear effects are indeed important in jammed packings, just as they are for crystals. Nevertheless, they are \emph{not} essential for understanding bulk response to leading order.

The primary result of this chapter is the perturbation argument presented in Sec.~\ref{sec:perturbation_argument}, which is valid for any potential of the form of Eq.~\eqref{pair_potential}. However, note that we only \emph{need} to invoke this argument for the case of Hookian repulsions ($\alpha=2$). The onset of contact nonlinearities is smooth when $\alpha>2$ and thus has the potential to cause problems only when $\alpha \leq 2$. This leads to the interesting and counterintuitive result that nonlinear pair potentials are more harmonic than one-sided linear springs. 

Our results are consistent with the recent work of van Deen {\it et al.}~\cite{vanDeen:2014vm}, who look at jammed sphere packings undergoing quasi-static shear. They measure the ratio of the shear modulus before and after a contact change, which they find to approach unity for $\Delta \phi N^2 \gg 10$.\footnote{This is reported in Ref.~\cite{vanDeen:2014vm} as $pN^2$ not $\Delta \phi N^2$, but these two scalings are equivalent for Hookian interaction, which they use. One would expect the scaling to remain $\Delta \phi N^2$ for $\alpha \neq 2$.} Note that this scaling was previously shown to control the finite-size behavior of the shear modulus, which only exhibits the canonical $G \sim \Delta \phi^{1/2}$ power law when $\Delta \phi N^2 \gg 10$~\cite{Goodrich:2012ck}. This suggests that if a system is large enough to exhibit this bulk scaling behavior, then it is large enough to be insensitive to individual contact changes. 

When an extensive number of contacts break, van Deen {\it et al.}~\cite{vanDeen:2014vm} also show that fluctuations in the shear modulus scale as $\Delta \phi^{-1/2-2\beta} N^{1-4\beta}$, where $\beta \approx 0.35$. As predicted, the shear modulus converges to a well-defined value in the thermodynamic limit but not in the limit $\Delta \phi \rightarrow 0$. Furthermore, Dagois-Bohy {\it et al.}~\cite{Dagois-Bohy:private:2014} study oscillatory rheology and find that the strain amplitude where linear response breaks down in large systems is independent of system size. This is also consistent with recent simulations by Tighe {\it et al.}~\cite{Tighe2014} that explicitly measure the extent of the linear regime as systems are sheared.

Our results also provide context for the work of Ikeda {\it et al.}~\cite{Ikeda:2013gu}, who studied nonlinearities that arise from thermal fluctuations. They find that nonlinearities begin to modify the linear vibrations when fluctuations in the distance between neighboring particles is comparable to the width of the first peak of the radial distribution function. Such fluctuations cause an extensive number of contacts to break and is therefore in complete agreement with our results.
References~\cite{Ikeda:2013gu} and \cite{Wang:2013gy} show that there is a well defined temperature scale $T^*$ that marks the breakdown of the harmonic approximation. For pair interactions of the form of Eq.~\eqref{pair_potential}, $T^*$ is proportional to $\Delta \phi^\alpha$, though the prefactor depends sensitively on the way one measures nonlinearities~\cite{Ikeda:2013gu,Wang:2013gy}.

Extending this result to experimental systems can be difficult because pair interactions are often not known precisely. 
Nevertheless, our results suggest that, for example, packings of soft colloidal micro-gels at room temperature should display harmonic behavior at high densities but cross over to nonlinear behavior as the density is lowered. In a recent experiment, Still {\it et al.}~\cite{Still:2014bd} measured the elastic moduli of a PNIPAM glass by calculating the dispersion relation from the displacement covariance matrix. They found clean harmonic behavior over a range of densities that agrees nicely with numerical calculations~\cite{Shundyak:2007ga,Somfai:2007ge} of frictional soft spheres.

Although expansion nonlinearities guarantee that the density of normal modes differs from the infinite-time spectral density of finite-amplitude vibrations, the harmonic approximation nonetheless provides the foundation from which we can understand such nonlinear behavior. Indeed, many aspects of nonlinear response are strongly correlated with linear-response properties.  For example, the Gruneisen parameter, an anharmonic property, depends on mode frequency, a harmonic property, in a way that is understood~\cite{Xu:2010fa}. The energy barrier to rearrangement in a given mode direction is strongly correlated with mode frequency as well~\cite{Xu:2010fa}, and the spatial location of particle rearrangements is strongly correlated with high-displacement regions in quasi-localized low-frequency modes~\cite{WidmerCooper:2009dp,Tanguy:2010dh,Manning:2011dk}.

Even at the onset of jamming, where the linear regime vanishes, it is essential to understand the linear response in order to approach the nonlinear response. This is illustrated by a recent analysis of shock waves in marginally jammed solids~\cite{Gomez:2012ji}.  The importance of linear quantities in the presence of a vanishingly small linear regime is not unique to jamming. 
In the Ising model, for example, the magnetic susceptibility diverges at the critical point, but the linear theory is still central to our understanding of the phase transition. Just as one must first understand phonons to understand phonon-phonon scattering, the density of normal modes and other linear response properties provide essential insight into the nature of jammed solids.

\section{Appendix\label{sec:linear_response_appendix}}
In this appendix, we take a crude look at how expansion nonlinearities depend on system size and the proximity to jamming. As has become common practice, for numerical convenience we use the pressure $p$ instead of the excess packing fraction $\Delta \phi$ to measure the distance to the jamming transition. The two are highly correlated and are related by $p \sim \Delta \phi^{\alpha-1}$, where $\alpha$ is the exponent in the interaction potential. Our approach will be to explicitly calculate the third order term in the energy expansion and compare it directly to the harmonic approximation to estimate the perturbation amplitude at which they become relevant. Of course, it is possible that fourth or higher order terms may become relevant before the third order term, so these results should be interpreted as an upper bound.

\begin{figure*}
	\centering	
	\epsfig{file=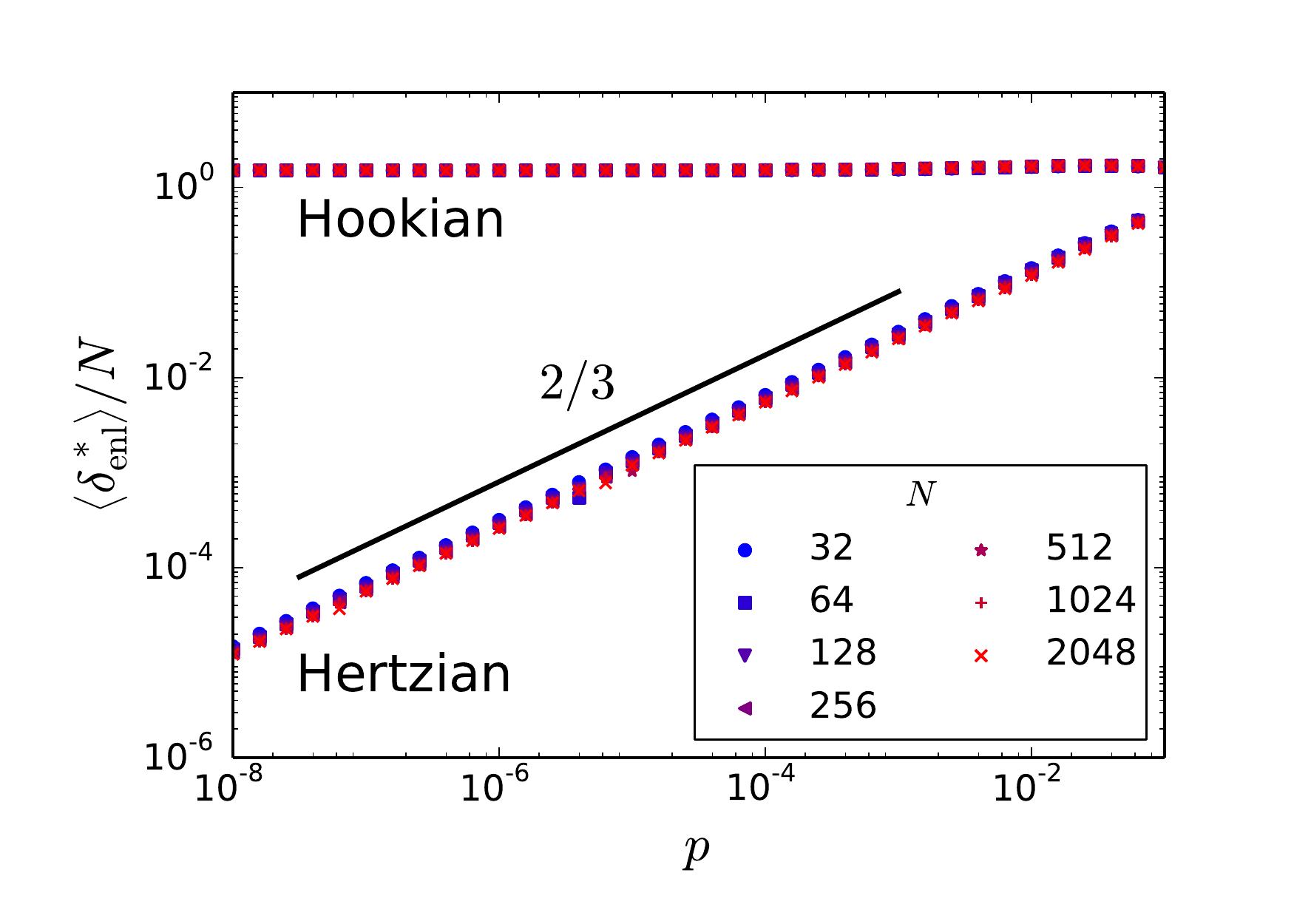,width=0.5\linewidth}%,viewport=30 20 500 330, clip}
	\epsfig{file=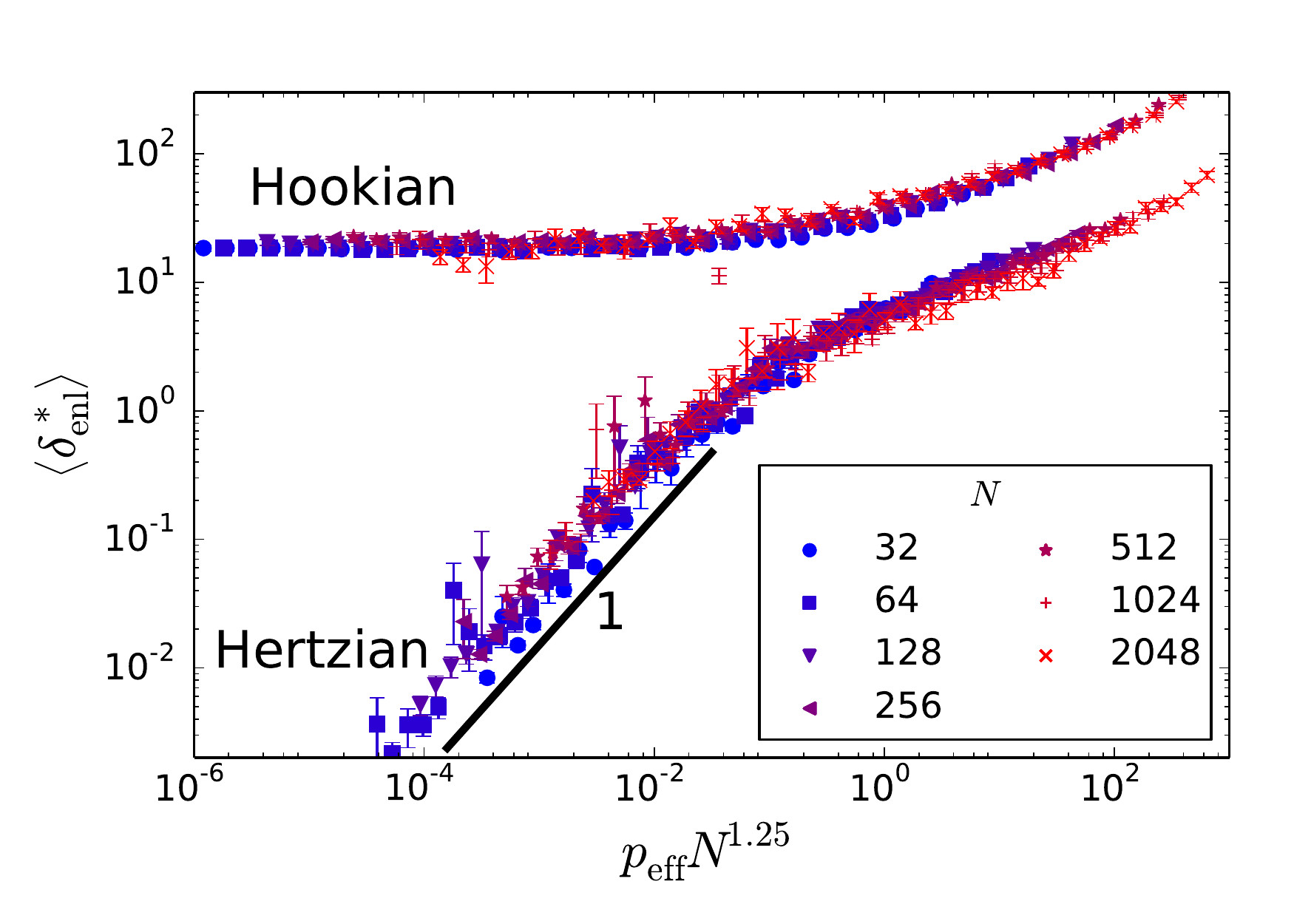,width=0.5\linewidth}%,viewport=15 15 480 325, clip}
	\caption[The onset of expansion nonlinearities.]{The average perturbation amplitude $\denl$ where the lowest order expansion nonlinearity equals the harmonic approximation of the total energy. Data is shown for two dimensional packings at different system sizes and pressures with Hookian ($\alpha=2$) and Hertzian ($\alpha=2.5$) interactions. For each configuration, $\denl$ was measured in 100 random directions (left) and the 5 lowest mode directions (right), and data from between 1000 and 5000 configurations were averaged for each system size, pressure, and potential. 
	}
	\label{fig:contact_nl_term3}
\end{figure*}

To be precise, for a given displacement of magnitude $\delta$ in the direction $\hat{u}$, 
we will define $\denl$ to be the displacement magnitude where the third order term equals the second order term. If $\hat u$ is properly normalized, this is given by $\denl \equiv \tfrac 12 D^0_{ij}\hat u_i \hat u_j / \left| \tfrac 1{3!}T^0_{ijk} \hat u_i\hat u_j\hat u_k\right|$. We now need to decide what direction $\hat u$ to look in. For randomly chosen directions, the left plot in Fig.~\ref{fig:contact_nl_term3} shows the average $\avg{\denl}$ as a function of pressure for different system sizes and for Hookian ($\alpha=2$) and Hertzian ($\alpha=2.5$) interactions. In arbitrary directions, the second and third order terms should be proportional to typical values of $D^0_{ij}$ and $T^0_{ijk}$. For Hookian interactions, both of these are dominated by the typical spring constant, which is proportional to $\Delta \phi^{\alpha-2}$, so the ratio should be constant in pressure, as observed. For Hertzian interactions, however, the third order term is dominated by $t\equiv \pd{^3V_{mn}(r)}{r^3}$, which is proportional to $\Delta \phi ^{\alpha-3}$ (but identically zero for Hookian interactions). Therefore, one would expect $\avg{\denl} \sim \Delta \phi \sim p^{2/3}$, as confirmed by the numerics. 

Near the jamming transition, however, there are special directions associated with the low frequency vibrational modes where the harmonic approximation is particularly small. As expansion nonlinearities are particularly dangerous in these directions, they deserve special treatment. The right hand plot in Fig.~\ref{fig:contact_nl_term3} shows $\avg{\denl}$ measured in the lowest five mode directions. 
We find that $\avg{\denl}$ can be written as a function of $p_\text{eff}N^w$, where $p_\text{eff} \equiv p^{1/(\alpha-1)} \sim \Delta \phi$ and $w = 1.25 \pm 0.2$. Note that $\avg{\denl}$ appears to diverge for fixed $\Delta \phi$ (fixed pressure) as $N\rightarrow \infty$. However, when $\Delta \phi \rightarrow 0$ at fixed $N$, $\avg{\denl}$ is constant for Hookian interactions and vanishes linearly with $\Delta \phi$ for Hertzian interactions.

\chapter{Diverging length scales at the jamming transition}
\label{chapter:LengthScales}
\section{The rigidity length scale}

Disordered solids exhibit many common features, including a characteristic temperature dependence of the heat capacity and thermal conductivity~\cite{Phillips:1981um} and brittle response to mechanical load~\cite{Maloney:2007en}.  A rationalization for this commonality is provided by the jamming scenario~\cite{OHern:2003vq,Liu:2010jx}, based on the behavior of packings of ideal spheres ({\it i.e.} soft frictionless spheres at zero temperature and applied stress), which exhibit a jamming transition with diverging length scales~\cite{OHern:2003vq,Silbert:2005vw,Wyart:2005wv,Wyart:2005jna,Drocco:2005ho,Ellenbroek:2006df,Ellenbroek:2009dp,Vagberg:2011fe,Goodrich:2012ck} as a function of packing fraction.  According to the jamming scenario, these diverging length scales are responsible for commonality, much as a diverging length near a critical point is responsible for universality.  

One of these length scales, the ``cutting length" $\lstar$, is directly tied to the anomalous low-frequency behavior that leads to the distinctive heat capacity and thermal conductivity of disordered solids~\cite{Phillips:1981um}, and is thus considered a cornerstone of our theoretical understanding of the jamming transition. This length arises from the so-called \emph{cutting argument} introduced by Wyart {\it et al.}~\cite{Wyart:2005wv,Wyart:2005jna} which is a counting argument that compares the number of constraints on each particle to the number of degrees of freedom in a system with free boundary conditions. Despite its importance, however, the connection between the cutting length derived by Wyart {\it et al.} and other physical length scales that diverge with the same exponent~\cite{Silbert:2005vw,Ellenbroek:2006df,Ellenbroek:2009dp} has not been understood. 

Here, we show that $\lstar$ is more robustly defined as a rigidity length. It is therefore relevant even for systems for which counting arguments are less useful, such as packings of frictional particles~\cite{Somfai:2007ge,Shundyak:2007ga,Henkes:2010kv} or ellipsoids~\cite{Donev:2007go,Zeravcic:2009wo,Mailman:2009ct}, or for experimental systems where it is not possible to count contacts. While this approach is applicable to these more general systems, we will use the traditionally employed soft sphere packings to motivate the rigidity length and illustrate its scaling behavior. We also show that $\lstar$ is directly related to a length scale identified by Silbert {\it et al.}~\cite{Silbert:2005vw} that arises from the longitudinal speed of sound.

\subsection{Model and numerical methods\label{sec:methods}}
{\it Generating mechanically stable packings.} 
We numerically generate packings of $N=4096$ frictionless disks in $d=2$ dimensions at zero temperature. Particles $i$ and $j$ interact with a harmonic, spherically symmetric, repulsive potential given by $V(r_{ij}) = \frac \epsilon 2 \left(1-r_{ij}/\sigma_{ij}\right)^2$ only if $r_{ij}<\sigma_{ij}$, where $r_{ij}$ is the center-to-center distance, $\sigma_{ij}$ is the sum of their radii and $\epsilon\equiv 1$ sets the energy scale. All lengths will be given in units of $\sigma$, the average particle diameter, and frequencies will be given in units of $\sqrt{\keff/m}$, where $\keff$ is the average effective spring constant of all overlapping particles and $m$ is the average particle mass.

Mechanically stable athermal packings were prepared with periodic boundary conditions by starting with randomly placed particles (corresponding to $T=\infty$) and then quenching the total energy to a local minimum. Energy minimization was performed using a combination of linesearch methods (L-BFGS and Conjugate gradient), Newton's method and the FIRE algorithm~\cite{Bitzek:2006bw} to maximize accuracy and efficiency. The distance to jamming is measured by the pressure, $p$, and the density of a system was adjusted until a target pressure was reached. Systems were discarded if the minimization algorithms did not converge. For reasons discussed in Sec.~\ref{sec_cutting_argument_review}, each packing was then replaced with a geometrically equivalent unstressed spring network.

The arguments we will present will concern the average number of contacts of each particle, $Z$, which approaches $2d$ in the limit of zero pressure. At positive pressure, the contact number is given for harmonic interactions by the relation $Z-2d \sim p^{1/2}$~\cite{Durian:1995eo,OHern:2003vq}.

{\it Creating a cut system.} We create a cut system by first periodically tiling the square unit cell, consistent with the periodic boundary conditions. We then remove all particles whose center is outside a box of length $L$.\footnote{A simulation box with periodic boundary conditions can be interpreted as a single unit cell that is then tiled to form a square lattice with a large basis. Elsewhere in this thesis, $L$ refers to the linear length of the simulation box or unit cell, but here $L$ refers to any linear length in the infinite tiled system.} By first tiling the system, we are able to take cuts that are larger than the unit cell, as well as cuts that are smaller. We have checked that our results are not dependent on the choice of $N=4096$ particles per unit cell.

{\it Calculating zero modes and rigid clusters.} To calculate the vibrational modes of the unstressed spring network, we diagonalize the $dN$ by $dN$ dynamical matrix $D_{ij}^{\alpha\beta}$, which is given by the second derivative of the total energy with respect to particle positions:
\eq{	D_{ij}^{\alpha\beta} = \sum_{\left<i,j\right>}k_{ij} \frac{\partial^2 r_{ij}}{\partial r_i^\alpha \partial r_j^\beta}, }
where $r_i^\alpha$ is the $\alpha$ component of the position of particle $i$, and $k_{ij} \equiv \frac{\partial^2V(r_{ij})}{\partial^2r_{ij}}$ is the stiffness of the bond.
The eigenvectors give the polarization of each mode, and the corresponding eigenvalues are the square of the mode frequency. Note that the dynamical matrix for sphere packings, as opposed to unstressed spring networks, has an additional term that is proportional to the stress.

Using the zero modes ({\it i.e.} modes with zero eigenvalues), one can easily calculate rigid clusters directly from their definition (see Sec.~\ref{sec_cluster_argument}). However, since only the zero modes are required to calculate rigid clusters, we use a pebble game algorithm developed by Jacobs and Thorpe~\cite{Jacobs:1995vi,Thorpe:1996uy} to understand the rigidity percolation transition in bond- and site-diluted lattices. This algorithm decomposes any network into distinct rigid clusters and can also be used to calculate the number of zero modes. We use the pebble game because its tremendous efficiency allows us to calculate rigid clusters for very large systems, although rigid clusters can always, in principle, be derived from modes of the dynamical matrix. Software for running the pebble game algorithm was obtained online at http://flexweb.asu.edu/.  

Note that zero modes, and thus rigid clusters, can be derived purely from the connectivity of the system without knowledge of the particular form of the interaction potential. Thus, our results are completely general for soft finite-ranged potentials; only the scaling between pressure and excess contact number needs to be adjusted, as described in Ref.~\cite{Liu:2010jx}, if other potentials were used. 

\subsection{Review of the cutting argument\label{sec_cutting_argument_review}} 

The cutting argument~\cite{Wyart:2005wv,Wyart:2005jna} addresses the origin of the low-frequency plateau in the density of vibrational modes in jammed packings~\cite{OHern:2003vq,Silbert:2005vw}.
Consider an infinite, mechanically stable packing of soft frictionless spheres in $d$ dimensions at zero temperature and applied stress. Two spheres repel if they overlap, {\it i.e.} if their center to center distance is less than the sum of their radii, but do not otherwise interact. ``Rattler" particles that have no overlaps should be removed. Since the remaining degrees of freedom must be constrained, the average number of contacts on each particle, $Z$, must be greater than or equal to $2d$, which is precisely the jump in the contact number at the jamming transition~\cite{OHern:2003vq,Goodrich:2012ck}.

It is instructive to study a simpler system, the ``unstressed" system, in which each repulsive interaction between pairs of particles in the system is replaced by a harmonic spring of equivalent stiffness $k$ at its equilibrium length.  The geometry of this spring network is identical to the geometry of the repulsive contacts between particles in the original system and the vibrational properties of the two systems are closely related~\cite{Wyart:2005jna}.
Now consider a square subsystem of linear size $L$ obtained by removing all the contacts between particles (or, in the language of the unstressed system, all springs) that cross the boundary between the subsystem and the rest of the infinite system.   Let the number of zero frequency modes in the cut system be $q$ and the number of these \emph{zero modes} that extend across the cut system be $q\p$.  Wyart {\it et al}.~\cite{Wyart:2005wv,Wyart:2005jna} used these modes to construct trial vibrational modes for the original infinite packing, as follows. If we restore the cut system with these $q\p$ extended zero modes back into the infinite system, the modes would no longer cost zero energy because of the contacts that connect the subsystem to the rest of the system. Trial modes are therefore created by deforming each extended zero mode sinusoidally so that the amplitudes vanish at the boundary. This deformation increases the energy of each mode to order $\omega_L^2$, where $\omega_L \sim 1/L$.

Note that if a mode is not extended, then it must be localized near the boundary, since the uncut system has no zero modes. However, the above procedure involves setting the mode amplitude to zero at the boundary, and so cannot be applied to such modes. It is therefore crucial to use only the $q\p$ extended modes to construct trial modes.

\begin{figure}[htb]
\centering
  \includegraphics[width=0.45\linewidth,angle=-90]{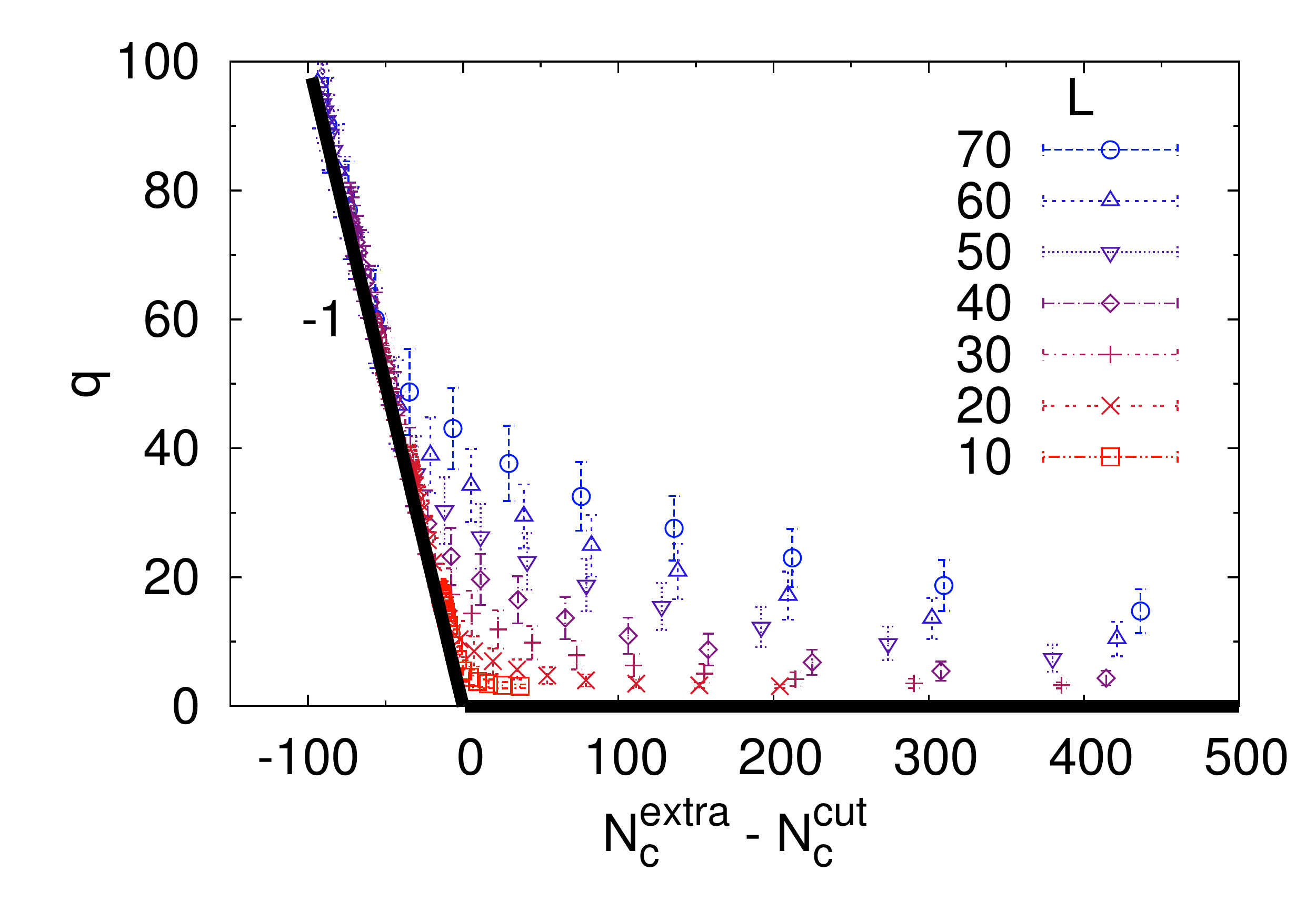}
  \caption[Number of excess zero modes as a function of the number of excess contacts after the cut.]{Number of excess zero modes as a function of the number of excess contacts after the cut. Each data point is an average of configurations at constant pressure.}
  \label{figa}
\end{figure}

The cutting argument now makes the assumption that $q\p = aq$, where $a$ is a constant independent of $L$. Before the cut, the number of extra contacts in the subsystem above the minimum required for stability is $\Ncextra \sim (Z-2d)L^d$. When the cut is made, we lose $\Nccut \sim L^{d-1}$ contacts. Naive constraint counting suggests that $q\p \sim q=\max\left(-\left(\Ncextra-\Nccut \right), 0\right)$, as shown by the solid black line in Fig.~\ref{figa}. Since $\Ncextra$ and $\Nccut$ both depend on $L$, we can define a length scale $\lstar$ by
\eq{
	\begin{array}{r l l }
		q\p\!\!\!\!\!\! &= 0 	& \mbox{if $L>\lstar$} \\
		q\p\!\!\!\!\!\! &> 0	& \mbox{if $L<\lstar$}.
	\end{array}
	 \label{lstardef}
}
The onset of zero modes is marked by $\Ncextra = \Nccut$, so
\eq{	\lstar \sim \frac{1}{Z-2d}.	\label{lstar_cuttingargument_def}}

\begin{figure}
	\centering
	\includegraphics[width=1.0\linewidth]{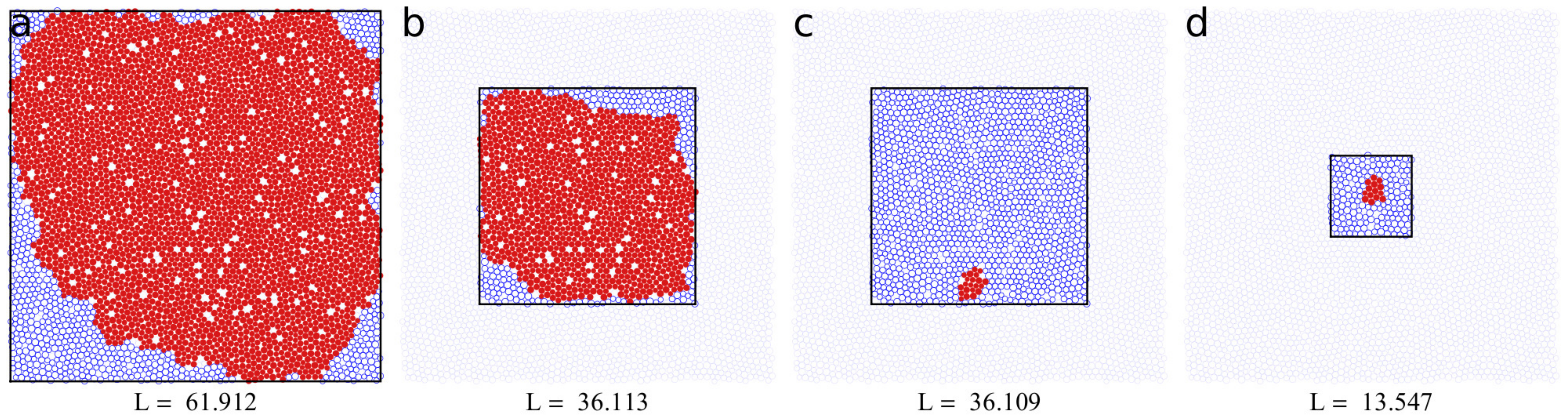}
	\caption[The breakup of macroscopic rigid clusters.]{The breakup of macroscopic rigid clusters. Subsystems are cut from a $N=4096$ particle packing at a pressure $p\approx2.5\times 10^{-4}$. a) A large subsystem with $q=60$ non-trivial zero modes. Only particles circled in blue participate in the zero modes. The solid red particles form a rigid cluster.
	b) A smaller subsystem with $q=35$ zero modes. 	c) A subsystem obtained by removing one additional particle from the system in (b). This added a single additional zero mode that extends across the entire system. The largest remaining rigid cluster only contains 21 particles. The breakup of the rigid cluster from (b) to (c), and the appearance of the corresponding extended zero mode, is the phenomenon associated with the cutting length.
	%{\bf The appearance of this extended mode and the corresponding breakup of the rigid cluster marks the length scale $\lstar$.}!!!!!!!!!!!!!!!!!!!!!!
	d) A small system below $\lstar$ with $q=33$ zero modes.  The largest rigid cluster contains $14$ particles.
	}
	\label{figb}
\end{figure}

The variational argument now predicts that at least $q\p/2$ of the total $L^d$ eigenmodes of the full system must have frequency less than order $\omega_L$, so the integral of the density of states from zero to $\omega_L$ must be
\eq{	\int_0^{\omega_L} \textrm{d}\omega D(\omega) \geq \frac{q\p}{2L^d}.	}
However, $D(\omega)$ is an intrinsic property of the infinite system and must be independent of $L$. Therefore, assuming no additional low frequency modes beyond those predicted by the variational argument, we can vary $L$ to back out the full density of states, as follows.

If $L>\lstar$, then $q\p=q=0$ and 
\eq{	\int_0^{\omega_L} \textrm{d}\omega D(\omega) = 0.\label{int_Domega_1}	}
For $L<\lstar$, we can write $q\p/2= a(\Nccut - \Ncextra)/2 \sim L^d\left(\omega_L - 1/\lstar\right)$, %where appropriate constants have been absorbed into $\omega_L$ and $\lstar$. 
which leads to 
\eq{	\int_0^{\omega_L} \textrm{d}\omega D(\omega) \sim \omega_L - 1/\lstar.\label{int_Domega_2}	}
Eqs.~\eqref{int_Domega_1} and \eqref{int_Domega_2} imply that 
\eq{	D(\omega) = \left\{ 
\begin{array}{l l}
	0 & \quad \mbox{if $\omega < \omega^*$}\\
	\text{const.} & \quad \mbox{if $\omega > \omega^*$,}\\ 
\end{array} \right. }
where $\omega^* \equiv 1/\lstar \sim Z-2d$ defines a frequency scale. Note that while $\lstar$ is potential independent, the units of frequency, and thus $\omega^*$, depend on potential~\cite{Liu:2010jx}. This argument predicts that the density of states has a plateau that extends down to zero frequency at the jamming transition, where $Z-2d=0$. Above the jamming transition, when $Z-2d>0$, the plateau extends down to a frequency $\omega^*$ before vanishing. This agrees well with numerical results on the unstressed system~\cite{Wyart:2005jna} (see also Fig.~\ref{fig:dos_intro}).  Note the importance of the length scale $\lstar$, which defines the frequency scale $\omega^*$ and is responsible for the excess low frequency modes.

\subsection{Too many zero modes\label{sec_lstar_problems}}

In the cutting argument, the length scale $\lstar$ is defined as the size of a cut region, $L$, where the number of extended zero modes, $q\p$, first vanishes (Eq.~\eqref{lstardef}). The argument then assumes that this coincides with the disappearance of all nontrivial zero modes, $q$, which is assumed to occure when the cut system is isostatic ({\it i.e.} when $\Nccut = \Ncextra$). Wyart {\it et al.} showed~\cite{Wyart:2005jna} numerically that this is true when $Z=2d$, but they do not provide such evidence for over-constrained systems. 

Fig.~\ref{figb}a shows a system that remains over-constrained after the cut ($\Ncextra>\Nccut$). The cutting argument would assert that the only zero modes are the trivial global translations and rotations, but we find that there are in fact $60$ non-trivial zero modes. This is generalized in Fig.~\ref{figa}, which shows that $q>0$ for all cut sizes $L$ and values of $\Ncextra-\Nccut$. Clearly, one cannot use the onset of zero modes to determine $\lstar$.

However, note that the zero modes in Fig.~\ref{figb}a exist only around the boundary (the particles depicted by blue circles), while \emph{none} of the non-trivial zero modes extend into the region of solid red particles. Since these zero modes are not fully extended, the system is above the cutting length.  As noted by Wyart et al.~\cite{Wyart:2005wv}, the scaling of the cutting argument would still be robust if the number of these excess boundary zero modes scales as $L^{d-1}$.  However, as can be seen in Fig.~\ref{figb}a, these modes penetrate a non-negligible distance into the bulk of the system and so this scaling is not obvious.

\section{The cluster argument\label{sec_cluster_argument}}
We now reformulate the cutting argument in a way that does not rely on the \emph{total} number of zero modes but is specifically designed to identify the onset of \emph{extended} zero modes, which are the ones needed to obtain $\omega^*$.
The mathematics will be similar to that in the cutting argument, but the setup and interpretation will be different. We will first introduce the idea of rigid clusters and illustrate the associated phenomenon that identifies the cutting length. We will then provide a rigorous derivation of the scaling of $\lstar$ in jammed packings.

Our argument is motivated by the simple fact that if none of the zero modes are extended, then by definition there must be a cluster of central particles that these modes do not reach. Since this cluster does not participate in any zero modes, any deformation to the cluster increases its energy. Thus, such clusters have a finite bulk modulus and we will refer to them as being \emph{rigid}. The solid red particles in Fig.~\ref{figb}a are an example of a rigid cluster. 
To be precise, a rigid cluster is defined as a group of particles (within an infinite $d$ dimensional system with average contact number $Z$) such that, if all other particles were removed, the only zero modes in the unstressed system would be those associated with global translation and rotation. This is purely a geometrical definition and is independent of potential. 

Figure~\ref{figb}b-d shows the same system as Fig.~\ref{figb}a, except with progressively smaller cut regions. Figure~\ref{figb}a and b are both dominated by a rigid cluster that covers approximately 84\% of the cut region. However, while the cut region in Fig.~\ref{figb}c differs from that in Fig.~\ref{figb}b by only a single particle, it has no rigid cluster larger than 21 particles (it is comprised of many small rigid clusters, the largest of which is shown in red). Apparently, the removal of a single particle introduced a zero mode that extends throughout the system and is precisely the type needed by the variational argument of Wyart {\it et al.}~\cite{Wyart:2005wv,Wyart:2005jna}.

This sudden breakup of the rigid cluster, which coincides with the onset of extended zero modes, is a non-trivial phenomenon that marks the length scale $\lstar$. We will now provide a formal derivation of this phenomenon, which leads to a clear physical definition of $\lstar$ and allows us to derive its scaling.

\begin{figure}[tpb]
	\centering
	\includegraphics[width=0.55\linewidth,viewport=80 583 280 725, clip]{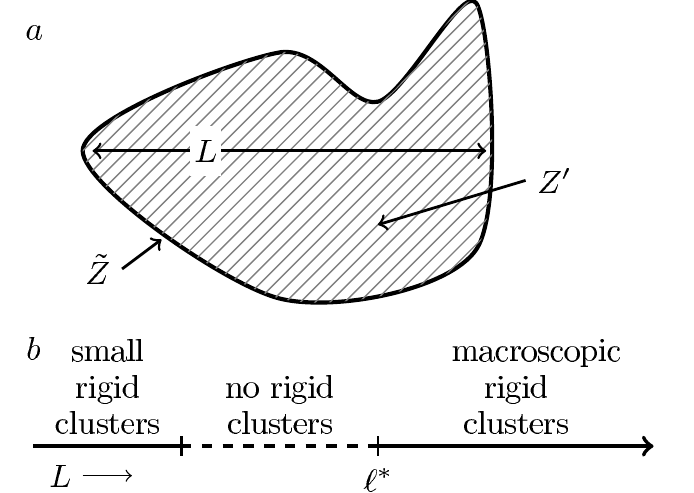}
%	\[\begin{tikzpicture}
%	\draw (-1.,1.2) node{a};
%	\draw[very thick] plot [smooth cycle] coordinates {(0,0) (2,1) (3, 0.5) (4, 1.5) (4,-1) (2,-1.5)};
%	\begin{scope}
%	\clip plot [smooth cycle] coordinates {(0,0) (2,1) (3, 0.5) (4, 1.5) (4,-1) (2,-1.5)};
%	\foreach \x in {-2,-1.85,...,4}
%	{
%		\draw[gray] (\x,-2) -- (\x+4,2);
%	}
%	\draw[<->,thick] (0.1,0) -- (4.1,0);
%	\filldraw[white] (1.1,0.25) rectangle (1.4,-0.25);
%	\draw (1.25,0) node{$L$};
%	\end{scope}
%	\draw[thick,->] (4.5,-0.3) node[right] {$Z^\prime$} -- (3,-0.75) ;
%	\draw[thick,->] (0.4,-1.2) node[left] {$\tilde Z$} -- (0.8,-0.9) ;	
%	\draw (-1.,-2) node{b};
%	\draw[very thick] (-0.5,-3) --  node[above]{\text{\parbox{37pt}{\centering small rigid clusters}}} (1,-3) ;
%	\draw[thick] (1,-2.9) -- (1,-3.1);
%	\draw[very thick,dashed] (1,-3)  -- node[above]{\text{\parbox{37pt}{\centering no rigid clusters}}}(3,-3);
%	\draw[thick] (3,-2.9) -- (3,-3.1) node[below] {$\ell^*$};
%	\draw[very thick,->] (3,-3) --  node[above]{\text{\parbox{37pt}{\centering macroscopic rigid clusters}}}(5.8,-3);
%	\draw[->](0,-3.3) node[left]{$L$} -- (0.5,-3.3);
%	\end{tikzpicture}\]
	\caption[Schematic of the cluster argument.]{Schematic of the cluster argument. a) An arbitrary surface (solid black line) of size $L$ and an enclosed rigid cluster (stripes). The rigid cluster has an average contact number of $Z\p$ in the bulk and $\tilde{Z}$ at the boundary. As $L$ becomes large, fluctuations in $Z\p$ and $\tilde{Z}$ vanish. b) Possible values of $L$ such that a rigid cluster fits within the surface. Rigid clusters can either be small or larger than some minimum value. This minimum value defines $\lstar$.}
	\label{fig_clstr_ex}
\end{figure}

Consider an arbitrary $d-1$ dimensional closed surface with characteristic size $L$ (for example, the solid black curve in Fig.~\ref{fig_clstr_ex}a). We will begin by asking whether or not it is \emph{possible} for all the particles within this surface to form a single rigid cluster.
For the cluster to be rigid, it must satisfy
\eq{	N_\text{c} - dN \ge -\frac 12 d(d+1) \label{clstr_ineq_1}}
where $N$ and $N_\text{c}$ are the number of particles and contacts in the cluster, respectively, and $\frac 12 d(d+1)$ is the number of global translations and rotations. This is a necessary but not sufficient condition for rigidity.
We can write $N_\text{c}$ as
\eq{	N_\text{c} = \frac 12 Z\p(N - N_\text{bndry}) + \frac 12 \tilde{Z}N_\text{bndry}	,	}
where $\tilde{Z}$ is the contact number of the $N_\text{bndry}$ particles on the boundary and $Z\p$ is the contact number of the particles not on the boundary (see Fig.~\ref{fig_clstr_ex}a). 
We can also define the positive constants $a$ and $b$ such that $N = 2aL^d$ and $N_\text{bndry} = 2bL^{d-1+\gamma}$, where $\gamma\ge 0$ depends on the shape of the surface, with $\gamma =0$ for non-fractal shapes.\footnote[3]{For now, we place no restrictions on the fractal dimension of the shape.} 
For shapes that have multiple characteristic lengths, {\it e.g.} a long rectangle, the choice of which length to identify as $L$ is irrelevant as it only leads to a change in the constants $a$ and $b$. For concreteness, we will always take $L$ to be the radius of gyration. 

Equation~\eqref{clstr_ineq_1} now becomes
\eq{	aL^{d-1+\gamma} \left( (Z\p-2d)L^{1-\gamma} - c \right) \ge -\frac 12 d(d+1), \label{clstr_ineq_2}	}
where $c=\frac ba(Z\p-\tilde{Z})>0$. Equation~\eqref{clstr_ineq_2} is trivially satisfied if $(Z\p-2d)L^{1-\gamma}-c > 0$, which implies
\eq{	L> L_\text{min}(Z\p,c,\gamma) \equiv \left(\frac{c}{Z\p-2d}\right)^{1/(1-\gamma)}. \label{L_bound}	} 
We will refer to clusters that satisfy Eq.~\eqref{L_bound} as \emph{macroscopic} clusters. However, it is also possible for $(Z\p-2d)L^{1-\gamma}-c < 0$, provided $L$ is very small, because the right hand side of Eq.~\eqref{clstr_ineq_2} is small and negative. 

It follows that it is only possible for the particles in our arbitrary surface to form a rigid cluster if the cluster is either very small or larger than $L_\text{min}$; rigid clusters of intermediate sizes cannot exist! 
Rigid clusters cannot exist below $L_\text{min}$ because the balance between the over constrained bulk and the under constrained boundary shifts towards the boundary as the cluster size decreases.  On the other hand, if a cluster is sufficiently small, then it can be rigid, as can be seen from the following constraint count for a triangular cluster of three particles.  For this cluster, there are six degrees of freedom, three constraints and three zero modes.  Because the three zero modes correspond to rigid translation in two directions and rigid rotation, they do not destroy the rigidity of the cluster.

Note that if $L$ is large, then fluctuations in $Z\p$ and $c$ vanish and $Z\p = Z$. $L_\text{min}$ is thus constant for all translations and rotations of the surface and is independent of $L$, depending only on the actual shape of the surface. 

Given our arbitrary shape parameterized by $c$ and $\gamma$, and the infinite packing parameterized by $Z$, $L_\text{min}(Z,c,\gamma)$ is the minimum possible size of any macroscopic rigid cluster in the $Z - 2d \ll 1$ limit.
However, we wish to find the minimum size of any rigid cluster \emph{regardless of shape}, which we do by finding $c^*$ and $\gamma^*$ that minimize $L_\text{min}$ and defining $\lstar \equiv L_\text{min}(Z,c^*,\gamma^*)$.
In the limit $Z \rightarrow 2d$, we immediately see that $\gamma^* = 0$ and
\eq{	\lstar = \frac {c^*}{Z-2d}. \label{lstar_scaling_prediction}}
As depicted in Fig.~\ref{fig_clstr_ex}b, we are left with the result that rigid clusters must either be very small or larger than $\lstar$, which we now interpret as a rigidity length.

\subsection{Estimating an upper bound}
We will now derive an upper bound for the magnitude of $\lstar$ in the $Z\rightarrow 2d$ limit. Since $c \sim LN_\text{bndry}/N$, $c$ is minimized when the shape is a $d$ dimensional hypersphere. We can approximate $N$ and $N_\text{bndry}$ to be $N \approx \phi V_d^L$ and $N_\text{bndry} \approx \phi S_{d-1}^L$, where $\phi$ is the packing fraction and $V_d^L$ and $S_{d-1}^L$ are the volume and surface area of a $d$ dimensional hypersphere with radius of gyration $L$. Using $S_{d-1}^L/V_d^L = w_dd/L$, where $w_d$ is the ratio of the radius of gyration of a hypersphere to its radius,\footnote[4]{$w_2 = \sqrt{1/2}$ and $w_3=\sqrt{3/5}$}
the $Z\rightarrow 2d$ limit of $\lstar$ becomes
\eq{	\lstar \approx \frac{w_dd (2d-\tilde{Z})}{Z-2d}. \label{lstar_derivation}	}
Equation~\eqref{lstar_derivation} is a quantitative derivation of $\lstar$ as a function of $Z$ that depends only on the value of $\tilde{Z}$, the average contact number at the boundary. 

We put an upper bound on $\lstar$ by obtaining a lower bound for $\tilde{Z}$. Note that any particle at the boundary of the rigid cluster cannot have $d$ or fewer contacts. Removing such a particle would remove $d$ degrees of freedom and at most $d$ constraints, and so the rigidity of the rest of the cluster would not be affected. Thus, $\tilde{Z} \geq d+1$ and 
\eq{	\lstar \leq \frac {w_dd(d-1)}{Z-2d}. \label{lstar_upper_bound}}

\begin{figure}[htpb]
	\centering
	\includegraphics[width=0.5\linewidth]{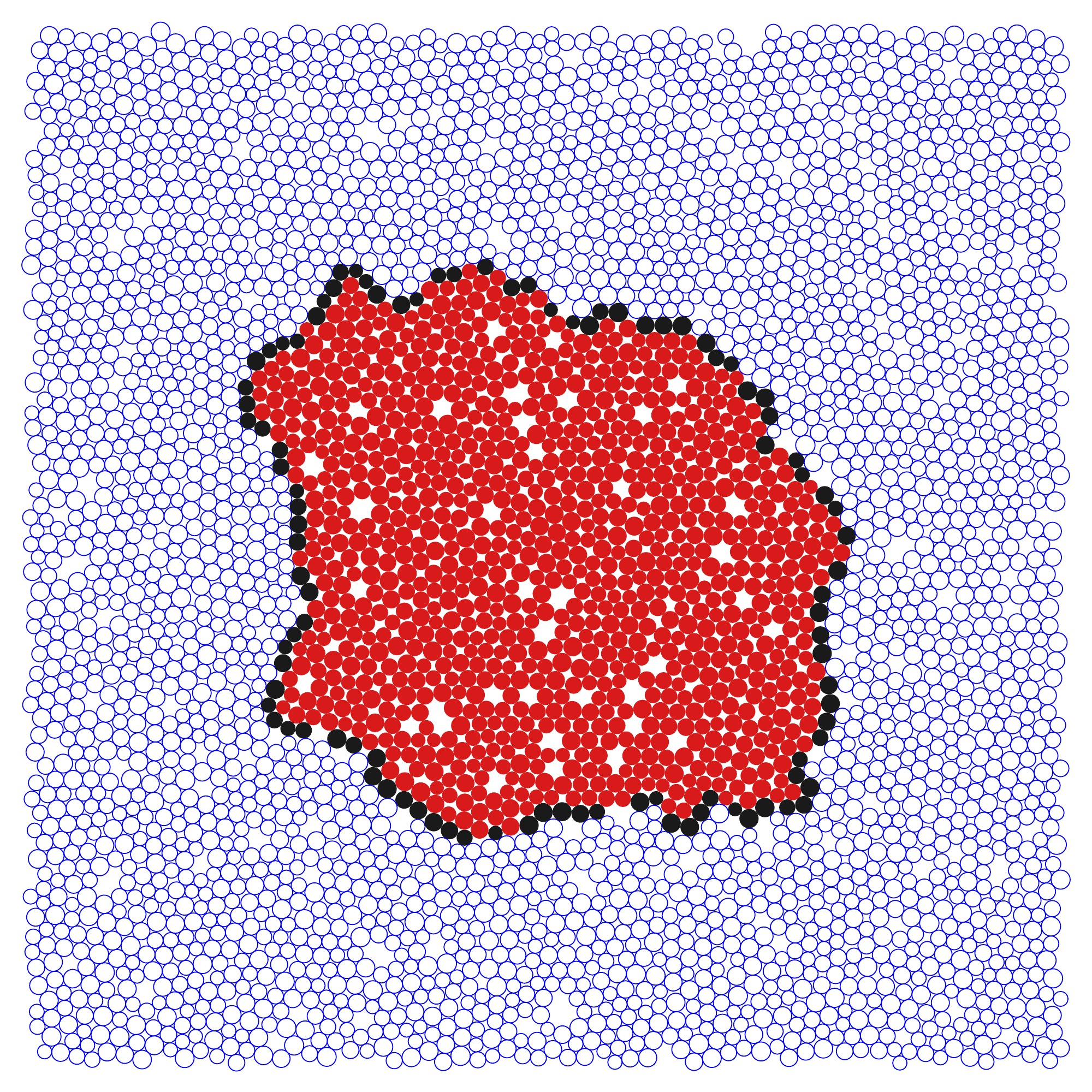}
	\caption[The smallest macroscopic rigid cluster.]{The smallest macroscopic rigid cluster for the system depicted in Fig.~\ref{figb}. The rigidity of the cluster formed by the solid red and black particles is destroyed if any of the black boundary particles are removed. None of the red particles make physical contact with the blue particles (which are not in the rigid cluster) and are not considered part of the boundary. The rigidity length, which is defined as the radius of gyration of the cluster, is $\lstar = 12.8$ (in units of the average particle diameter). }
	\label{fig:min_cluster}
\end{figure}

\subsection{Numerical verification\label{sec:numerical_verification}}
We will now use the cluster argument to calculate $\lstar$ numerically. Note that the rigid cluster in Fig.~\ref{figb}b is not necessarily the \emph{smallest} rigid cluster. The cluster breaks apart when the particle closest to the edge is removed (Fig.~\ref{figb}c), but it is possible that other particles at the edge of the rigid cluster can be removed without destroying the rigidity. The minimum rigid cluster that defines $\lstar$ has the property that rigidity is lost if any boundary particle is removed.

We calculate $\lstar$ by taking a large cut system (see Sec.~\ref{sec:methods}) and finding the smallest macroscopic rigid cluster.
To do this, we remove a particle that is randomly chosen from the boundary and decompose the remaining particles into rigid clusters. If there is no longer a macroscopic rigid cluster, then the boundary particle was necessary for rigidity and is put back. If the rigid cluster remains then the particle was not necessary for rigidity and we do not replace it. This process is repeated with another randomly chosen boundary particle until all the particles at the boundary of the rigid cluster are deemed necessary for rigidity. %See the Electronic Supplementary Information for a video that demonstrates this process. 
The resulting rigid cluster ({\it e.g.} see Fig.~\ref{fig:min_cluster}) cannot be made any smaller and so its radius of gyration measures $\lstar$.

\begin{figure}[htpb]
	\centering
	\includegraphics[width=0.75\linewidth]{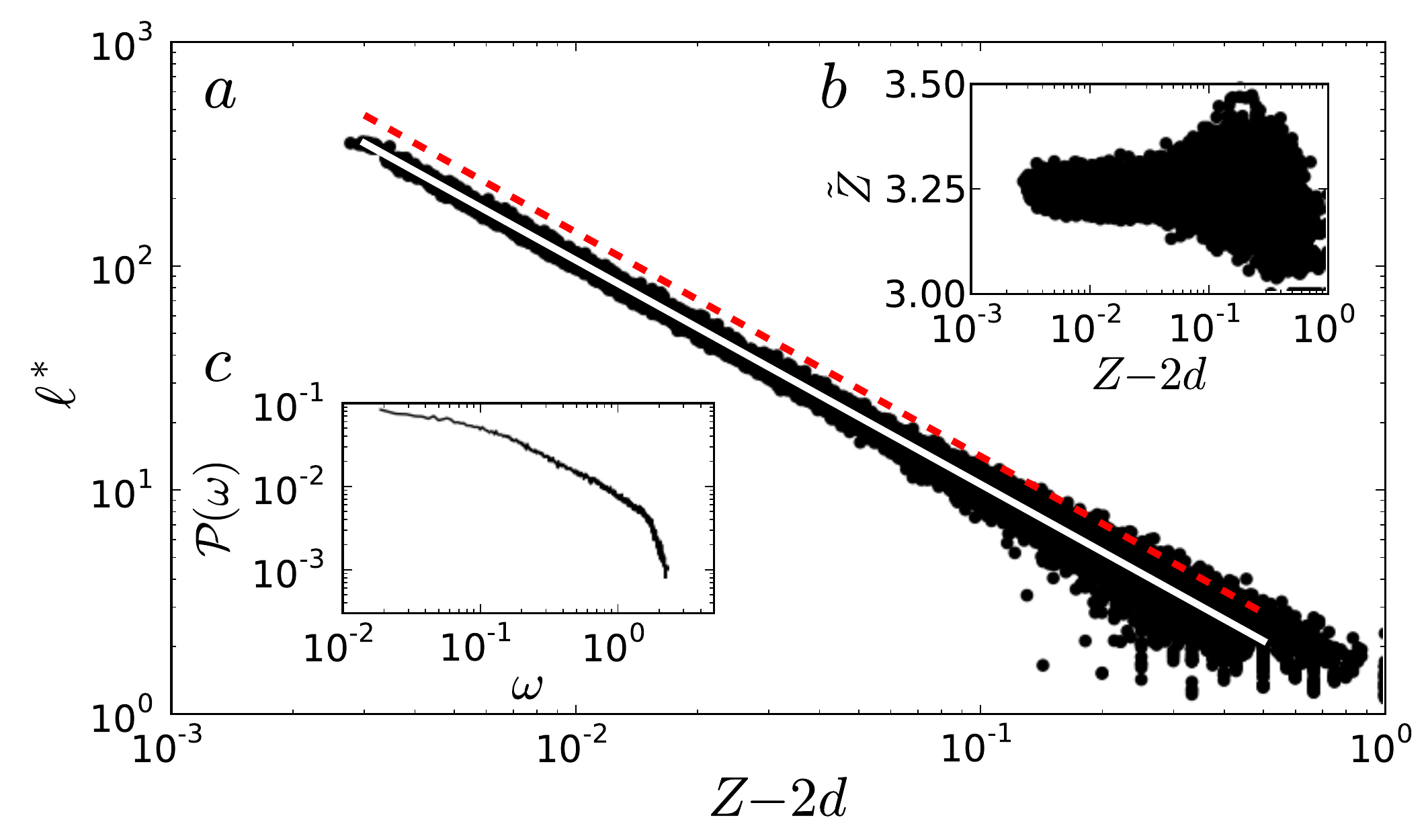}
	\caption[Numerical verification of the rigidity length.]{Numerical verification of the rigidity length. a) $\lstar$ as a function of $Z-2d$, measured for individual systems as described in the text. b) $\tilde{Z}\approx 3.25$ in the limit $Z\rightarrow 2d$, close to the predicted bound. The solid white line in a) is the quantitative prediction of Eq.~\eqref{lstar_derivation} using $\tilde{Z} = 3.25$, while the dashed red line is the upper bound obtained from $\tilde Z = 3$. c) The projection, $\mathcal{P}(\omega)$, of the single extended zero mode just below $\lstar$ onto the modes of frequency $\omega$ in the uncut system, averaged over many realizations.}
	\label{fig_lstar_scatter}
\end{figure}

Figure~\ref{fig_lstar_scatter}a shows that $\lstar$ diverges as $(Z-2d)^{-1}$, consistent with the cutting argument and our reformulation. In the small $Z-2d$ limit, $\lstar$ is just below the theoretical upper bound of Eq.~\eqref{lstar_upper_bound} (red dashed line). Figure~\ref{fig_lstar_scatter}b shows that $\tilde Z$, the contact number of boundary particles, is approximately $3.25$ as $Z \rightarrow 2d$, slightly above the lower bound of $3$. The solid white line in Fig.~\ref{fig_lstar_scatter}a shows the quantitative prediction from Eq.~\eqref{lstar_derivation} using $\tilde{Z} = 3.25$, which agrees extremely well with the data.

According to Ref.~\cite{Wyart:2005wv}, the extended zero modes of the cut system should be good trial modes for the low frequency modes of the system with periodic boundaries. Consider a system just below $\lstar$ so that there is only one extended zero mode. The global translations and rotations, as well as the boundary zero modes, can be projected out of the set of zero modes by comparing them to the modes of the system just above $\lstar$. Figure~\ref{fig_lstar_scatter}c shows the projection of that single extended zero mode onto the $dN$ modes of the full uncut system as a function of the frequency of the uncut modes. This mode projects most strongly onto the lowest frequency modes, implying that it is, in fact, a good trial mode from which to extract the low frequency behavior, as assumed~\cite{Wyart:2005wv}. Along with the first direct measurement of $\lstar$, our results provide the first numerical verification that the trial modes of the variational argument are highly related to the low frequency modes of the periodic system. 

\subsection{Advantages of the cluster argument over the counting argument}
Along with adequately dealing with the excess zero modes in Fig.~\ref{figa}, the cluster argument has a few additional advantages.
In it, $\lstar$ is defined as the smallest rigid cluster, regardless of shape, whereas the cutting argument has to specify a flat cut. This is a potential issue because the value of $\lstar$ is sensitive to the shape of the cut.  For example, if one were to consider a shape with a non-trivial fractal dimension, then $\Nccut$ would no longer scale as $L^{d-1}$, resulting in a length with entirely different scaling. Wyart {\it et al.}~\cite{Wyart:2005wv,Wyart:2005jna} argue that a flat cut is a reasonable choice for the purposes of their variational argument, but a physical length scale with relevance beyond the variational argument should be more naturally defined. The cluster argument not only provides such a physical definition, it explains unambiguously why a flat, non-fractal cut was the correct choice in the cutting argument. 

Furthermore, defining $\lstar$ in terms of the number of zero modes can be problematic. For example, rattlers must be removed and internal degrees of freedom like particle rotations must be suppressed. For packings of ellipsoidal particles, to take one example, the choice of degrees of freedom is critical.  Jammed packings of ellipsoids lie below isostaticity~\cite{Donev:2007go} and their unstressed counterparts can have an extensive number of extended zero modes. Despite this, when the aspect ratios of the ellipsoids are small, there is a band of modes similar to those for spheres, with a density of states that exhibits a plateau above $\omega^*\sim Z - 2d$~\cite{Zeravcic:2009wo}. One would thus expect a length scale $\lstar \sim 1/\omega^*$, but constraint counting of the cutting argument does not predict this. While the cluster argument also relies on zero modes and thus cannot be applied directly in this case, the intuition that $\lstar$ is a rigidity length scale should carry over. Packings of ellipsoids can have zero modes and still be rigid, and the cluster argument would predict that there is a length scale below which a packing with free boundaries loses its rigidity.

Experimental systems present a similar challenge because the contact network is often difficult to determine. However, our result that $\lstar$ marks a rigidity transition suggests that the elastic properties of a system could be used to measure $\lstar$. Such a measurement should be experimentally tractable, would not require knowledge of the vibrational properties, and would not require specification of the degrees of freedom of the system.

\subsection{Additional comments}
As in the cutting argument, the cluster argument assumes that spatial fluctuations in $Z$ are negligible.
Wyart {\it et al.} argue~\cite{Wyart:2005jna} that fluctuations in $Z$ are negligible in $d>2$ dimensions, and that the condition of local force balance suppresses such fluctuations even in $d=2$ in jammed packings. We have applied our procedure from Sec.~\ref{sec:numerical_verification} to bond-diluted hexagonal lattices where these fluctuations are not suppressed. Although these systems display a global rigidity transition~\cite{Jacobs:1995vi,Thorpe:1996uy} when they have periodic boundary conditions, they do not exhibit an abrupt loss of rigidity at some length scale that could be interpreted as $\lstar$ when they have free boundary conditions. It remains to be seen if $\lstar$ exists in this sense for bond-diluted 3 dimensional lattices.

Finally, our result that rigid clusters cannot exist on length scales below $\lstar$ appears to be consistent with results of Tighe~\cite{Tighe:2012gm}, as well as that of D\"uring {\it et al.}~\cite{During:2012bsa}, for floppy networks below isostaticity.  There, they find that clusters with free boundaries replaced by pinned boundaries cannot be rigid for length scales above $1/|Z-2d|$. The use of pinning boundary particles has also been used by Mailman and Chakraborty~\cite{Mailman:2011hz} to calculate a point-to-set correlation length above the transition that appears to scale as $\lstar$.

\subsection{Discussion\label{sec_discussion}}
We have reformulated the cutting argument in terms of rigidity instead of constraint counting. Networks derived form sphere packings can only be rigid when they have free boundaries if they are larger than a characteristic length $\lstar$, which diverges at the jamming transition. Systems just smaller than this rigidity length exhibit extended zero modes that are highly correlated with the anomalous low-frequency modes of the periodic systems, confirming the variational argument of Wyart {\it et al.}~\cite{Wyart:2005wv,Wyart:2005jna}. In contrast to the original counting argument, the generalized definition of $\lstar$ does not depend on the nature of an arbitrary cut. The insight that $\lstar$ marks a rigidity transition extends the relevance of the length to systems where constraint counting is either non-trivial (such as packings with internal degrees of freedom) or not practical (such as experimental systems where determining contacts is often difficult).

The new rigidity interpretation of $\lstar$ makes it transparently clear that the cutting length $\lstar$ is equivalent to the length scale $\ell_\text{L}$, identified by Silbert {\it et al.}~\cite{Silbert:2005vw}.  For systems with periodic boundaries, the anomalous modes derived from the zero modes swamp out sound modes at frequencies above $\omega^*$.  Thus, the minimum wavelength of longitudinal sound that can be observed in the system is $\ell_\text{L}=c_\text{L}/\omega^*$, where $c_\text{L}=\sqrt{B/ \rho}$ is the longitudinal speed of sound, $B \sim (Z-2d)^0$ is the bulk modulus, and $\rho$ is the mass density of the system.  

For systems with free boundaries that are smaller than $\lstar$, rigid clusters cannot exist so the bulk modulus and speed of sound vanish. The minimum wavelength of longitudinal sound that can be supported is therefore given by the minimum macroscopic cluster size, $\lstar$. From the scalings of $B$ and $\omega^*$, we see that $\ell_\text{L} \sim  \left(Z - 2d \right)^{-1} \sim \lstar$. Our definition of $\lstar$ implies that the two length scales not only have the same scaling but have the same physical meaning.  

Silbert {\it et al.}~\cite{Silbert:2005vw} also identified a second smaller length scale $\ell_\text{T}$ from the transverse speed of sound, which depends on the shear modulus.  For systems with free boundaries to be rigid, they must support both longitudinal and transverse sound, and so while our reasoning applies to both $\ell_\text{L}$ and $\ell_\text{T}$, $\lstar$ should be the larger of the two, so that the condition for rigidity for a cluster of size $L$ is $L \gtrsim \lstar=\ell_\text{L}$.  Note that systems with periodic boundary conditions of size $L \gg \ell_\text{T}$ are stable to infinitesimal deformations of the shape of the boundary~\cite{Schoenholz:2013jv,DagoisBohy:2012dh}.

Ideal sphere packings have the special property that the number of contacts in a packing with periodic boundary conditions is exactly isostatic at the jamming transition in the thermodynamic limit~\cite{OHern:2003vq,Goodrich:2012ck}.  Here, we have shown that the number of contacts in such a system with \emph{free} boundary conditions is exactly isostatic (Eq.~\eqref{clstr_ineq_1} is satisfied with a strict equality) in the cluster of size $\lstar$.   This simplicity makes ideal sphere packings a uniquely powerful model for exploring the marginally jammed state.

\section{The transverse length scale}

At the jamming transition of ideal spheres, the removal of a single contact causes the rigid system to become mechanically unstable~\cite{OHern:2003vq,Liu:2010jx,Goodrich:2012ck}.  Thus at the transition, the replacement of periodic-boundary conditions with free-boundary conditions destroys rigidity even in the thermodynamic limit~\cite{Wyart:2005wv,Wyart:2005jna,Goodrich:2013ke}. Recognizing that packings at the jamming threshold are susceptible to boundary conditions, Torquato and Stillinger~\cite{Torquato:2001bm} drew a distinction between collectively jammed packings, which are stable when the confining box is not allowed to deform, and strictly jammed packings, which are stable to arbitrary perturbations of the boundary.  Indeed, Dagois-Bohy {\it et al.}~\cite{DagoisBohy:2012dh} have shown that jammed packings with periodic-boundary conditions can be linearly unstable to shearing the box.  

At densities greater than the jamming transition there are more contacts than the minimum required for stability~\cite{OHern:2003vq,Liu:2010jx}. In this regime one would expect packings that are sufficiently large to be stable to changes in the boundary.  How does the characteristic size for a stable system depend on proximity to the jamming transition?  %The finite-size scaling of quantities such as the contact number~\cite{Goodrich:2012ck} is governed not by the linear system size, $L$, but by the total number of particles in the system, $N=L^d$, for dimension $d \ge 2$; such scaling is expected for systems at or above their upper critical dimension.   In contrast, 
The sensitivity to free- versus periodic-boundary conditions is governed by a length scale, $\ell^*$, that diverges at jamming transition. For $L \gg \ell^*$, the system is stable even with free boundaries~\cite{Wyart:2005wv,Wyart:2005jna,Goodrich:2013ke}.

Our aim in this section is to identify the range of system pressures and sizes over which the system is likely to be unstable to a more general class of boundary perturbations in which particle displacements violate periodic boundary conditions.  We will show that stability is governed by a competition between transverse plane waves and the anomalous modes that are unique to disordered systems. Thus, we show that stability for a large class of boundary perturbations is governed by a 
length scale, $\ell_\text{T}$, that also diverges at the jamming transition.  Packings with $L \gg \ell_\text{T}$ are linearly stable with respect to these boundary perturbations. We understand this as a competition between jamming transition physics at low pressures/system sizes, and transverse acoustic wave physics at high pressures/system sizes. The two lengths, $\ell^*$ and $\ell_\text{T}$, are the same as the longitudinal and transverse lengths associated with the normal modes of jammed sphere packings~\cite{Silbert:2005vw}. We will discuss the physical meaning of these length scales in more detail in Sec.~\ref{sec:two_lengths}.%Physically, $\ell^*$ and $\ell_\text{T}$ give the lowest wavelengths at which longitudinal and transverse sound modes, respectively, are able to cleanly propagate. Sound modes at smaller wavelengths are overwhelmed by the enormous number of excess anomalous modes that are unique to disordered systems.

%\section{Model and numerical methods}
We analyze athermal, frictionless packings with periodic boundary conditions composed of equal numbers of small and large spheres with diameter ratio 1:1.4 all with equal mass, $m$. The particles interact via the repulsive finite-range harmonic pair potential
\begin{equation}\label{eq:potential}V(r_{ij}) = \frac{{\varepsilon}}2 \left(1-\frac{r_{ij}}{\sigma_{ij}}\right)^2 \end{equation}
if $r_{ij}<\sigma_{ij}$ and $V(r_{ij}) = 0$ otherwise. Here $r_{ij}$ is the distance between particles $i$ and $j$, $\sigma_{ij}$ is the sum of the particles' radii, and $\varepsilon$ determines the strength of the potential. Energies are measured in units of $\varepsilon$, distances in units of the average particle diameter $\sigma$, and frequencies in units of $\sqrt{{\varepsilon}/m \sigma^2}$.  We varied the total number of particles from $N= 32$ to $N=512$ at 36 pressures between $p=10^{-1}$ and $p=10^{-8}$. Particles are initially placed at random in an infinite temperature, $T=\infty$, configuration and are then quenched to a $T=0$ inherent structure using a combination of linesearch methods, Newton's method and the FIRE algorithm~\cite{Bitzek:2006bw}. The resulting packing is then compressed or expanded uniformly in small increments until a target pressure, $p$, is attained.  After each increment of $p$, the system is again quenched to $T=0$.

\subsection{Symmetry-breaking perturbations}
\begin{figure}[h!tb]
\centering
\includegraphics[width=0.45\textwidth]{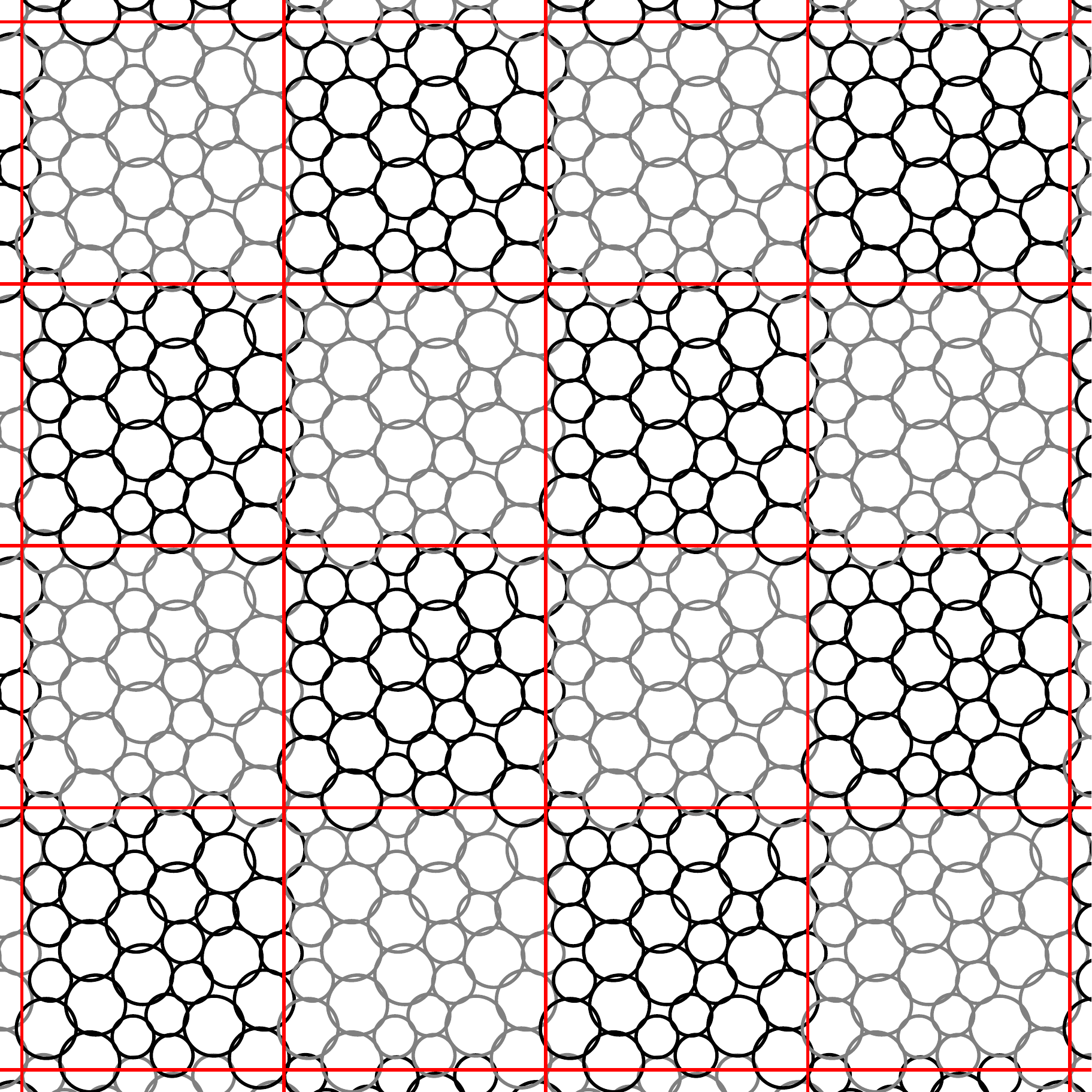}
\caption[A 32 particle system with periodic boundary conditions interpreted as a tiling in space.]{A 32 particle system with periodic boundary conditions interpreted as a tiling in space. Here a 4$\times$4 section of the tiling is shown. Shading is used to provide contrast between adjacent copies of the system.}
\label{fig:checker}
\end{figure}

The boundary conditions can be perturbed in a number of ways. The dramatic change from periodic to free boundaries has been studied in Refs.~\cite{Wyart:2005wv,Wyart:2005jna,Goodrich:2013ke}. Dagois-Bohy {\it et al.}~\cite{DagoisBohy:2012dh} considered infinitesimal ``shear-type" deformations to the \emph{shape} of the periodic box, such as uniaxial compression, shear, dilation, etc. Here, we relax the periodic boundary conditions by considering a third class of perturbations that allow particle displacements that violate periodicity. To do this, we treat our system with periodic boundary conditions as a tiling of identical copies of the system over all space (see Fig.~\ref{fig:checker}). Thus, an $N$-particle packing in a box of linear size $L$ with periodic boundary conditions can be viewed as the $N$-particle unit cell of an infinite hypercubic lattice.

\begin{figure}[h!tb]
%\centering
a: Stable dispersion \qquad \qquad \qquad \qquad c: Example stable mode\\
\includegraphics[width=1\textwidth]{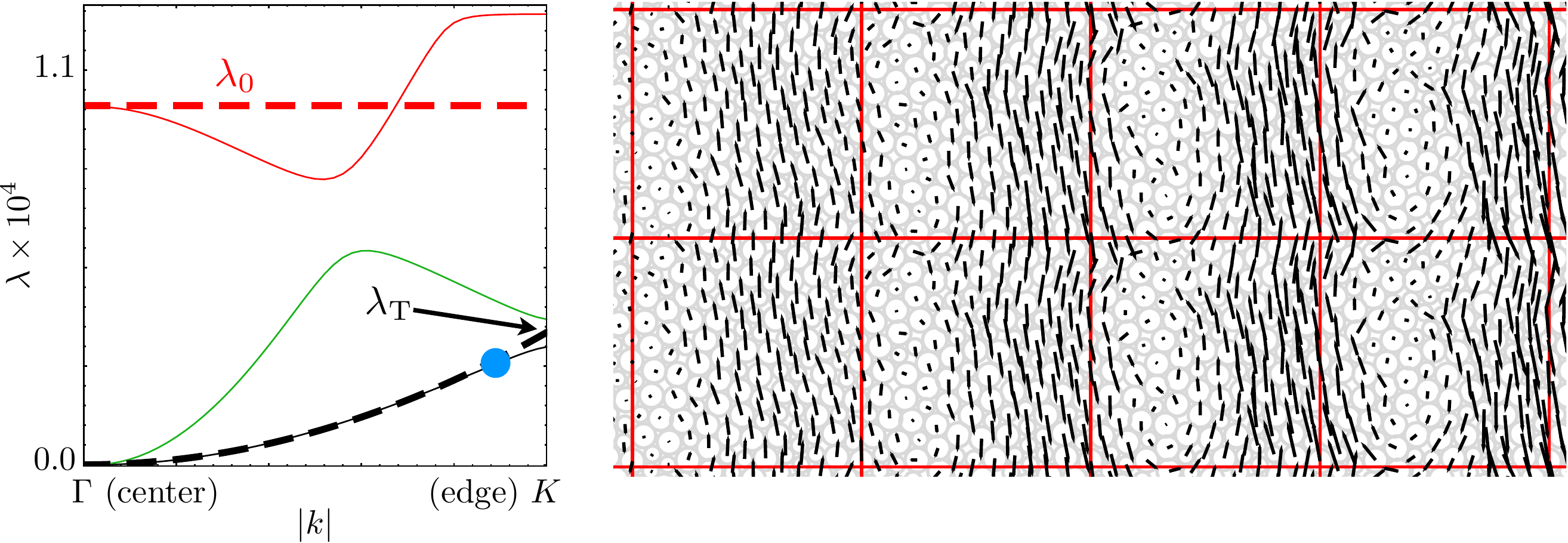}\\
b: Unstable dispersion \quad \qquad  \qquad \qquad d: Example unstable mode \\
\includegraphics[width=1\textwidth]{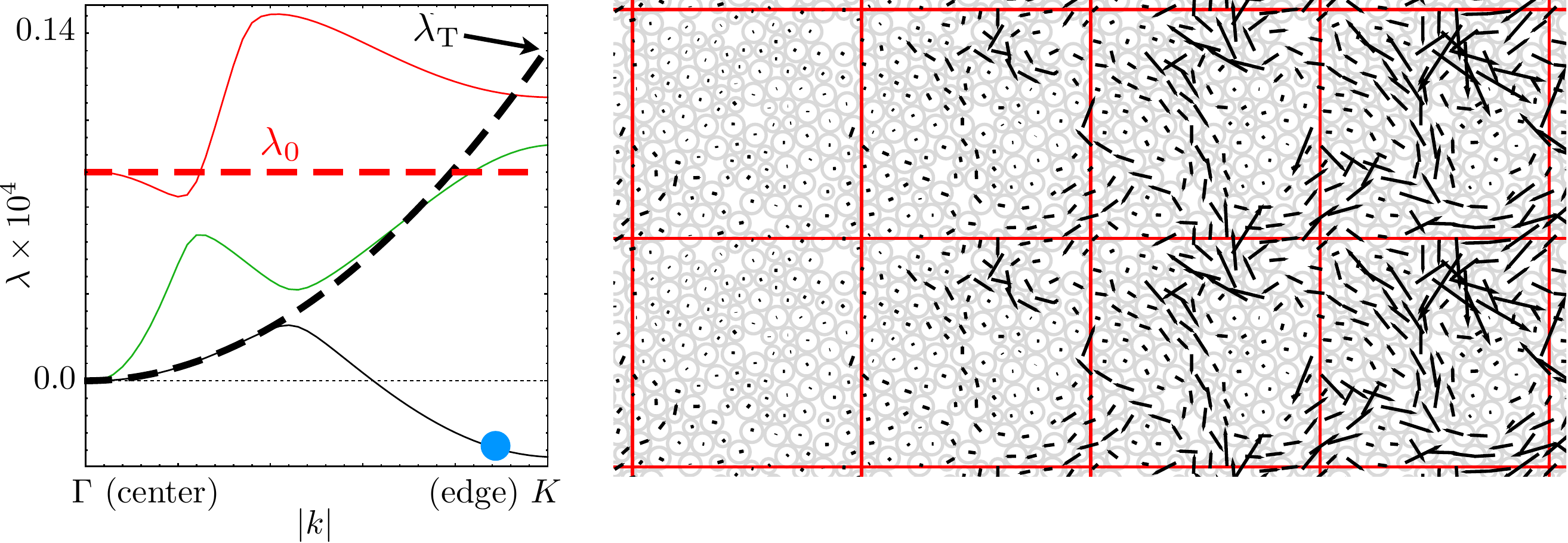}
\caption[Stable and unstable modes.]{Stable and unstable modes. (a) The dispersion relation for the transverse-acoustic mode (black), the longitudinal-acoustic mode (green), and the lowest anomalous mode (red) along the horizontal axis for a system of 128 particles in two dimensions at a pressure of $p=10^{-1}$. The black dashed line shows the quadratic approximation to the transverse branch and has a value of $\lambda_\text{T}$ at the zone edge. The dashed horizontal red line shows the flat approximation of the lowest anomalous mode and has a value of $\lambda_0$. The system is stable because the lowest branch is never negative. (b) Similar dispersion relations for a system at $p=10^{-3}$. This system has a lattice instability because the lowest mode is negative at large $k$. (c)-(d) A $2\times 4$ section of the periodically tiled systems from (a)-(b), respectively. Overlaid are the displacement vectors $\bm u_{i\mu}$ for the lowest modes near the zone edge ($\bm k\approx0.9\pi/L\bm {\hat x}$, see the blue dot in (a)-(b)). Note that the mode in (c) has strong transverse plane-wave character, while the mode in (d), which is unstable because $\lambda<0$, has developed strong anomalous character.}
\label{fig:pw-anom}
\end{figure}

The normal modes of vibration can be found by solving the equations of motion to linear order.  To do this, we assume the particles begin in mechanical equilibrium at positions specified by $\bm r_{i\mu}^0$, where $i$ indexes particles in each cell and $\mu$ indexes unit cells, so that $\bm r_{i\mu}^0$ is the equilibrium position of particle $i$ in cell $\mu$. 
The energy of the system to lowest order in particle displacements about its minimum value, $\bm u_{i\mu} = \bm r_{i\mu} - \bm r_{i\mu}^0$, can generically be written as
\begin{equation}\label{eq:linearization}U = \sum_{{\left<i\mu,j\nu\right>}}V(r_{i\mu j\nu}) \sim U_0 + \sum_{{\left<i\mu,j\nu\right>}}\bm u_{i\mu}^\text{T} \frac{\partial^2 U}{\partial \bm r_{i\mu}\partial \bm r_{j\nu}}\bigg|_{\bm r= \bm r^0}\bm u_{j\nu},
\end{equation}
where the sums are over all pairs of particles $i\mu$ and $j\nu$ that are in contact.
The equations of motion resulting from Eq.~\eqref{eq:linearization} can be solved by a plane-wave ansatz,
$\bm u_{i\mu} = \mathrm{Re}\left\{\bm \epsilon_i \exp\left[i(\bm k\cdot \bm R_\mu-\omega t)\right]\right\}$.
Here $\bm \epsilon_i$ is a $dN$-dimensional polarization vector, $\bm k$ is a $d$-dimensional wavevector and $\bm R_\mu$ is the $d$-dimensional vector corresponding to the position of cell $\mu$. This gives the eigenvalue equation
\begin{equation}\label{eq:eigenvalue}\lambda_n(\bm k)\bm \epsilon_{ni}(\bm k) = \sum_{j}\bm D_{i j}(\bm k)\bm \epsilon_{nj}(\bm k),\end{equation}
where
\begin{equation}\label{eq:dynamical}\bm D_{ij}(\bm k) = \sum_{\mu\nu}\frac{\partial^2  U}{\partial \bm r_{i\mu}\partial \bm r_{j\nu}}\bigg|_{\bm r= \bm r^0}e^{i\bm k\cdot(\bm R_\mu - \bm R_\nu)}\end{equation}
is the dynamical matrix of dimension $dN \times dN$, and $n$ labels the eigenvalues and eigenvectors. From Eq.~\eqref{eq:linearization}, the frequency of the modes in the $n$th branch are $\omega_n(\bm k)=\sqrt{\lambda_n(\bm k)}$ with eigenvector $\bm \epsilon_{ni}(\bm k)$. Figure~\ref{fig:pw-anom} shows $\lambda_{n}(\bm k)$ as a function of $\bm k$ for two example systems, as well as examples of $\bm u_{i\mu}$ that solve the eigenvalue equation. The behavior of these examples will be discussed in detail below.

With $\bm k$ allowed to vary over the first Brillouin zone, the eigenvectors comprise a complete set of states for the entire tiled system. It follows that any displacement of particles at the boundary of the unit cell can be written as a Fourier series,
\begin{equation}\label{eq:fourierseries}\bm u_{i\mu} = \sum_{\bm k,n}A_{\bm k,n}\bm \epsilon_{ni}(\bm k) \exp\left[i(\bm k\cdot \bm R_\mu)\right],\end{equation}
which is valid for all particles $i$.
Therefore, the system will be unstable to some collective perturbation of its boundary if and only if there is some $\bm k$ and $n$ for which $\lambda_n(\bm k)< 0$. This procedure allows us to characterize boundary perturbations by wavevector.\footnote[3]{Note that shear-type deformations can be considered concurrently by adding the term $\Lambda \cdot \bm r_{i\mu}^0$ to Eq.~\eqref{eq:fourierseries}, where $\Lambda$ is a global deformation tensor. This term represents the affine displacement and is neglected for our purposes.} If we find a wavevector whose dynamical matrix has a negative eigenvalue, it follows that the system must be unstable with respect to the boundary perturbation implied by the corresponding eigenvector. We will show that stability is governed by a competition between transverse plane waves and anomalous modes, examples of which are shown in Fig.~\ref{fig:pw-anom}.

The gist of our argument can be understood as follows.  The lowest branch of the normal mode spectrum is composed of transverse plane waves; it is these modes that are the ones most easily perturbed by disorder to produce a negative eigenfrequency and therefore a lattice instability.  The largest perturbation comes from an interaction with the higher-frequency anomalous modes which have a characteristic frequency of $\omega^*$.  If $\omega^*$ is large, which would occur at high pressure, or if the highest-frequency transverse mode ({\it i.e.}, at the zone boundary) is small, which would occur for large systems, then there is unlikely to be a strong perturbation and all the modes will remain positive and there will be no instability.  %Thus we see that it a competition between the transverse plane waves and the anomalous modes that govern the stability to these perturbations. 

To illustrate, we show in Fig.~\ref{fig:pw-anom}a the lowest three eigenvalue branches along the horizontal axis of the Brillouin zone for an example system that is shown in Fig.~\ref{fig:pw-anom}c. At low $k$, the lowest two branches correspond to transverse and longitudinal plane waves, respectively, and increase quadratically in $k$. The third branch corresponds to the lowest anomalous mode in the system. When the longitudinal branch approaches the anomalous mode, band repulsion pushes the branch down to lower energies. However, the transverse branch maintains its quadratic behavior all the way to the zone edge. Near the zone edge, modes of this lowest branch have very strong plane-wave character, as shown by the displacement vectors in Fig.~\ref{fig:pw-anom}c.

Contrast this to the dispersion curves shown in Fig.~\ref{fig:pw-anom}b, %which correspond to the lowest three branches for the system shown in Fig.~\ref{fig:pw-anom}b. 
for a system at much lower pressure. Here, the lowest anomalous branch is roughly an order of magnitude lower than in the previous example, and therefore has a more drastic effect on the lowest two branches. Specifically, the transverse branch experiences band repulsion about half way to the zone edge, and the quadratic extrapolation (the dashed black line) is a very poor predictor of the eigenvalue at the edge. In this second example, modes near the edge have negative eigenvalues, meaning they are unstable. Furthermore, as illustrated in Fig.~\ref{fig:pw-anom}d, these unstable modes do not resemble plane-waves, but instead appear disordered and closely resemble anomalous modes. %are anomalous instead of plane waves %(see Fig.~\ref{fig:pw-anom}b) 

This scenario is generic. 
We define $\lambda_\text{T}$ to be the energy of the transverse branch extrapolated to the zone boundary and $\lambda_0$ to be the energy of the lowest anomalous branch (see Fig.~\ref{fig:pw-anom}). When $\lambda_0$ is much larger than $\lambda_\text{T}$, then the lowest mode maintains its transverse plane-wave character throughout the Brillouin zone. However, when $\lambda_0$ is less than $\lambda_\text{T}$, then the lowest mode at the zone edge is anomalous and is more likely to be unstable. Importantly, plane waves and anomalous modes are governed by different physics. By understanding the scaling of $\lambda_0$ and $\lambda_\text{T}$, we will show that stability is controlled by the transverse length scale, $\ell_\text{T}$.

Our strategy will be to first consider the so-called ``unstressed system," which replaces the original system of spheres with an identical configuration of particles connected by unstretched springs with stiffness given by the original bonds. In the unstressed system, eigenvalues can never be negative and so all systems are stable.  We will then construct scaling relations for $\lambda_0$ and $\lambda_{\text T}$ of the unstressed system and conclude that the lowest eigenvalue will be $\min\{\lambda_0,\lambda_{\text T}\}$. We then construct additional scaling relations for the accompanying shift in the eigenvalues upon reintroducing the stress. By finding the pressures and system sizes where these two quantities are comparable, we will thus determine the scaling of the susceptibility of packings to perturbations described by Eq.~\eqref{eq:fourierseries}.

\subsection{The unstressed system}
The dynamical matrix in Eq.~\eqref{eq:dynamical} is a function of the second derivative of the pair potentials of Eq.~\eqref{eq:potential} with respect to particle positions. To construct the dynamical matrix of the unstressed system, we rewrite this as as
\begin{align}
\label{eq:stressed}
\frac{\partial^2V}{\partial r_{\alpha}\partial r_{\beta}} &= \frac{\partial^2 V}{\partial r^2}\frac{\partial r}{\partial r_{\alpha}}\frac{\partial r}{\partial r_{\beta}} + \frac{\partial V}{\partial r}\frac{\partial^2 r}{\partial r_{\alpha}\partial r_{\beta}},
\end{align}
where $V$ is the potential between particles $i\mu$ and $j\nu$, $\bm r \equiv \bm r_{j\nu} - \bm r_{i\mu}$ and $\alpha$ and $\beta$ are spatial indices. 
The second term is proportional to the negative of the force between particles. If this term is neglected, there are no repulsive forces and the system will be ``unstressed"~\cite{Wyart:2005jna,Alexander:1998vc}.
The dynamical matrix of the unstressed system, obtained from {\it just} the first term in Eq.~\eqref{eq:stressed}, has only non-negative eigenvalues.
 %because it represents a system of unstretched springs
We will use a subscript ``$\text{u}$" (as in, {\it e.g.}, $\lambda_{\text{u}}$) to refer to quantities corresponding to the unstressed system.

\begin{figure*}
\centering
\includegraphics[width=1.0\textwidth,viewport=110 10 1180 345,clip]{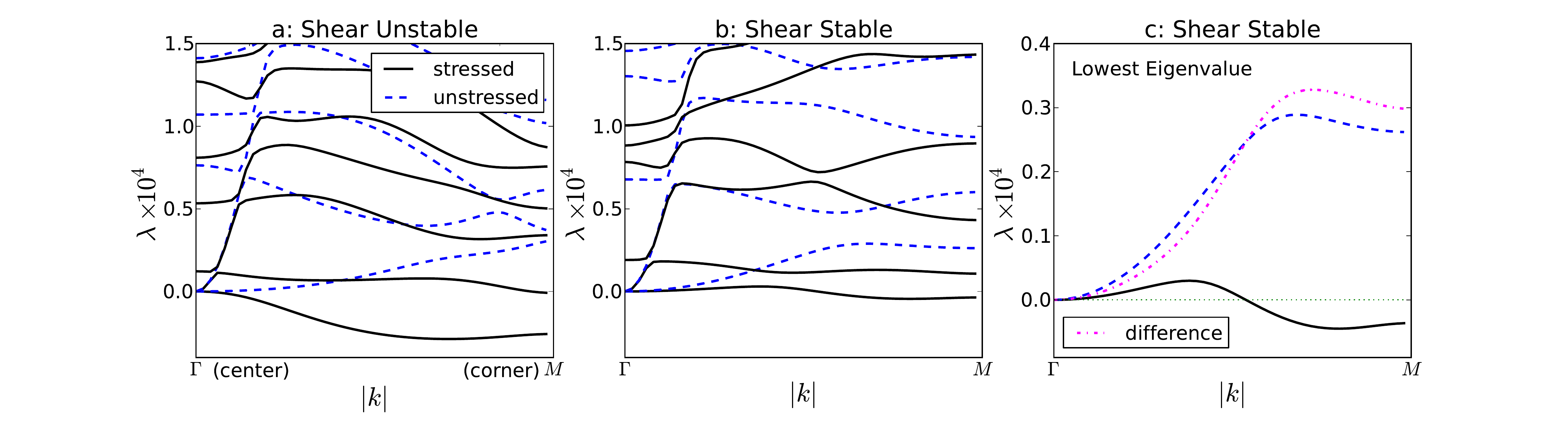}
\caption[Dispersion relations along the $\Gamma-M$ line.]{Dispersion relations along the $\Gamma-M$ line for the lowest few branches of 2 different 2-dimensional packings of $N=1024$ particles.  The $\Gamma$ point is at the Brillouin zone center ($\bm k = \bm 0$) and the $M$ point is at the zone corner where the magnitude of $\bm k$ is greatest. (a)  A ``shear-unstable'' packing (\emph{i.e.} a sphere packing that is unstable at low $k$) at $p = 10^{-4}$ (black) and the dispersion relation for the corresponding unstressed system (blue). (b) A ``lattice-unstable'' sphere packing (\emph{i.e.} a packing that is stable near $k=0$ but is unstable at higher $k$) at $p=10^{-4}$ (black) and the dispersion relation for the corresponding unstressed system (blue). (c) Comparison between the lowest branch in the sphere packing (black) and its unstressed counterpart (blue) for the same system as in (b).  The dashed magenta line is the difference between the two eigenvalue branches.}
\label{fig:examplemodes}
\end{figure*}

The black curves in Fig.~\ref{fig:examplemodes}a show the six lowest eigenvalue branches, $\lambda({\bm k})$, for a 2-dimensional packing of $N=1024$ disks at a pressure $p = 10^{-4}$.
Also shown (dashed blue curves) are the lowest eigenvalue branches for the corresponding unstressed system, $\lambda_\text{u}({\bm k})$, where the second term in Eq.~\eqref{eq:stressed} has been omitted.  Here, the lowest branch of the stressed system has negative curvature at $k=0$ implying that the system has a negative shear modulus.  Such shear-type instabilities have been analyzed previously~\cite{DagoisBohy:2012dh}.  In the remainder of this chapter, we restrict our attention to shear-stable packings that are stable near $k=0$ but potentially unstable at higher wave vectors. We refer to this type of instability as a ``lattice instability."
% ``$k > 0$ instability.''

Figure~\ref{fig:examplemodes}b compares the 6 lowest eigenvalue branches of a sphere packing (black) with a lattice instability to those of its unstressed counterpart (dashed blue).  The lowest branch for the sphere packing has positive curvature at $k=0$, but becomes negative at higher $k$.  This implies that the system is unstable to boundary perturbations corresponding to Eq.~\eqref{eq:fourierseries} over a range of wavevectors. In contrast, the unstressed system, by construction, can have only positive (or possibly zero) eigenvalues.   In Fig.~\ref{fig:examplemodes}c, the dotted magenta line shows the difference between the lowest eigenvalue branch of the unstressed system (blue) and of the sphere packing (black).

\subsection{The lowest eigenvalue of the unstressed system}  
We first evaluate the eigenvalues in the unstressed system and then consider the effect of the stress term (\emph{i.e.} the second term in Eq.~\eqref{eq:stressed}). In order to obtain the scaling of the lattice instabilities, we first estimate the eigenvalues at the $M$ point, which is located at the corner of the Brillouin zone. The wave vector $\bf k_M$ at the corner has the smallest wavelength anywhere in the Brillouin zone and thus corresponds to the most drastic perturbation. 
%at the largest wavevector, $\bm k_M$, \emph{i.e.} at the $M$ point located at the corner of the Brillouin zone.  
We then extend the argument to the rest of the Brillouin zone. %(In two dimensions the $M$ point, at the zone corner, is at $\bm k = (\pi/L,\pi/L)$).

For the unstressed system, the lowest eigenvalue at the corner of the Brillouin zone can be estimated as follows.  As Fig.~\ref{fig:examplemodes}b suggests, the mode structure is fairly straightforward. 
Low-frequency vibrations are dominated by two distinct classes of modes: plane waves and the so-called ``anomalous modes" that are characteristic of jammed systems~\cite{Wyart:2005wv,Liu:2010jx}.
%\cite{Goodrich:2013ke,Liu:2010jx}. 
The lowest plane-wave branch is transverse and parabolic at low $k$ (see Fig.~\ref{fig:examplemodes}b): $\omega_{\text{T},\text{u}} \approx c_{\text{T},\text{u}}k$ or equivalently
\begin{equation}
\lambda_{\text{T},\text{u}} \approx c_{\text{T},\text{u}}^2 k^2 \sim G_\text{u} k^2, \label{eq:parabola}
\end{equation}
where $G_\text{u}$ is the shear modulus of the unstressed system.

The eigenvalue of the lowest anomalous mode can be understood as follows. Wyart {\it et al.}~\cite{Wyart:2005wv,Wyart:2005jna} showed that the density of vibrational states at $k=0$ for unstressed systems, $D_\text{u}(\omega)$, can be approximated by a step function, so that $D_\text{u}(\omega) \approx 0$ for $\omega<\omega_\text{u}^*$ while $D_\text{u}(\omega) \approx \mathrm{const}$ for $\omega>\omega_\text{u}^*$.  As suggested by Fig.~\ref{fig:examplemodes}, the anomalous modes are nearly independent of $\bm k$, so this form for $D_\text{u}(\omega)$ is a reasonable approximation not only at $k=0$ but over the entire Brillouin zone. %Assuming that optical bands are flat is simply the Einstein approximation,\cite{Ashcroft:1976ud} and is frequently used in the study of solids. 
Thus, the eigenvalue of the lowest anomalous mode is approximated by $\omega_\text{u}^*$ at any $\bm k$.

Note that if $\omega_{\text{T},\text{u}} \ll \omega_\text{u}^*$, the lowest branch will maintain its transverse-acoustic-wave character and hence will remain parabolic in $k$ all the way to the zone corner.  
However, when $\omega_{\text{T},\text{u}} \gg \omega_\text{u}^*$, the lowest mode at the corner no longer has plane-wave character because the transverse acoustic mode will hybridize with anomalous modes and will develop the character of those modes.  It follows that there is a crossover between jamming physics and plane-wave physics when $\omega_{\text{T},\text{u}} \approx \omega_\text{u}^*$ or, equivalently, when the system size is
\begin{equation}\label{eq:crossoverlength}L \approx \ell_\text{T}\equiv c_{\text{T},\text{u}}/\omega_\text{u}^*.\end{equation}
Here $\ell_\text{T}$ is the transverse length identified by Silbert {\it et al.}~\cite{Silbert:2005vw}.

Near the jamming transition, many properties scale as power laws with the excess contact number above isostaticity, $\Delta Z \equiv Z - Z^N_\mathrm{iso}$, where $Z$ is the average number of contacts per particle and $Z^N_\mathrm{iso} = 2d \left(1-1/N\right) \approx 2d$ for large systems~\cite{Goodrich:2012ck}. In particular, for the harmonic potentials we consider here, $\omega_\text{u}^*\sim \Delta Z$,  $G_\text{u} \sim \Delta Z$, and $p\sim\Delta Z^2$ for dimensions $d \ge 2$. (These results are easily generalized to potentials other than the harmonic interactions used here~\cite{Liu:2010jx}.)
Therefore, Eqs.~\eqref{eq:parabola} and \eqref{eq:crossoverlength} predict that, for $d \ge 2$, $\ell_\text{T} \sim p^{-1/4}$ and the crossover will occur at 
\begin{equation}\label{eq:crossover} pL^4 \sim\text{const.}\end{equation}

\begin{figure}
\centering
\includegraphics[width=0.7\textwidth,viewport=0 75 500 650,clip]{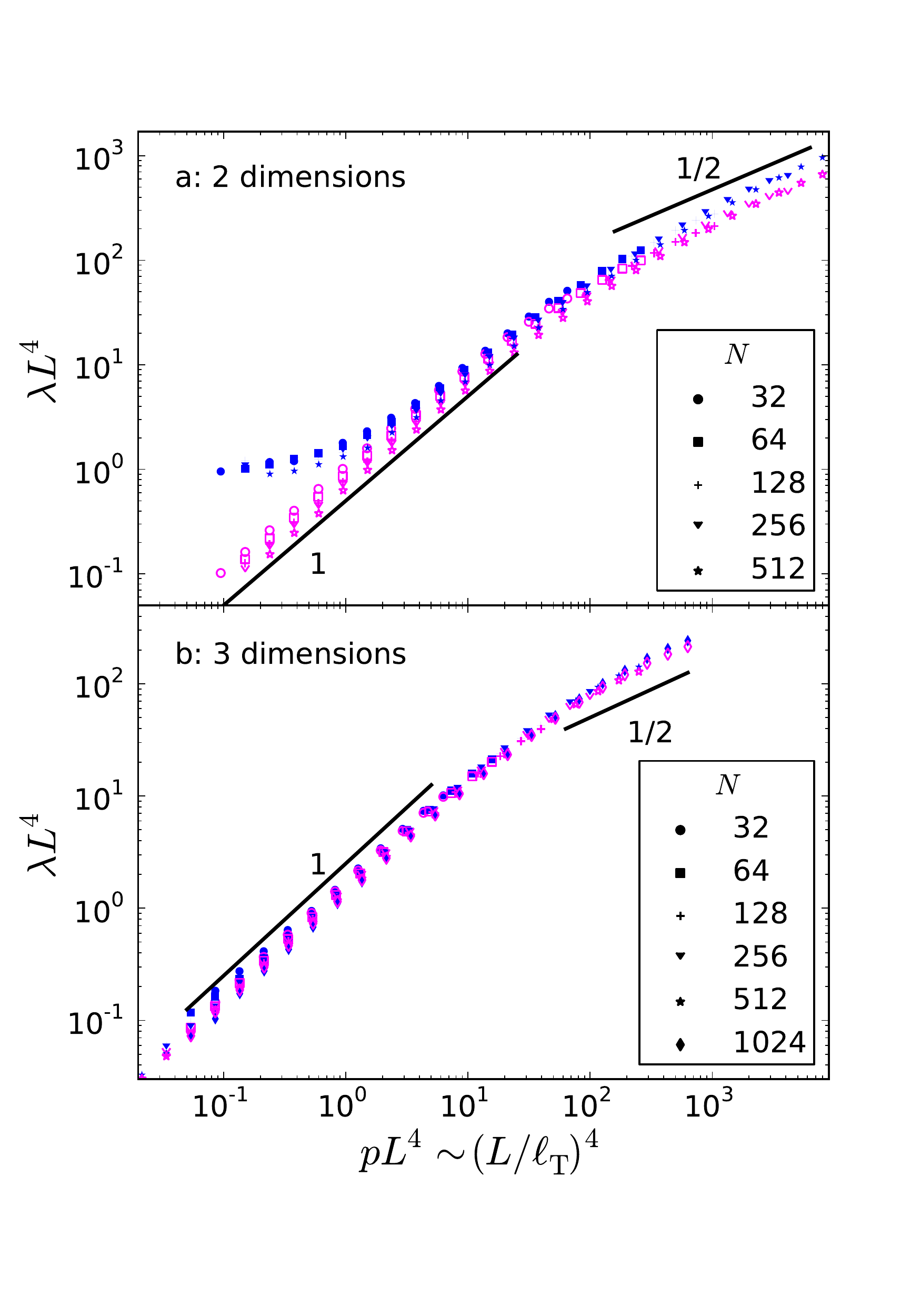}
\caption[Scaling of the lowest eigenvalue.]{Scaling of the lowest eigenvalue. The average eigenvalue of the lowest mode at the corner of the Brillouin zone of the unstressed system (blue), as well as the average difference between the unstressed system and the original packing (magenta), in (a) two and (b) three dimensions. The blue and magenta data exhibit collapse as predicted by Eqs.~\eqref{eq:lambda0} and \eqref{eq:stressdiff}, with the caveat that we were unable to reach the low pressure regime in three dimensions where we expect the scaling to be different. Data is only shown when at least 20 shear stable configurations were obtained.}
\label{fig:stressedunstressed}
\end{figure} %wouter for G in stressed vs unstressed systems. Europhysics letter

A second crossover occurs at very low pressures and is due to finite-size effects~\cite{Goodrich:2012ck} that change the scaling of $\Delta Z$ to $\Delta Z\sim1/N\sim L^{-d}$ in $d$ dimensions, independent of $p$.  In this regime, $\omega_\text{u}^*$ and $G_\text{u}$ remain proportional to $\Delta Z$, and thus also scale as $L^{-d}$. We therefore expect the lowest eigenvalue of the unstressed system at the corner of the Brillouin zone ($k_{M} = \sqrt{d}\pi/L$) to feature three distinct regimes.% at the $M$ point, where the wavevector magnitude is $k_{M} = \sqrt{d}\pi/L$:
\begin{equation}
	\begin{array}{ r r l}
		\text{low pressure:} &  		\lambda_\text{u} & \sim \;\;  1/N^2 \sim L^{-2d} \\
		\text{intermediate pressure:} &	\lambda_\text{u}  & \sim \;\; \omega^{*2}_\text{u} \sim p  \\
 		\text{high pressure:} &		\lambda_\text{u}  & \sim \;\; c_\text{T}^2 k_M^2 \sim p^{1/2} L^{-2}. \label{eq:lambda0}
	\end{array}
\end{equation}
In two dimensions, we expect that $\lambda_\text{u}L^4$ will collapse in all three regimes as a function of $pL^4$.

This prediction is verified in Fig.~\ref{fig:stressedunstressed}. The blue symbols in Fig.~\ref{fig:stressedunstressed}a, corresponding to $\lambda_\text{u} L^4$ of the unstressed system in $d=2$, exhibit a plateau at low pressures/system sizes. At intermediate $pL^4$, the blue symbols have a slope of 1 and at high $pL^4$ a slope of 1/2, as predicted by Eq.~\eqref{eq:lambda0}. 
In three dimensions (Fig.~\ref{fig:stressedunstressed}b), we observe the two higher pressure regimes, with a crossover between them that scales with $pL^{4}$, as expected. We did not reach the low-pressure plateau regime in three dimensions because it is difficult to generate shear stable configurations at low pressures.  Note, however, we do not expect the crossover to the low pressure regime to collapse in $d=3$ with $pL^4$.  

\subsection{The effect of stress on the lowest eigenvalue}
With the behavior of the lowest eigenvalue of the unstressed system in hand, we now turn to the effects of stress to explore the behavior of actual sphere packings at the zone corner, $k_{M} = \sqrt{d}\pi/L$.  The second term in Eq.~\eqref{eq:stressed} shifts the shear modulus to smaller values without affecting the scaling with pressure~\cite{Ellenbroek:2009to}, $G \sim \sqrt{p}$.   It therefore follows that  $G_\text{u}-G\sim\sqrt p$. In the high-pressure regime, where the lowest mode is the transverse plane wave, we therefore expect that $\lambda_\text{u}-\lambda \sim G_\text{u}-G \sim \sqrt{p}$.   Likewise, it was shown~\cite{Goodrich:2013ke,Wyart:2005vu,Xu:2007dd} that the second term in Eq.~\eqref{eq:stressed} lowers ${\omega_\text{u}^*}^2$ by an amount proportional to $p$, so at intermediate and low pressures it follows that $\lambda_\text{u}-\lambda \approx{\omega_\text{u}^*}^2-{\omega^*}^2 \sim p$.

The difference in the lowest eigenvalues of the sphere packing and the unstressed system will therefore feature two distinct regimes: % at low and intermediate pressures, and at high pressures, respectively:
\begin{equation}
	\begin{array}{ r r l}
		\text{low and int. $p$:} &	\left(\lambda_\text{u}-\lambda\right) & \sim p  \\
 		\text{high $p$:} &		\left(\lambda_\text{u}-\lambda\right) & \sim p^{1/2} L^{-2}. \label{eq:stressdiff}
	\end{array}
\end{equation}
 
\subsection{The lowest eigenvalue and stability of the original packing}
\begin{figure}
\centering
\includegraphics[width=1\textwidth,viewport=75 75 950 650,clip]{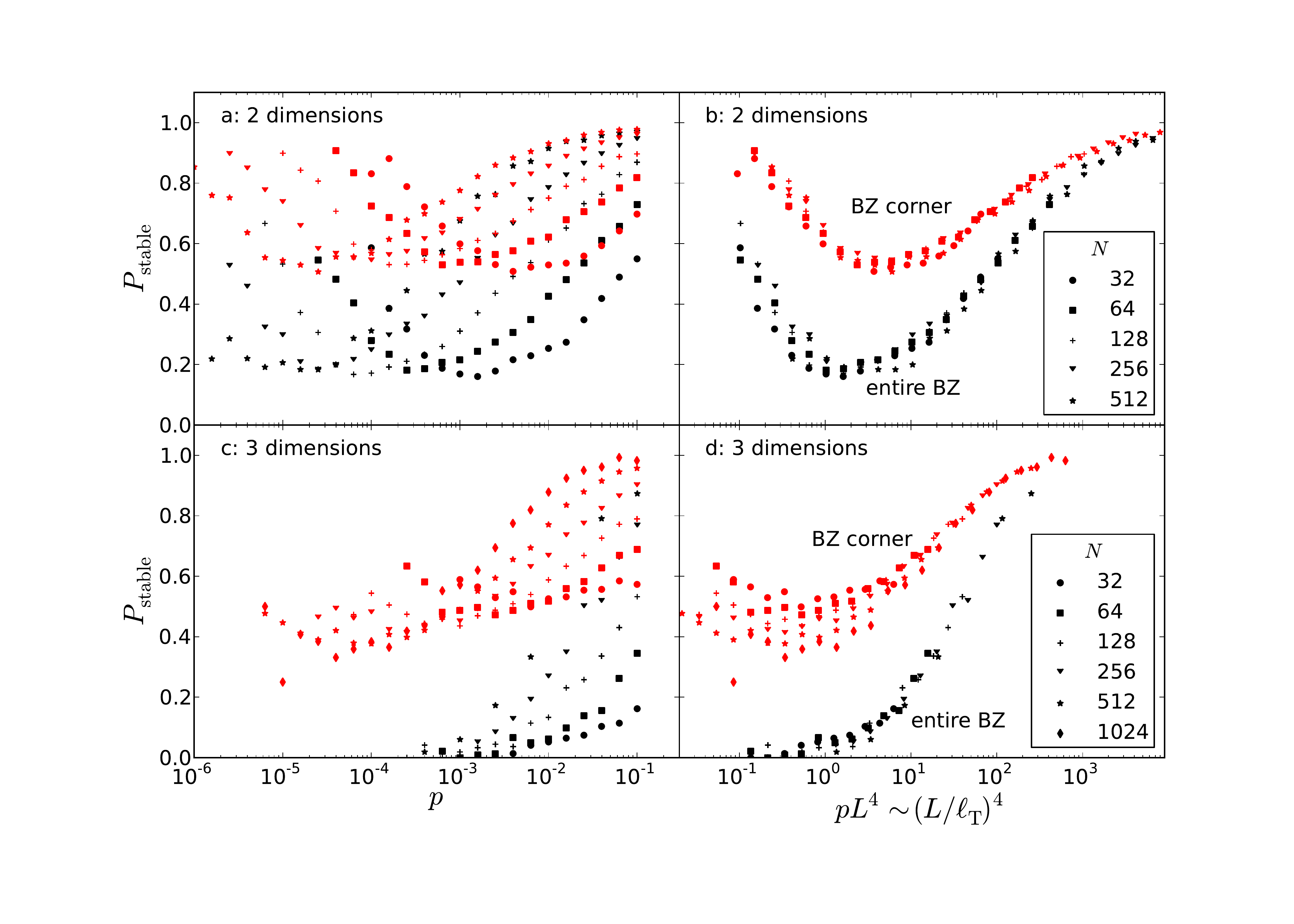}
\caption[The fraction of shear-stable systems in two and three dimensions.]{The fraction of shear-stable systems in two and three dimensions that are also $k>0$-stable. In red we plot the fraction of systems that are unstable at the $M$ point while in black we plot the fraction of systems that are stable everywhere. We see that both collapse with $pL^4$, or equivalently $L/\ell_\text{T}$, with the expected exception of the low pressure regime in three dimensions.}
\label{fig:goodapples}
\end{figure}

Comparing these results to Eq.~\eqref{eq:lambda0}, we see that the lowest eigenvalue of the packing $\lambda$ should be positive in the low pressure limit where $\lambda_\text{u}-\lambda \ll \lambda_\text{u}$.  At high and medium pressures, $\lambda_\text{u}$ and $\lambda_\text{u}-\lambda$ are comparable and obey the same scaling. One would therefore expect instabilities to arise in this regime.  At high pressures, however, we know that $\lambda$ should be positive since $\lambda \sim G k_M^2$ and shear stability implies $G>0$~\cite{DagoisBohy:2012dh}. Therefore, fluctuations about the average scaling behavior are most likely to drive the system unstable at intermediate pressures.
Since this regime collapses with pressure and system size as $pL^4$, we expect the fraction of systems that are unstable to obey this scaling.  This prediction is corroborated in Fig.~\ref{fig:goodapples}, where we see that the fraction of systems that are stable at the $M$ point (red data) depends on pressure $p$ and system size $L$ as expected.

We have thus far illustrated our reasoning for wavevectors at a zone corner, which correspond to the smallest wavelengths and thus the most drastic perturbations. Systems should therefore be more likely to go unstable at the zone corner than anywhere else. However, our arguments apply equally well to any point in the Brillouin zone. 

This argument is confirmed in Fig.~\ref{fig:goodapples}. To investigate the stability over the entire zone, we computed the dispersion relation over a fine mesh in $k$ space and looked for negative eigenvalues.  Any configuration with a negative eigenvalue at any value of $\bm k$ was labeled as unstable.  The black data in Fig.~\ref{fig:goodapples} shows the fraction of systems that are stable over the entire Brillouin zone. It exhibits the same qualitative features as at just the zone corner (red data), but with fewer stable configurations overall. 

Note that our scaling arguments for the unstressed system apply equally well to the original packing. However, scaling alone cannot tell us the {\it sign} of the lowest eigenvalue. Treating the unstressed and stressed terms of the dynamical matrix separately, and exploiting the fact that the unstressed matrix is non-negative, is thus necessary for understanding the stability of the packing and emphasizes that ``lattice instabilities" are caused by stress, not the geometry of the packing.

%\section{Discussion}
\subsection{The two length scales\label{sec:two_lengths}}
Dagois-Bohy {\it et al.}~\cite{DagoisBohy:2012dh} found that in two dimensions, the fraction of states that are shear stable collapses with $pL^4$ and approaches $1$ at large $pL^4$. 
We have shown that shear stable packings can be unstable to a class of boundary perturbations that correspond to particle motion at non-zero $\bm k$, and that the susceptibility of packings to such perturbations also collapses with $pL^4$ (except for the $N$-dependent finite-size effects at low pressure). 
The combination of these two results suggests that a system should be stable to all infinitesimal boundary perturbations for $L \gg %\max{\left\{\ell_0,\ell_\text{T}\right\}}=
\ell_\text{T}$, where $\ell_\text{T}\sim p^{-1/4}$ (for harmonic interactions) is the so-called transverse length scale.  This implies that the closer the system is to the jamming transition the larger the system must be in order to be insensitive to changes in the boundaries.  It also implies that the distinction between collectively and strictly jammed~\cite{Torquato:2001bm} packings decreases as $L \gg \ell_\text{T}$ and vanishes in the thermodynamic limit for any nonzero pressure.

While the sensitivity of jammed packings to infinitesimal changes to the boundary is controlled by the diverging length scale $\ell_\text{T}$, Wyart {\it el al.}~\cite{Wyart:2005wv,Wyart:2005jna} argued that the stability to a more drastic change of boundary conditions, in which the periodic boundaries are replaced with free ones, is governed by the larger length $\ell^*\sim p^{-1/2}$: jammed packings with \emph{free} boundaries are stable only for $L \gg \ell^*$~\cite{Wyart:2005wv,Wyart:2005jna,Goodrich:2013ke}.
%$L \gg \ell_\text{L}$.~\cite{Goodrich:2013ke} 
%Goodrich {\it et al.}~\cite{Goodrich:2013ke} have shown 
We saw earlier in this chapter that this length is equivalent, not only in scaling behavior but also in physical meaning, to the longitudinal length $\ell_\text{L} =\omega^*/c_\text{L}$. % that was proposed alongside $\ell_\text{T}$ by Silbert {\it et al.}~\cite{Silbert:2005vw}% and is the wavelength of the \emph{longitudinal} plane wave with frequency $\omega^*$.

$\ell_\text{L}$ was proposed alongside $\ell_\text{T}$ by Silbert {\it et al.}~\cite{Silbert:2005vw}, and both length scales can be understood from a competition between plane-wave physics and jamming physics. Silbert {\it et al.} showed that at large wavelengths, there are well defined transverse and longitudinal sound modes; the Fourier transform $f_\text{T,L}(\omega,k)$ is sharp in $\omega$ at low $k$~\cite{Silbert:2005vw,Silbert:2009iw}. However, at small wavelengths, the sound modes mix with the anomalous modes and $f_\text{T,L}$ gets very broad. The crossover between these two regimes occurs when the frequency of the sound mode is $\omega^*$, the lowest frequency at which anomalous modes exist. Since the frequency of the sound modes is given by $\omega = c_\text{T,L}k$, the wave length at the crossover is $\ell_\text{T,L} \sim c_\text{T,L}/\omega^*$.
%there are well-defined, large wavelength transverse modes with frequency $\omega$ and wavevector $|\bm k| \approx \omega/c_\text{T}$, provided that $|\bm k| < \omega^*/c_\text{T}$. At higher wavevectors, and thus smaller wavelengths, the transverse sound mode mixes with the anomalous modes and the Fourier transform $f_\text{T}(\omega,\bm k)$ becomes very broad.
Thus, the two length scales $\ell_\text{T}$ and $\ell^*=\ell_\text{L}$ are both related to the wavelength where the density of anomalous modes overwhelms the density of plane-wave sound modes~\cite{Silbert:2005vw}. Our results indicate that they also both control the stability of jammed packings to different classes of boundary perturbations. The two lengths can therefore be viewed as equally central to the theory of the jamming transition.

\subsection{Discussion}
To understand why some boundary perturbations are controlled by $\ell_\text{T}$ while others are controlled by $\ell_\text{L}$, we must understand the nature of the transverse and longitudinal deformations allowed.  In the case of systems with free boundaries, the system size of $\ell_\text{L}$ needed to support compression is also sufficient to support shear since $\ell_\text{L}>\ell_\text{T}$.  The minimum system size needed to support both is therefore $\ell_\text{L}$. Under periodic boundary conditions, jammed systems necessarily support compression regardless of system size and the issue is whether the system is stable to shear as well.  Our arguments show that the minimum system size needed to support both longitudinal and transverse perturbations is $\ell_\text{T}$.  

The same reasoning can be applied to understanding the stability with respect to a change from periodic to fixed boundary conditions.  In that case, neither longitudinal nor transverse deformations are allowed and it follows that jammed systems of any size will be stable to this change. Mailman and Chakraborty~\cite{Mailman:2011hz} have studied systems with fixed boundary conditions to calculate point-to-set correlations. Their analysis, as well as arguments based on the entropy of mechanically stable packings, reveal a correlation length that scales like $\ell^*$.

The fact that the diverging length scales control the response to boundary changes but do not enter into the finite-size scaling of quantities such as the contact number and shear modulus~\cite{Goodrich:2012ck} is consistent with the behavior of a system that is at or above its upper critical dimension.  The fact that power-law exponents do not depend on dimensionality for $d \ge 2$ is also consistent with this interpretation~\cite{OHern:2003vq,Parisi:2010uu}.  However, critical systems are generally controlled by only one diverging length.  Jamming is thus a rare example of a phase transition that displays two equally important diverging length scales. 
It remains to be seen whether other diverging lengths that have been reported near the jamming transition~\cite{OHern:2003vq,Drocco:2005ho,Olsson:2007df,Vagberg:2011fe,Mailman:2011hz} can also be associated with boundary effects.

\chapter{Jamming in finite systems: stability, anisotropy, fluctuations and scaling}
\label{chapter:finite_size}

\newcommand{\Nc}[0]{\ensuremath{N_\text{c}}}
\newcommand{\Nciso}[0]{\ensuremath{N_\text{c}^\text{iso}}}
\newcommand{\Nbndry}[0]{\ensuremath{N_\text{dof}^\text{bndry}}}
\newcommand{\Ndof}[0]{\ensuremath{N_\text{dof}}}
\newcommand{\Nzm}[0]{\ensuremath{N_\text{zm}}}

\newcommand{\He}[0]{\ensuremath{\hat K}}
\newcommand{\Zmin}[0]{\ensuremath{Z^N_\text{min}}}
\newcommand{\Gdc}[0]{\ensuremath{G_{DC}}}

\newcommand{\Rc}[0]{\ensuremath{\mathcal R_{comp}}}
\newcommand{\Ra}[0]{\ensuremath{\mathcal R_{all}}}
\newcommand{\Rap}[0]{\ensuremath{\mathcal R_{all}^+}}
\newcommand{\Ec}[0]{\ensuremath{\mathcal E_{comp}}}
\newcommand{\Ea}[0]{\ensuremath{\mathcal E_{all}}}
\newcommand{\Eap}[0]{\ensuremath{\mathcal E_{all}^+}}

%\let\vaccent=\v % rename builtin command \v{} to \vaccent{}
%\renewcommand{\v}[1]{\ensuremath{\mathbf{#1}}} % for vectors
%\newcommand{\gv}[1]{\ensuremath{\mbox{\boldmath$ #1 $}}}
%% for vectors of Greek letters
%\newcommand{\uv}[1]{\ensuremath{\mathbf{\hat{#1}}}} % for unit vector
%\newcommand{\guv}[1]{\ensuremath{\mathbf{\hat{\mbox{\boldmath$#1$}}}}} % for greek unit  vector
%\newcommand{\abs}[1]{\left| #1 \right|} % for absolute value
%\newcommand{\avg}[1]{\left< #1 \right>} % for average
%\newcommand{\T}[0]{^\text{T}}
%
%\newcommand{\ket}[1]{\left| #1 \right>} % for Dirac bras
%\newcommand{\bra}[1]{\left< #1 \right|} % for Dirac kets
%\newcommand{\braket}[2]{\left< #1 \vphantom{#2} \right| \left. \! #2 \vphantom{#1} \right>} % for Dirac brackets
%\newcommand{\matrixel}[3]{\left< #1 \vphantom{#2#3} \right|
% #2 \left| #3 \vphantom{#1#2} \right>} % for Dirac matrix elements

\section{Introduction and conclusions}

The theory of jammed amorphous solids has been largely based on packings at zero temperature of frictionless spheres with finite-range repulsions. Over the past decade, numerous studies have characterized the transition of such systems from an unjammed ``mechanical vacuum'' in which no particles interact at low packing fraction, $\phi$, to a jammed, rigid structure at high $\phi$ (see~\cite{Liu:2010jx,vanHecke:2009go} and references therein). %\cite{OHern:2003vq,Liu:2010jx}\cite{grr}. 
The scenario that has emerged is that the jamming transition is a rare example of a random first-order transition.\footnote{Note that the jamming transition appears to be a random first-order transition in dimensions $d \ge 2$, and is distinct from the glass transition, which is a random first-order transition in infinite dimensions~\cite{Parisi:2010uu}.} %In the large system limit, the bulk modulus $B$ and shear modulus $G$ both become non-zero for finite pressure --- 
%$G$ grows smoothly with pressure, whereas
%$B$ is discontinuous at zero pressure.
%Crucially, both moduli become non-zero simultaneously
%and concurrent with the onset of finite pressure --- thus making the transition both clear and well-defined.  
At the jamming transition, the average number of contacts per particle, $Z$, jumps discontinuously from zero to the value given by the rigidity criterion proposed originally by Maxwell.  Power-law scaling over many decades in confining pressure has been observed near the transition for the bulk modulus, shear modulus, energy, non-affinity, a characteristic frequency scale, various length scales 
and the excess contact number~\cite{Liu:2010jx,Durian:1995eo,OHern:2003vq,Silbert:2005vw,Wyart:2005vu,Wyart:2005jna,Wyart:2005wv,Silbert:2006bd,Ellenbroek:2006df,Ellenbroek:2009dp,Goodrich:2013ke}.
%Moreover, the excess contact number and shear modulus have recently been shown to exhibit finite-size scaling, consistent with the critical nature of the jamming transition~\cite{Goodrich:2012ck}.

For ordinary critical phase transitions, singularities are rounded in finite systems but the nature of the transition remains qualitatively the same as it is in infinite ones.  
However, because the particle interactions in a jammed packing are purely repulsive and the force on every particle has to be balanced, a jammed packing must have a rigid structure that is system-spanning.  As a result, the nature of the boundary conditions is inextricably linked with the onset of rigidity, and boundary conditions play a particularly important role in finite jammed systems~\cite{Torquato:2001bm}.  For example, systems prepared in the standard way, in a fixed simulation box with periodic boundary conditions (that is, with the repeated zone of constant volume with {\em fixed} angles), can be unstable to shear even though they can support a pressure~\cite{DagoisBohy:2012dh}. 

Even for configurations that are stable to both shear and compression, the definition of the rigidity onset in terms of the development of nonzero bulk and shear moduli requires attention.  This is because jammed systems are only truly isotropic in the thermodynamic limit. Any finite system should properly be described by six elastic constants in 2 dimensions, or 21 in 3 dimensions, rather than the two elastic constants, the bulk and shear moduli, that describe isotropic systems.  Finally, the mechanical response of a finite system depends not only on the boundary conditions, but on whether or not the configuration has residual shear stress.  These considerations necessitate a careful reevaluation of jamming in finite systems. 

In this chapter, we take all of these potential complications into account to develop a comprehensive finite-size analysis of compressed, athermal sphere packings with periodic-boundary conditions. 
We recast the $6$ ($21$) elastic constants needed in $2$ ($3$) dimensions in terms of (i) two combinations that are finite in the thermodynamic limit: the bulk modulus, $B$, and $G_{DC}$ (which approaches the shear modulus in the thermodynamic limit) and (ii) three combinations that measure anisotropic fluctuations and vanish in that limit.  Despite the complications alluded to above, for all of the ensembles studied and independent of the criteria used to identify the jamming transition, we show that $pN^2$ (where $p$ is the pressure and $N$ is the system size) is the correct scaling variable for the key quantities of excess contact number, $B$ and $G_{DC}$.  
%This is consistent with earlier results for one of these ensembles~\cite{Goodrich:2012ck}.  
(In the case of two dimensions, our results are consistent with the presence of logarithmic corrections to scaling, supporting the conjecture~\cite{Liu:2010jx,Charbonneau:2012fl,Goodrich:2012ck,Wyart:2005vu} that the upper critical dimension for jamming is $d=2$.)

One of the three elastic constants (defined above) that vanish in the thermodynamic limit also collapses with $pN^2$ and vanishs in the limit of $pN^2 \rightarrow \infty$ as $1/\sqrt{N}$. This is consistent with the central-limit theorem. The remaining two exhibit this behavior only for ensembles that have zero residual shear stress.  Thus, for the ensembles with no shear stress, we observe scaling collapse with $pN^2$ for all variables studied.

We note that one consequence of the scaling collapse with $pN^2$ is that one needs larger and larger systems as the jamming transition is approached to be in the thermodynamic limit. If the limit is properly taken, however, our results show that the bulk modulus, $B$, the shear modulus, $G$, and the ratio of the two, $G/B$, all become nonzero simultaneously\footnote{This simultaneity can be misunderstood because $G$ increases continuously while $B$ experiences a discontinuous jump, meaning that $G$ can be arbitrarily small while $B$ remains finite. The precise statement is that there exists a $\phi_c$ such that, for all $\epsilon>0$, $B = G = 0$ when $\phi = \phi_c - \epsilon$ and $B>0$ and $G>0$ when $\phi = \phi_c + \epsilon$.} at the jamming transition, consistent with earlier claims~\cite{OHern:2003vq}.

The location of the jamming transition depends on both system size~\cite{OHern:2003vq,Vagberg:2011fe} and protocol~\cite{Chaudhuri:2010jg}.  Thus, the packing fraction at the transition fluctuates from state to state.  Several studies have focused on finite-size effects associated with this distribution of packing fractions at the onset of jamming~\cite{OHern:2003vq,Vagberg:2011fe, Chaudhuri:2010jg,Liu:2014gu}.  In contrast, we concentrate on finite-size scaling in bulk quantities above the transition, and bypass the effects of the distribution of jamming onsets by looking at behavior as a function of pressure, or equivalently, $\phi-\phi_c$, where $\phi_c$ is the packing fraction at the jamming onset for a given state.  

In Sec.~\ref{sec:jamming_def}, we introduce the three ensembles based on the different jamming criteria and review the constraint counting arguments for each one~\cite{Goodrich:2012ck,DagoisBohy:2012dh}.  We introduce the $\tfrac 18 d(d+1)(d^2+d+2)$ independent elastic constants in $d$ dimensions, and use them to find the conditions required for mechanical stability.  We then recast them in terms of combinations that either approach the bulk and shear moduli or vanish in the thermodynamic limit.  Section~\ref{sec:results} contains the numerical results for the excess contact number and the elastic constant combinations versus pressure and system size.  We also present results for statistical fluctuations of the excess contact number, bulk modulus, and $G_{DC}$.

\section{Jamming, ensembles and constraint counting in finite systems\label{sec:jamming_def}}

\begin{table}
\centering
\begin{tabular}{| c | c | c | c |}
\hline
Index & Meaning & Range & {\it e.g.}  \\ \hline
$\alpha$, $\beta$ & particle position DOF & $[1,dN]$ & $r_\alpha$ \\
$\bar\alpha$, $\bar\beta$ & position and boundary DOF & $[1,dN+\Nbndry]$ & $q_{\bar\alpha}$ \\
$b$ & simulation box shape DOF & $[1, d(d+1)/2 -1]$ & $L_b$\\
$i$, $j$, $k$, $l$ & dimension & $[1,d]$ & $\epsilon_{ij}$\\
$n$ & mode number & $[1,dN]$ & $\lambda_n$ \\ \hline
\end{tabular}
\caption[List of indices used in this chapter and their meaning.]{\label{table:indices} List of indices used in this chapter and their meaning. Note that $d$ is the dimensionality and $N$ is the total number of particles.}
\end{table}

\subsection{Jamming criteria and ensembles\label{sec:JammingCriteria}}

We will consider athermal ($T=0$) packings of $N$ soft spheres that interact only when they overlap with a purely repulsive spherically symmetric potential in $d$ dimensions. For now, we will not be concerned with the specific form of the interaction potential and only require that it has a finite range that defines the particle diameter.
What does it mean for such a packing to be jammed? The answer to this is clear in the thermodynamic limit.
At sufficiently low packing fractions, $\phi$, there is room for the spheres to avoid each other so that none of them overlap, and the number of load-bearing contacts vanishes. The potential-energy landscape is locally flat and the pressure and elastic moduli, which are respectively related to the first and second derivatives of the energy, are zero; 
in no way should the system be considered a solid.
At high $\phi$, however, there is no longer room for the particles to avoid each other and they are forced to overlap, and 
the system possesses enough contacts for rigidity.
It no longer sits at zero energy and develops a non-zero stress tensor
with positive pressure. Moreover, the shear modulus $G$ and bulk modulus $B$ are positive. Such a system possesses all the characteristics of a solid and is therefore jammed.

When we are not in the large system limit, the onset of rigidity is more complex. In this section, we will discuss the behavior of three quantities -- the average contact number, the pressure and the elastic constants  -- in finite systems at the jamming transition.

{\em I: Connectivity ---} It has long been known that there is a connection between the jamming transition and the contact number $Z$ ({\it i.e.}, the average number of load-bearing contacts per non-rattling particle), which is given by $Z\equiv 2\Nc/N_0$, where $\Nc$ is the total number of contacts and $N_0$ is the number of particles that are not rattlers~\cite{BOLTON:1990uy,Durian:1995eo,Alexander:1998vc,Moukarzel:1998vn}.  $Z=0$ below the jamming transition because there are no overlapping particles. (Note, it is possible for two particles to \emph{just} touch, but such a contact cannot bear any load.) At the transition, $Z$ jumps to a finite value and increases further as the system is compressed. This finite jump has been understood from the Maxwell criterion, which is a mean-field argument stating that a rigid network of central-force springs must have an average contact number of at least $\Ziso$. 
When $Z=\Ziso$, the number of contacts just balances the number of degrees of freedom.

However, the use of constraint counting and isostaticity as a measure of jamming has some serious drawbacks. For example, packings of ellipsoids jam well below isostaticity~\cite{Donev:2007go,Zeravcic:2009wo,Mailman:2009ct}.
Also, as contacts in frictional packings are able to constrain multiple degrees of freedom, the contact number at jamming depends sensitively on the strength of the frictional part of the interactions and lies below $2d$~\cite{Shundyak:2007ga,Somfai:2007ge,Henkes:2010uu,Henkes:2010kv,Papanikolaou:2013fa}.
Furthermore, the Maxwell criterion assumes that as a system approaches isostaticity, none of the contacts are redundant (in a manner that can be defined precisely for certain networks). Although we will show below that this assumption is often correct, it is not a generic feature of sphere packings.

For example, consider a 50/50 mixture of large and small particles in two dimensions just above the jamming transition. Such bidisperse packings are quite common in the study of jamming because a monodisperse mixture leads to local crystallization. Even for bidisperse mixtures, however, there is a non-negligible probability that a particle is surrounded by 6 particles of exactly the same size. It is easy to see that these 7 particles have a redundant contact even at the transition, but this extra contact does not contribute to the global stability of the rest of the packing. Therefore, the contact number at the transition will be slightly greater than the isostatic value.\footnote{We find the difference to be small, of order $10^{-3}$.} A corollary of this is that a packing might have $Z>\Ziso$ and still be unjammed. 
(As discussed in Appendix~\ref{sec:numerical_procedures}, our numerical calculations use a polydisperse distribution of particle sizes in two dimensions to avoid this issue.) Therefore, we see that constraint counting is not a robust indicator of whether or not a system is jammed. 

{\em II: Positive pressure ---}
For packings of purely repulsive particles, positive pressure is clearly a necessary condition for jamming. If a particle is trapped by its neighbors, then there must be a restoring force to counteract any small displacement. Such forces can only come from particle-particle interactions which, when integrated over the system, lead to non-zero pressure. If the pressure is zero, then there cannot be any particle-particle interactions and the system is not jammed, regardless of system size. Therefore, positive pressure is a necessary condition for jamming.

{\em III: Mechanical rigidity ---} A solid must resist global deformations such as compression and shear. We first consider the response to compression. As we saw above, particle-particle overlaps in a jammed system push outward and lead to non-zero pressure. Upon compression, these forces must increase to linear order, implying that the bulk modulus, $B$, is positive. 

\begin{figure}[htp]
\begin{center}
\includegraphics[width=.6\linewidth]{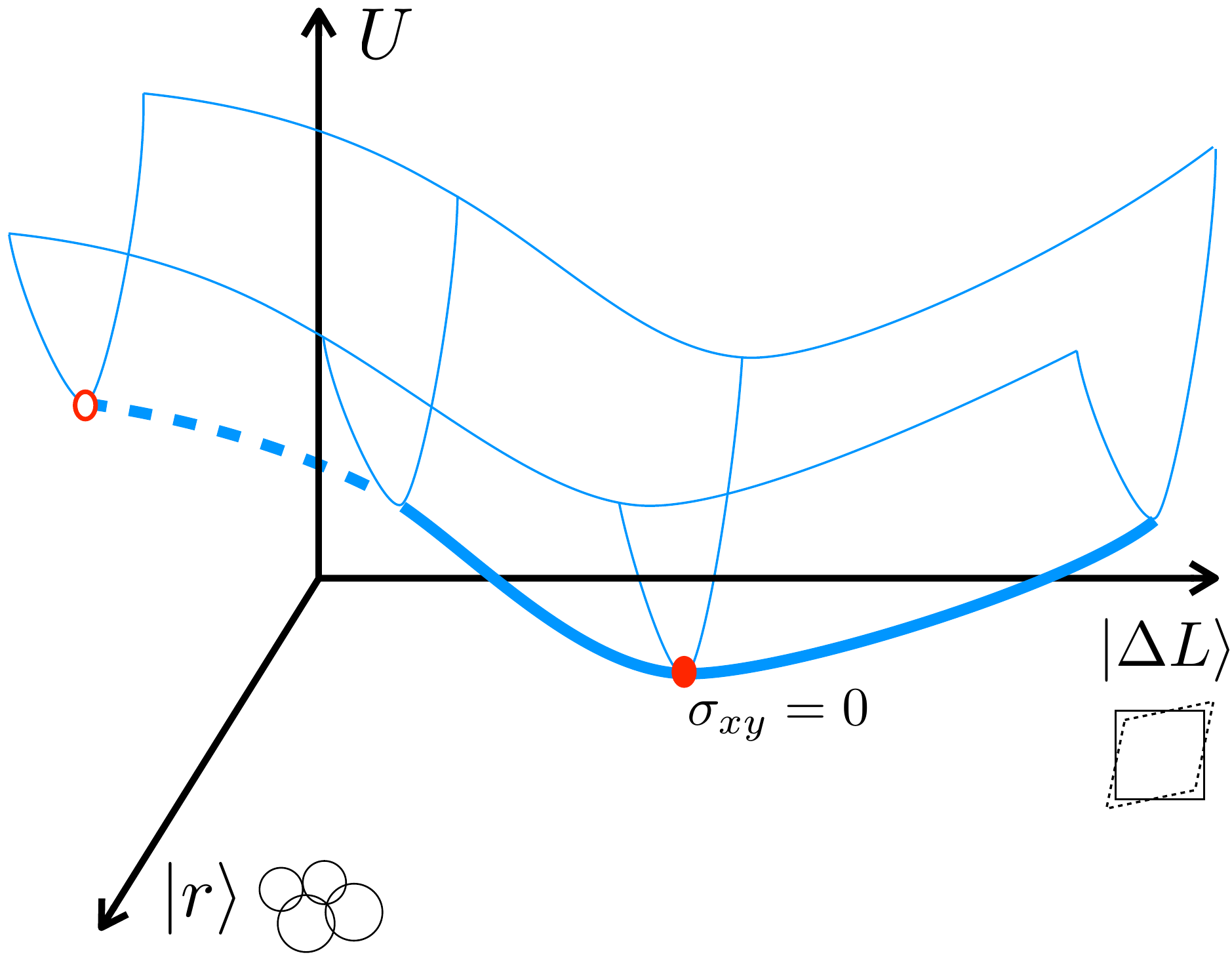}
\end{center}
\caption[Schematic energy landscape.]{\label{fig:SS_energy_landscape}
Schematic energy landscape where $|r\rangle$ denotes the particle degrees of freedom and $|\Delta L \rangle$ all possible shear deformations of the box. 
For packings in the $\Ec$ ensemble, $|\Delta L\rangle$ is fixed and the system is $\Rc$ stable if it sits at a minimum of $U$ with respect to $|r\rangle$, {\it i.e.} the open circle. 
$\Ra$ stability is governed by the curvature of $U$ along the global shear degrees of freedom. 
Thus, $\Rc$ stable states can be $\Ra$ unstable if the curvature of $U$ is negative along any of the $|\Delta L\rangle$ directions (thick dashed curve). Such states can and do occur.
If the curvature of $U$ is positive along all global shear directions (thick solid curve), the packing is $\Ra$ stable. Such $\Ra$ stable packings can have finite shear stresses (non-zero gradient along global shear directions). Finally, packings that are at a local minimum of $U$ with respect to the $|r\rangle$ \emph{and} $|\Delta L\rangle$ directions (filled circle) have zero residual shear stress in addition to being $\Ra$ stable, and thus satisfy the $\Rap$ requirement.
}
\end{figure}

The situation for shear deformations is more subtle, and various jamming criteria can be defined depending on the boundary conditions~\cite{DagoisBohy:2012dh}. 
Consider the potential energy landscape as a function of (1) the $dN$ particle positions $r_\alpha$, (2) the $d(d+1)/2 -1$ degrees of freedom $\Delta L_b$ associated with the shape of the box, and (3) the volume $V$. Common jamming algorithms fix the shape and size of the box and generate packings at a minimum of $U$ with respect to $\left| r \right> = \{r_\alpha\}$ (see Fig.~\ref{fig:SS_energy_landscape}).  In this case, no further constraints are necessary beyond those needed for the system to resist compression. 

The criterion that the system resist compression will be referred to as the $\Rc$, or ``Rigid to Compression,'' requirement, and the ensemble of systems that satisfy this requirement will be referred to as the $\Ec$ ensemble.  Experimental examples are when particles are placed in a rigid container or when the shape of the container is externally controlled.  Note that when the boundary is not allowed to deform, residual shear stresses and shear moduli correspond to the first and second derivatives, respectively, of $U$ along a strain direction without permitting the shape to equilibrate.  As a result, both the macroscopic residual stress and shear modulus are uncontrolled.  Figure~\ref{fig:SS_energy_landscape} illustrates that systems that are $\Rc$ stable do not need to be stable to shear, as pointed out by Dagois-Bohy {\it et al.}~\cite{DagoisBohy:2012dh}. 

The criterion that the system resists all global deformations, including shear and compression, will be referred to as the $\Ra$, or ``Rigid to All,'' requirement.  Previous work showed that the fraction of $\Ec$ packings that are $\Ra$ unstable becomes of order one for finite systems at sufficiently low pressure~\cite{DagoisBohy:2012dh}.

We can also consider the situation where all volume-conserving shear deformations are allowed. In this case, the degrees of freedom associated with the \emph{shape} of the container or simulation box can be considered dynamic degrees of freedom along with the particle positions~\cite{DagoisBohy:2012dh,Torquato:2010hb}. These additional variables expand the dimensionality of the energy landscape and allow the particles to relax further.  We have developed an algorithm for generating states that are not only $\Ra$ stable but also have zero residual shear stress~\cite{DagoisBohy:2012dh}.  In short, two-dimensional packings are generated by finding minima of $U$ with respect to both $\left| r \right>$ and the two shear degrees of freedom (labeled $\left| \Delta L \right> = \{\Delta L_b\}$ in Fig.~\ref{fig:SS_energy_landscape}). Because derivatives of $U$ with respect to shear degrees of freedom give shear stresses, the packings generated by this algorithm have a purely hydrostatic stress tensor.  We will refer to these combined criteria ($\Ra$ stable plus zero residual shear stress) as the $\Rap$ requirement.

The ensemble of systems that satisfy the $\Ra$ and $\Rap$ requirements will be referred to as the $\Ea$ and $\Eap$ ensembles, respectively. As illustrated in Fig.~\ref{fig:SS_energy_landscape}, these three jamming conditions have a simple interpretation in terms of the energy landscape. Furthermore, the ensembles have the hierarchical structure: $\Eap \subset \Ea \subset \Ec$. % (see Fig.~\ref{fig:diagram}). 

In the remainder of this chapter we study three different ensembles of packings, the $\Ec$, $\Ea$ and $\Eap$ ensembles described above.
The standard $\Ec$ packings dominate the jamming literature; we study them in both two and three dimensions. We will refer to these as the ``2d \Ec" and ``3d \Ec" ensembles, respectively.  We will also study two dimensional packings that are $\Rap$ stable (stable to shear deformations in all directions \emph{and} have no residual shear stress), which make up the ``\Eap" ensemble. Finally, to compare these two ensembles, we consider the two-dimensional $\Ea$ ensemble, which is a ``filtered $\Ec$" ensemble where we include only the $\Ec$ configurations that happen to be $\Ra$ stable. Like the $\Eap$ states, $\Ea$ states have positive shear modulus; unlike the $\Eap$ states, $\Ea$ states have generically non-zero residual shear stress. 
The essential scenario is %depicted in Fig.~\ref{fig:diagram}: whereas 
that for small $p N^2$ the packings in these different ensembles are significantly different, but for large $p N^2$ these differences become smaller and vanish when $p N^2 \rightarrow \infty$.
For further details and numerical procedures, see Appendix~\ref{sec:numerical_procedures}.

\begin{table}
\centering
\begin{tabular}{ | c | c | c | c |}
	\hline
	Ensemble & Criteria  & Preparation algorithm  & Dimensionality \\ \hline
	$\Ec$ & $\Rc$  &standard jamming algorithm & $2d$, $3d$\\ \hline
	$\Ea$ & $\Ra$ & filtered $\Ec$ ensemble & $2d$ \\ \hline
	$\Eap$ & $\Rap$ & new shear-stabilized algorithm & $2d$ \\ \hline
\end{tabular}
\caption[List of ensembles, the jamming criteria they satisfy, the algorithm used, and the dimensionality in which we studied them.]{\label{table:ensembles}List of ensembles, the jamming criteria they satisfy, the algorithm used, and the dimensionality in which we studied them: $\Eap \subset \Ea \subset \Ec$.  The distinction between these ensembles vanishes in the large-system limit.}
\end{table}

\subsection{Jamming criteria in terms of the extended Hessian\label{sec:precise_formulation}}
Here we show that the jamming criteria introduced in Sec.~\ref{sec:JammingCriteria} can be formulated in terms of an extended Hessian that includes the boundary degrees of freedom~\cite{DagoisBohy:2012dh,Tighe:2011fq}. 
By defining jamming in terms of global deformations, we avoid requiring that individual particles be constrained.  Assumptions about the existence of zero modes are also not required.  This formulation therefore avoids the ambiguities of previous definitions based on counting zero modes.
In practice, zero modes can be present in jammed systems, such as those associated with rattlers and the extended quartic modes in the zero pressure limit of jammed packings of ellipsoids~\cite{Donev:2007go,Zeravcic:2009wo,Mailman:2009ct} --- as long as they are decoupled from the boundary degrees of freedom, they do not prevent the packing from being jammed.

We will begin by considering the $\Ra$ requirement that the system be stable with respect to all possible boundary deformations, 
and then show how the less strict $\Rc$ requirement can be deduced in the same framework.
We start with the Taylor expansion of the potential energy $U$ about a reference state with energy $U^0$, volume $V^0$, and particles positions $r_\alpha^0$. We restrict our attention to reference states in which the sum of forces on each particle is zero. The goal will be to determine if the reference state is jammed.

To test the $\Ra$ requirement, we need to include the $\Nbndry = d(d+1)/2$ degrees of freedom associated with boundary deformations in the energy expansion. It will be convenient to represent these variables as a symmetric strain tensor, $\epsilon_{ij}$. By differentiating the energy with respect to $\epsilon_{ij}$, we get the stress tensor of the reference state:
\eq{	\sigma^0_{ij} = \frac{1}{V^0} \left( \frac{\partial U}{\partial \epsilon_{ij}}\right)_0 \, .	}
$\sigma^0_{ij}$ represents prestress in the system and the trace of $\sigma_{ij}^0$ is proportional to the pressure.

Now consider the set of $dN$ particle displacements $\{u_\alpha\}$ about the reference state, $u_\alpha \equiv r_\alpha - r^0_\alpha$. The net force on each particle is given by the derivative of the energy with respect to $u_\alpha$, but this must be identically zero to satisfy force balance. To treat the boundary deformations and particle displacements together, let $\{q_{\bar \alpha}\} = \{u_\alpha,\epsilon_{ij}\}$ be the combination of the $dN$ particle displacements and the $\Nbndry$ independent components of the strain tensor. The first order term in the energy expansion is $\left( \frac{\partial U}{\partial q_{\bar\alpha}} \right )_0 q_{\bar\alpha}$, but this reduces to $\sigma^0_{ij}\, \epsilon_{ji} \, V^0$ due to the presence of force balance.

If the boundary was held fixed, then the second order term in the expansion would be obtained from the Hessian matrix $\He_{\alpha\beta}^0$, which is given by 
\eq{	\He^0_{\alpha \beta} \equiv \left( \frac{\partial^2 U}{\partial u_\alpha \partial u_\beta} \right)_{\!0} \, ,	\label{eq:reduced_hessian}}
where the derivatives are evaluated at the reference state. $\He^0_{\alpha \beta}$ is also the well-studied dynamical matrix of a packing where every particle has unit mass; its eigenvectors give the normal modes of vibration. For perturbations that include the boundary, however, we instead need the ``extended Hessian" matrix $\hat K$ \cite{DagoisBohy:2012dh,Tighe:2011fq},
\eq{	\He_{\bar\alpha \bar\beta} \equiv \left( \frac{\partial^2U}{\partial q_{\bar\alpha}\partial q_{\bar\beta}} \right)_{\!0} \,.	
\label{eq:hessian}}
We refer to $\hat K$ as an extended Hessian due to the inclusion of the global degrees of freedom. 

To second order in $q$, the change in energy $\Delta U = U - U^0$ associated with a deformation is
\eq{
	\Delta U &\approx  \left( \frac{\partial U}{\partial q_{\bar\alpha}} \right )_0 q_{\bar\alpha}
		+ \frac{1}{2} \left( \frac{\partial^2 U}{\partial q_{\bar\alpha} \, \partial q_{\bar\beta}} \right )_0 q_{\bar\alpha} q_{\bar\beta} \nonumber \\
	    &\approx \sigma^0_{ij}\, \epsilon_{ji} \, V^0 + \frac{1}{2}  \hat K_{\bar\alpha \bar\beta} \, q_{\bar\alpha} q_{\bar\beta}  \,,
	\label{eq:expansion}
}
where the strain tensor $\epsilon_{ij}$ is determined from the last $\Nbndry$ components of $q_{\bar \alpha}$.  
The linear term represents work done against the pre-stress. Only the strain degrees of freedom contribute to the linear term; all other contributions sum to zero as a result of force balance in the reference state.

Two observations follow directly from the energy expansion of Eq.~(\ref{eq:expansion}). First, the presence of a linear term indicates that packings where force balance is satisfied on every particle do not generically sit at a minimum of their energy $U$ with respect to boundary deformations (Fig.~\ref{fig:SS_energy_landscape}). Instead, gradients of the enthalpy-like quantity $H \equiv U -   \sigma^0_{ij} \, \epsilon_{ji} \, V^0$ vanish, $({\partial H}/{\partial q_{\bar\alpha}} )_0 = 0$: this requirement serves as a mechanical equilibrium condition.
Second, packings that are in {\em stable} $\Rap$ mechanical equilibrium under fixed confining stress must minimize $H$; this constrains the curvature of $\Delta H = H - H^0$, which is determined by the eigenvalues of the real and symmetric matrix $\hat K_{\bar \alpha \bar\beta}$. Packings that are only $\Ra$ stable do not minimize $H$ but still have the same constraints on the curvature of $\Delta H$. Defining $e_n$ and $\lambda_n$ to be the $n$th eigenvector and eigenvalue of $\He_{\bar\alpha\bar\beta}$, respectively,  we can write
\eq{
\Delta H = \frac{1}{2}  \hat K_{\bar\alpha \bar\beta} \, q_{\bar\alpha} q_{\bar\beta}  = \frac{1}{2} (q_{\bar\alpha} \, e_{n,\bar\alpha})^2 \lambda_n \,.
}
If $\lambda_n < 0$ for any mode, then the system is linearly unstable to perturbations along that mode. In this case, the system does not sit at a local energy minima and therefore is not jammed. 
In principle, zero modes ($\lambda_n=0$) are allowed, but if a zero mode has a non-zero projection onto any of the $\Nbndry$ boundary variables, then the system is unstable to that global deformation and again is not jammed. 
%If $\lambda_n = 0$, then 
%If such an unstable mode has any non-zero projection on any of the $\Nbndry$ boundary variables, then the system is unstable to that global deformation.

Therefore, for a system to be jammed according to the $\Ra$ requirement, it must satisfy
\eq{	\lambda_n \geq 0 \quad \forall n, 	\label{eq:jamming_def2}	}
and
\eq{	e_{n,\bar\alpha\p} &= 0 \quad \mbox{whenever $\lambda_n = 0$}, 	 \label{eq:jamming_def}	}
where $\bar\alpha\p$ runs only over the set of degrees of freedom associated with boundary deformations. Note that this definition automatically accounts for the presence of rattlers and the $d$ global translational zero modes.
%Thus, a system is jammed if $\He_{\bar\alpha \bar\beta}$ has no negative modes and if none of the zero modes couple to the boundary degrees of freedom.

For systems where the $\Rc$ requirement is the appropriate condition, jamming can be determined in much the same way. The only difference is in the relevant boundary variables and therefore the definition of the extended Hessian. Instead of considering all $d(d+1)/2$ boundary degrees of freedom, we only include isotropic compression/expansion. $\Nbndry = 1$ and the extended Hessian is thus a $dN+1$ by $dN+1$ matrix, but Eqs.~\eqref{eq:expansion}-\eqref{eq:jamming_def} follow identically.

For finite systems, the $\Ra$ requirement is significantly more strict than the $\Rc$ requirement. Packings made by standard jamming algorithms, which
are jammed according to the $\Rc$ requirement, can still have negative modes if shear deformations are included in the extended Hessian.  The fraction of states in the $\Ec$ ensemble that are also in the $\Ea$ ensemble is a function of $pN^2$ --- this fraction vanishes for small $pN^2$ but approaches 1 for large $pN^2$~\cite{DagoisBohy:2012dh}. %This is depicted schematically in Fig.~\ref{fig:diagram}.

We stress that the definition in Eqs.~\eqref{eq:jamming_def2} and \eqref{eq:jamming_def} considers the eigenvalues and vectors of the extended Hessian defined in Eq.~(\ref{eq:hessian}). Although it is possible to calculate elastic moduli, and thus the stability, from the usual ``reduced'' Hessian of Eq.~\eqref{eq:reduced_hessian}~\cite{Maloney:2006dt}, the eigenvalues of the reduced Hessian are not sufficient to determine if a system is jammed. Indeed, a packing can be unstable to global deformations even when the reduced Hessian is positive semi-definite because positive (or zero) modes can become negative when they are allowed to couple to the boundary. 

\subsection{Jamming criteria in terms of elastic constants\label{sec:jamming_from_elastic_constants}}
The $\Rc$ and $\Ra$ requirements that a system be stable to boundary deformations are equivalent to placing restrictions on the elastic moduli.
For isotropic systems, where the elasticity is described by the bulk modulus, $B$, and the shear modulus, $G$, the connection between stability requirements and
elastic moduli is simple: the $\Rc$ requirement is satisfied when the bulk modulus is positive, while the $\Ra$ requirement is satisfied when both the bulk and shear moduli are positive.

However, finite-sized systems are not isotropic. As a result, individual packings with periodic boundary conditions should be treated as crystals with the lowest possible symmetry. In this section, we will discuss the elastic constants of such systems.

A global affine deformation is given to lowest order by a specific strain tensor $\epsilon_{ij}$, which transforms any vector $r_i$ according  to
\eq{ r_i \rightarrow r_i + \sum_j \epsilon_{ij} r_j. \label{strain_tensor_transformation}	}
Note that in $d$ dimensions, the strain tensor has $d(d+1)/2$ independent elements. Now, when a mechanically stable system is subject to an affine deformation, it usually does not remain in mechanical equilibrium. Instead, there is a secondary, non-affine response, which can be calculated within the harmonic approximation from the Hessian matrix discussed above. Details of this calculation are presented in Refs.~\cite{Ellenbroek:2006df,Ellenbroek:2009dp}.

The change in energy can be written as
\eq{	
	\frac {\Delta U}{V^0} =  \sigma^0_{ij}  \epsilon_{ji}  +  \frac 12 c_{ijkl}\epsilon_{ij}\epsilon_{kl},	
	\label{elastic_modulus_tensor_def}
 }
where $c_{ijkl}$ is the $d \times d \times d \times d$ elastic modulus tensor and $V^0$ is again the volume of the initial reference state. 
The symmetries of $\epsilon_{ij}$ imply:
\eq{	c_{ijkl} = c_{jikl} = c_{ijlk} = c_{klij}.	}
When no further symmetries are assumed, the number of independent elastic constants becomes $\tfrac 18 d(d+1)(d^2+d+2)$, which is 6 in 2 dimensions and 21 in 3 dimensions.

It is convenient to express Eq.~\eqref{elastic_modulus_tensor_def} as a matrix equation by writing the elastic modulus tensor as a symmetric $d(d+1)/2$ by $d(d+1)/2$ dimensional matrix $\tilde c$ and the strain tensor as a $d(d+1)/2$ dimensional vector $\tilde \epsilon$. In 2 dimensions, for example, these are\footnote{While it is not necessary for our discussion here, the factors of 2 and 4 that appear in $\tilde c$ usually are included instead in $\tilde \epsilon$.}
\eq{	
	\begin{array}{c c}
\tilde c = \left( \begin{array}{ccc}
 c_{xxxx} &  c_{xxyy} & 2 c_{xxxy} \\
. &  c_{yyyy} & 2 c_{yyxy} \\
. & . & 4 c_{xyxy}
\end{array} \right),	&
\tilde \epsilon = \left( \begin{array}{c}
\epsilon_{xx} \\ \epsilon_{yy} \\ \epsilon_{xy}
\end{array} \right).
	\end{array}
	\label{C_colors_2d}
}
We can now rewrite Eq.~\eqref{elastic_modulus_tensor_def} as a matrix equation for the enthalpy-like quantity $\Delta H$:
\eq{	\frac{\Delta H}{V^0} = \frac 12 \tilde{\epsilon}^T \tilde{c}\,\tilde{\epsilon}. \label{DeltaU_matrixForm}}

We can now state the $\Rc$ and $\Ra$ requirements in terms of the anisotropic elastic moduli. The $\Rc$ requirement is that the system is stable against compression. This is measured by the bulk modulus, which can be written in terms of the elements of $c_{ijkl}$:
\eq{	B \equiv \frac 1{d^2} \sum_{k,l}c_{kkll}.	\label{B_def} }
The $\Rc$ requirement is satisfied if and only if $B>0$, which can be tested using Eqs.~\eqref{elastic_modulus_tensor_def} and \eqref{B_def}.

Unlike the bulk modulus, the shear modulus is not uniquely defined for anisotropic systems. Any traceless strain tensor constitutes pure shear, and to test the $\Ra$ requirement, we take a direct approach. The $\Ra$ requirement is satisfied if and only if $\Delta H > 0$ for all strain directions, {\it i.e.} for any $\epsilon_{ij}$. From Eq.~\eqref{DeltaU_matrixForm}, we see that this is the case if all the eigenvalues of $\tilde c$ are positive. Thus, the $\Ra$ requirement is satisfied if and only if $\tilde c$ is positive definite.

Note that the $\Rc$ and $\Ra$ requirements place different restrictions on the rank of $\tilde c$. For the $\Rc$ requirement, $\tilde c$ can have as few as one non-zero eigenvalue, while all $d(d+1)/2$ eigenvalues must be positive for the $\Ra$ requirement. This fact will be important in Sec.~\ref{sec:constraint_counting}.

\subsubsection{Useful elastic constant combinations\label{sec:usefulelasticconstants}}

Given the multitude of elastic constants, especially in higher dimensions, it is useful to divide them into 5 distinct types, based on their symmetry, as illustrated in Table~\ref{table:elastic_constant_types}. The most familiar are Types 1 and 2, which correspond to uniaxial compression and pure shear, respectively. 
For anisotropic systems, each elastic constant is independent and (generically) nonzero. However, our systems are \emph{prepared} under isotropic conditions; there is no {\it a priori} difference between any two axes, as there can be for crystals. Since the reference axes are arbitrary, we can rotate our coordinate system so that the elastic constant $c_{xxxx}$, for example, \emph{becomes} $c_{yyyy}$ in the new reference frame. The groups outlined in Table~\ref{table:elastic_constant_types} are defined so that any elastic constant can be rotated into another of the same type. They are thus conceptually equivalent, although of course their actual values will differ.

We will now exploit the conceptual distinction between the various types of elastic constants to define three orientation dependent moduli. A general description of this process is given in Appendix~\ref{AppendixA}, but for brevity we simply quote the results here. Let $\hat \theta$ be the set of generalized Euler angles that represent rotations in $d$ dimensions. The three $\hat \theta$-dependent moduli are the generalized shear modulus $G(\hat \theta)$, the modulus of uniaxial compression $U(\hat \theta)$, and the dilatancy modulus $D(\hat \theta)$. 
One could also construct an orientation dependent moduli for the Type 5 constants, but these only exist in three dimensions and will not be discussed here.
The bulk modulus is independent of orientation and is given by Eq.~\eqref{B_def}.

\begin{table}
\centering
\begin{tabular}{ | c | c | c | c |}
	\hline
	Type & Definition ($i \neq j \neq k$) & \# of constants & Example(s) \\ \hline
	1 & $c_{iiii}$ & $d$ & $c_{xxxx}$ \\ \hline
	2 & $c_{ijij}$ & $d(d-1)/2$ & $c_{xyxy}$ \\ \hline
	3 & $c_{iijj}$ & $d(d-1)/2$ & $c_{xxyy}$ \\ \hline
	4 & $c_{iiij}$, $c_{iijk}$ & $d^2(d-1)/2$ & $c_{xxxy}$, $c_{yyxz}$ \\ \hline
	5 & $c_{ijik}$ &$d(d-2)(d^2-1)/8$ &  $c_{xyxz}$ \\ \hline
\end{tabular}
\caption{\label{table:elastic_constant_types}Classification of elastic constants.}
\end{table}

As an example, consider the generalized shear modulus $G(\theta)$ in two dimensions. The set of symmetric, traceless strain tensors can be parameterized by the \emph{shear angle} $\theta$:
\eq{	\epsilon(\theta) = \frac \gamma 2 \left( \begin{array}{cc} \sin(2\theta) & \cos(2\theta) \\ \cos(2\theta) & -\sin(2\theta) \end{array} \right),	\nonumber}
where $\gamma \ll 1$ is the magnitude of the strain. When $\theta=0$, the response is given by $c_{xyxy}$, but when $\theta=\pi/4$, the response is $\tfrac 14(c_{xxxx}+c_{yyyy}-2c_{xxyy})$. For arbitrary angles, the response $G(\theta)$ is a sinusoidal function of $\theta$~\cite{DagoisBohy:2012dh} (see Appendix~\ref{AppendixA}).

\begin{figure}[htp]
	\centering
	\epsfig{file=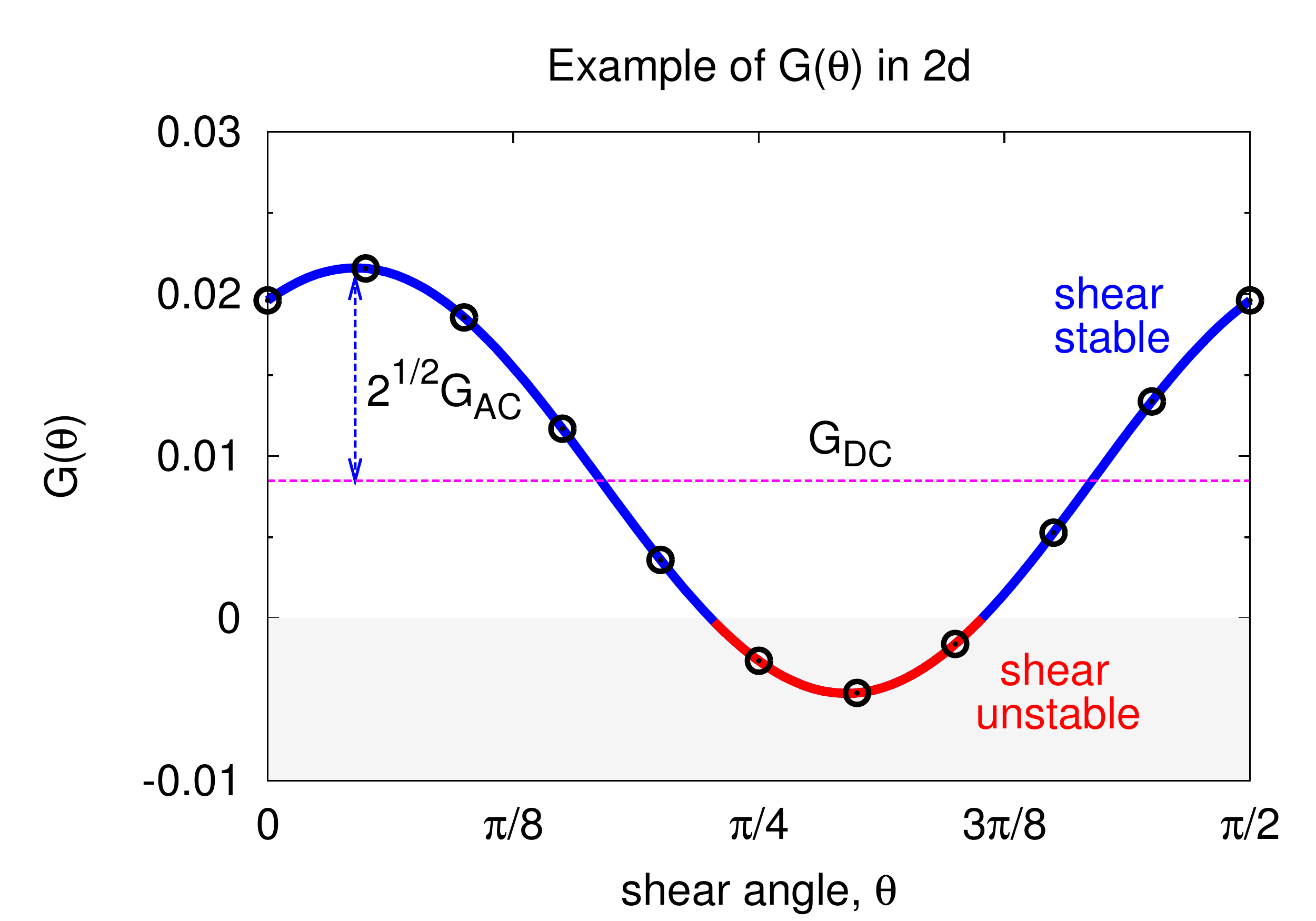,height=0.75\linewidth,angle=-90}
	\caption[Example of the sinusoidal function $G(\theta)$.]{\label{fig:G_theta}
	Example of the sinusoidal function $G(\theta)$ for a two-dimensional system of $N=256$ particles at a pressure of $p=10^{-3}$. The black circles show direct numerical calculations using various $\epsilon(\theta)$, while the solid line shows the prediction using the elastic modulus tensor at $\theta=0$ (see Appendix~\ref{AppendixA}). The system is stable to shear when $G(\theta)>0$ (blue) and unstable when $G(\theta)\leq 0$ (red). The data agree with the prediction. The horizontal dashed line shows the average $\Gdc$; $G_{AC}$ is obtained from the amplitude of the sinusoidal curve. 
	}
\end{figure}

An example of $G(\theta)$ for a two-dimensional $\Ec$ packing is shown in Fig.~\ref{fig:G_theta}. Notice that there is a range of angles for which $G(\theta)<0$, implying that the system is unstable to that set of shear deformations. By construction, this does not occur for systems in the $\Ea$ and $\Eap$ ensembles. We define the angle-averaged shear modulus $\Gdc$ to be (see Fig.~\ref{fig:G_theta})
\eq{	G_{DC} &\equiv \frac{1}{\pi} \int_0^{\pi} G(\theta) d\theta.	}
We can also define $G_{AC}$ to characterize the variation of $G(\theta)$ about this average:
\eq{	G_{AC}^2 &\equiv \frac{1}{\pi} \int_0^{\pi} \left( G(\theta) - G_{DC} \right)^2 d\theta. \label{eq:G_AC_def}}

Note that for an isotropic system, $\Gdc = G$ ({\it i.e.}, the usual shear modulus) and $G_{AC}=0$. In three dimensions, the generalized shear modulus is no longer a simple sinusoidal function and instead depends on the three Euler angles. Nevertheless, we can still define $\Gdc$ and $G_{AC}$ to be the mean and standard deviation of the response to shear. This is discussed in detail in Appendix~\ref{AppendixA}.

In a similar manner, $U(\hat \theta)$ measures the response to uniaxial compression along an axis determined by $\hat \theta$. The full expression for $U(\hat \theta)$ is more complicated than $G(\hat \theta)$, but we can still define $U_{DC}$ and $U_{AC}^2$ to be the average and variance of $U(\hat \theta)$, respectively. However, since $U_{DC}$ can be expressed in terms of the bulk modulus and average shear modulus,
\eq{	U_{DC} = B + G_{DC}, \label{eq:Udc_from_B_and_G}}
it is redundant and will not be considered further. Finally, the Type 4 dilatancy constants can be generalized to $D(\hat \theta)$, and the average and variance defined as $D_{DC}$ and $D_{AC}^2$. One important result is that $D_{DC} = 0$ for any individual system (see Appendix~\ref{AppendixA}) and therefore will not be discussed further. 

In summary, we will consider the elastic constant combinations $B$, $G_{DC}$, $G_{AC}$, $U_{AC}$, and $D_{AC}$.  Expressions for these quantities in terms of the original elastic constants, $c_{ijkl}$, are provided in Appendix~\ref{AppendixA}.  Note that of these five quantities, $B$ and $G_{DC}$ reduce to the bulk and shear modulus, respectively, in the thermodynamic limit, which is isotropic.  As expected, we will see in Sec.~\ref{mvh2} that the remaining combinations,  $G_{AC}$, $U_{AC}$, and $D_{AC}$, vanish in the thermodynamic limit.

\subsection{Constraint counting and isostaticity\label{sec:constraint_counting}}
Earlier, we indicated that the contact number $Z$ is not an ideal metric for determining whether a system is jammed. However, the value of $Z$ at the jamming transition is of considerable importance. In this subsection, we %review arguments from Ref.~\cite{Goodrich:2012ck} that 
derive the exact value of $Z$ at the jamming transition for packings of frictionless spheres in finite-sized systems in the $\Ec$ ensemble. We also generalize the arguments to include the $\Ea$ and $\Eap$ ensembles and find that the contact number at the transition for these ensembles is slightly different~\cite{DagoisBohy:2012dh,Goodrich:2012ck}. This difference in contact number is easily understood from the additional degrees of freedom associated with boundary deformations that need to be constrained in the $\Ea$ and $\Eap$ ensembles. Furthermore, we will see in Sec.~\ref{sec:finite_size_scaling} that once this slight difference is taken into account, the increase in contact number with pressure is identical for the various ensembles. % (see Figs.~\ref{fig:finite_size_scaling} and \ref{fig:corrections_to_scaling}).

As discussed above, a system is isostatic when the number of constraints equals the number of degrees of freedom. Such a statement hides all subtleties in the definition of the relevant constraints and degrees of freedom.
For example, for a system with periodic boundary conditions in $d$ dimensions, particle-particle contacts cannot constrain global translational motion. Therefore, the isostatic number of contacts is
\eq{	\Nciso \equiv dN_0 - d,	}
where $N_0$ is the number of particles in the system after the rattlers have been ignored.  The isostatic contact number is therefore $\Ziso \equiv 2d - 2d/N_0$, which approaches $2d$ in the thermodynamic limit.

We now revisit the relationship between isostaticity and the jamming transition for packings of frictionless spheres. Suppose that $\Nzm$ of the total $dN+\Nbndry$ vibrational modes of the extended Hessian are zero modes, meaning they have zero eigenvalue. As before, $\Nbndry$ depends on the boundary conditions: $\Nbndry = 1$ in the $\Ec$ ensemble and $\Nbndry = d(d+1)/2$ in the $\Ea$ and $\Eap$ ensembles. A particle-particle contact has the potential  to constrain at most one degree of freedom, and every unconstrained degree of freedom results in a zero mode. Therefore, the number of contacts must satisfy
\eq{\Nc \geq dN + \Nbndry - \Nzm. \label{Nc_inequality} }
 Equation~\eqref{Nc_inequality} is an inequality because some contacts might be redundant, meaning they could be removed without introducing a zero mode. Such redundancies correspond to states of self stress, and Eq.~\eqref{Nc_inequality} can be written as $\Nc = dN + \Nbndry - \Nzm + S$, where $S$ is the number of states of self stress~\cite{Lubensky:2015tz}.
 %\cite{StenullLubensky}.

The $d$ global translations, as well as every rattler, each lead to $d$ trivial zero modes.
We will now use the numerical result that the only zero modes observed in jammed sphere packings are those associated with global translation and rattlers~\cite{Goodrich:2012ck}.  
Thus, the total number of zero modes in a jammed system is $\Nzm = d + d(N-N_0)$, and $\Nc$ and $Z$ must satisfy
\eqs{	
	\Nc & \geq N_\text{c,min} \equiv  \Nciso + \Nbndry, \\
	Z &\ge \Zmin \equiv \Ziso + \frac 2{N_0} \Nbndry.
	\label{Zmin_prediction}
}

If a system is exactly isostatic, then it has enough contacts to constrain the position of every particle, but it does \emph{not} have enough contacts to constrain the global degrees of freedom, and thus cannot be jammed. Since the $\Rc$ and $\Ra$ requirements do not explicitly forbid nontrivial zero modes, it is possible for the global variables to become constrained \emph{before} all the positional degrees of freedom. While this indeed occurs for ellipsoid packings, the fact that this is never observed for sphere packings implies that zero modes associated with translations of the spheres are extended and inevitably interact with the boundary.

\begin{figure}[htp]
	\centering
	\epsfig{file=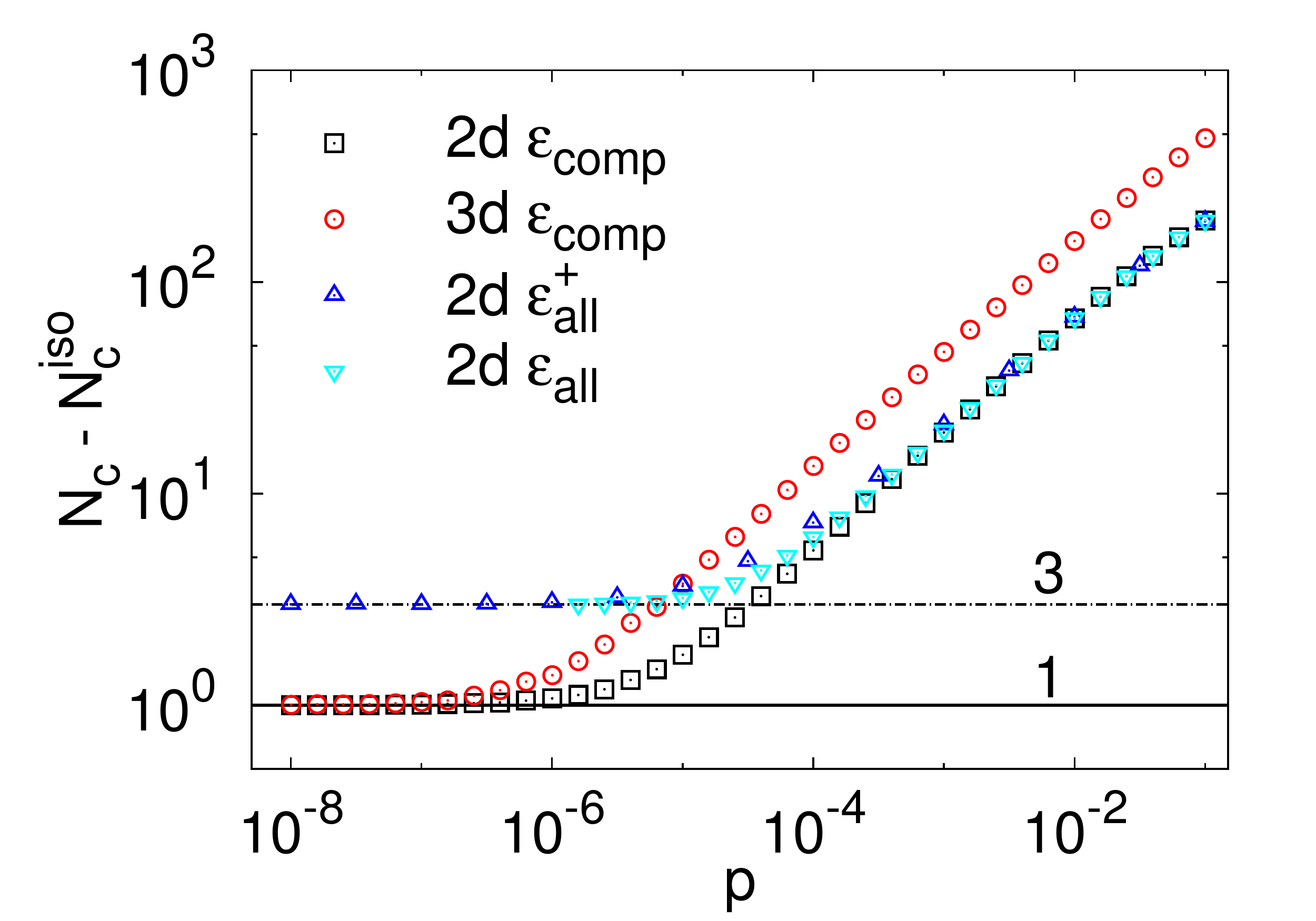,width=0.4\linewidth,angle=-90}
	\caption[The total number of contacts above isostaticity.]{The total number of contacts $\Nc$ above the isostatic number $\Nc^\text{iso}$ as a function of pressure for systems of $N=256$ particles. The solid horizontal line is at $\Nc-\Nciso=1$ and the dashed horizontal line is at $\Nc-\Nciso=3$.}
	\label{FigureNcmNciso}
\end{figure}

While Eq.~\eqref{Zmin_prediction} states that a system can only be jammed if $\Nc \geq \Nciso + \Nbndry$, this is clearly not a sufficient condition for jamming because some of the contacts could be redundant and not contribute to the overall rigidity of the system. However, we find numerically that Eq.~\eqref{Zmin_prediction} is indeed an equality as the transition is approached (provided the system is sufficiently disordered, recall the discussion in Sec.~\ref{sec:JammingCriteria}I regarding bidisperse packings in two dimensions). This is demonstrated in Fig.~\ref{FigureNcmNciso}, which shows that the number of contacts above isostaticity, in the limit of zero pressure, approaches $\Nc - \Nciso \rightarrow 1$ for the $\Ec$ ensembles (where $\Nbndry = 1$) and $\Nc - \Nciso \rightarrow 3$ for the $\Ea$ and $\Eap$ ensembles (where $\Nbndry = 3$). Importantly, we do not find \emph{any} systems that are jammed ({\it i.e.} satisfy Eqs.~\eqref{eq:jamming_def2} and~\eqref{eq:jamming_def}) but do not satisfy Eq.~\eqref{Zmin_prediction}.

Finally, note that the $\Nbndry$ additional contacts required for jamming can also be understood in terms of the normal reduced hessian and the matrix $\tilde c$ discussed in Sec.~\ref{sec:jamming_from_elastic_constants}. $\Nciso$ contacts are needed to remove any nontrivial zero modes from the reduced hessian. However, the $\Rc$ and $\Ra$ requirements necessitate $\Nbndry$ positive eigenvalues of $\tilde c$, leading to the additional $\Nbndry$ contacts in Eq.~\eqref{Zmin_prediction}.

\section{Numerical results\label{sec:results}}

In this section we examine the finite-size scaling behavior of the contact number and the elastic constants as a function of system size, $N$, and proximity to the jamming transition, which we quantify by the pressure, $p$, which vanishes at the transition. We will focus on soft-sphere potentials that have harmonic interactions (see Appendix~\ref{sec:numerical_procedures} for details), but extending our results to other soft-sphere potentials is straightforward~\cite{Liu:2010jx}. 

In Sec.~\ref{sec:finite_size_scaling} we present results for the excess contact number, $Z-\Ziso$, as well as for the two elastic constant combinations, $B$ and $G_{DC}$, that approach the bulk and shear moduli, respectively, in the thermodynamic limit (see Sec.~\ref{sec:usefulelasticconstants}).  Section~\ref{mvh2} contains the finite-size scaling results for the three ``$AC$" elastic constant combinations that vanish in the thermodynamic limit (again defined in Sec.~\ref{sec:usefulelasticconstants}).  Finally, Sec.~\ref{sec:statistical_fluctuations} examines the standard deviation of the distributions of the nonvanishing quantities, $Z-\Ziso$, $B$ and $G_{DC}$, namely $\sigma_Z$, $\sigma_B$ and $\sigma_{\Gdc}$.  These standard deviations must also vanish in the thermodynamic limit relative to the mean.  We note that when a single measurement of the response to shear, for example, is performed on a finite packing, both the angular variation and statistical fluctuations play a role ---
in earlier work we have shown examples where the angular and statistical fluctuations are taken together~\cite{DagoisBohy:2012dh}.

The results presented below can be summarized as follows.  First, we find subtle differences in $Z-\Ziso$, $B$ and $G_{DC}$ between the $\Ec$, $\Ea$ and $\Eap$ ensembles. These differences vanish as $pN^2 \rightarrow \infty$.  In addition, $G_{AC}$, $U_{AC}$, and $D_{AC}$ all vanish in the thermodynamic limit, as expected, and the fluctuations, $\sigma_Z$, $\sigma_B$ and $\sigma_{\Gdc}$, all vanish as $1/\sqrt{N}$ relative to the mean. All 6 quantities that vanish in the thermodynamic limit ($G_{AC}$, $U_{AC}$, and $D_{AC}$, $\sigma_Z$, $\sigma_B$ and $\sigma_{\Gdc}$) collapse with $pN^2$ in all 3 ensembles, with the exception of $U_{AC}$ and $D_{AC}$, which only collapse in the $\Eap$ ensemble, where there is no residual shear stress.  We will discuss these two exceptions further below.  In all, these results show that the thermodynamic limit is well defined for any $p$, although the number of particles needed to observe this limit diverges as the jamming transition is approached.  

Second, we find non-trivial finite-size corrections to the scaling of $Z-\Ziso$, $B$ and $G_{DC}$, in all three ensembles. %, as found for the $\Ec$ ensemble earlier~\cite{Goodrich:2012ck}. 
These corrections scale with the total system size, $N$, rather than the system length, $L$, in 2 and 3 dimensions. %, consistent with Ref.~\cite{Goodrich:2012ck}.  
In addition, we find that the two-dimensional results can be better described when logarithmic corrections to scaling are included.  These results therefore reinforce the conclusion that  jamming is a phase transition with an upper critical dimension of two. 

\begin{figure*}
	\centering
	\epsfig{file=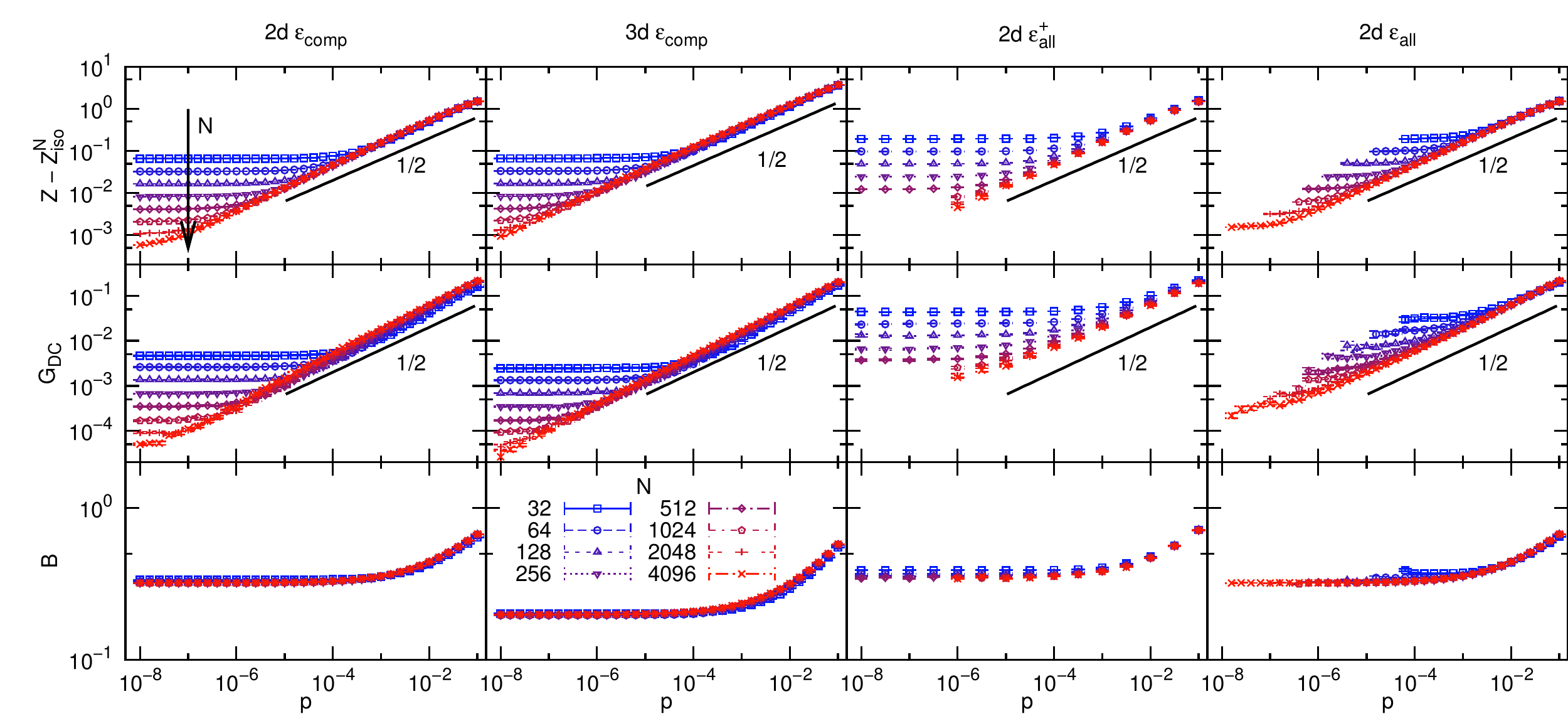,height=0.98\linewidth,angle=-90}
	\caption[The excess contact number, shear modulus, and bulk modulus as a function of pressure and system size.]{The excess contact number ($Z - \Ziso$, top row), shear modulus ($G_{DC}$, middle row) and bulk modulus ($B$, bottom row) as a function of pressure for system sizes ranging from $N=32$ (blue) to $N=4096$ (red). 
	The four columns show, from left to right, the $2d$ $\Ec$ ensemble, the $3d$ $\Ec$ ensemble, the $2d$ $\Eap$ ensemble, and the $2d$ $\Ea$ ensemble. The solid black lines all have slope $1/2$.}
	\label{FigureRawData}
\end{figure*}

\subsection{Finite-size scaling: $Z-\Ziso$, $G_{DC}$ and $B$\label{sec:finite_size_scaling}}

In this section we probe the finite-size scaling of the ensemble-averaged values of the angle-independent quantities that do not vanish in the thermodynamic limit:
the contact number above isostaticity, $Z-\Ziso$, the shear modulus, $G_{DC}$, and the bulk modulus, $B$.  We study these for all three ensembles defined earlier.

\subsubsection{Finite-size plateau}

Figure~\ref{FigureRawData} shows the excess contact number, $Z- \Ziso$, average shear modulus,
$G_{DC}$, and bulk modulus, $B$, as a function of pressure for different system sizes and ensembles. At high pressures we measure the scaling relationship $Z-\Ziso \sim p^{1/2}$ that has previously been observed~\cite{Durian:1995eo,OHern:2003vq,Liu:2010jx} for harmonic interaction potentials. However, at low pressures the excess contact number plateaus to $2\Nbndry/N_0$. As expected, this correction to the excess contact number due to stabilizing the boundaries is a finite-size effect: as the system size increases, the onset pressure of this plateau decreases so that $Z-\Ziso \sim p^{1/2}$ is valid for all pressures in the thermodynamic limit. 

Similar to the excess contact number, the shear modulus has a high-pressure regime that conforms to the known scaling of $G_{DC} \sim p^{1/2}$ and a low-pressure plateau that scales as $1/N$. This plateau also vanishes in the thermodynamic limit and is a finite-size effect. 
The nearly constant behavior of the bulk modulus as a function of pressure is consistent with previous results and persists for large systems.

The fact that $\Gdc/B$ in the limit of zero pressure is proportional to $1/N$ and thus vanishes for large systems is a key feature of the jamming transition. In random spring networks, which are often used to model disordered solids, both the shear and bulk moduli vanish when the system approaches isostaticity such that the ratio of the two remains finite~\cite{Ellenbroek:2009to}. The only model system we are aware of that exhibits this jamming-like behavior in $\Gdc/B$ is the set of ``generic" rational approximates to the quasi-periodic Penrose tiling. In recent work~\cite{Stenull:2014hv}, Stenull and Lubensky show that such networks near isostaticity have constant bulk modulus (for sufficiently large $N$) and a shear modulus that vanishes with $1/N$. Their results are also consistent with our discussion in Sec.~\ref{sec:constraint_counting}.

\subsubsection{Finite-size scaling of excess contact number, bulk and shear moduli}

\paragraph{Contact number:}
If jamming is a phase transition, then quantities like the excess contact number, $Z-\Ziso$, must be analytic for finite $N$. However, the bulk scaling of $Z-\Ziso \sim p^{1/2}$ that has been known for over a decade~\cite{Durian:1995eo,OHern:2003vq} is clearly not analytic at $p=0$. Thus, there must be finite-size rounding of this singular behavior if jamming is to be considered critical. We already saw that finite-size effects in $Z-\Ziso$ emerge in the limit of zero pressure, resulting in a plateau that is proportional to $1/N$. Here, we predict additional finite-size effects that cannot be understood from constraint counting alone. Instead, we will use our theoretical understanding of the low-pressure plateau and the high-pressure scaling to infer the scaling of more subtle finite-size effects that arise from the assumption that $Z-\Ziso$ is analytic for finite $N$.  %These arguments were presented in an abbreviated form in Ref.~\cite{Goodrich:2012ck}, and are included with more detail here.

\begin{figure*}
	\centering
	\epsfig{file=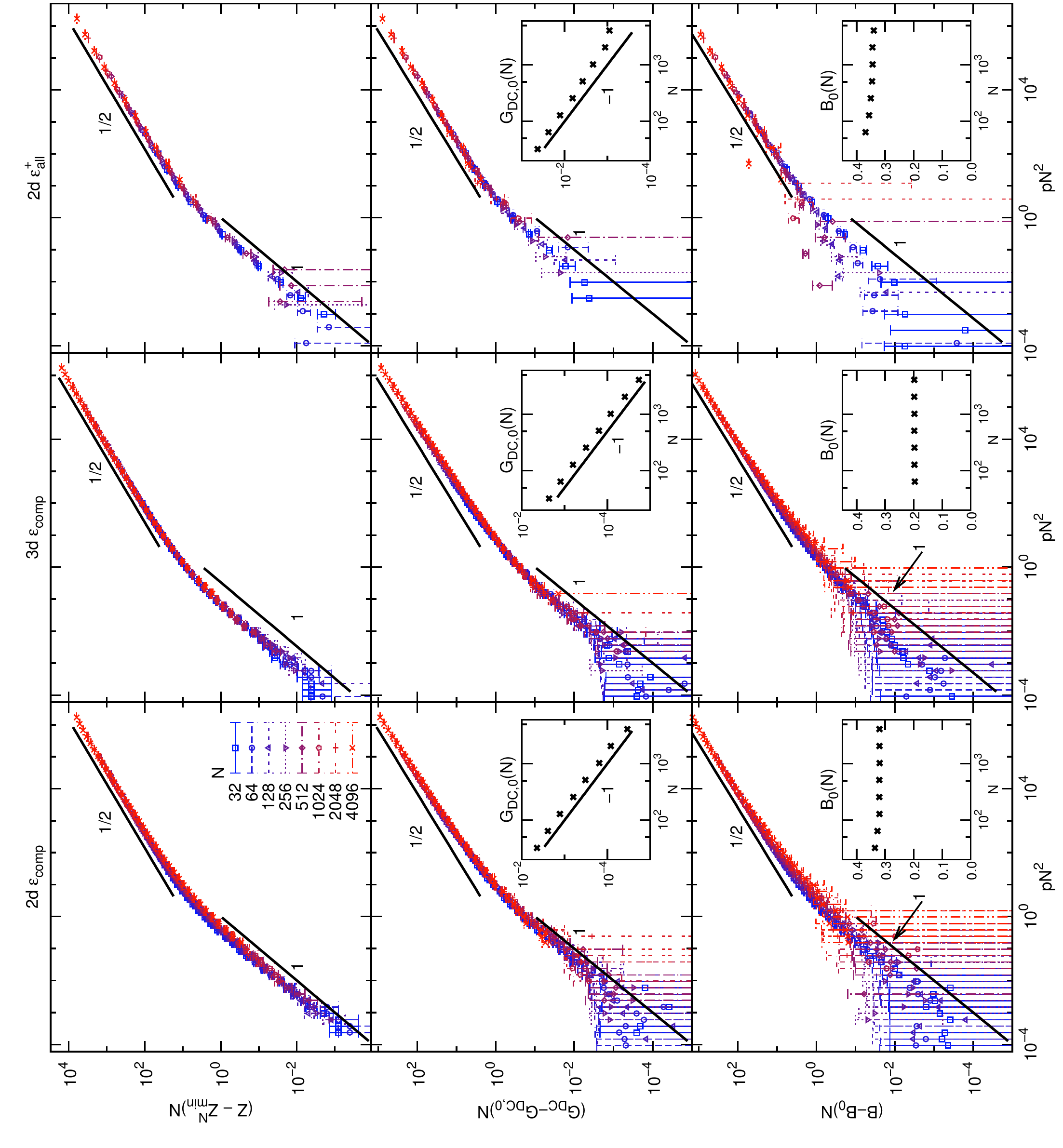,height=1\linewidth,angle=-90}
	\caption[Finite-size scaling collapse of the excess contact number and average elastic moduli.]{\label{fig:finite_size_scaling}Finite-size scaling collapse of the excess contact number and average elastic moduli. Top row: $Z$ minus the theoretical minimum $\Zmin$ (see Eq.~\eqref{Zmin_prediction}). $\left(Z-\Zmin\right)N$ collapses as a function of $pN^2$ for the 2d $\Ec$ (left), 3d $\Ec$ (middle) and 2d $\Eap$ (right) ensembles. Note that at low $pN^2$, most of the $\Ec$ packings are $\Ra$ unstable and our filtered, 2d $\Ea$ ensemble does not have many states at low $pN^2$. This is why data is not shown for this ensemble. At large $pN^2$, $Z-\Zmin \sim Z-\Ziso \sim p^{1/2}$ (Eq.~\eqref{ZmZiso_p_scaling}), while $Z-\Zmin \sim pN$ at low $pN^2$ (Eq.~\eqref{ZmZmin_predicton}). The crossover between these scalings occurs when the total number of extra contacts is of order 10.
	Middle row: $\Gdc$ minus the measured $p\rightarrow 0$ plateau. $\left(\Gdc - G_{DC,0}\right)N$ collapses as a function of $pN^2$ and has the same crossover behavior as $\left(Z-\Zmin\right)N$. The insets show that $G_{DC,0}$ is proportional to $N^{-1}$.
	Bottom row: $B$ minus the measured $p\rightarrow 0$ plateau. Note that the plateau $B_0$ of the bulk modulus is much larger than for the shear modulus. Therefore, uncertainties in $B$ lead to the large error bars in $\left(B- B_0\right)N$ at low $pN^2$. The insets show that $B_0$ is roughly constant in $N$, as expected for particles with harmonic interactions. It is not clear from the data whether there is an additional $N^{-1}$ contribution to the plateau ({\it i.e.} $B_0(N) = B_0(\infty) + a N^{-1}$).
	The colors and symbols are the same as in Fig.~\ref{FigureRawData}.
	}
\end{figure*}

The assertion that jamming is a phase transition leads to a falsifiable prediction for the form of the scaling function.  First we summarize the three main ingredients of the argument. {\em(i)} The low pressure plateau in $Z-\Ziso$ derives from the extra contact(s) needed to satisfy the jamming criteria and is proportional to $1/N$. {\em(ii)} In the limit of large $N$ and at sufficiently large pressures, $Z-\Ziso$ exhibits power-law scaling with a known exponent of $1/2$:
\eq{	Z-\Ziso \sim p^{1/2}.	\label{ZmZiso_p_scaling} }
{\em(iii)} $Z$ is analytic in $p$ for finite $N$.

From the first two assertions, we see that if finite-size scaling is obeyed, it must be of the form
\eq{	Z-\Ziso = \frac 1N F\left( pN^2\right),	\label{scaling_form} }
where $F(x)$ is a scaling function that must satisfy, first, that
$F(x)\sim 1$ for small $x$, second, that $F(x) \sim x^{1/2}$ for large $x$, and third, that $F(x)$ is analytic in $x$ at $x=0$.

The third requirement regarding analyticity implies that
the expansion of the contact number for small $p$ takes the form
\eq{	\left(Z-\Ziso\right)N = c_0 + c_1 pN^{2} + ..., \label{ZmZiso_expansion}}
where $c_0 = 2\Nbndry$ gives the zero pressure plateau and $c_1$ is a constant.
Although the leading terms in the expansion clearly fail to describe the $Z-\Ziso \sim p^{1/2}$ scaling at large pressure, they should be valid at small pressure.
Our reasoning thus predicts that as the pressure vanishes, the contact number should approach its limiting value $\Zmin$ as
\eq{	Z-\Zmin \approx c_1 pN \qquad \text{for $p \ll 1$},	\label{ZmZmin_predicton} }
where the constant $c_1$ is independent of system size. Furthermore, there should be a crossover between this low-pressure regime and a high-pressure regime where $Z-\Zmin \sim Z-\Ziso \sim p^{1/2}$.

This is verified in the top row of Fig.~\ref{fig:finite_size_scaling}, which shows $\left(Z-\Zmin\right)N$ as a function of $pN^2$. The scaling with exponent $1/2$ at high $pN^2$ is consistent with Eq.~\eqref{ZmZiso_p_scaling}, while the slope of $1$ at low  $pN^2$ is consistent with Eq.~\eqref{ZmZmin_predicton}. Since $\left(Z-\Zmin\right)N$ is exactly twice the total number of contacts above the minimum ({\it i.e.}, $\Nc - N_\text{c,min}$), our data shows that the crossover to the low-pressure regime occurs when the total number of extra contacts in the system is of order 10, regardless of the system size. Importantly, the low-pressure scaling is not predicted from constraint counting arguments and data collapse in this region is not trivial. However, both follow immediately from the notion that jamming is a phase transition.

\paragraph{Shear modulus:}
We now turn our attention to the average shear modulus $\Gdc$. We saw in Fig.~\ref{FigureRawData} that the behavior of $\Gdc$ is strikingly similar to that of $Z-\Ziso$. Specifically, the shear modulus deviates from the canonical $\Gdc \sim p^{1/2}$ scaling at low pressure and instead exhibits a plateau that decreases with system size. As we discussed above, this plateau is due to the $\Rc$ and $\Ra$ requirements that there are at least $\Nbndry$ constraints above the isostatic value.

Since $Z -\Ziso \sim N^{-1}$ in the zero-pressure limit, one would also expect the plateau in $\Gdc$ to be proportional to $N^{-1}$. Using the same reasoning as above, if finite-size scaling exists in the shear modulus it must be of the form $\Gdc N \sim  F(pN^2)$, where again $F(x)\sim 1$ for small $x$ and $F(x) \sim x^{1/2}$ for large $x$.
Also, the assertion that $\Gdc$ is analytic for finite $N$ implies that the low-pressure limit of the shear modulus is of the form
\eq{	\Gdc N = g_0 + g_1 pN^{2} + ...}
where $g_0$ and $g_1$ are constants.

The middle row of Fig.~\ref{fig:finite_size_scaling} confirms this scaling. For each ensemble and system size, we first calculated the plateau value $G_{DC,0}$ of $\Gdc$, and then plotted $\left(\Gdc-G_{DC,0}\right)N$ as a function of $pN^2$. The values of $G_{DC,0}$ are shown in the insets and are proportional to $N^{-1}$, confirming that $g_0$ is indeed constant.
$\Gdc$ increases from this plateau at low pressures with $pN$ before crossing over to the known $p^{1/2}$ scaling.

\paragraph{Bulk modulus:} The same reasoning as above can also be applied to the scaling of the bulk modulus. As the bottom row of Fig.~\ref{fig:finite_size_scaling} shows, our data appear consistent with $(B-B_0)N$ scaling linearly with $pN^2$ close to the transition.  However, the error bars are very large as the plateau value for the bulk modulus is orders of magnitude larger than that of the shear modulus so the bulk modulus does not supply nearly as strong support for the existence of nontrivial scaling as the shear modulus and coordination number.

The finite-size effects presented in Figs.~\ref{FigureRawData} and \ref{fig:finite_size_scaling} clearly depend on the pressure, which is a useful measure of the distance to jamming for an individual system. 
A recent paper~\cite{Liu:2014gu}, however, claims to see finite-size scaling of the contact number and shear modulus with $(\phi-\phi_{c,\infty})L^{1/\nu}$, where $\nu \approx 0.8$, which is the same scaling that controls the mean of the distribution of critical packing fractions~\cite{OHern:2003vq,Vagberg:2011fe}.
To understand this, note that there are two different finite-size effects that come into play: 1) the corrections to $\phi_c$ that scale with $(\phi-\phi_{c,\infty})L^{1/\nu}$, and 2) the rounding shown in Figs.~\ref{FigureRawData} and \ref{fig:finite_size_scaling} that scale with $pN^2\sim pL^{2d}$. Since $1/\nu<2d$, one would expect the corrections to $\phi_c$ to influence the contact number and shear modulus over a broader range of $\phi$, leading to the observations of Ref.~\cite{Liu:2014gu}. However, the true behavior of these quantities as a function of $\phi$ is a convolution of the two finite-size effects. Thus, given their different scaling, finite-size collapse does not exist as a function of $\phi$. The appearance of scaling collapse observed in Ref.~\cite{Liu:2014gu} is because their data is not sufficiently sensitive at low pressures.

\begin{figure}
	\centering
	\epsfig{file=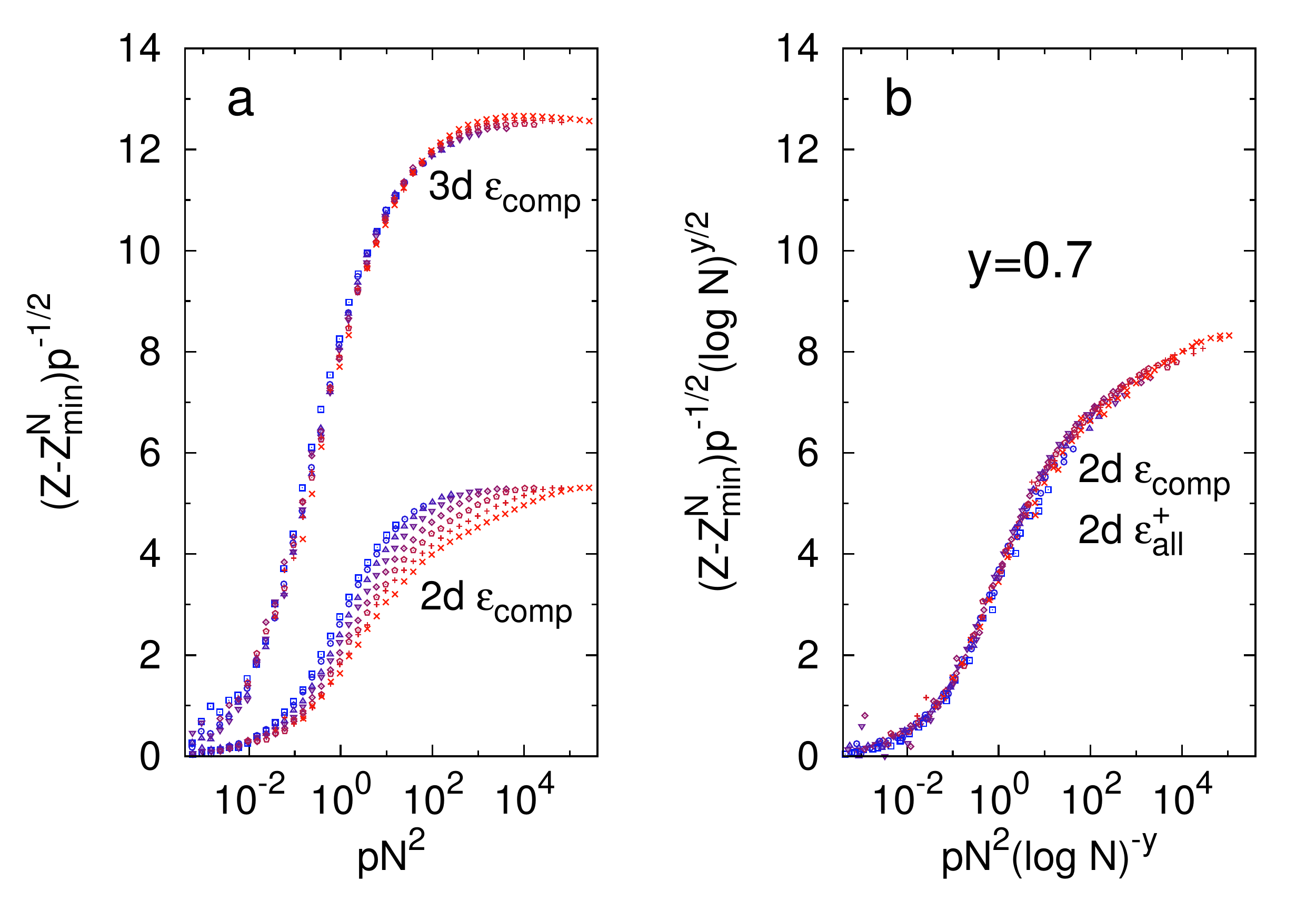,height=0.7\linewidth,angle=-90}
	\caption[Logarithmic corrections to scaling.]{\label{fig:corrections_to_scaling}Logarithmic corrections to scaling. a) $\left(Z-\Zmin\right)p^{-1/2}$ as a function of $pN^2$ on a linear scale for the two and three dimensional $\Ec$ ensembles. The $3d$ data shows good collapse but there is a system size dependence in $2d$ that was not as clear in Fig.~\ref{fig:finite_size_scaling}. The two dimensional $\Eap$ data (not shown) is indistinguishable from the 2d $\Ec$ data. b) $\left(Z-\Zmin\right)p^{-1/2}\left(\log N\right)^{y/2}$ as a function of $pN^2\left(\log N\right)^{-y}$, with $y=0.7$. The 2d $\Ec$ and 2d $\Eap$ ensembles are both shown and collapse perfectly onto each other. This shows that the system size dependence in a) can be accounted for by introducing a logarithmic correction of the form of Eq.~\eqref{scaling_form_log_corrections}. The colors and symbols are the same as in Fig.~\ref{FigureRawData}.}
\end{figure}

\subsubsection{Corrections to scaling in two dimensions}

We return now to the scaling for the contact number, and note the quality of the data collapse in three dimensions, which spans over 8 decades in $pN^2$ and over 5 decades in $\left(Z-\Zmin\right)N$ (see Fig.~\ref{fig:finite_size_scaling}). In both of the two-dimensional ensembles, however, there is a very slight systematic trend at intermediate $pN^2$. This can be seen more clearly by dividing $\left(Z-\Zmin\right)N$ by $p^{1/2}N$ and showing the data on a linear scale. Figure~\ref{fig:corrections_to_scaling}a shows that the collapse of the $3d$ data remains extremely good while there are clear deviations in the $2d$ data.

These deviations can be interpreted as corrections to scaling, which are often observed in critical phenomena at the upper critical dimension. 
One would expect potential corrections to scaling to be logarithmic and lead to scaling of the form
\eq{	Z-\Ziso = \frac 1N F\left( pN^2/\left(\log N\right)^y \right),	 \label{scaling_form_log_corrections} }
with some exponent $y$. Figure~\ref{fig:corrections_to_scaling}b shows both the 2d $\Ec$ data and the 2d $\Eap$ data scaled according to Eq.~\eqref{scaling_form_log_corrections}. We find that including a logarithmic correction with $y = 0.7 \pm 0.1$ leads to very nice data collapse in two dimensions.

The finite-size scaling that we observe depends on the total number of particles $N$ rather than the linear size of the system $L\sim N^{1/d}$. Such scaling is typically associated with first-order transitions and with second-order transitions above the upper critical dimension~\cite{BINDER:1985vl,Dillmann:1998ty}. Along with the corrections to scaling that we see in $d=2$, this is consistent with the notion that jamming is a mixed first/second order phase transition with an upper critical dimension of $d_\text{c} = 2$, in accord with previous results
~\cite{Liu:2010jx,Charbonneau:2012fl,Goodrich:2012ck,Wyart:2005vu}.

Unlike $\left(Z-\Ziso\right)N$, which approaches the same small pressure plateau in every individual system, the plateaus in $\Gdc$ vary from system to system. It is only when averaged over many systems that $G_{DC,0}$ has a clear $N^{-1}$ scaling. This explains why $(\Gdc - G_{DC,0})N$ is much more noisy at low $pN^2$ than $(Z-\Zmin)N$, which makes it impossible to see from our data whether or not there are corrections to scaling in $\Gdc$ in two dimensions.

\begin{figure*}[htp]
	\centering
	 \epsfig{file=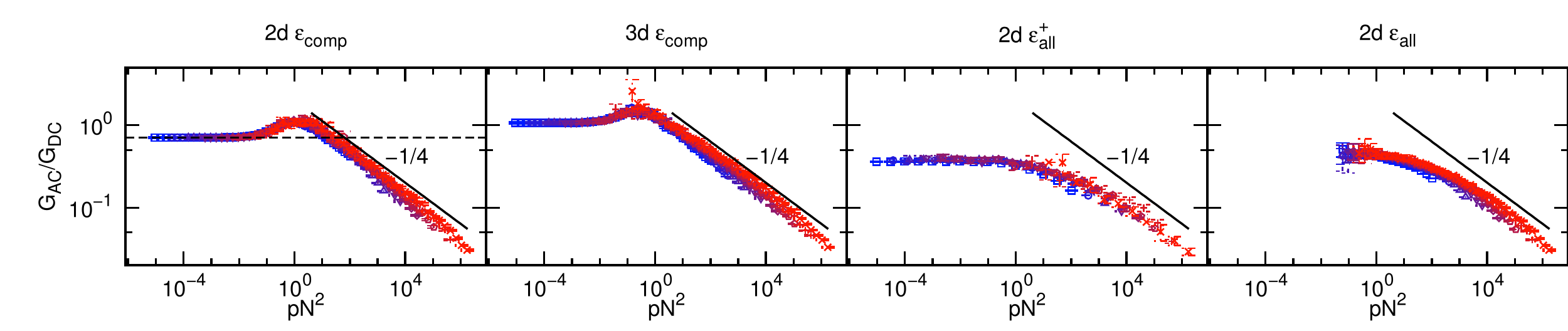,height=1\linewidth,angle=-90}\\
	\epsfig{file=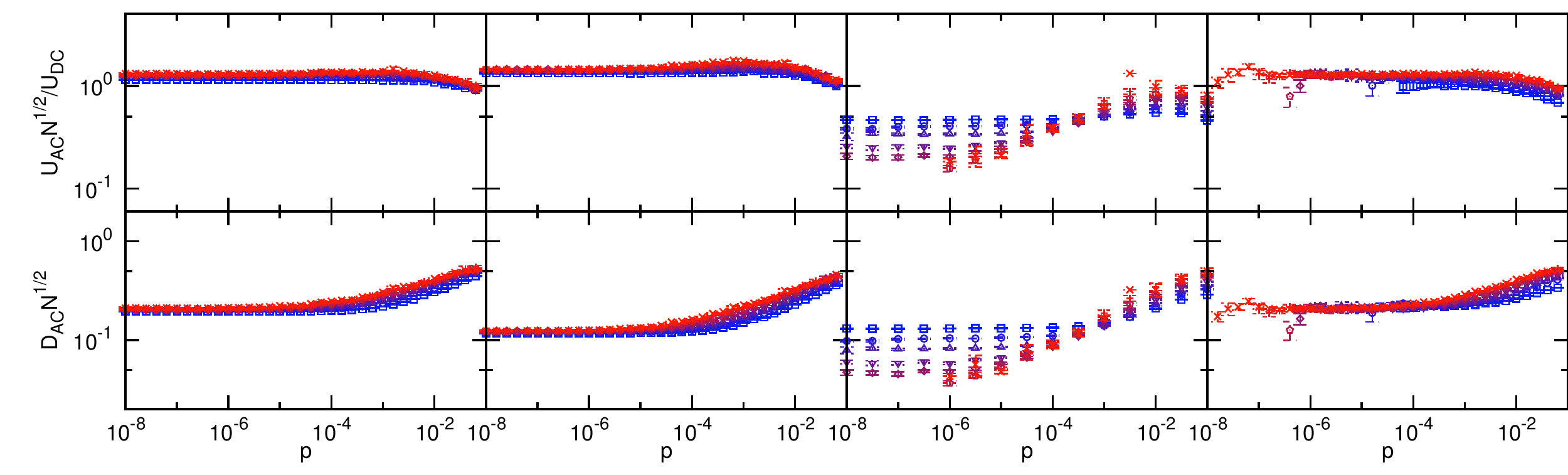,height=1\linewidth,angle=-90}
	\caption[The average ``$AC$" quantities.]{\label{fig:anisotropic_AC_scaled}
	The average ``$AC$" quantities, which are defined in Appendix~\ref{AppendixA} and discussed in the text. The colors and symbols are the same as in Fig.~\ref{FigureRawData}.}
\end{figure*}

\begin{figure}[htp]
	\centering
	 \epsfig{file=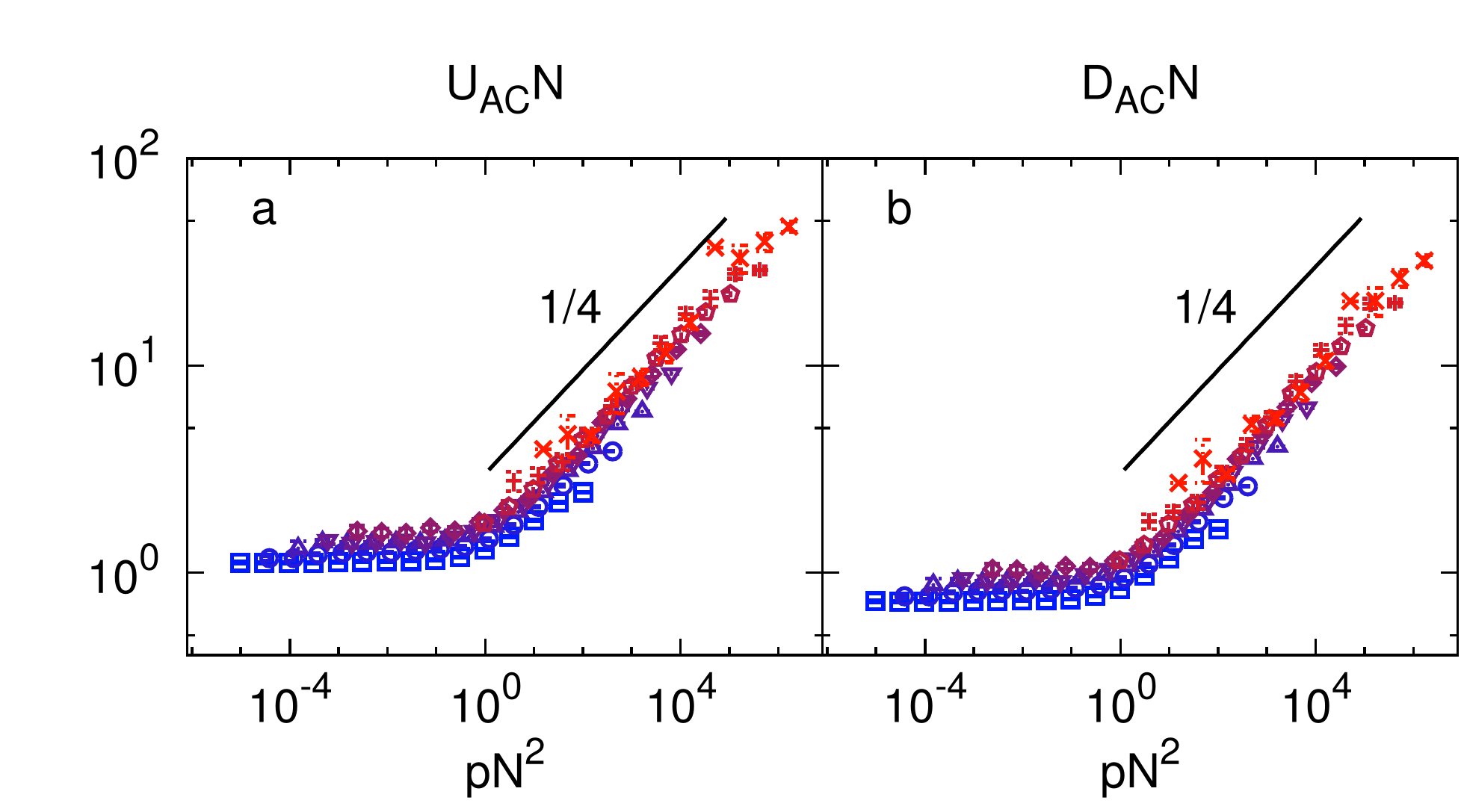,height=0.6\linewidth,angle=-90}
	\caption[Scaling collapse of $U_{AC}$ and $D_{AC}$ for the 2d $\Eap$ ensemble.]{\label{fig:anisotropic_AC_scaled_SS}
	Scaling collapse of $U_{AC}$ and $D_{AC}$ for the 2d $\Eap$ ensemble. The scaling of these two quantities is the only unexpected difference between the three ensembles that we have found. The colors and symbols are the same as in Fig.~\ref{FigureRawData}.}
\end{figure}

\begin{figure}[htp]
	\centering
	 \epsfig{file=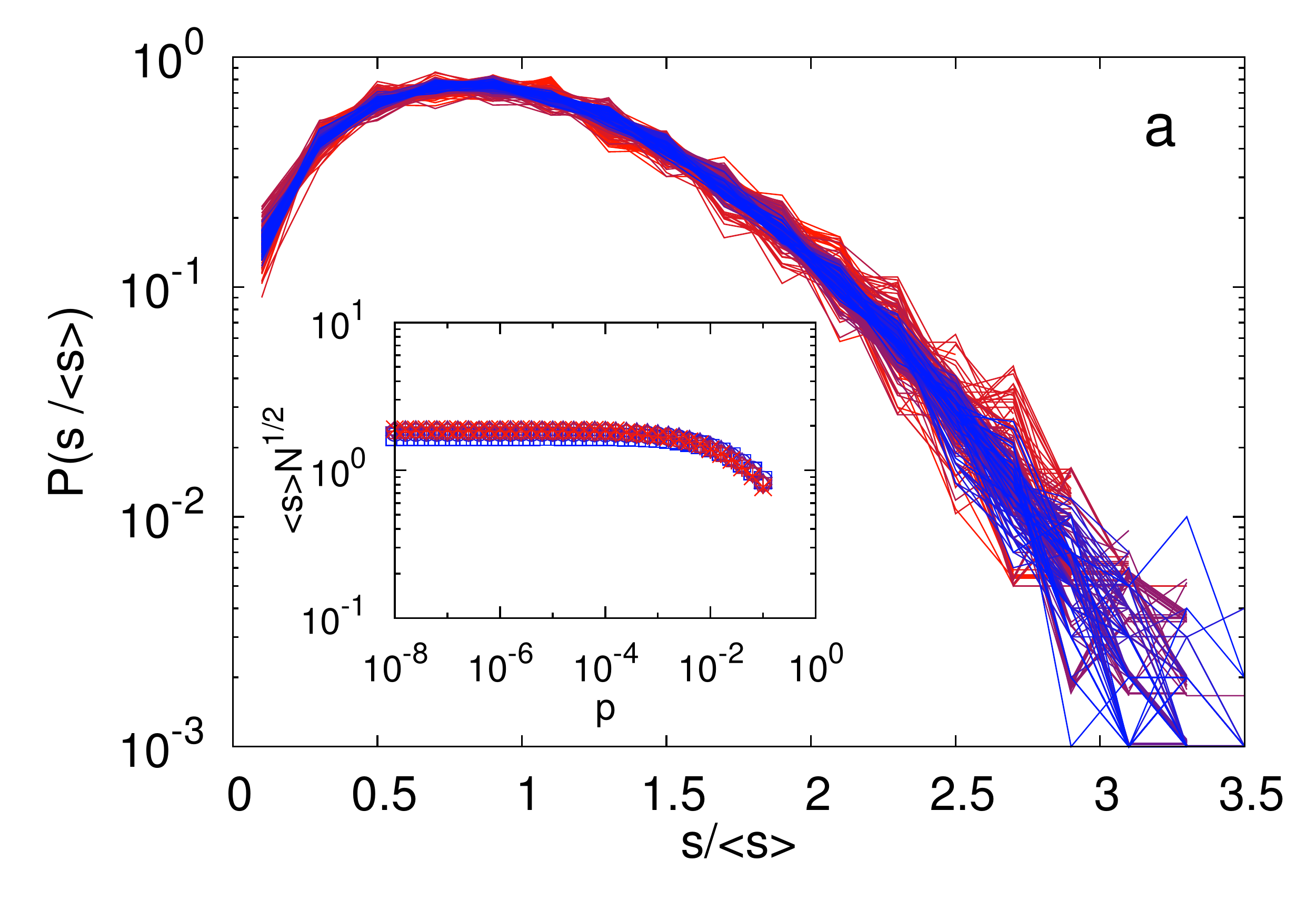,height=0.63\linewidth,angle=-90} \\
	 \epsfig{file=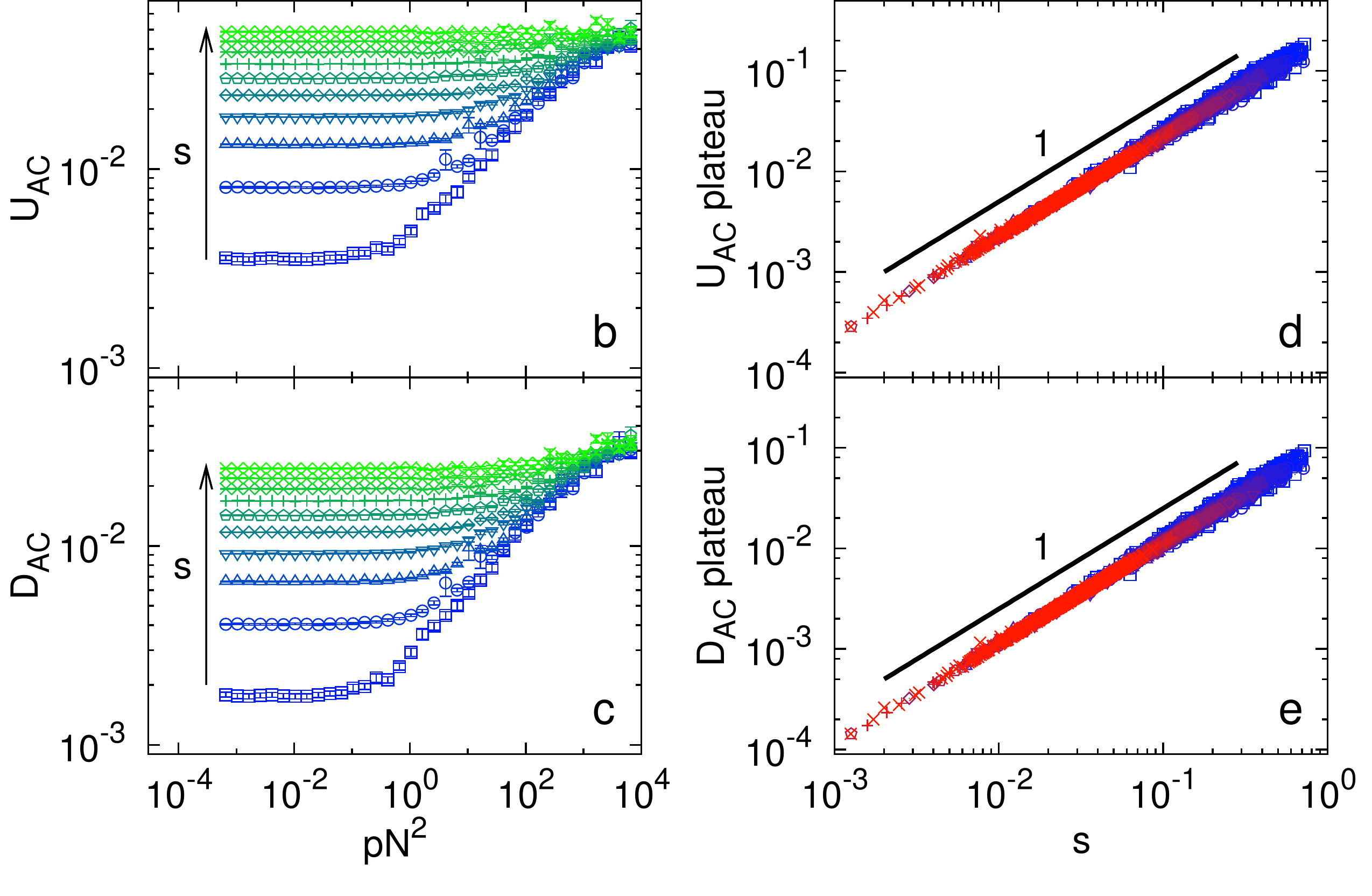,height=0.7\linewidth,angle=-90}
	\caption[The residual shear stress.]{\label{fig:impact_of_residual_stress}The residual shear stress.
		a) The probability distribution $P(s/\!\avg{s})$ of the residual shear stress divided by the ensemble average. $P(s/\!\avg{s})$ collapses onto a single curve and is independent of system size and pressure. Inset: $\avg{s} N^{1/2}$ as a function of pressure.
		b) $U_{AC}$ as a function of $pN^2$ for $N=256$ systems from the 2d $\Ec$ ensemble. Systems are binned according to the residual shear stress before averaging. For high $s$ (green data), $U_{AC}$ is roughly constant but for low $s$ (blue data), $U_{AC}$ is similar to the $\Eap$ data in Fig.~\ref{fig:anisotropic_AC_scaled_SS}.
		c) $D_{AC}$ displays the same behavior as $U_{AC}$.
		d) and e) Scatter plot of the lowest pressure values of $U_{AC}$ and $D_{AC}$, both of which show a remarkable linear dependence on the shear stress.
		The colors and symbols in a), d) and e) are the same as in Fig.~\ref{FigureRawData}.
	}
\end{figure}

\subsection{Anisotropy\label{mvh2}}

In this section we characterize the anisotropic modulations of the elastic constants. 

\subsubsection{Finite-size scaling of anisotropic elastic constant combinations}

As discussed above in Sec.~\ref{sec:usefulelasticconstants}, the elasticity of a jammed packing can be conveniently (though not completely\footnote{Since the full elasticity of an anisotropic system is described by 6 (21) independent constants in two (three) dimensions, the 5 quantities $B$, $G_{DC}$, $G_{AC}$, $U_{AC}$ and $D_{AC}$ are not sufficient to completely characterize a system's elastic properties. Unlike the elements of the elastic modulus tensor, however, they provide an intuitive description that conveniently isolates anisotropic fluctuations.}) described by the five quantities $B$, $G_{DC}$, $G_{AC}$, $U_{AC}$ and $D_{AC}$. The first two of these represent the average response to compression and shear, while the final three represent anisotropic fluctuations. Since anisotropy in jamming is a finite-size effect, one would expect the three ``$AC$" values to vanish in the thermodynamic limit. Here we explore their nontrivial dependence on system size and pressure, {\it i.e.}, proximity to the jamming transition. 
\begin{figure*}
	\centering
	\epsfig{file=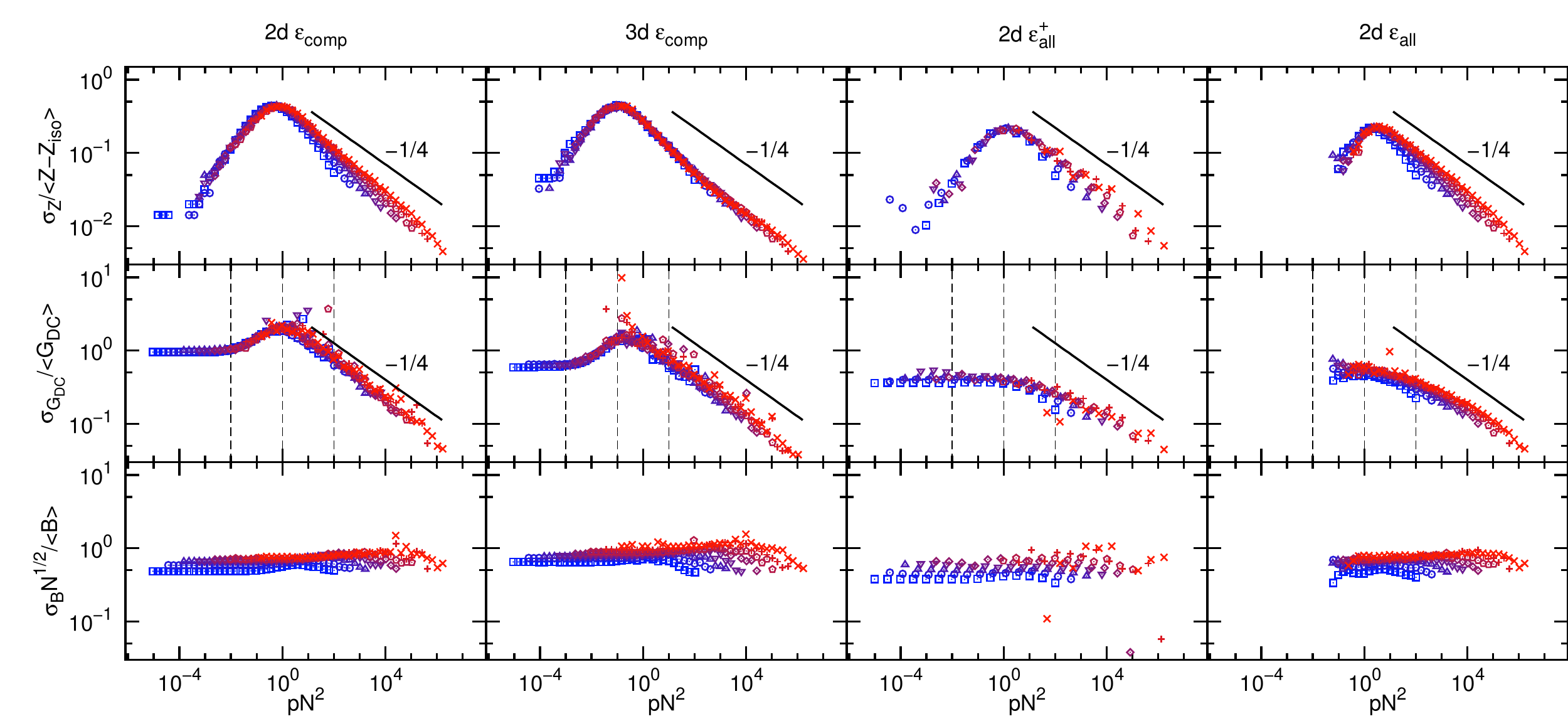,height=1\linewidth,angle=-90}
	\caption[The relative fluctuations in $Z$, $\Gdc$ and $B$.]{\label{fig:anisotrpoic_sigma_scaled}
	The relative fluctuations in $Z$, $\Gdc$ and $B$. The colors and symbols are the same as in Fig.~\ref{FigureRawData}.}
\end{figure*}

\begin{figure*}
    \centering
    \begin{tabular}{c}
    	\epsfig{file=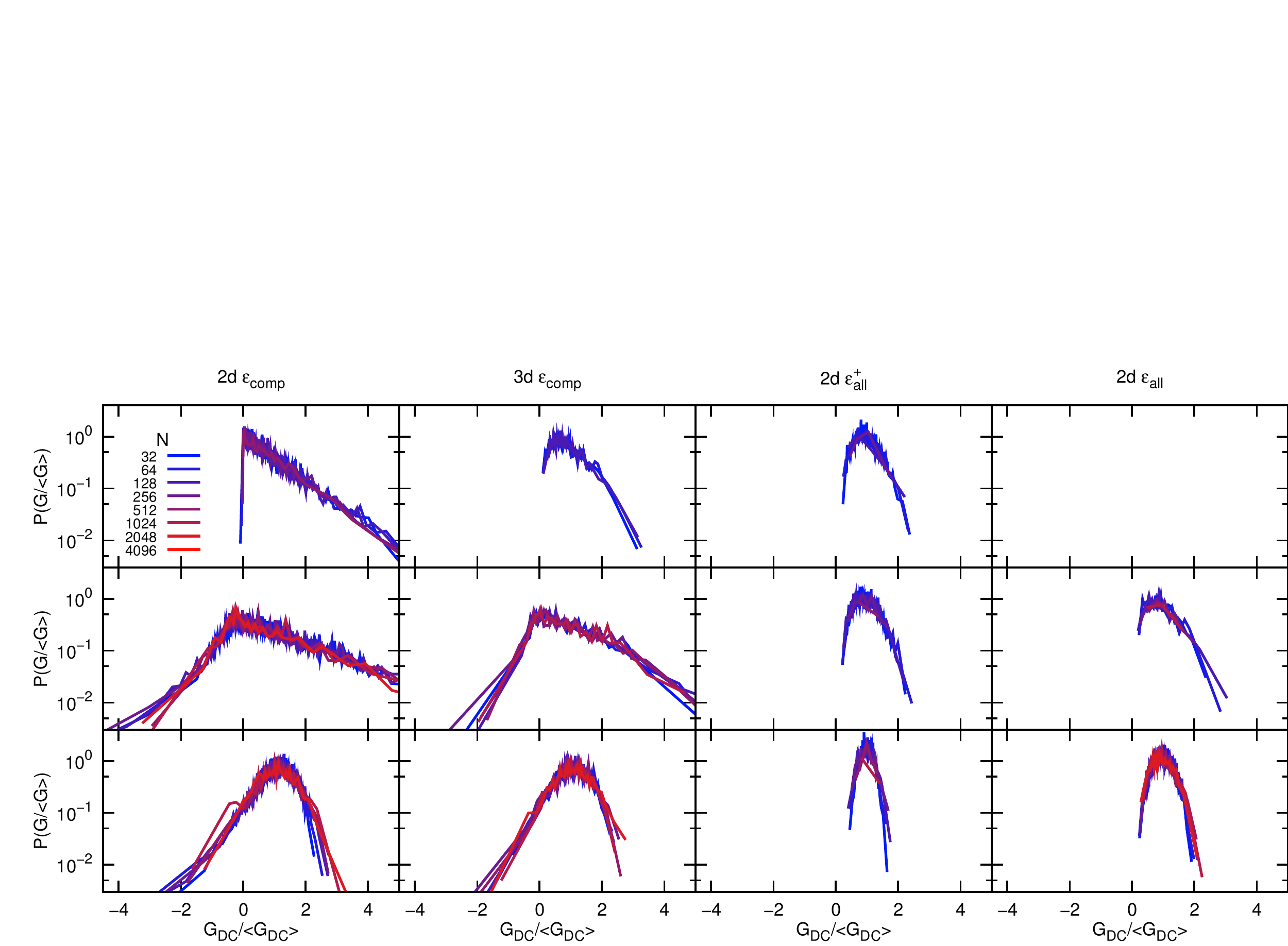,width=0.7\linewidth,angle=-90}
    \end{tabular}
    \caption[The distribution of $G_{DC}$ for the different ensembles.]{\label{fig:Gdist}
    The distribution of $G_{DC}$ for the different ensembles. The shape of the distribution function is different for low $pN^2$ (top row), medium $pN^2$ (middle row) and high $pN^2$ (bottom row), but collapses for systems at similar $pN^2$. The precise values of $pN^2$ correspond to the vertical dashed lines in Fig.~\ref{fig:anisotrpoic_sigma_scaled} ($10^{-2}$, $10^0$ and $10^2$ for the $2d$ ensembles; $10^{-3}$, $10^{-1}$ and $10^{1}$ for the $3d$ ensemble). }
\end{figure*}

The top row in Fig.~\ref{fig:anisotropic_AC_scaled} shows the anisotropic fluctuations of the shear modulus, $G_{AC}$, normalized by the average $\Gdc$ for all four ensembles. 
When plotted as a function of $pN^2$, the data collapse nicely onto a single curve, consistent with the finite-size scaling of Sec.~\ref{sec:finite_size_scaling}. We can distinguish three regimes, depending on the magnitude of $pN^2 $.

{\em (i)  $pN^2 \ll 1$:} Close to jamming, both $\Gdc$ and $G_{AC}$ are constant in pressure. For the two-dimensional $\Ec$ ensemble, the ratio $G_{AC}/\Gdc$ is approximately $1/\sqrt{2}$ (see the black dashed line). 
To understand this, first note that $G_{AC}$ is proportional to the peak height of the sinusoidal function $G(\theta)$ (see Fig.~\ref{fig:G_theta}), and the minimum of $G(\theta)$ is $G_\text{min} = \Gdc - \sqrt{2}G_{AC}$. Also note that $G_\text{min}$ is bounded by $-p$ at low pressures because a negative response can only arise from the pre-stress between contacts~\cite{Wyart:2005jna}. For the 2d $\Ec$ ensemble, we find that $G_\text{min}$ does indeed vanish as $p\rightarrow 0$, implying that $G_{AC}/\Gdc \rightarrow 1/\sqrt{2}$.
The fact that $G(\theta)$ reaches 0 (instead of remaining positive for all $\theta$) indicates that low $pN^2$ packings in this ensemble are on the edge of stability.  We note that while $G(\theta)$ is non-negative on average, it can nevertheless be negative for individual configurations included in the ensemble, either over a range of $\theta$ or even for all $\theta$, as noted earlier in Ref.~\cite{DagoisBohy:2012dh}.

{\em (ii)  $pN^2 \approx 1$:} In the crossover regime,
the minimum of $G(\theta)$ becomes negative for $\Ec$ packings, which implies that $G_{AC}/\Gdc > 1/\sqrt{2}$, leading to the characteristic ``bump'' in the $\Ec$ curves. However, this cannot happen for $\Ea$ or $\Eap$ packings because $G(\theta)$ must always be positive, and this bump is clearly absent there. 

{\em (iii)  $pN^2 \gg 1$:} At large pressures and system sizes, our results are consistent with the scaling $G_{AC}/\Gdc \sim \left(pN^2\right)^{-1/4}$. The $N$ dependence of this scaling is what one would expect from the central-limit theorem: relative fluctuations should be proportional to $1/\sqrt{N}$. The origin of the $p^{1/4}$ pressure dependence is not {\it a priori} obvious, but does follow if one assumes finite-size scaling with $pN^2$. Thus, the combination of the collapse in all three regimes with the non-trivial pressure dependence is strong evidence that finite-size scaling at the jamming transition is not a coincidence. Just as it is for classical phase transitions, finite-size scaling is a fundamental feature of jamming. 

The second row of Fig.~\ref{fig:anisotropic_AC_scaled} shows $U_{AC}$, normalized by the average $U_{DC}N^{-1/2}$. $U_{DC}$ itself is not shown but is given by Eq.~\eqref{eq:Udc_from_B_and_G}, and is constant at low pressures. The bottom row of Fig.~\ref{fig:anisotropic_AC_scaled} shows $D_{AC}$, which is normalized only by $N^{-1/2}$ because $D_{DC}=0$. For the $\Ec$ and $\Ea$ ensembles, $U_{AC}$ and $D_{AC}$ are constant at low and intermediate pressures, and deviate slightly at large pressures. They are also both proportional to the square root of the system size, again consistent with the central-limit theorem. 

From the data presented in Fig.~\ref{fig:anisotropic_AC_scaled} it is not clear if $U_{AC}$ and $D_{AC}$ collapse (note that the abscissa on these plots is $p$, not $pN^2$).  As we show below, there is solid evidence that these quantities have no single parameter scaling in the $\Ec$ and $\Ea$ ensembles. For the $\Eap$ ensemble (third column of Fig.~\ref{fig:anisotropic_AC_scaled}), $U_{AC}$ and $U_{DC}$ are qualitatively different. Interestingly, $U_{AC}N$ and $D_{AC}N$ in the $\Eap$ ensemble behave similarly to  $G_{AC}N$; as shown in Fig.~\ref{fig:anisotropic_AC_scaled_SS}, they are constant at low $pN^2$ and are proportional to $(pN^2)^{1/4}$ at high $pN^2$.  In the $\Eap$ ensemble, there is therefore clear evidence that $U_{AC}$ and $D_{AC}$ scale as $N^{-1/2}$ in the large $pN^2$ limit, consistent with expectations from the central limit theorem.

The discrepancy between the $\Eap$ and the other ensembles is due to the presence of residual shear stress in $\Ec$ and $\Ea$ packings. Figure~\ref{fig:impact_of_residual_stress}a shows that the distribution $P(s/\!\avg{s})$, where $s$ is the residual shear stress and $\avg{s}$ is the ensemble average, is independent of pressure and system size. In the inset, we see that $\avg{s}$ is roughly constant in pressure and is proportional to $N^{-1/2}$. To see the effect of the residual stress on $U_{AC}$ and $D_{AC}$, we bin systems according to $s$ and recalculate the average AC values. The results, which are shown in Fig.~\ref{fig:impact_of_residual_stress}b-c, clearly demonstrate the effect of residual stress on the low $pN^2$ behavior. For low $s$, $U_{AC}$ and $D_{AC}$ are similar to the $\Eap$ results in Fig.~\ref{fig:anisotropic_AC_scaled_SS}, where $s=0$ exactly. However, for high $s$, $U_{AC}$ and $D_{AC}$ are roughly flat. When considered together, the large $s$ data dominates the average leading to the lack of collapse seen in Fig.~\ref{fig:anisotropic_AC_scaled}.

\subsection{Statistical fluctuations in $Z-\Ziso$, $B$ and $\Gdc$ \label{sec:statistical_fluctuations}} 

In addition to $G_{AC}$, $U_{AC}$ and $D_{AC}$,  anisotropy effects can also be characterized by the distributions of contact number, bulk modulus and shear modulus. The simplest way to characterize these distributions is by their standard deviation. However, since the average quantities themselves change by many orders of magnitude, we normalize the standard deviations by the mean.

We begin with the distribution of the average number of contacts. The top row of Fig.~\ref{fig:anisotrpoic_sigma_scaled} shows the standard deviation $\sigma_Z$ of this distribution, normalized by the average of $Z-\Ziso$, which collapses as a function of $pN^2$. In the high and low $pN^2$ limits, the width of the distribution vanishes relative to the average. At intermediate $pN^2$, however, $\sigma_Z$ is of order $Z-\Ziso$. The second row of Fig.~\ref{fig:anisotrpoic_sigma_scaled} shows $\sigma_{\Gdc}$, which is almost identical to $G_{AC}$ (top row of Fig.~\ref{fig:anisotropic_AC_scaled}). Similarly, $\sigma_B$ is shown in the bottom row of Fig.~\ref{fig:anisotrpoic_sigma_scaled}. Interestingly, $\sigma_B$ is qualitatively similar to the high $s$ data for $U_{AC}$ in the $\Ec$ ensembles: $\sigma_B/B$ is proportional to $N^{-1/2}$ but roughly independent of pressure. The distinctive behavior of $U_{AC}$ in the $\Eap$ ensemble is not observed in $\sigma_B$.

One can also look at the full distributions of these quantities. We will focus on the shear modulus $\Gdc$. Fig.~\ref{fig:Gdist} shows the distribution of $\Gdc$, normalized by the average, for the four ensembles. The top, middle and bottom rows correspond to systems with low, intermediate and high $pN^2$, respectively, the precise values of which are given in the caption and depicted by vertical dashed lines in Fig.~\ref{fig:anisotrpoic_sigma_scaled}.

For a given ensemble, both the average of $\Gdc$ and $\sigma_{\Gdc}$ are independent of system size provided that $pN^2$ is held constant. Fig.~\ref{fig:Gdist} shows that this is true for the entire distribution of $\Gdc$. Indeed, the distribution can be considered a one parameter family of functions. Note that at low $pN^2$ (top row), the distribution vanishes very close to $\Gdc=0$ because, as discussed above, negative responses can only arise from stresses, which vanish with pressure.  At higher $pN^2$, however, $G_{DC}$ can be negative for the $\Ec$ ensemble.

%\appendix
\section{Appendix}

\subsection{Numerical procedures\label{sec:numerical_procedures}}
A $d$ dimensional packing of $N$ spheres with equal mass $M$ is described by the position vectors $\vec r_{m}$ and radii $R_m$. Here, the index $m$ goes over the $N$ particles.
We will consider a simulation box with periodic boundaries made from the lattice vectors $\vec L_i$, where $i$ again indicates the dimension.
The center-center distance between particles $m$ and $m\p$ is given by
\eq{	r_{mm\p} = |\vec r_m-\vec r_{m\p} + \sum_b n_{mm\p}^i \vec L_i|,	} 
where $n_{mm\p}^i \in \{-1,0,1\}$ accounts for interactions across the periodic boundaries. 
The spheres interact via the harmonic soft-sphere potential
\eq{	U_{mm\p} = \frac \varepsilon 2 \left(1 - \frac{r_{mm\p}}{R_m+R_{m\p}}\right)^2	}
only when they overlap, {\it i.e.} when $r_{mm\p} < R_m + R_{m\p}$.
The units of length, mass, and energy are $D_\text{avg}$, $M$, and $\varepsilon$ respectively, where $D_\text{avg}\equiv N^{-1}\sum_m 2R_m$ is the average particle diameter.

\subsubsection{Generating sphere packings in the $\Ec$ ensemble}
To generate packings that satisfy the $\Rc$ requirement, we fix the lattice vectors:
\eq{	\vec L_i = L \vec e_i,}
where $\vec e_i$ is the unit vector in the $i$th direction. In other words, we use a standard cubic simulation box whose length $L$ is determined by the packing fraction $\phi$. 

In two dimensions, we choose the particles' radii to be uniformly distributed between 1 and 1.4 to prevent the issue discussed in Sec.~\ref{sec:JammingCriteria}I. In three dimensions, we use a 50/50 bidisperse mixture with ratio 1.4.
We begin by placing the particles at random at a very high packing fraction. We then quench the system to a zero temperature configuration by minimizing the total energy. We do this with a combination of line-search methods (L-BFGS and the Pollak-Ribi\`ere variant of Conjugate Gradient), the Newton-Rhapson method,\footnote{When calculating the inverse of the Hessian matrix in the Newton-Raphson method, we add to it $\lambda_0 \v I$, with $\v I$ the identity and $\lambda_0$ small, to suppress the global translations.} and the FIRE algorithm~\cite{Bitzek:2006bw}. This combination of minimization algorithms was chosen to maximize accuracy and efficiency. However, given its speed, ease of implementation, and sensitivity to shallow features in the energy landscape, we would now recommend the exclusive use of the FIRE algorithm.

We then incrementally adjust the packing fraction, minimizing the energy after each iteration, until we are within $1\%$ of a desired pressure $p_\text{target}^1 = 10^{-1}$. Starting now with this configuration, we repeat this process with a slightly lower target pressure, $p_\text{target}^2 = 10^{-1.2}$. We continue lowering the target pressure incrementally until we reach $p_\text{target}^{36} = 10^{-8}$. Thus, for each initial random configuration, we obtain 36 states at logrithmically spaced pressures.

For each system size and dimension, we repeat this process for at least 1000 different initial random configurations. For small $N$ in two dimensions, we generate up to 5000 configurations to improve statistics. We do not consider systems for which the minimization algorithms fails to converge. This gives us the full two and three dimensional $\Ec$ ensembles. Finally, we can consider only the subset of systems that satisfy the $\Ra$ requirement to form the $\Ea$ ensemble.

\subsubsection{Generating sphere packings in the $\Eap$ ensemble}
To generate two dimensional packings that satisfy the $\Rap$ requirement, we also let the lattice vectors $\vec L_i$ vary. To separate the total volume from the shear degrees of freedom (and to suppress global rotations), we make the following change of variables:
\eqs{
	\vec{L}_1 &= L\left(\frac{1}{1+b},0\right) \\
	\vec{L}_2 &= L(a, 1+b).
}
The degrees of freedom of the system are thus the $dN$ components of the particle positions as well as $L$, $a$, and $b$. 
We then minimize the enthalpy-like potential introduced in Sec.~\ref{sec:precise_formulation}, 
\eq{H = U + p_\text{target} L^2, \label{eq:H_target_pressure}}
with respect to these $dN+3$ degrees of freedom.
This produces a system that 1) satisfies force balance at each particle, 2) has no residual shear stress, and 3) is at a pressure given precisely by $p_\text{target}$~\cite{DagoisBohy:2012dh}. 

Since minimizing Eq.~\eqref{eq:H_target_pressure} brings the system directly to the target pressure, we do not need to adjust the packing fraction manually. We also only use the Conjugate Gradient and FIRE~\cite{Bitzek:2006bw} algorithms. Note that in the FIRE algorithm, we set the effective mass of the boundary degrees of freedom to be $\sqrt{N}$.

\subsection{Elastic constants in two and three dimensions \label{AppendixA}}
Consider the symmetric, two dimensional strain tensor
\eq{ \overleftrightarrow{\epsilon}	 = \left( \begin{array}{cc} \epsilon_{xx} & \epsilon_{xy} \\  \epsilon_{xy}  &  \epsilon_{yy}  \end{array} \right).	}
We will consider the three dimensional case below.
This deformation is imposed on the system in accordance with Eq.~\eqref{strain_tensor_transformation}. After the system is allowed to relax, we define the response to be $R\equiv 2\frac{\Delta U}{V^0}$, where $\Delta U$ is the change in energy of the system and $V^0$ is the volume. To linear order, this is given in terms of the elastic modulus tensor:
\eqs{
	R = {}& c_{ijkl}\epsilon_{ij}\epsilon_{kl} \\
	= {} & c_{xxxx} \epsilon_{xx} ^2 + c_{yyyy} \epsilon_{yy}^2 \\
	&+ 4c_{xyxy} \epsilon_{xy}^2 + 2c_{xxyy}\epsilon_{xx}\epsilon_{yy} \\
	&+ 4c_{xxxy}\epsilon_{xx}\epsilon_{xy}+ 4c_{yyxy}\epsilon_{yy}\epsilon_{xy}. \label{R_from_cijkl}
}
Thus, if the 6 elastic constants $c_{xxxx}$, $c_{yyyy}$, $c_{xyxy}$, $c_{xxyy}$, $c_{xxxy}$, and $c_{yyxy}$ are known, then the linear response to any small deformation is easily obtained.

Although we are assuming that the system is not isotropic, there is no fundamental difference between the various directions -- the choice of axes is arbitrary. 
For a particular strain tensor, we can rotate the deformation by an angle $\theta$:
\eq{
	\overleftrightarrow{\epsilon}(\theta) &=
	\left( \begin{array}{cc} \cos\theta & \sin\theta \\ -\sin\theta & \cos\theta \end{array} \right)
	\left( \begin{array}{cc} \epsilon_{xx} & \epsilon_{xy} \\ \epsilon_{xy} & \epsilon_{yy} \end{array} \right)
	\left( \begin{array}{cc} \cos\theta & -\sin\theta \\ \sin\theta & \cos\theta \end{array} \right). \nonumber 
}
This results in a new deformation with a response $R(\theta)$.
Using the components of the rotated strain tensor,
\eq{	
	\epsilon_{xx}(\theta) &= \tfrac 1 2 (\epsilon_{xx}+\epsilon_{yy}) + \tfrac 1 2(\epsilon_{xx}-\epsilon_{yy})\cos2\theta + \epsilon_{xy} \sin2\theta \nonumber \\
	\epsilon_{yy}(\theta) &= \tfrac 1 2 (\epsilon_{xx}+\epsilon_{yy}) -   \tfrac 1 2(\epsilon_{xx}-\epsilon_{yy})\cos2\theta - \epsilon_{xy}\sin2\theta \nonumber \\
	\epsilon_{xy}(\theta) &=-\tfrac 1 2 (\epsilon_{xx}-\epsilon_{yy})\sin2\theta + \epsilon_{xy}\cos2\theta. \nonumber
}
The new response can be calculated from Eq.~\eqref{R_from_cijkl}.
Note that given the symmetry of Eq.~\eqref{R_from_cijkl}, $\theta$ can always be taken to be in the interval $[0,\pi]$.

By considering deformations that are rotations of each other, $R(\theta)$ is a convenient way to observe anisotropic fluctuations -- in an isotropic system, $R(\theta)$ is always independent of $\theta$. 
The first quantity of interest is the average response,
\eqs{
	R_{DC} &\equiv \avg{R(\theta)} \\
	& = \frac 1 {\pi}\int_0^{\pi}d\theta R(\theta), \label{Rbar}
}
which integrates out the anisotropic fluctuations. We can then characterize the anisotropy by the variance of the response:
\eqs{
	R_{AC}^2 &\equiv \avg{(R(\theta)- R_{DC})^2} \\
	&= \frac 1 {\pi} \int_0^{\pi}d\theta (R(\theta)- R_{DC})^2.  \label{sigmaR2}
}

Equations~\eqref{Rbar} and \eqref{sigmaR2} are generic in that we have not yet specified the initial strain tensor. Our strategy going forward will be to choose physically relevant strain tensors, {\it e.g.} corresponding to pure shear, calculate the response as a function of $\theta$, and use Eqs.~\eqref{Rbar} and \eqref{sigmaR2} to characterize the mean response as well as the fluctuations.
In doing so, it will be convenient to make the following definitions:
\eqs{
	G_0  &=c_{xyxy},\\
	G_{\tfrac{\pi}{4}} & = \frac 14 \left(c_{xxxx}+c_{yyyy} - 2c_{xxyy}\right) \\
	A_2 & = \sqrt{\frac 14 \left(c_{xxxx}-c_{yyyy}\right)^2 + \left(c_{xxxy}+c_{yyxy}\right)^2} \\
	\phi_2 &= \tan^{-1}\left(-2\left(c_{xxxy}+c_{yyxy}\right),c_{xxxx}-c_{yyyy}\right) \\
	A_4 &= -\frac 12 \sqrt{\left(c_{xxxy}-c_{yyxy}\right)^2 + \left(G_0 - G_{\tfrac{\pi}{4}}\right)^2} \\
	\phi_4 &= \tan^{-1}\left(c_{xxxy}-c_{yyxy},G_0 - G_{\tfrac{\pi}{4}}\right).
}

\subsubsection{Uniform Compression}
Uniform compression is obtained from the strain tensor
\eq{ \overleftrightarrow{\epsilon}	 = \frac \gamma 2 \left( \begin{array}{cc} 1 & 0 \\ 0 & 1 \end{array} \right),	}
 where we are interested in the limit $\gamma \ll 1$. This does not change under rotation and so the response, {\it i.e.} the bulk modulus $B$, can be calculated directly from Eq.~\eqref{R_from_cijkl}:
 \eq{	B = \frac 14 \left(c_{xxxx} + c_{yyyy} + 2c_{xxyy}\right).}

\subsubsection{Shear}
Pure shear can be obtained by setting $\epsilon_{xx} = \epsilon_{yy} = 0$ and $\epsilon_{xy} = \gamma/2$, resulting in the strain tensor
\eq{	\overleftrightarrow{\epsilon}(\theta) = \frac \gamma 2 \left( \begin{array}{cc} \sin(2\theta) & \cos(2\theta) \\ \cos(2\theta) & -\sin(2\theta) \end{array} \right),	}
where $\theta$ is the angle of shear. We will define $G(\theta)$ to be the response, which can be written as (see Fig.~\ref{fig:G_theta})
\eq{	G(\theta) = \frac 12 \left( G_0 + G_{\tfrac{\pi}{4}}  \right) - A_4 \sin \left( 4\theta + \phi_4 \right).	}
Note that although the generic period of $R(\theta)$ is $\pi$, $G(\theta)$ is periodic over the interval $[0,\pi/2]$.
Note also that $G(0)=G_0$ and $G(\pi/4) = G_{\tfrac{\pi}{4}}$.
From Eqs.~\eqref{Rbar} and \eqref{sigmaR2}, we see that
\eqs{
	G_{DC} &= \frac 12 \left( G_0 + G_{\tfrac{\pi}{4}}  \right) \\
	G_{AC} &= \frac {A_4}{\sqrt{2}}.
}

\subsubsection{Uniaxial Compression}
Uniaxial compression can be obtained by setting $\epsilon_{xx} = \gamma$ and $\epsilon_{yy} = \epsilon_{xy} = 0$, resulting in the strain tensor
\eq{	\overleftrightarrow{\epsilon}(\theta) = \frac \gamma 2 \left( \begin{array}{cc} 1+\cos(2\theta) & -\sin(2\theta) \\ -\sin(2\theta) & 1-\cos(2\theta) \end{array} \right).	}
We will define $U(\theta)$ to be the response, which can be written as
\eqs{	
	U(\theta) ={}& B + G_{DC} \\
	& + A_2 \sin(2\theta + \phi_2) \\
	&+ A_4 \sin(4\theta + \phi_4).
}
Note that $U(0) = c_{xxxx}$ and $U(\pi/2) = c_{yyyy}$.
From Eqs.~\eqref{Rbar} and \eqref{sigmaR2}, we see that
\eqs{
	U_{DC} &= B + G_{DC} \\
	U_{AC} &= \sqrt{ \frac 12 \left( A_2^2 + A_4^2 \right)}.
}

\subsubsection{Dilatancy}
Linear dilatancy can be understood from setting $\epsilon_{xx} = \epsilon_{xy} = \gamma/2$ and $\epsilon_{yy} = 0$, resulting in the strain tensor
\eq{	\epsilon(\theta) = \frac \gamma2 \left(
\begin{array}{cc}
 1+ \cos (2 \theta)+2\sin (2 \theta) & 2\cos (2 \theta)- \sin (2 \theta) \\
 2\cos (2 \theta)- \sin (2 \theta) & 1-\cos (2 \theta)-2\sin (2 \theta)
\end{array}
\right).
}
If the response of such a deformation is $R(\theta)$, then the dilatent response is
\eqs{	
	D(\theta) &= R(\theta) - \frac 14 U(\theta) - G(\theta) \\
	& = -\frac {A_2}2 \cos\left(2\theta + \phi_2\right) - A_4 \cos(4\theta + \phi_4).
}
When $\theta=0$, for example, we have from Eq.~\eqref{R_from_cijkl} that
\eq{	
	R(0) &= \frac 14 c_{xxxx} + c_{xyxy} + c_{xxxy} \\
	&= \frac 14 U(0) + G(0) + c_{xxxy}
}
so
\eq{	D(0) &= R(0) - \frac 14 U(0) - G(0) = c_{xxxy}.	}
Similarly, $D(\pi/2) = -c_{yyxy}$.
From Eqs.~\eqref{Rbar} and \eqref{sigmaR2}, we see that
\eqs{
	D_{DC} & = 0\\
	D_{AC} & = \sqrt{\frac 18 \left( A_2^2 + 4A_4^2 \right)}.
}
\\\\
\subsubsection{Three dimensions}
Extending the above definitions to three dimensions is straight forward. We begin with the strain tensor
\eq{	\overleftrightarrow{\epsilon} = \left(\begin{array}{ccc}\epsilon_{xx} & \epsilon_{xy} & \epsilon_{xz} \\ \epsilon_{xy} & \epsilon_{yy} & \epsilon_{yz} \\\epsilon_{xz} & \epsilon_{yz} & \epsilon_{zz}\end{array}\right) }
and the rotation matrix
\eq{	\mathcal{R}(\theta_1, \theta_2, \theta_3) = \mathcal{R}^\#(\theta_3) \cdot \mathcal{R}^*(\theta_2) \cdot \mathcal{R}^\#(\theta_1)}
where $\theta_1$, $\theta_2$ and $\theta_3$ are Euler angles and $\mathcal{R}^\#$ and $\mathcal{R}^*$ are given by
\eq{
	\mathcal{R}^\#(\theta) &=
		\left(	\begin{array}{ccc}
 			\cos \theta & -\sin \theta & 0 \\
			\sin \theta & \cos \theta & 0 \\
 			0 & 0 & 1
		\end{array} \right), \\
	\mathcal{R}^*(\theta) &=
		\left(	\begin{array}{ccc}
 			\cos \theta & 0 & \sin \theta \\
			0 & 1 & 0 \\
			-\sin \theta & 0 & \cos \theta
		\end{array} \right).
}
The rotated strain tensor,
\eq{	\overleftrightarrow{\epsilon} (\theta_1, \theta_2, \theta_3) = \mathcal{R}^{-1}(\theta_1, \theta_2, \theta_3) \cdot \overleftrightarrow{\epsilon} \cdot \mathcal{R}(\theta_1, \theta_2, \theta_3),}
and the response, $R(\theta_1, \theta_2, \theta_3)$, are functions of the three Euler angles.
Finally, the average response $R_{DC}$ and variance $R_{AC}^2$ are obtained from properly integrating over the three angles:
\eq{	
	R_{DC} &= \mathcal{I}^3\; R(\theta_1, \theta_2, \theta_3), \\
	R_{AC}^2 &= \mathcal{I}^3 \left[ R(\theta_1, \theta_2, \theta_3) - R_{DC} \right]^2,
}
where $\mathcal{I}^3$ stands for $ \frac{1}{32\pi^2}\int_0^{4\pi} d \theta_3 \int_0^\pi d \theta_2  \sin \theta_2 \int_0^{4\pi}  d \theta_1$.

\chapter{The crystal and the anticrystal}
\label{chapter:anticrystal}

\section{Introduction}

For over a century, physicists have been taught to describe real solids in terms of perturbations about perfect crystalline order~\cite{Ashcroft:1976ud}.  Such an approach only takes us so far: a glass, another ubiquitous form of rigid matter, cannot be described in any meaningful sense as a defected crystal~\cite{Phillips:1981um}.  Is there an opposite extreme to a crystal, a solid with Òcomplete disorder,Ó that would be an alternate starting point for understanding real materials? In this Chapter we argue that the solid made of particles with finite-ranged interactions at the jamming transition~\cite{Liu:1998up,OHern:2003vq,Liu:2010jx} constitutes such a limit. It has been shown that the physics associated with this transition can be extended to interactions that are long ranged~\cite{Xu:2007dd}. Here, we demonstrate that jamming physics is not restricted to amorphous systems but dominates the behavior of solids with surprisingly high order. Much as the free-electron and tight-binding models represent two idealized poles from which to understand electronic structure~\cite{Ashcroft:1976ud}, we identify two extreme poles of mechanical behavior. Thus, the physics of jamming can be set side-by-side with the physics of crystals to provide an organizing structure for understanding the mechanical properties of solids over the entire spectrum of disorder.

The jamming transition can be studied in its purest form at zero temperature in disordered packings of spheres interacting via finite-range, repulsive interactions~\cite{OHern:2003vq,Liu:2010jx}. At the transition, the packing fraction, $\phi$, is just sufficient to cause unavoidable contact between particles. 
The marginally-jammed state at this transition represents an extreme limit of solids -- the epitome of disorder -- in several ways.  (1) For a perfect crystal, the ratio of the shear to bulk modulus, $G/B$, is of order unity while for a liquid, $G/B=0$~\cite{Ashcroft:1976ud}. 
At the jamming transition, the response to shear is infinitely weaker than the response to compression~\cite{OHern:2003vq}. That $G/B$ vanishes at the jamming transition implies that the marginally-jammed solid lies at the extreme edge of rigidity~\cite{OHern:2003vq,Liu:2010jx}. 
(2) Any crystalline solid supports sound modes at sufficiently low frequency; since the wavelength of sound is long enough to average over microscopic details, the crystal, even with defects, is well approximated as an elastic medium~\cite{Ashcroft:1976ud}.  At the jamming transition, however, there are diverging length scales that exceed the wavelength of sound, even at arbitrarily low frequencies. This leads to qualitatively new physics: a new class of vibrational modes that overwhelms plane-wave behavior~\cite{OHern:2003vq,Liu:2010jx,Silbert:2005vw,Wyart:2005wv,Xu:2007dd}.

The marginally-jammed solid also sits at a critical point, the jamming phase transition. This suggests that properties of this state, including the two distinguishing characteristics of vanishing $G/B$ and a plateau in the density of states, might  be reflected in systems away from the transition. For instance, although the jamming transition is inaccessible to particles with Lennard-Jones interactions,  features like the boson peak in Lennard-Jones glasses can be understood by treating the long-range attractions as a correction to jamming physics~\cite{Xu:2007dd}. Similarly, numerous other studies indicate that the jamming transition influences the behavior of systems with  (a) 3-body interactions~\cite{Wyart:2008jg,Phillips:1979wh,Phillips:1981vy,Thorpe:1983tw,Boolchand:2005bi}, (b) friction~\cite{Song:2008jb,Henkes:2010kv,Papanikolaou:2013fa}, (c) temperature~\cite{Zhang:2066bv}, and (d) aspherical particles~\cite{Donev:2004fr,Zeravcic:2009wo,Mailman:2009ct}. Here, we show that the physics of jamming is relevant over a surprisingly broad range of disorder.   Even when a solid looks crystalline to the eye, it may nevertheless manifest mechanical properties better described by jamming than by crystalline physics. 

\section{Between order and disorder}

We focus on the simplest systems that can capture both jamming and crystalline physics, namely packings of frictionless spheres of mass $m_\mathrm{s}$ and diameter $\sigma$. Spheres $i$ and $j$ interact only if they overlap, in which case they feel a linear repulsive force of magnitude $f_{ij} = \epsilon \left( 1 - \frac{r_{ij}}{\sigma} \right)$, where $r_{ij}$ is their separation. All quantities are in units of $m_\mathrm{s}$, $\sigma$ and $\epsilon$.  
We generate packings using an unbiased algorithm that can tune continuously between a crystal and the marginally-jammed solid.  We begin with $N$ spheres arranged in a perfect crystal and introduce point defects in three ways. In variant A, we begin with $N=4000$ spheres in a face-centered cubic (fcc) lattice and introduce vacancy defects by randomly removing some fraction, $m$, of the spheres. The size of the box is then adjusted to obtain a desired volume fraction $\phi$. The system is then relaxed to a local energy minimum using the FIRE algorithm~\cite{Bitzek:2006bw}. Variant B is identical to A except that we begin with a body-centered cubic (bcc) lattice with $N=4394$. Variant C again begins with an fcc lattice with $N=4000$ but defects are introduced as vacancy-interstitial pairs ({\it i.e.}, for every sphere that is removed, another is randomly inserted).

Clearly, these algorithms generate perfect crystals when $m=0$. When $m=1$, variant C is identical to the algorithm commonly used for jamming~\cite{Liu:2010jx}. For all variants, $m\approx 1/2$ 
leads to configurations indistinguishable from jammed ones; thus the two end points of all three variants coincide respectively with the perfect crystal and jamming. We adjust $m$ and $\phi$ to generate states that span the two extreme cases. We then characterize the order using $F_6$ to measure neighbor correlations of bond-orientational order~\cite{Auer:2004db,Russo:2012km} (see Sec.~\ref{sec:apndx_f6_methods_summary}).  $F_6 =1$ for a perfect crystal and $F_6\approx0$ for a disordered system. 

\begin{figure}
	\centering
	\includegraphics[width=1.09\linewidth]{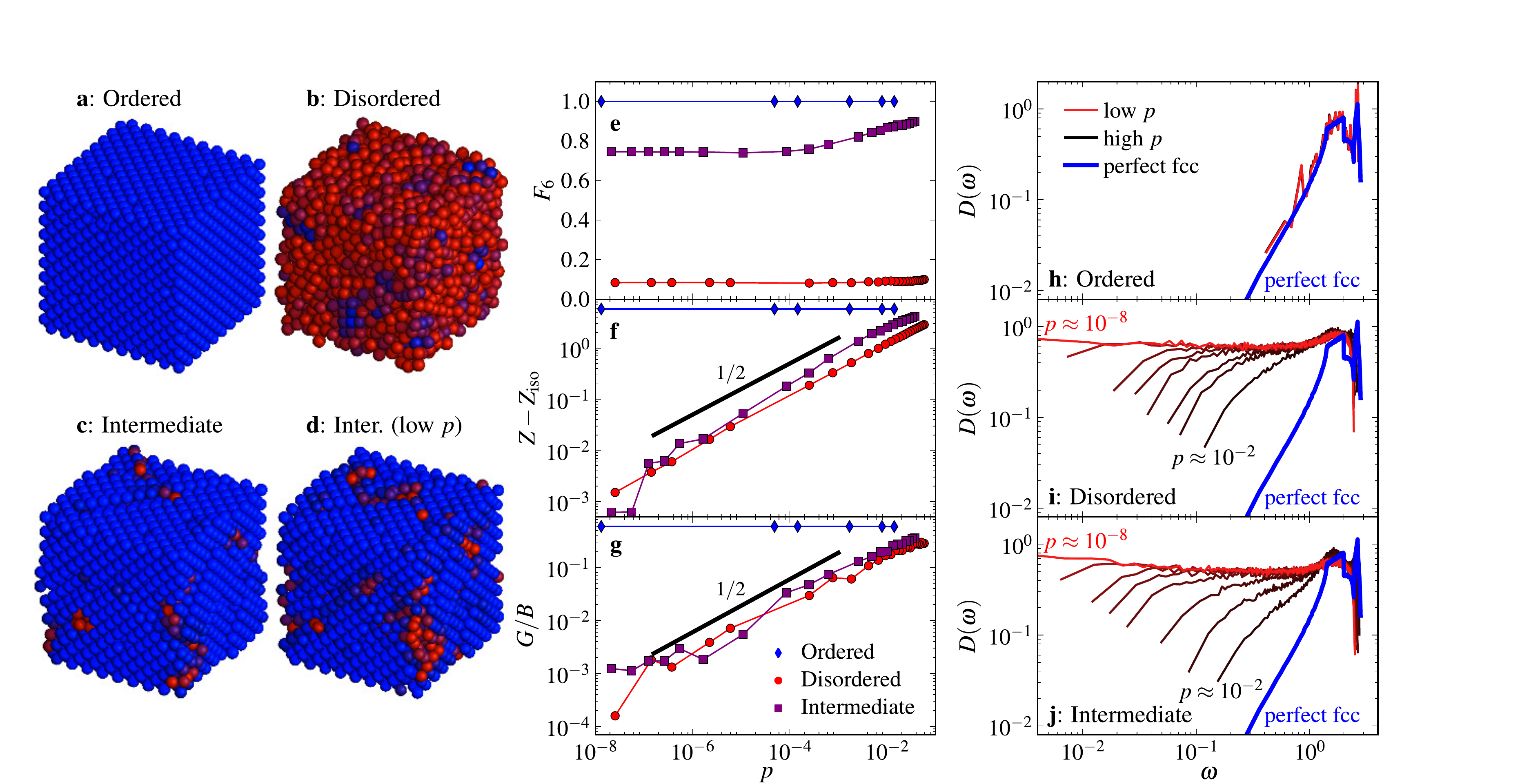}
	\caption[The behavior of three example systems.]{\label{fig:example_systems}The behavior of three example systems.
	(a-c) Images of three example systems generated using Variant A of our algorithm. (a) A nearly perfect crystal with $F_6 = 1.0$, (b) a disordered sample with $F_6 = 0.1$, and (c) a defected crystal with $F_6 = 0.9$. Blue (red) particles have high (low) local order (see Sec.~\ref{sec:apndx_f6_methods_summary}). 
	(d) The same system as (c) after the pressure has been lowered by about seven orders of magnitude. 
	(e-g) The global order parameter $F_6$, excess contact number $\ZmZiso$, and ratio of the shear modulus to bulk modulus $G/B$ as a function of pressure for the three systems. 
	(h-j) The density of vibrational modes $D(\omega)$ for the three systems at different pressures. $D(\omega)$ for a perfect fcc crystal is provided for comparison.
	Note that the intermediate system, though very highly ordered, behaves in its mechanical and vibrational properties much more like the disordered sample than like the nearly perfect crystal.}
\end{figure}

We study the mechanical response of packings as a function of the pressure, $p$. Three example zero-temperature configurations at high pressure are shown in Fig.~\ref{fig:example_systems}a-c.  The first is a slightly defected crystal with a very small density of vacancies. The second is disordered, indistinguishable from a jammed system. 
The third ``intermediate" system is {\em structurally} a defected crystal with $F_6 \approx 0.9$, but, as we shall see, behaves {\em mechanically} like a jammed solid.  As $p$ decreases, Fig.~\ref{fig:example_systems}e shows that for the first two systems $F_6$ remains constant; for the third $F_6$ decreases but stays high, $F_6>0.75$. Even as $p\rightarrow 0$, its structure remains predominantly crystalline (Fig.~\ref{fig:example_systems}d).

For the perfect fcc crystal, each particle has exactly 12 neighbors at any $p>0$.  Likewise, the slightly defected crystal of Fig.~\ref{fig:example_systems}a has an average number of contacts per particle, $Z$, that is independent of pressure (see Fig.~\ref{fig:example_systems}f).   The disordered system of Fig.~\ref{fig:example_systems}b decreases towards the isostatic value, $Z_\text{iso}\approx 2d$, as $p$ is lowered, consistent with the well-established scaling of $\ZmZiso \sim p^{1/2}$ for jammed spheres with harmonic interactions.\cite{Durian:1995eo,OHern:2003vq,Liu:2010jx} Surprisingly, as the pressure is lowered in the system with intermediate order, we find that $Z$ also approaches $Z_\text{iso}$ as a power law with roughly the same exponent. Apparently, despite the high degree of crystallinity, this system becomes marginally coordinated at low pressures.

The mechanical responses of these three systems are shown in Fig.~\ref{fig:example_systems}g.  The slightly defected crystal shows the expected response, with $G/B$ remaining constant as $p \rightarrow 0$, while for the disordered system, $G/B \sim p^{1/2}$ as expected near the jamming transition~\cite{OHern:2003vq,Liu:2010jx}.
For the intermediate sample, 
not only is $\ZmZiso$ similar to that of a jammed solid, but Fig.~\ref{fig:example_systems}g shows that the mechanical properties are also more like a jammed solid than a crystalline one: $G/B$ vanishes as $p\rightarrow 0$ with the \emph{same power law} as in jamming.  

We also measure the density of normal modes of vibration, $D(\omega)$, which is directly related to the mechanical and thermal properties of the solid within the harmonic approximation~\cite{Goodrich:2014jw}. Low-frequency modes in crystals correspond to plane waves and follow Debye scaling: $D(\omega)\sim \omega^{d-1}$ (see Fig.~\ref{fig:example_systems}h). 
Disordered systems, in addition, possess a class of ``anomalous" modes that result in a low-frequency plateau in $D(\omega)$ (see Fig.~\ref{fig:example_systems}i)~\cite{Silbert:2005vw}, related to $\ZmZiso$~\cite{Wyart:2005wv}. Figure~\ref{fig:example_systems}j shows that the intermediate sample, which \emph{looks} highly crystalline, has a low-frequency density of states that closely resembles that of the disordered system~\cite{Mari:2009ci}.

The system in Fig.~\ref{fig:example_systems}c indeed behaves mechanically as if it were jammed, despite its extremely ordered structure. To establish the generality of this observation, we have generated hundreds of thousands of packings with varying amounts of order, $F_6$, using the three versions of our algorithm.

\begin{figure}
	\centering
	\includegraphics[width=0.6\linewidth]{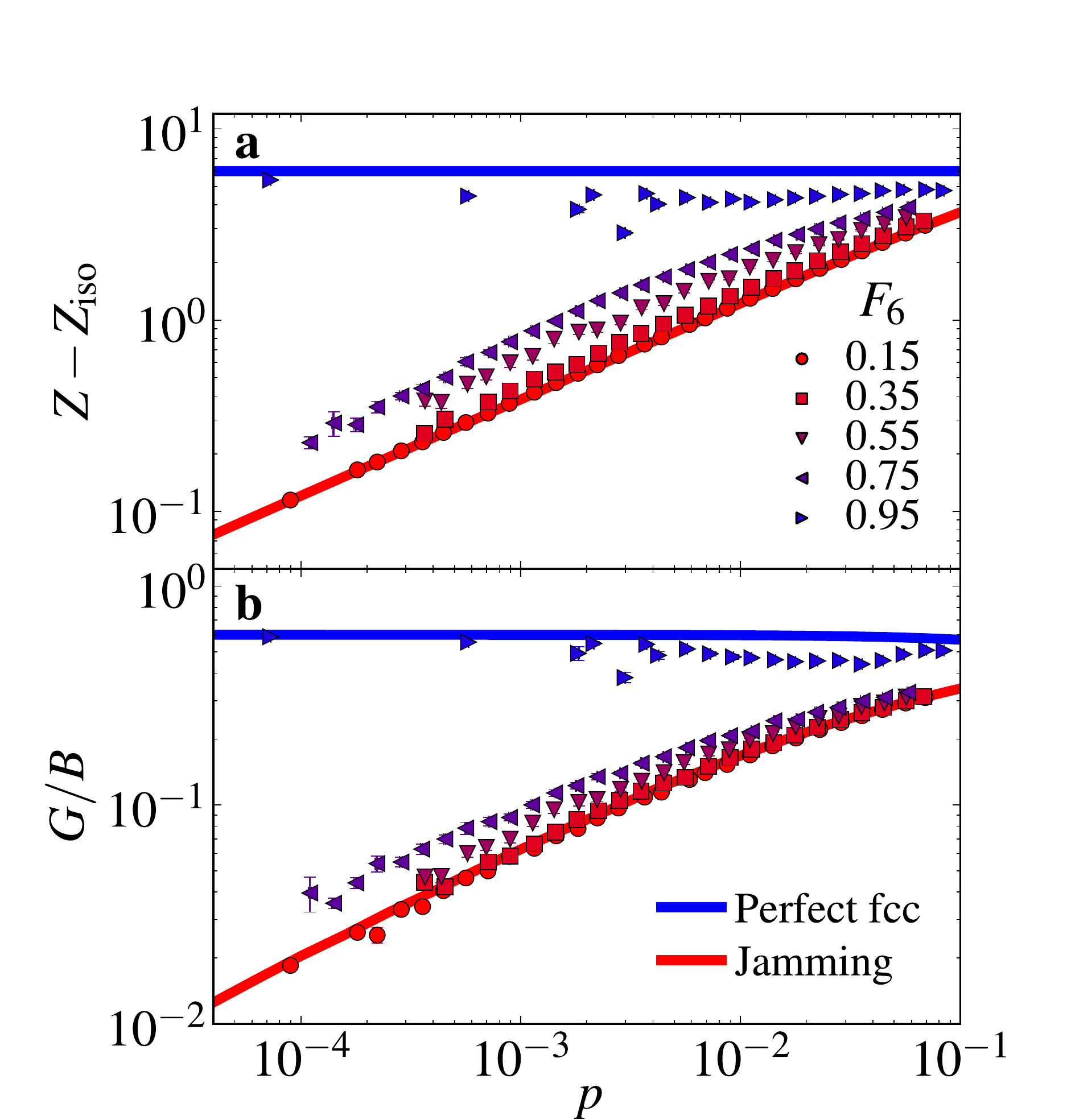}
	\caption[The crossover from jamming physics to crystalline physics.]{\label{fig:F6binned_plots}
	The crossover from jamming physics to crystalline physics.
	(a) The excess contact number as a function of pressure at different values of $F_6$ averaged over an ensemble of systems. The blue line shows the constant $\ZmZiso$ behavior for a perfect fcc crystal, while the red line shows the behavior at jamming. 
	(b) The ratio of the shear modulus to bulk modulus for the same systems. Note that $G/B$ for a perfect crystal is constant at low pressures but decreases very slightly at higher pressures. 
	The error bars represent the standard deviations.
	Only for systems with very high values of $F_6$ do the properties have the pressure independence expected for a crystal.
	}
\end{figure}

\begin{figure}
	\centering
	\includegraphics[width=0.55\linewidth]{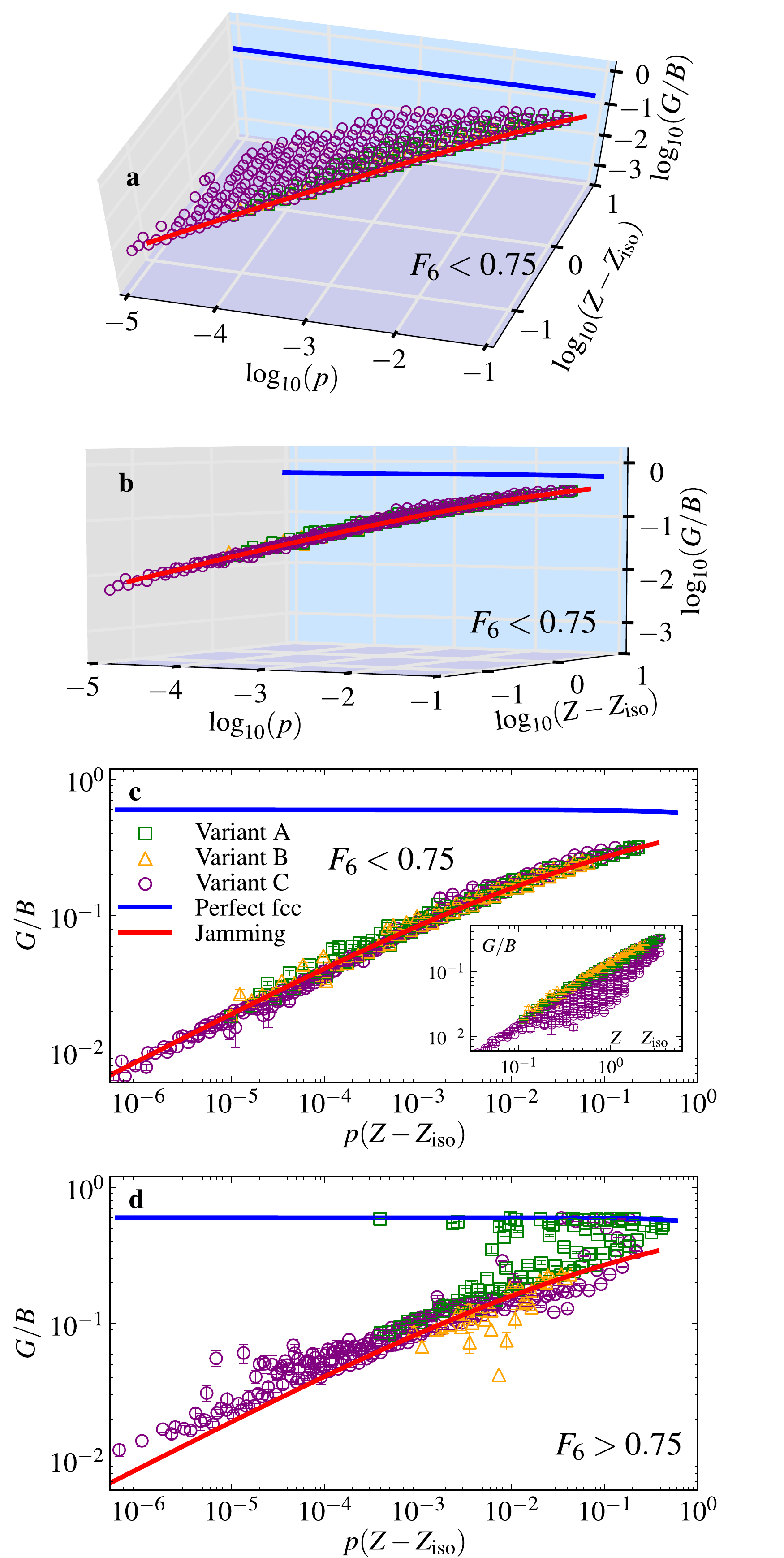}
	\caption[Robustness of jamming behavior.]{\label{fig:3d_plots}Robustness of jamming behavior.
	(a) A three-dimensional plot of $G/B$ as a function of pressure and $\ZmZiso$ for systems with $F_6 < 0.75$, generated with Variant A (green squares), B (orange triangles) and C (purple circles). The solid blue and red lines give the behavior of a perfect fcc crystal and a jammed solid, respectively. 
	(b) Same data but shown from a different point of view to emphasize that all the points lie on a well-defined plane. 
	(c) As a function of $p\left(\ZmZiso\right)$, $G/B$ collapses onto a single curve for all $F_6 < 0.75$. This curve coincides with the one for jamming. The inset shows the data plotted as a function of just $\ZmZiso$.
	(d) For $F_6 > 0.75$, $G/B$ is not cleanly described by jamming physics. 
	The error bars in (c) and (d) represent the standard deviations.
	In the high $F_6$ regime, many systems are better described as a defected crystal whereas all states with $F_6 < 0.75$ are better described from the jamming scenario.
	}
\end{figure}

We first separate packings into bins with different degrees of order. Figure~\ref{fig:F6binned_plots} shows, for different values of $F_6$, the average of $\ZmZiso$ and $G/B$ versus $p$. Note that before averaging, packings were also binned in  pressure. For the systems in Fig.~\ref{fig:example_systems}, we lowered pressure by gradually decreasing $\phi$ and reminimizing energy. For the packings in Figs.~\ref{fig:F6binned_plots}-\ref{fig:3d_plots}, we did not lower $p$ but simply varied $\phi$ to obtain packings over a range of pressures, making it more difficult to produce low-$p$ packings. 
Data are only shown for systems generated with variant A, but the other ensembles show similar features. The solid lines show the two limiting extremes corresponding to the perfect fcc crystal (blue) and jammed packings (red).
At a given pressure, both $\ZmZiso$ and $G/B$ increase as systems become more ordered, but they retain the scaling with pressure associated with jamming, {\em not} the constant value associated with crystalline behavior, up to relatively high values of $F_6$.  Thus the system with intermediate order in Fig.~\ref{fig:example_systems}c is indeed a representative example of this trend.

In jamming, the value of $\ZmZiso$ at a given pressure is well-defined.  However, for our systems there is a large spread in $\ZmZiso$ at fixed pressure. Therefore, we have binned systems according to both $p$ and $\ZmZiso$ and then separated them into whether they do, or do not, behave mechanically like jammed systems.  The separation between these two groups of configurations in terms of $F_6$ is not sharp: some packings with $F_6>0.95$ behave like jammed ones and a few packings with $F_6$ as low as $0.75$ respond more like a crystal.  This suggests that there is crossover from crystalline behavior to jamming behavior that occurs at remarkably high order.

To demonstrate this more clearly, Fig.~\ref{fig:3d_plots}a shows that for $F_6<0.75$, $G/B$ falls cleanly onto a single ``jamming surface."   This includes all the data from all three ensembles. The collapse of the data onto a single plane is made more clear in Fig.~\ref{fig:3d_plots}b, where we have rotated the point of view to look at the surface edge on.  
Figure~\ref{fig:3d_plots}c shows that the jamming surface can be scaled onto a single curve that matches the result expected for jammed packings (red line). Note that given the smoothness of the data, we can choose different scaling on the two axes and get similar collapse. However, as shown by Fig.~\ref{fig:F6binned_plots}a and the inset in Fig.~\ref{fig:3d_plots}c, $G/B$ is \emph{not} simply a function of just $p$ or $\ZmZiso$, as one might have expected.
As $F_6 \rightarrow 1$, the manner in which systems approach the crystalline limit depends on the protocol -- as expected, perturbations about a crystal are sensitive to the details of the perturbation. Figure~\ref{fig:3d_plots}d shows $G/B$ averaged over all systems with $F_6>0.75$. In this region there is a mixture of states that behave like crystalline configurations with those that behave like jammed packings. 

\section{Discussion}

It is important to understand the generality of our results. Our algorithms correspond to introducing ordinary vacancies or vacancy-interstitial pairs to an otherwise pefect crystal. As more of these simple defects are added, the mechanical behavior changes character; at some point the solid loses all semblance of crystalline behavior even though its structure appears exceptionally well ordered. We have shown that in this regime the physical properties can be profitably described as perturbations about the {\em disordered limit} of a solid. Our model is also simplified because we use finite-range potentials. This choice makes the relevance of the jamming transition clear but does not limit its applicability: as emphasized above, the persistence of key features of the jamming scenario in systems with 3-body interactions, friction, temperature, and aspherical particles implies that our results should be generally relevant. In particular, long-range attractions, as in Lennard-Jones systems, can be treated as merely a perturbation around a jamming description of the mechanical properties even though the jamming transition itself is not accessible~\cite{Xu:2007dd}. 
Recently, it was suggested that hard sphere systems below the jamming transition might have somewhat different physics~\cite{Zargar:2014uj}.

Of particular note, crossovers from ordered to disordered behavior have been observed for highly crystalline systems in other contexts.  When crystalline KBr is doped with KCN, the impurities form an ``orientational glass" without destroying the crystalline structure. De Yoreo {\it et al.}~\cite{DeYoreo:1983ul} found a crossover in the thermal conductivity from crystalline ($T^3$) to glassy ($T^2$) scaling  when the concentration of KCN is only around $1\%$. Moreover, for concentrations of $25\%$ and $50\%$, the specific heat and thermal conductivity appeared equally glassy.
In sheared quasi-2$d$ foams, Katgert {\it et al.}~\cite{Katgert:2008bk,Katgert:2009eo} found a crossover in the flow profile to that of disordered systems in highly ordered foams with only $2\%$ area defects. 
Finally, Mari {\it et al.}~\cite{Mari:2009ci} showed that introducing a very slight polydispersity (.003\%) to an fcc crystal of hard spheres just below jamming causes the vibrational properties to be indistinguishable from a hard-sphere glass. It is not clear in this case if there exists a crossover to crystalline behavior as the polydispersity is decreased, or if that limit is singular.

Our results provide a new vantage point for understanding the mechanical response of solids.  Starting from the jamming scenario, we can profitably describe the behavior of even highly-ordered materials. For example, many polycrystals might be better described in terms of corrections to jamming behavior than in terms of defected crystals. Although perfect order is a well-defined concept, it has been less clear what perfect disorder means.  Our results suggest that instead of using a structural quantity like $F_6$, one can characterize order in terms of mechanical properties. A system {\emph{behaves}} ordered when, for example, $G/B$ is roughly constant in pressure and {\emph{behaves}} disordered when it decreases with decreasing pressure. The spectrum of order in solids therefore has two well-defined limits: a perfect crystal, with constant $G/B$ and $\ZmZiso$, and a perfect anti-crystal, with $G/B$ and $\ZmZiso$ vanishing at zero pressure.

\section{Appendix}
\subsection{Methods summary\label{sec:apndx_f6_methods_summary}}
We measure the degree of order in a given configuration by calculating the parameter $F_6$. For each pair of neighboring spheres $i$ and $j$, the function $S_6(i,j)$ measures correlation of neighbor orientation: 
\begin{align}
S_6(i,j) \equiv \frac{\sum_{m=-6}^6 q_{6m}(i)\cdot q_{6m}^*(j)}{\left| \sum_m q_{6m}(i)\right| \left|\sum_m q_{6m}(j)\right|},
\end{align} 
where $q_{lm}(i)$ is the standard bond-orientational order parameter~\cite{STEINHARDT:1983uh}.
Summing over the $N_c(i)$ neighbors of $i$, we define~\cite{Auer:2004db,Russo:2012km}
\begin{align}
f_6(i) \equiv \frac 1{N_c(i)} \sum_{j \text{ nn } i} \Theta \left(S_6(i,j) - S_6^0\right),
\end{align}
where $\Theta$ is the step function and $S_6^0$ is a threshold that is typically taken to be $0.7$.  $f_6(i)$ measures the fraction of sphere $i$'s neighbors with highly correlated neighbor orientations. In a perfect crystal, $f_6=1$ and in a disordered state it is small. Finally, we average $f_6(i)$ over the system to obtain a global order measure, $F_6$.

\chapter{The principle of independent bond-level response}
\label{chapter:independent_response}

\newcommand{\dZ}[0]{\Delta Z} 
\newcommand{\dZi}[0]{\Delta Z_\text{initial}} 
\newcommand{\muBm}[0]{{\mu_{B_-}}}
\newcommand{\muBp}[0]{{\mu_{B_+}}}
\newcommand{\muGp}[0]{{\mu_{G_+}}}

%\title{The principle of independent bond-level response: tuning by pruning to exploit disorder for global behavior}%response}

\section{Introduction}
The properties of amorphous solids are essentially and qualitatively different from those of simple crystals~\cite{Goodrich:2014fl}.  In a crystal, identical unit cells are interminably and symmetrically repeated, ensuring that all cells make identical contributions to the global response of a solid to an external perturbation~\cite{Ashcroft:1976ud,kittel2004introduction}.  Unless a crystal's unit cell is very complicated, all particles or inter-particle bonds contribute nearly equally to any global quantity, so that each bond plays a similar role in determining the physical properties of the solid. For example, removing a single bond from a perfectly ordered array or network decreases the overall elastic strength of the system, but in such a way that the resistance to shear and the resistance to compression drop in tandem~\cite{Feng:1985vr}, leaving their ratio nearly unaffected.
Disordered materials are not similarly constrained. We will show that as a consequence, one can exploit disorder to achieve a unique, varied, textured and tunable global response.

A tunable global response is a corollary to a new principle that emerges for disordered matter: independent bond-level response.  
This independence refers not only to 1) the significant variation in the response at the individual bond level, but also, and more importantly, to 2) the dearth of strong correlations between the responses of any specific bond to different perturbations.  To illustrate this principle, we consider the specific perturbations of compression and shear.  We construct networks in which individual bonds are successively removed to drive the overall system into different regimes of behavior characterized by the ratio $G/B$ of the shear modulus, $G$, to the bulk modulus, $B$.  
Starting from the same initial network, we can remove as few as 2\% of the bonds to produce a network with a value of $G/B$ that is either nearly zero (incompressible limit where the Poisson ratio is $\nu = 1/(d-1)$ in $d$ dimensions) or nearly infinite (maximally auxetic with $\nu = -1$~\cite{Greaves:2011ku}) merely by removing different sets of bonds. Moreover, by using different algorithms or starting with different networks, one can confine the region within which the bonds are removed to strips of controllable size, ranging from a few bond lengths to the size of the entire sample~\cite{Driscoll:2015vf}. This has the practical consequence that one can achieve precise spatial control in tuning  properties from region to region within the network -- as is needed for creating origami~\cite{Witten:2007cq,Mahadevan:2005hr} or kirigami~\cite{Castle:2014jg} materials.

\section{Response at the bond level}

We construct networks numerically by starting with a configuration of particles produced by a standard jamming algorithm~\cite{OHern:2003vq,Liu:2010jx}.  
We place $N$ soft repulsive particles at random in a box of linear size $L$ and minimize the total energy until there is force balance on each particle.  
We work in either two or three dimensions and start with a packing fraction that is above the jamming density.  After minimizing the energy of a configuration, we create a network by replacing each pair of interacting particles with an unstretched spring of unit stiffness between nodes at the particle centers~\cite{Wyart:2005jna}. We characterize the network by the excess coordination number $\dZ \equiv Z - Z_\text{iso}$, where $Z$ is the average number of bonds at each node and $Z_\text{iso} \equiv 2d - 2d/N$ is the minimum for a system to maintain rigidity in $d$ dimensions~\cite{Goodrich:2012ck}. We note that networks produced this way have no long-range order~\cite{OHern:2003vq}, unlike networks constructed by randomly displacing cites on a lattice~\cite{GARBOCZI:1989wr,Thorpe:1990wb}.

\begin{figure}[htpb]
	\centering
	\includegraphics[width=0.6\linewidth]{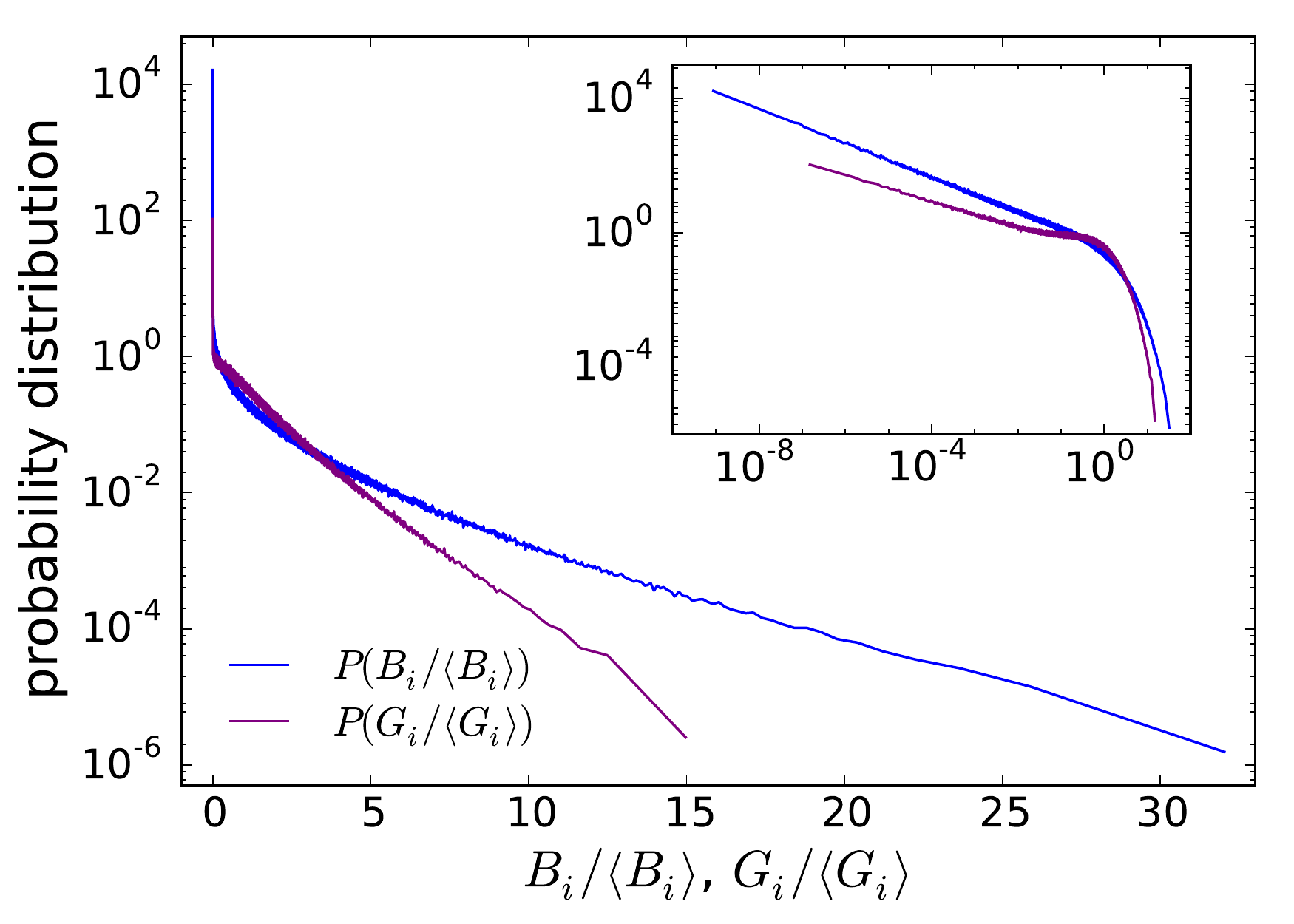}
	\caption[Variation in bond-level response.]{\label{fig:Ri_distributions_3d}Variation in bond-level response. Distribution on a log-linear scale (inset: log-log scale) of the contribution of each bond to the macroscopic bulk and shear moduli, $B_i$ and $G_i$, for $3d$ networks with $\dZi \approx 0.13$. Here $i$ indexes bonds. 
	%The distributions are normalized by their means. 
	At low $B_i$ or $G_i$, the distributions follow power-laws with exponents $-0.51$ and $-0.38$, respectively.  At high values, the distributions decay over a range that is broad compared to their means, $\avg{B_i}$ and $\avg{G_i}$.} 
\end{figure}

For each network, we use linear response to calculate the contribution $B_i$ of each bond $i$ to the bulk modulus, $B=\sum_i B_i$. ($B_i$ is proportional to the change in energy of bond $i$ when the system is uniformly compressed, see Sec.~\ref{sec:apndx_bondlevel_response_calculations} for details). 
The distribution of $B_i$ in three dimensions is shown in blue in Fig.~\ref{fig:Ri_distributions_3d}. In all plots, data is averaged over 500 networks that, unless explicitly stated, have approximately 4000 nodes and an initial excess coordination number $\dZi \approx 0.13$ (corresponding to a total number of bonds that is about 2\% above the minimum needed for rigidity).

Similarly, we can start with the same initial network and calculate $G_i$, the contribution of each bond to the shear modulus, $G=\sum_i G_i$.  (As shown in Chapter~\ref{chapter:finite_size}, a finite system is not completely isotropic, so the shear modulus varies with direction~\cite{DagoisBohy:2012dh}; we calculate the angle-averaged shear modulus, which approaches the isotropic shear modulus in the infinite system-size limit~\cite{Goodrich:2014iu}.)   The resulting distribution for $G_i$ is shown in purple in Fig.~\ref{fig:Ri_distributions_3d}.  Note that the distributions of the bond contributions to $B$ and $G$ are continuous, broad, and non-zero in the limit $B_i,G_i \rightarrow 0$.  That is, some bonds have nearly zero contribution to the bulk or shear modulus while others contribute disproportionately. 
For both $B$ and $G$, the distribution forms a power law at low values of $B_i$ or $G_i$, which is then terminated above $\avg{B_i}$ and $\avg{G_i}$ by approximately exponential cut-offs. 
Such significant variation in bond-level response is consistent with previous observations~\cite{Ellenbroek:2006df,Ellenbroek:2009dp}
and is in stark contrast to a perfect crystal where the distributions would be composed of discrete delta functions.

\begin{figure}[htpb]
	\centering
	\includegraphics[width=0.6\linewidth]{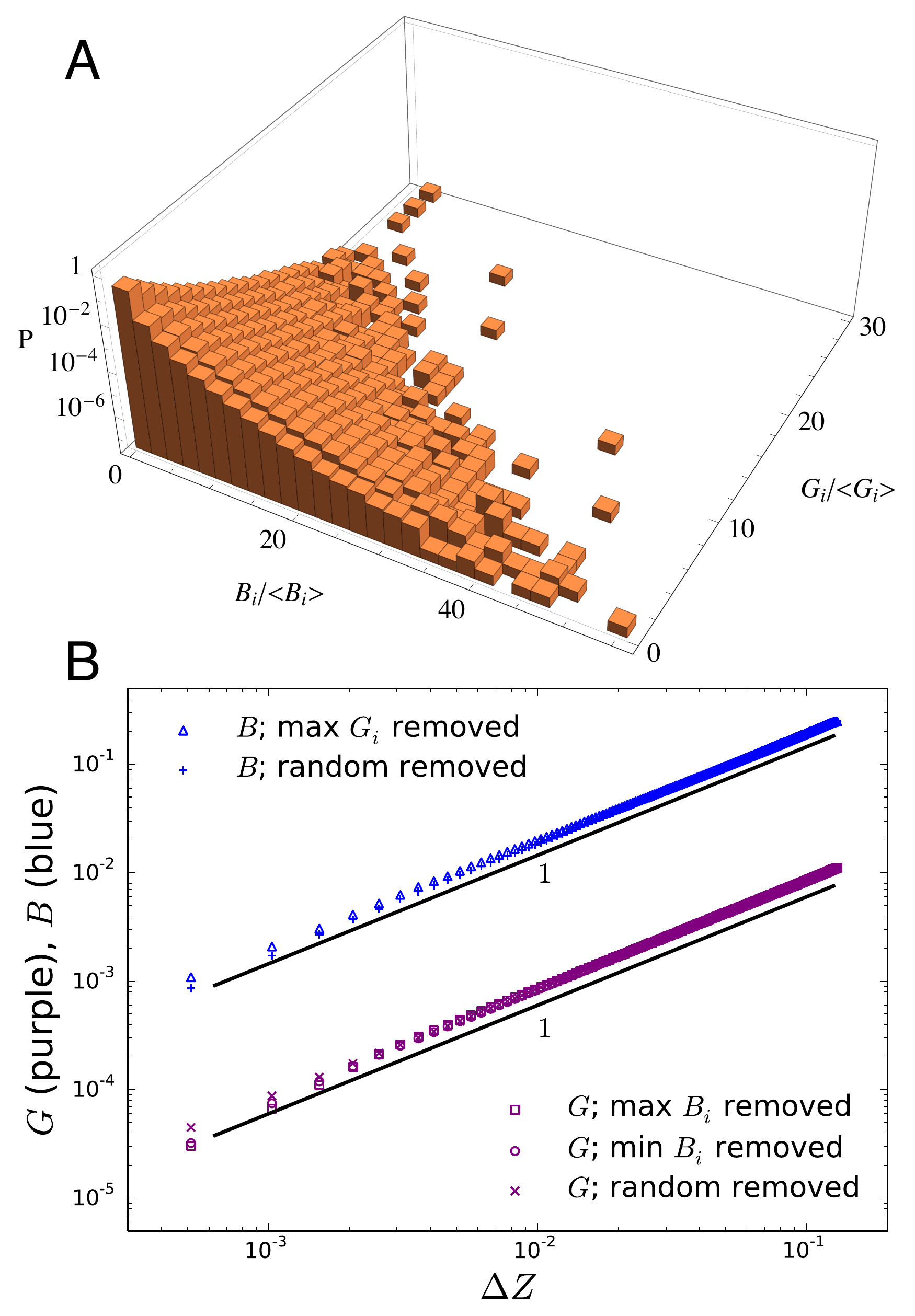} 
	\caption[Independence of bond-level response.]{\label{fig:bond_level_independence}Independence of bond-level response. ({\bf A}) Joint probability distribution of $B_i$ and $G_i$ for $3d$ networks with $\dZi \approx 0.13$. There is little apparent correlation between the response to compression ($B_i$) and to shear ($G_i$) for a given bond $i$.  ({\bf B}) The value of $G$ when bonds with the largest (purple squares) and smallest (purple circles) $B_i$ are removed is nearly indistinguishable from $G$ when bonds are removed at random (purple crosses). Similarly, $B$ is very similar whether bonds with the largest $G_i$ (blue triangles) are removed or bonds are removed at random (blue pluses).}
\end{figure}

We next ask if there is a correlation between how an individual bond responds to shear and how it responds to compression.  Do bonds with a large contribution to the bulk modulus also have a proportionately large contribution to the shear modulus?  Figure~\ref{fig:bond_level_independence}a shows the joint probability distribution $P(B_i ,G_i)$.  A strong positive correlation between $B_i$ and $G_i$ would produce a linear trend on this graph, which is clearly not observed. %While the correlations are not negligible (see Supplementary Material) 
We conclude that the correlations are weak, although we note that they are also not vanishingly small (see Sec.~\ref{sec:apndx_correlations_analysis} for more details).  %The residual correlations that we do find occur for those bonds with very small values of $G_i$.
This lack of strong correlation between $B_i$ and $G_i$ is again qualitatively different from what one would find for a simple crystal. Thus, Figs.~\ref{fig:Ri_distributions_3d} and~\ref{fig:bond_level_independence}a illustrate a previously-unrecognized property that is well obeyed by our disordered networks: independent bond-level response.

\section{Tuning by pruning}

This new property suggests that one can tailor the behavior of the network by selectively removing (pruning) those bonds that contribute more or less than the average to one of the moduli.  By so doing, one can decrease one modulus with respect to the other.  

\begin{figure}[htpb]
	\centering
	\includegraphics[width=0.6\linewidth]{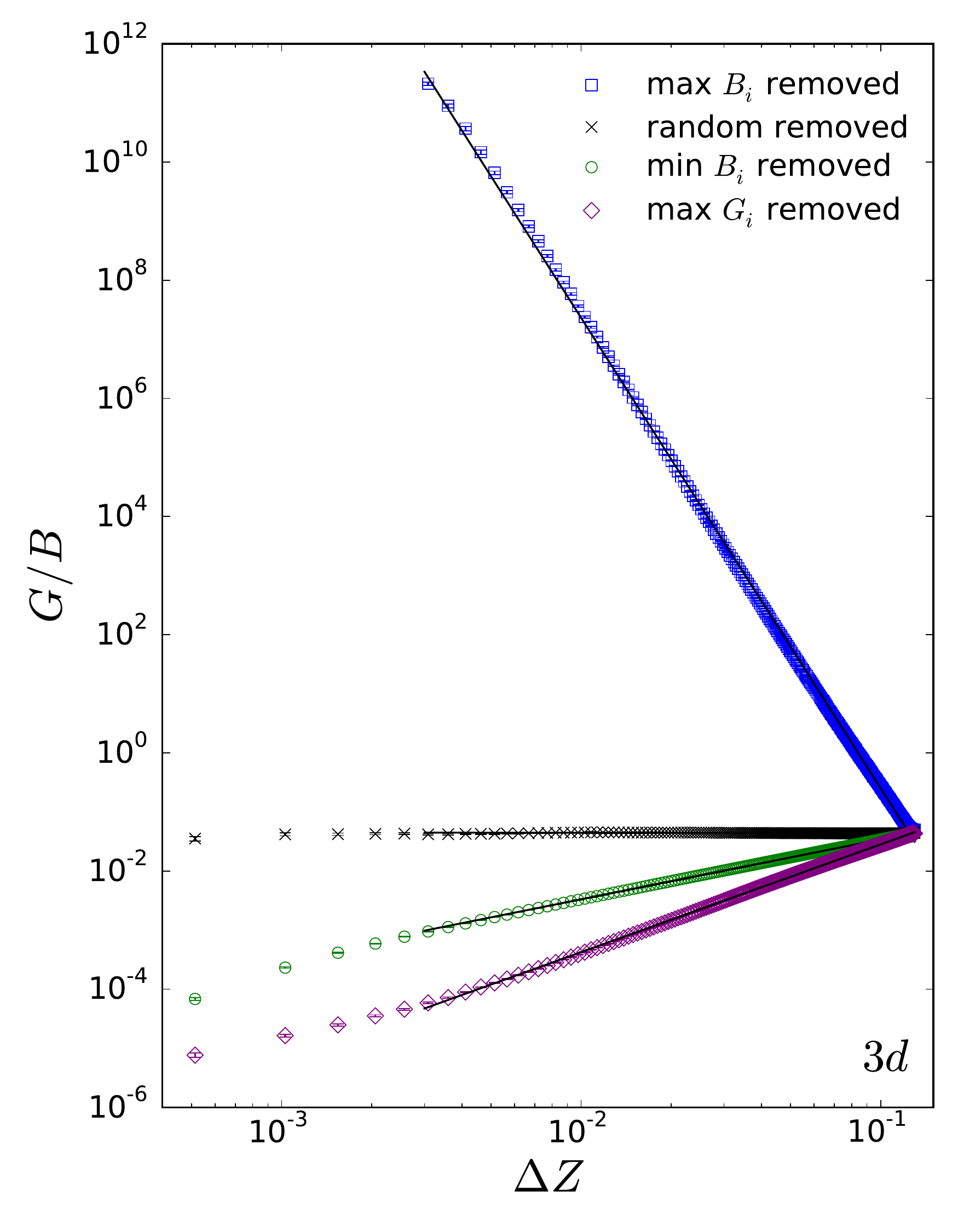} 
	\caption[Tuning global response in three dimensions.]{\label{fig:Global_response}Tuning global response in three dimensions. The ratio of shear to bulk modulus, $G/B$, for  four pruning algorithms. Error bars (included) are smaller than the symbols. Lines are fits to the data over the indicated range and have slopes, from top to bottom, of -7.96, -0.01, 1.01, and 1.82. Starting with the same initial conditions, we can tune global response by 16 orders of magnitude by pruning of order 2\% of the bonds.}
\end{figure}

First, we consider the known case of \emph{rigidity percolation}~\cite{Feng:1985vr,Ellenbroek:2009to,Ellenbroek:2014uh}, where a bond is picked at random and removed. This pruning is repeated until the system becomes unstable at $\dZ = 0$. We have implemented a slight variation of this procedure: at each step, a bond is removed only if each node it connects has at least $d+1$ remaining bonds in $d$ dimensions.  This is the condition for local stability of a particle in the original jammed packing~\cite{Levine2001}.
As the excess coordination number decreases, the bulk and shear moduli vanish together, so that $G \sim B \sim \dZ$~\cite{Feng:1985vr,Ellenbroek:2009to,Ellenbroek:2014uh}\footnote{Higher order corrections to the elastic constants are negligible for the range of $\Delta Z$ discussed here.} (see Fig.~\ref{fig:bond_level_independence}b).
Therefore, as shown in Fig.~\ref{fig:Global_response}, $G/B$ is independent of $\dZ$.

We now implement the idea of \emph{selected}-bond removal in a variety of ways.   
First we remove the bond with the smallest $B_i$, namely the weakest contribution to the bulk modulus (provided, as above, that each node connected to this bond has at least $d+1$ remaining bonds). Since the distribution $P(B_i)$ is continuous and nonzero as $B_i \rightarrow 0$, bond removal has almost no effect on the bulk modulus. However, since there is little correlation between the contribution of each bond to the bulk and shear moduli, there is a much larger effect on the shear modulus. 
Once the bond has been removed, the contributions $B_i$ and $G_i$ of the remaining bonds to the moduli must be recalculated because they depend on the connectivity of the entire network. This process of removing the bond with the smallest $B_i$ and then recalculating the values of $B_i$ for the remaining bonds is then repeated many times. 
Figure~\ref{fig:bond_level_independence}b shows that when bonds with the smallest $B_i$ are successively removed, the \emph{shear} modulus is linearly proportional to $\Delta Z$. Furthermore, it is quantitatively indistinguishable from the case where bonds are removed at random. 
The ability to alter the behavior of $B$ without affecting the behavior of $G$ is a clear demonstration of the principle of independent bond-level response. 

Since removing bonds with the smallest $B_i$ has little effect on the bulk modulus, we would expect $G/B \rightarrow 0$ as $\dZ \rightarrow 0$. As shown in Fig.~\ref{fig:Global_response}, we indeed find that $G/B \sim \dZ^\muBm$, with $\muBm = 1.01 \pm 0.01$. This behavior is identical to the scaling found in the original jammed sphere packings, where $\dZ$ is lowered by decompressing the system.

We can drive the same initial network to the opposite limit, $G/B \rightarrow \infty$, by successively removing bonds with the \emph{largest} contribution to $B$. 
As before, independent bond-level response predicts that the shear modulus will again decrease linearly with $\dZ$, as we indeed find (see Fig.~\ref{fig:bond_level_independence}b). However, the bulk modulus will decrease more quickly, as prescribed by the high $B_i$ tail of the distribution, suggesting that the ratio $G/B$ should \emph{increase}. 
The result of this successive bond-removal algorithm is shown by the blue squares in Fig.~\ref{fig:Global_response}.  We find that $G/B \sim \dZ^{\muBp}$, where $\muBp = -7.96 \pm 0.01$.  Thus, the increase in $G/B$ occurs with a \emph{much} steeper power law than the decrease of $G/B$ when the bond with the smallest contribution to $B$ is removed.  

The algorithms mentioned above can be extended in a number of ways. As a further example, 
one can remove the bond with the largest contribution to the shear modulus to drive $G/B$ towards zero. In this case, independent bond-level response implies that the bulk modulus will respond as if bonds were removed randomly, so that $B \sim \dZ$ (see Fig.~\ref{fig:bond_level_independence}b). However, the shear modulus decreases more rapidly; we find $G/B \sim \dZ^{\muGp}$, where $\muGp = 1.82 \pm 0.01$ (purple diamonds in Fig.~\ref{fig:Global_response}).

Note that the presence of a non-trivial zero-frequency vibrational mode (which our bond-cutting procedure does not explicitly forbid) would herald an instability in the structure. We look for such modes by diagonalizing the dynamical matrix, but do not observe them until the system is at or extremely close to isostaticity.\footnote{For the cases where the largest $B_i$ or $G_i$ bonds are removed, zero modes do not appear until $\Delta Z = 0$. For the random case and the case where the bond with the smallest $B_i$ is removed, the first zero mode can appear when there are still a few bonds above the isostatic value.} If we remove bonds with the {\it smallest} $G_i$, 
however, we find that zero modes appear when $\Delta Z$ is still quite large, preventing $G/B$ from diverging as we would expect. A variant of our procedure could prevent this ({\it e.g.} by including a constraint that a removed bond not create any zero modes).

\begin{figure}[htpb]
	\centering
	\includegraphics[width=0.6\linewidth]{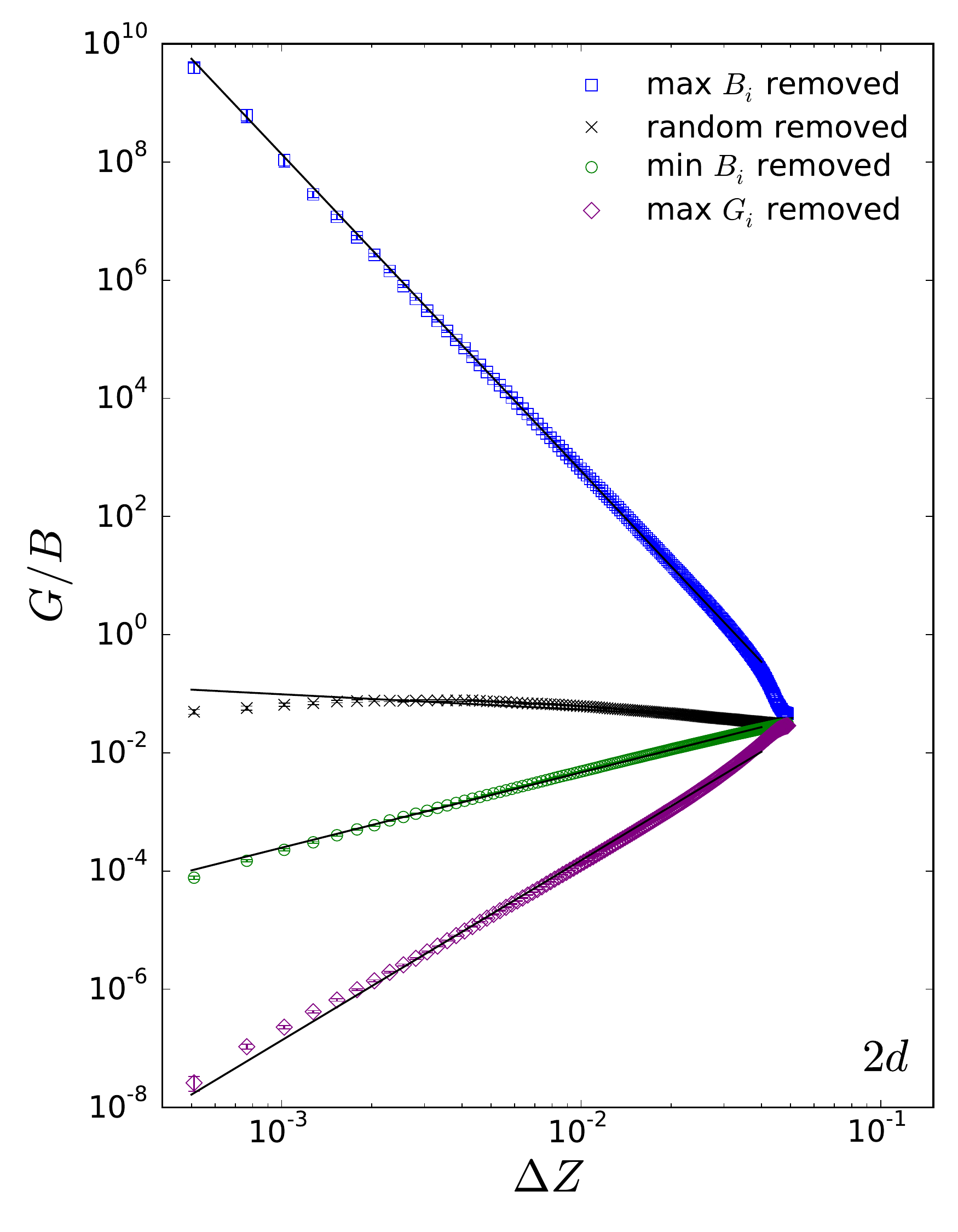} 
	\caption[Tuning global response in two dimensional networks.]{\label{fig:Global_response_2d}Tuning global response in two dimensional networks with $\dZi \approx 0.047$. The ratio of shear to bulk modulus, $G/B$, for  four pruning algorithms. Error bars (included) are smaller than the symbols. Lines are fits to the data over the indicated range and have slopes, from top to bottom, of -5.36, -0.26, 1.27, and 3.05. Starting with the same initial conditions, we can tune global response by 17 orders of magnitude by pruning of order 1\% of the bonds.}
\end{figure}

We can tune two-dimensional networks with equal ease. We construct spring networks in two dimensions with approximately 8000 nodes and an initial coordination number of $\dZi \approx 0.047$, which is about 1\% above the minimum needed for rigidity.  
As shown in Fig.~\ref{fig:Global_response_2d}, the behavior of $G/B$ is qualitatively similar to Fig.~\ref{fig:Global_response}.
When bonds with the smallest $B_i$ are removed, we find that $G/B \sim \dZ^\muBm$, with $\muBm = 1.27 \pm 0.01$. This is close to the behavior known for jammed packings ($G/B \sim \dZ^1$), though it is certainly not as clean as in three dimensions. %, where $\muBm = 1.01 \pm 0.01$. 
When we prune bonds that resist compression the most (largest $B_i$), we find that $G/B \sim \dZ^\muBp$, where $\muBp = -5.36 \pm 0.01$. 
At the smallest $\dZ$, $G/B \sim 10^{10}$.
Finally, when bonds with the largest $G_i$ are removed we find that $G/B \sim \dZ^\muGp$, with $\muGp = 3.05 \pm 0.01$.
Although $G/B$ diverges/vanishes with slightly different power laws in two and three dimensions, the overall effect is no less dramatic.

\section{Discussion}

Note that our procedures are remarkably efficient in tuning $G/B$.  Figure~\ref{fig:Global_response} shows that by removing about 2\% of the bonds in three-dimensional networks we can obtain a difference of more than 16 orders of magnitude in the tuned value of $G/B$, depending on which bonds we prune.  In two dimensions, we are able to obtain differences in $G/B$ that span over 17 orders of magnitude by pruning only $\sim 1\%$ of the bonds (Fig.~\ref{fig:Global_response_2d}). 
This is even more surprising given the fact that $B_i$ and $G_i$ are somewhat correlated. 
The fact that we can still tune the ratio of the moduli so drastically demonstrates the robustness of the principle of independent bond-level response.

The limit $G/B \rightarrow 0$ corresponds to the incompressible limit of a solid where the Poisson ratio, $\nu= (d - 2 G/B)/ [d (d-1)+ 2 G/B ]$ in $d$ dimensions, reaches its maximum value of  $\nu=+1$ (in $2d$) or $+1/2$ (in $3d$).   The limit $G/B \rightarrow \infty$ corresponds to the auxetic limit where the Poisson ratio reaches its minimum value of $\nu=-1$.  By using these different pruning algorithms, we can tailor networks to have any Poisson ratio between these two limits.  This ability provides great flexibility in the design of network materials.

For many materials~\cite{Greaves:2011ku} the Poisson ratio decreases with increased connectivity of the constituent particles and increases with packing density.  We note that neither of these correlations hold for the algorithms we have introduced for tuning the Poisson ratio (or ratio of shear and bulk moduli).  We can reach $G/B \rightarrow \infty$ (minimum Poisson ratio)  or $G/B \rightarrow 0$ (maximum Poisson ratio) by removing the same number of bonds from the same starting configuration. Neither the overall connectivity nor the overall density is different in the two final states.  Thus, our procedures for producing tunable Poisson ratio materials are fundamentally different from correlations considered in the literature.

We turn now to spatial correlations between cut bonds.  Driscoll {\it et al.}~\cite{Driscoll:2015vf} have conducted numerical simulations and experiments in which they removed bonds with the {\it largest} strain under uniaxial or isotropic compression or shear. They showed that the cut bonds form a damage zone whose width increases and diverges as the initial excess coordination number, $\dZi \rightarrow 0$; for sufficiently small $\dZi$, the pruned bonds are homogeneously distributed throughout the entire system. Outside this zone, they found that the network is essentially unaffected.

\begin{figure}[htpb]
	\centering
	\includegraphics[width=0.9\linewidth]{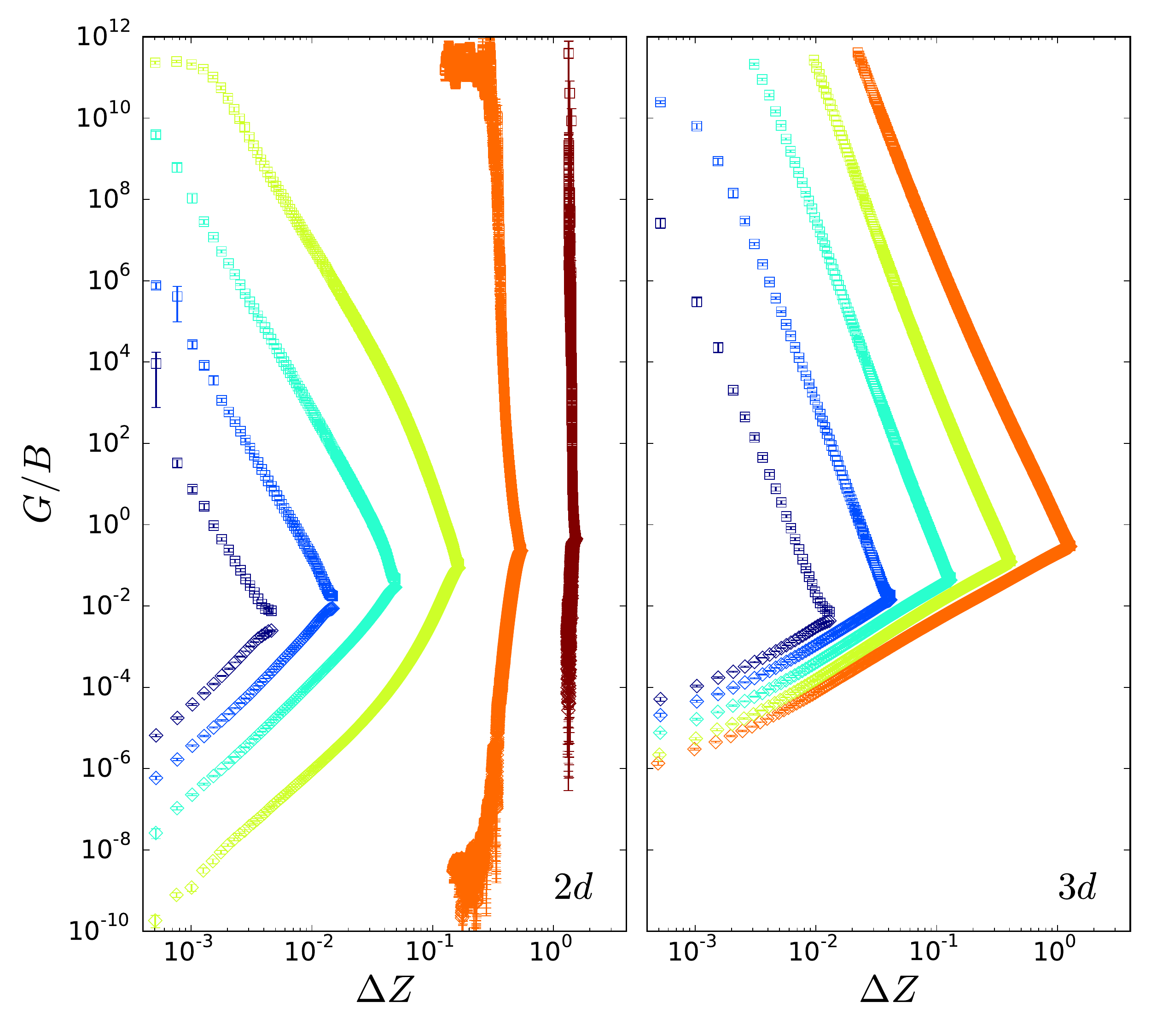} 
	\caption[Tuning the global response with different starting networks.]{\label{fig:different_deltaZ_initial}Tuning the global response with different starting networks in $2d$ (left panel) and $3d$ (right panel). In two dimensions, the data correspond to systems with approximately 8000 nodes and $\dZi$ of, from right to left, 1.5, 0.53, 0.16, 0.047, 0.014, and 0.0041. In three dimensions, the data correspond to systems with approximately 4000 nodes and $\dZi$ of, from right to left, 1.3, 0.40, 0.13, 0.039, and 0.012.}
\end{figure}

Since $B_i$ (or $G_i$) in our simulations is proportional to the strain squared, our procedure is identical to that of Driscoll {\it et al.}~\cite{Driscoll:2015vf} if we remove bonds with the {\it largest} contribution to the relevant elastic constant. 
So far, all the data we have presented are for systems with a sufficiently small $\dZi$ so that the distribution of the cut bonds appears homogeneous.  However, we find that $G/B$ diverges/vanishes regardless of $\dZi$ (see Fig.~\ref{fig:different_deltaZ_initial}), demonstrating that the ability to drastically tune $G/B$ does not depend on the spatial distribution of removed bonds or system size. When we remove the bond with the {\it smallest} contribution to $B$ or $G$, the bonds are removed homogeneously throughout the system, independent of $\dZi$. 
Our results, combined with the work of Driscoll {\it et al.}, means that elastic properties can be tuned not only globally but also on a local scale controlled by the initial connectivity -- one region may be highly incompressible while a nearby region may be highly auxetic.  This offers tremendous flexibility in the design of new and interesting materials.

We have presented a number of ways of tuning $G/B$ in disordered networks by using the principle of independent bond-level response. However, these ideas may be extended to other global properties as well. For example, one can imagine controlling thermal expansion by tuning nonlinear terms. One can even consider different classes of systems, such as a disordered resistor network~\cite{DEARCANGELIS:1985uh,DEARCANGELIS:1986vk} where one may be able to independently adjust the components of the conductivity tensor to design a highly anisotropic device. 
In general, to tune two properties relative to each other, one first must be able to quantify contributions at the single-bond level. The principle of independence holds if 1) there is a sufficient variation in the bond-level contributions ({\it i.e.} Fig.~\ref{fig:Ri_distributions_3d}), and 2) the contribution of a bond to one property is not strongly correlated with its contribution to the other ({\it i.e.} Fig.~\ref{fig:bond_level_independence}a). One could then independently tune these properties by removing bonds that contribute disproportionally to one property or the other.

Our results demonstrate that disordered networks provide particularly elegant opportunities for constructing mechanical metamaterials with tunable, flexible and spatially textured response.  However, the algorithms we have presented are not restricted to artificially constructed materials. 
Compressing a real network composed of springs that fail when stressed past a given threshold would lead to the same network as removing springs with the largest $B_i$, provided that the threshold is sufficiently small. 
It is also not beyond imagination that one could selectively break bonds at the nano-scale level in response to global perturbations in complex solids.  Indeed, biology appears to be able to target structures in networks that are under particularly high stress and to enhance their strength (such as in  trabecular bone~\cite{Keyak:2013ga}). Alternatively, there may be mechanisms to buckle or sever strongly stressed fibers (such as in actin networks~\cite{Lenz:2012df}).  It is interesting to ask if such selective repair or destruction of biological structures changes 
ratios of different mechanical responses such as the Poisson ratio.

\section{Appendix}
\subsection{Calculation of bond-level elastic response\label{sec:apndx_bondlevel_response_calculations}}
We consider networks of nodes connected by unstretched central-force springs with stiffness $k=1$.  Let $\vec{\delta r}_i$ be the total strain on bond $i$ when the system is deformed according to some strain tensor $\epsilon_{\alpha\beta}$. The change in energy of the network is then given to lowest order by
\eq{	\Delta E = \sum_{i} k_i \delta r_{i,\parallel}^2, \label{eq:energy_change}}
where $\delta r_{i,\parallel}$ is the component of $\vec {\delta r}_i$ that is parallel to the bond direction.  Thus, the bond that contributes the most (least) to the response to a given boundary deformation is the one with the largest (smallest) $\delta r_{i,\parallel}^2$. To remove the bond that contributes the most to the bulk modulus, for example, one would remove the bond with the largest $\delta r_{i,\parallel}^2$ under compression.  This procedure can be implemented in either a simulation or an experiment.  

In practice, for our computations, we use linear algebra to calculate the response of each bond more efficiently, as follows. 
The bulk elasticity of a system is described to linear order by the elastic modulus tensor $c_{\alpha\beta\gamma\delta}$, so that if the system is distorted by the symmetric strain tensor $\epsilon_{\alpha\beta}$, the change in energy is given to leading order by
\eq{	\Delta E/V = \frac 12 \epsilon_{\alpha\beta}c_{\alpha\beta\gamma\delta}\epsilon_{\gamma\delta}, }
where $V$ is the volume of the system.
In general, there are 6 (21) independent components of the elastic modulus tensor in two (three) dimensions, but in the isotropic limit this reduces to just the bulk modulus $B$ and the shear modulus $G$. 

The components of $c_{\alpha\beta\gamma\delta}$ are calculated from the change in energy of the system under various boundary deformations using Eq.~\ref{eq:energy_change}. %$\vec {\delta r}_i$ is composed of an affine strain and a non-affine response. 
The strain $\vec{\delta r}_i$ can be decomposed into two distinct parts. First there is an affine strain set directly by the strain tensor. However, this results in a nonzero net force, $\vec f_m$, on each node $m$, leading to a secondary non-affine response. This non-affine response is calculated by solving the following system of equations
\eq{	\mathcal{M}_{mn} \vec u^\text{NA}_{m} = \vec f_n,}
where %the indices $m$ and $n$ run over the $Nd$ degrees of freedom associated with node displacements, 
$\mathcal{M}_{mn}$ is the Hessian matrix and $\vec u^\text{NA}_m$ is the non-affine displacement of each node. The total strain $\vec {\delta r}_i$ of bond $i$ is calculated from the sum of the affine and non-affine displacements of the two nodes that the bond connects.
Since $\Delta E$ can be written as a sum over bonds, so too can the elastic modulus tensor:
\eq{	c_{\alpha\beta\gamma\delta} = \sum_i c_{i,\alpha\beta\gamma\delta}. }
Under the deformation $\epsilon_{\alpha\beta}$, the change in energy of bond $i$ is
\eq{	\Delta E_i = \frac 12 \epsilon_{\alpha\beta}c_{i,\alpha\beta\gamma\delta}\epsilon_{\gamma\delta}. }
$c_{i,\alpha\beta\gamma\delta}$ thus completely describes the bond-level elastic response for bond $i$, and can be used to calculate the quantities $B_i$ and $G_i$ considered in the main text. 

The global bulk and shear moduli are linear combinations of the components of the elastic modulus tensor. In two dimensions, they are
\eq{
	B &= \tfrac 14 \left( c_{xxxx} + c_{yyyy} + 2c_{xxyy}\right) \\
	G &= \tfrac 18 \left( 4c_{xyxy} + c_{xxxx} + c_{yyyy} - 2c_{xxyy}\right),
}
while in three dimensions they are
\eq{
	B ={}& \tfrac 19 \left( c_{xxxx} + c_{yyyy} + c_{zzzz} + 2c_{yyzz} + 2c_{xxzz} + 2c_{xxyy}\right) \\
	G ={}& \tfrac 1{15} \left( 3c_{yzyz} + 3c_{xzxz} + 3c_{xyxy} \right. \nonumber \\
	& \left. +\, c_{xxxx} + c_{yyyy} + c_{zzzz} - c_{yyzz} - c_{xxzz} - c_{xxyy}\right).
}
Finite disordered systems are never perfectly isotropic, so the shear modulus always has some dependence on the angle of shear. The above expressions for $G$ represent the angle-averaged shear modulus, which reduces to the shear modulus in the isotropic limit of infinite system size. We calculate the contribution of bond $i$ to the bulk and shear moduli in exactly the same way:
\eq{
	B_i &= \tfrac 14 \left( c_{i,xxxx} + c_{i,yyyy} + 2c_{i,xxyy}\right) \\
	G_i &= \tfrac 18 \left( 4c_{i,xyxy} + c_{i,xxxx} + c_{i,yyyy} - 2c_{i,xxyy}\right),
}
in two dimensions, and
\eq{
	B_i ={}& \tfrac 19 \left( c_{i,xxxx} + c_{i,yyyy} + c_{i,zzzz} + 2c_{i,yyzz} + 2c_{i,xxzz} + 2c_{i,xxyy}\right) \\
	G_i ={}& \tfrac 1{15} \left( 3c_{i,yzyz} + 3c_{i,xzxz} + 3c_{i,xyxy} \right. \nonumber \\
	& \left. +\, c_{i,xxxx} + c_{i,yyyy} + c_{i,zzzz} - c_{i,yyzz} - c_{i,xxzz} - c_{i,xxyy}\right)
}
in three dimensions.

%\clearpage
\subsection{Correlations in bond-level response\label{sec:apndx_correlations_analysis}}
Figure~\ref{fig:bond_level_independence}a shows that $B_i$ and $G_i$ are not strongly correlated, but it is also apparent that they are not completely independent. Here, we will attempt to characterize the extent of this correlation. The simplest thing we can do is measure the correlation coefficient $r$, defined as
\eq{ r = \frac{\avg{B_i - \avg{B_i}}\avg{G_i - \avg{G_i}}}{\sigma_{B_i} \sigma_{G_i}}, }
where the averages are over all $i$, and $\sigma_{B_i}$ and $\sigma_{G_i}$ are the standard deviations of the $B_i$ and $G_i$ distributions. $r$ is often referred to as the ``sample Pearson correlation coefficient," and has extreme values of $r=1$ (perfect linear correlation) and $r=-1$ (perfect linear anti-correlation), with $r=0$ when there is a complete lack of correlation. We find that $r \approx 0.325$ for the $3d$ data in Fig.~\ref{fig:bond_level_independence}a and $r \approx 0.171$ for the equivalent data in two dimensions. 

This level of correlation is \emph{statistically} significant given the amount of data averaged over, meaning that it could not have arisen from chance. (In other words, we can reject the null hypothesis of a complete lack of correlation with exceedingly high probability.) However, interpreting the meaning of these correlations is particularly difficult due to the extremely non-Gaussian nature of the underlying distributions. %and it is not clear if these correlations are \emph{physically} significant. 

It is possible to have a distribution with correlations that are statistically significant but physically unimportant because the elastic moduli $B$ and $G$ are dominated by the large $B_i$ and $G_i$ tails of the distributions. If correlations existed only at small $B_i$ and $G_i$, one would not expect them to have any effect on our bond-pruning algorithm to tune the relative values of $B$ and $G$. Furthermore, since the densities of $B_i$ and $G_i$ are much higher at low values (Fig.~\ref{fig:Ri_distributions_3d}), such correlations would have a large effect on $r$ without affecting the physics.

\begin{figure}[htpb]
	\centering
	\begin{tabular}{cc}
	\includegraphics[width=0.47\linewidth]{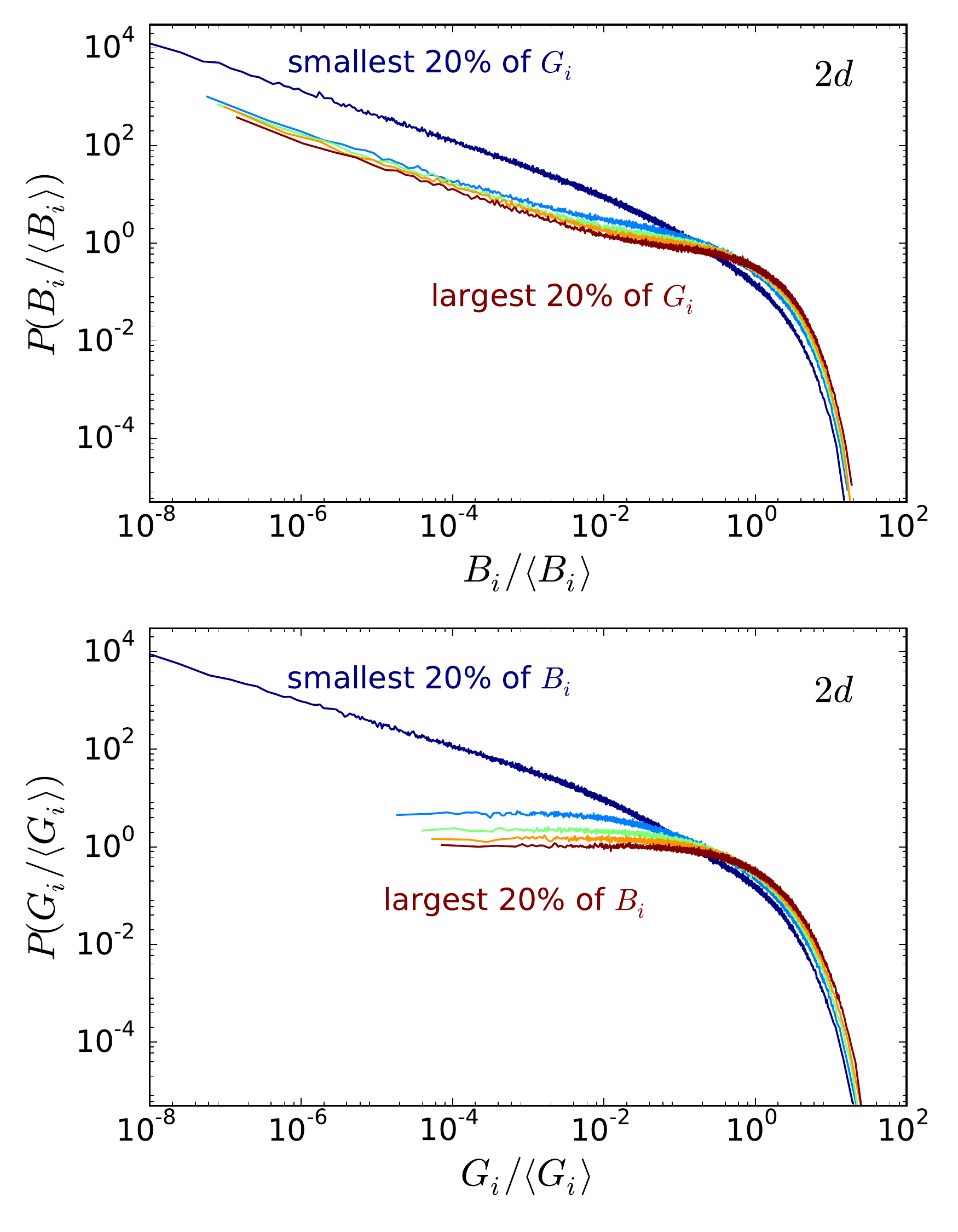} &
	\includegraphics[width=0.47\linewidth]{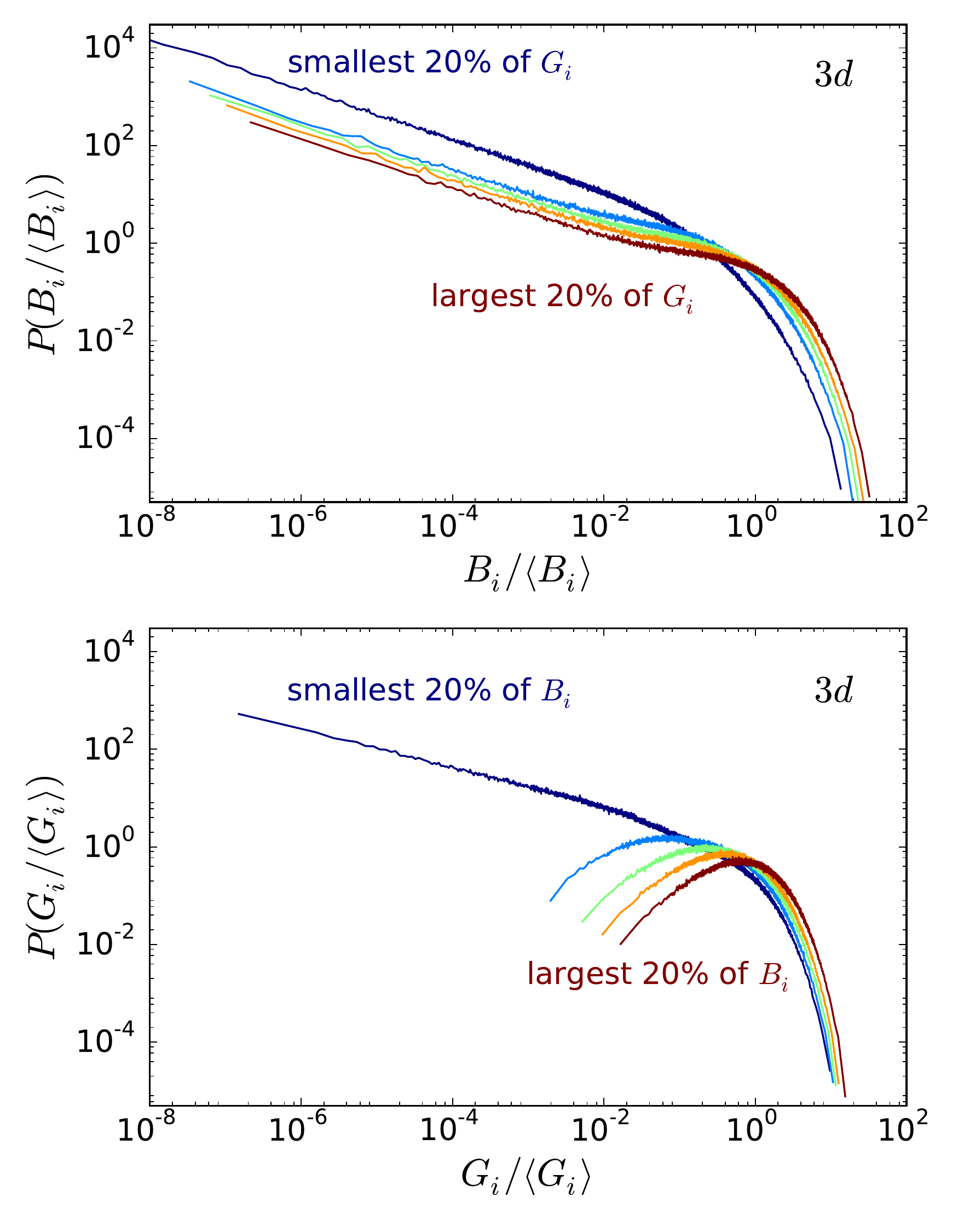} 
	\end{tabular}
	\caption[Conditional probability distributions of bond-level response.]{\label{fig:correlations}Conditional probability distributions of $B_i$ and $G_i$ for two dimensional (left column) and three dimensional (right column) data. The top row shows $P(B_i/\avg{B_i})$ as a function of $G_i$, and the bottom row shows $P(G_i/\avg{G_i})$ as a function of $B_i$.}
\end{figure}

Instead of focusing on the correlation coefficient, it is more instructive to look at the conditional probability distributions, which are shown in Fig.~\ref{fig:correlations} for both two and three dimensional data. The top row shows $P(B_i)$ for given $G_i$. Data was first sorted according to the value of $G_i$, and then divided into five even bins. Each curve represents one fifth of the data: the dark blue curve corresponds to bonds whose $G_i$ is in the smallest $20\%$, the next curve corresponds to bonds whose $G_i$ is in the second $20\%$, and so on until the dark red curve, which corresponds to bonds whose $G_i$ is in the largest $20\%$. The data is normalized by the average, $\avg{B_i}$, over \emph{all} bonds. The bottom row shows the conditional probability distribution of $G_i$ where data is similarly grouped according to $B_i$. 

For completely uncorrelated data, the curves in Fig.~\ref{fig:correlations} would collapse. This is clearly not the case, and the proximity of the curves to each other shows us where correlations arise. For example, the dominant source of correlations can be seen at small values of $G_i$: $P(G_i)$ is qualitatively different for data with small $B_i$ than for data with large $B_i$. However, as discussed above, it is the other end of the distributions that dominate the global elastic constants. While there is still some correlation at large $B_i$ and large $G_i$, this is clearly much smaller than at low values. Ultimately, the proof that the correlation between $B_i$ and $G_i$ is ``sufficiently small" for the idea of independent bond-level response to apply comes from Fig.~\ref{fig:bond_level_independence}b. Despite the correlation seen in Fig.~\ref{fig:correlations}, removing the largest $B_i$ bond has the same quantitative effect on the shear modulus as removing the smallest $B_i$ bond or a random bond.

\chapter{Summary and conclusion}
It has been exactly 100 years since Sir William Henry Bragg and his son William Lawrence Bragg shared the Nobel Prize in Physics ``for their services in the analysis of crystal structure by means of X-rays"~\cite{nobelprize1915}. Just two years earlier, Bragg and his son had demonstrated experimentally that crystals contain long-range periodic order, and ever since the perfect crystal has served as the starting point for understanding ordered solids~\cite{Ashcroft:1976ud}.
%Ever since the discovery by Bragg and Bragg just over a century ago that crystals have periodic order~\cite{Ashcroft:1976ud}, the perfect crystal has served as the starting point for understanding ordered solids. 
In an attempt to simplify the study of disordered systems, much progress has been made since the turn of the century in developing a unified ``jamming" framework for amorphous materials. The theory of jamming focuses on athermal packings of ideal soft spheres, which undergo a rigidity transition as a function of density that displays characteristics of a mixed first/second order phase transition. The motivating idea here is that if we understand the physics underlying this critical point, then maybe it can serve as a starting point, much like the perfect crystal, but for understanding disordered solids. 

Much of this thesis deals with the mechanical response of disordered solids, in particular linear elasticity and low-frequency vibrations. Our approach is to initially focus on the ideal jamming transition, and then to take a step back and consider whether the ideas developed there can be extended more generally. We started in Chapter~\ref{chapter:linear_response} by demonstrating the validity of linear response in the thermodynamic limit. A valid linear regime is necessary for quantities like the elastic constants (including the bulk and shear moduli) as well as the density of states to carry any physical meaning, but the existence of such a linear regime for packings of soft spheres had previously been called into question~\cite{Schreck:2011kl}. Therefore, the results of this chapter are essential for the rest of the dissertation. 

We then turned our attention to the critical nature of the jamming transition. Building off of previous work, Chapter~\ref{chapter:LengthScales} redefined the length scale $\lstar$, which was originally proposed in order to understand the unique behavior of the density of states near the transition~\cite{Wyart:2005wv,Wyart:2005jna}. This new understanding 1) showed that the length scale had a natural interpretation in terms of rigid clusters, 2) rationalized some assumptions made in the original work, 3) allowed us to make a semi-quantitative prediction for the length scale, and 4) enabled us to measure $\lstar$ numerically. We also saw that this length is intimately related to a separate ``transverse" length scale that also diverges at the jamming transition. This smaller length can also be interpreted as a rigidity scale, but it governs rigidity against a more subtle class of deformations.

Numerical simulations of particulate systems are unavoidably limited to a finite number of particles, and it has long been recognized in the context of phase transitions that this limitation can be exploited~\cite{Fisher:1972hr} to yield insight into the nature of the transition.
%Numerical simulations are necessarily always of finite-size systems, and so whenever a system possesses a diverging length scale it is important to identify and understand any finite-size effects. 
In Chapter~\ref{chapter:finite_size}, we thoroughly investigated the impact of finite size on jammed packings. We first saw that the very definition of jamming requires that one specify the class of boundary deformations allowed, allowing us to define various configurational ensembles. We then saw that non-trivial finite-size effects exist in the excess contact number and the shear modulus. These corrections can be understood within the context of finite-size scaling, providing further evidence that jamming can indeed be considered a real phase transition with an upper critical dimension of two. We also saw precisely how anisotropic fluctuations vanish with system size.

With the theory of jamming starting to approach maturity, due not only to the work presented here but also that of many other groups, we then saw that jamming might in fact have the potential to complete the picture started by the Braggs over 100 years ago. Chapter~\ref{chapter:anticrystal} investigated systems with structural order between the two extremes of the perfect crystal and the marginally jammed solid. We argued that the \emph{physics} associated with the jamming transition of soft spheres should be thought of as a mechanical extreme, which we dub the acticrystal. Instead of asking where a system falls on the spectrum of structural order, one should ask where the system falls on the spectrum of mechanical behavior. At least for soft sphere packings, the spectrum of mechanical behavior has two well defined endpoints, the crystal and the anticrystal, and the expectation is that real materials with more complicated interactions can be understood in terms of one of these two poles.

Finally, Chapter~\ref{chapter:independent_response} showed that the dependence of the linear elastic behavior on the excess contact number in jammed sphere packings is a special case of a much more general physical principle. By focusing on disordered spring networks, where bonds can be removed more easily than in sphere packings, we saw that the ratio of the shear modulus to the bulk modulus could be tuned by over 16 orders of magnitude by removing a very small fraction of bonds. This amazing flexibility is a result of the principle of independent bond-level response, which corresponds to the lack of strong correlations between the way an individual bond contributes to different global properties. The fact that a disordered material is not constrained in its properties in the same way as a crystalline one presents significant and yet untapped potential for novel material design.

It has now been 17 years since my advisors originally proposed the jamming phase diagram~\cite{Liu:1998up} and 12 years since the seminal work on soft sphere packings~\cite{OHern:2003vq} that jumpstarted the field. By now, the theory of ideal packings at and above the jamming transition is well understood and we are starting to see that this intellectual foundation can have a significant and dramatic effect on the way we study disordered materials. 
For example, the theory of ``soft spots" was first developed by Manning and Liu~\cite{Manning:2011dk} to identify ``defects" in disordered sphere packings that are particularly susceptible to rearrangement, and this notion has since been shown to apply to a vast array of numerical and experimental systems~\cite{Chen:2011faa,Rottler:2014ik,Schoenholz:2014fy,Cubuk:2015cd,Smessaert:2014gx,Sussman:2015if}. Another example is the work of Driscoll {\it et al.}~\cite{Driscoll:2015vf} that describes failure in marginal materials.

%Possibly the most glaring example of this is the idea of ``soft spots," which refer to local regions in a material that are particularly susceptible to rearrangement. This theory was initially developed by Manning and Liu~\cite{Manning:2011dk} for soft sphere packings, but has since been shown to apply to a vast array of numerical and experimental systems~\cite{Chen:2011faa,Rottler:2014ik,Schoenholz:2014fy,Cubuk:2015cd,Smessaert:2014gx,Sussman:2015if}. 

As I see it, there are two major directions going forward. First, using the foundation of the jamming transition, what more can be learned about sphere packings? Soft spots are a perfect example of this, as is the work presented in Chapters~\ref{chapter:anticrystal} and \ref{chapter:independent_response}. There is then the separate question of connecting the behavior of ideal soft spheres to more complicated and realistic systems. An elegant example of this is that soft spots can be identified robustly from local structural signatures using a machine learning approach~\cite{Cubuk:2015cd}. Extending ideas from soft sphere packings to real materials can be straight forward at times and extremely non-trivial at other times as it depends sensitively on both the phenomenon being studied and the system to which the idea may apply. While this thesis largely leaves such work to future studies, establishing this connection is nevertheless essential for the theory of jamming to realize its full potential. 

% the full potential of the theory of jamming to be realized. 

%\bibliographystyle{plainnat}
%\bibliography{CPG_bibtex_library.bib,additional_bibtex.bib}

%\fullpage{1-4}{Teo10a}

\end{document}